\documentclass[12pt,a4paper]{article}

\usepackage{amsmath,amssymb,amsfonts,bm,graphicx,subfiles}

\newlength{\abstwidth}
\renewcommand{\b}{\mathrm{b}}
\renewcommand{\c}{\mathrm{c}}
\renewcommand{\d}{\mathrm{d}}
\newcommand{\e}{\mathrm{e}}

\newcommand{\g}{\mathrm{g}}
\newcommand{\hrm}{\mathrm{h}}

\newcommand{\n}{\mathrm{n}}
\newcommand{\p}{\mathrm{p}}
\newcommand{\q}{\mathrm{q}}
\newcommand{\s}{\mathrm{s}}

\renewcommand{\u}{\mathrm{u}}
\newcommand{\A}{\mathrm{A}}
\newcommand{\B}{\mathrm{B}}

\newcommand{\K}{\mathrm{K}}

\newcommand{\M}{\mathrm{M}}
\newcommand{\N}{\mathrm{N}}

\newcommand{\W}{\mathrm{W}}

\newcommand{\Z}{\mathrm{Z}}

\newcommand{\dbar}{\overline{\mathrm{d}}}

\newcommand{\nbar}{\overline{\mathrm{n}}}
\newcommand{\pbar}{\overline{\mathrm{p}}}
\newcommand{\qbar}{\overline{\mathrm{q}}}
\newcommand{\sbar}{\overline{\mathrm{s}}}

\newcommand{\ubar}{\overline{\mathrm{u}}}
\newcommand{\Bbar}{\overline{\mathrm{B}}}

\newcommand{\Kbar}{\overline{\mathrm{K}}}

\newcommand{\Deltabar}{\overline\Delta}
\newcommand{\Sigmabar}{\overline\Sigma}

\newcommand{\mT}{m_{\perp}}
\newcommand{\pT}{p_{\perp}}
\newcommand{\pTo}{p_{\perp 0}}
\newcommand{\rT}{r_{\perp}}

\newcommand{\CM}{\mathrm{CM}}
\newcommand{\lab}{\mathrm{lab}}
\newcommand{\AQM}{\mathrm{AQM}}
\newcommand{\lsum}{\sum\limits}
\newcommand{\lint}{\int\limits}
\newcommand{\total}{\mathrm{tot}}
\newcommand{\nondiff}{\mathrm{nondiff}}
\newcommand{\elastic}{\mathrm{el}}
\newcommand{\diff}{\mathrm{diff}}
\newcommand{\annihilation}{\mathrm{ann}}

\newcommand{\resonance}{\mathrm{res}}

\newcommand{\figref}[1]{Figure~\ref{#1}}
\newcommand{\secref}[1]{Section~\ref{#1}}
\newcommand{\tabref}[1]{Table~\ref{#1}}

\begin{document}
\sloppy
 
\pagestyle{empty}
 
\begin{flushright}
LU TP 20-11\\
MCnet-20-12\\
May 2020
\end{flushright}

\vspace{\fill}

\begin{center}
{\Huge\bf A Framework for Hadronic}\\[4mm]
{\Huge\bf Rescattering in pp Collisions}\\[10mm]
{\Large Torbj\"orn Sj\"ostrand and Marius Utheim} \\[3mm]
{\it Theoretical Particle Physics,}\\[1mm]
{\it Department of Astronomy and Theoretical Physics,}\\[1mm]
{\it Lund University,}\\[1mm]
{\it S\"olvegatan 14A,}\\[1mm]
{\it SE-223 62 Lund, Sweden}
\end{center}

\vspace{\fill}

\begin{center}
\begin{minipage}{\abstwidth}
{\bf Abstract}\\[2mm]
In this article, a framework for hadronic rescattering in the general-purpose
\textsc{Pythia} event generator is introduced. The starting point is
the recently presented space--time picture of the hadronization process.
It is now extended with a tracing of the subsequent motion of the primary
hadrons, including both subsequent scattering processes among them and
decays of them. The  major new component is cross-section parameterizations
for a range of possible hadron--hadron combinations, applicable from threshold
energies upwards. The production dynamics in these collisions has also
been extended to cope with different kinds of low-energy processes.
The properties of the model are studied, and some first comparisons with
LHC $\p\p$ data are presented. Whereas it turns out that approximately
half of all final particles participated in rescatterings, the net
effects in $\p\p$ events are still rather limited, and only striking in a
few distributions. The new code opens up for several future studies, 
however, such as effects in $\p$A and AA collisions.
\end{minipage}
\end{center}

\vspace{\fill}

\phantom{dummy}

\clearpage

\pagestyle{plain}
\setcounter{page}{1}

\section{Introduction}
One of the most unexpected discoveries at the LHC is that
high-multiplicity $\p\p$ events bear a striking resemblance to
heavy-ion AA events. The first example was the observation
of a ``ridge'', i.e.\ an enhanced particle production around the
azimuthal angle of a trigger jet, stretching away in (pseudo)rapidity
\cite{Khachatryan:2010gv,Aad:2015gqa,Khachatryan:2016txc}.
Even more spectacular is the smoothly increasing fraction of strange
baryon production with increasing charged multiplicity, a trend that
lines up with pA data before levelling out at the AA results
\cite{Adam:2015qaa,ALICE:2017jyt}. Further examples include
non-vanishing $v_2$ azimuthal flow coefficients 
\cite{Aad:2015gqa,Khachatryan:2016txc,Acharya:2019vdf},
strong peaks in hadron ratios such as $\Lambda^0 / \K^0_S$ 
at around $\pT \approx 2$~GeV \cite{Khachatryan:2011tm},
and an $\langle \pT \rangle$ strongly increasing with particle mass
\cite{Abelev:2014qqa}, all suggesting some form of collective flow.
A recent overview of relevant observations and related theoretical 
ideas and challenges can be found in Ref. \cite{Adolfsson:2020dhm}.

One possible explanation for these phenomena is that a
quark--gluon plasma (QGP) can be created in $\p\p$ collisions.
This runs counter to the conventional wisdom that, unlike in 
AA collisions, the $\p\p$ environment does not offer sufficiently
large volumes and long time scales for a QGP to form, see
e.g.~\cite{BraunMunzinger:2007zz,Busza:2018rrf,Nagle:2018nvi}.
Nevertheless, such models have been developed, for instance the
core--corona model implemented in EPOS \cite{Pierog:2013ria}.
In it a lower-density corona of colour strings can hadronize
independently, whereas in a higher-density core the strings can melt
into a QGP that hadronizes collectively. In its simplest form, 
a string here represents the colour confinement field between
a separated colour triplet--antitriplet pair, typically formed
in the collision and thereafter expanding mainly along the collision
axis. More central $\p\p$ collisions correlate both with a higher core
fraction and a higher multiplicity, thus offering a mechanism for 
multiplicity-dependent event properties that can be continued on
to $\A\A$ collisions.

Alternatively, the similarity between $\p\p$ and AA could be viewed
as incentive to explore what phenomena could be explained without
recourse to QGP formation. As examples, the formation
of ropes with a higher colour charge than the string may explain a
changed particle composition \cite{Bierlich:2014xba}, while the
shoving of overlapping strings can give collective flow
\cite{Bierlich:2016vgw}. Strings squeezed into a smaller transverse
area could also offer a higher string tension and thereby a changed
particle composition \cite{Fischer:2016zzs}.

Whatever approach is taken, one issue is
that both strings and particles are produced very closely packed,
in fact physically overlapping to a large extent. This is nothing
new, but is already a consequence e.g.\ of the 
\textsc{Pythia} model for MultiParton Interactions (MPIs)
\cite{Sjostrand:1987su,Sjostrand:2017cdm} and the Lund 
string model view of particle production \cite{Andersson:1983ia}.
The former assumes that several strings are drawn out from a 
collision area of a typical proton size, and the latter that each 
of these strings individually has about the same transverse size. 
Even allowing for the transverse expansion of the string systems,
the overlap of fragmenting strings and of primary produced hadrons 
in $\p\p$ collisions is alarmingly high \cite{Ferreres-Sole:2018vgo}. 
This opens up for the above-mentioned modifications of the string
properties, and would also suggest that hadrons can interact with 
each other (elastically or inelastically) on the way out from the 
production region surrounding the primary ``scattering''. This is
what is referred to as hadronic rescattering.

So why has this overlap not attracted attention in traditional high-energy
$\p\p$ generators, such as Herwig \cite{Bahr:2008pv,Bellm:2015jjp},
\textsc{Pythia} \cite{Sjostrand:2006za,Sjostrand:2014zea} or
\textsc{Sherpa} \cite{Gleisberg:2008ta, Bothmann:2019yzt}?
One practical reason is that close-packing corrections did not seem 
necessary to describe $\p\p/\p\pbar$ data up to Tevatron energies, 
either because they were not there or (more likely) because nobody 
looked. Concerning rescattering in particular, another is that hadrons produced in 
a given space--time region of an event also tend to move in the same 
direction. The most obvious example of this is the ordering in rapidity with
respect to the collision axis. This implies that hadronic rescattering
tends to occur between pairs of rather low invariant mass and therefore
should not upset the overall structure of the event, in particular
if hadrons of different species are not distinguished. Furthermore, 
in high-$\pT$ jets the parton-shower evolution spreads out the colour
strings, such that overlaps are far less frequent than in the
low-$\pT$ region \cite{Fischer:2016zzs}. As we will see,
rescattering indeed only appears to have a noticeable impact on a select few
distributions in $\p\p$ collisions. 

The situation is different in heavy-ion physics, where the hadronic
densities could be even higher, and the
density drops slower per unit time for a larger expanding system, so 
there are more opportunities for rescattering on the way out.
Several rescattering frameworks have been developed 
as part of the description of AA collisions, see e.g.\ the overview and 
comparison in Ref. \cite{Zhang:2017esm}. The best 
known probably is UrQMD \cite{Bass:1998ca}, which much of our current work is based upon. SMASH 
\cite{Weil:2016zrk} is a recent addition still being actively developed. 
\textsc{Luciae} \cite{Sa:1995fj} / \textsc{Paciae} \cite{Sa:2011ye}
has its roots in Lund, even if now disconnected.
Many of these programs make use of Lund string fragmentation.

With the recent implementation of an explicit space--time picture for
the hadronization in \textsc{Pythia} \cite{Ferreres-Sole:2018vgo}, it
becomes possible to use e.g.\ UrQMD to simulate rescattering on
\textsc{Pythia} generated events. This was recently done \cite{daSilva:2019and},
with interesting results. Unavoidably it is a kludge, however: while
\textsc{Pythia}~8 is written in C++, information has to be transferred
to the UrQMD Fortran code, and then UrQMD in turn relies on the
older \textsc{Pythia}~6 Fortran version for some tasks. Interfacing
SMASH would have the advantage of being able to stay with C++, but
again SMASH in its turn makes use of \textsc{Pythia}.

We therefore believe it would be worthwhile to develop and provide
a purely internal implementation of hadronic rescattering. 
In this article we will present such a new framework, and show some of 
the first results obtained with it. This does not preclude the usage of 
and comparison with other packages, but rather that interfacing with such packages could 
be simplified. For instance, one could imagine implementing alternative
cross section parameterizations while still retaining the underlying
space--time tracing. As part of developing this framework, our work includes implementations of low energy hadron-hadron interactions. This means event generation in \textsc{Pythia} becomes available for beam energies all the way down to the mass threshold, a feature which may have other applications not related to rescattering.

The outline of this article is as follows.
\secref{sec:spacetime} reviews the space--time hadron production
picture that provides the starting point for the subsequent
rescattering. It also describes the algorithm for finding hadronic rescattering vertices and the evolution of the event through the rescattering phase. \secref{sec:rescattering} describes the dynamics of low energy 
processes. This includes how such processes are implemented, and how total, 
partial and differential cross sections are modelled for the different
processes. It represents the bulk of the new features that have
been included into \textsc{Pythia} as a result of this work.
Then \secref{sec:modelTests} presents some model tests and
model features, while \secref{sec:comparisons} shows some
comparisons with experimental data of relevance for the model.
Finally \secref{sec:summary} gives a summary and outlook.

Natural units are assumed throughout the article, i.e.\ $c = \hbar = 1$.
Energy, momentum and mass are given in GeV, space and
time in fm, and cross sections in mb.

\section{The space--time model}
\label{sec:spacetime}

In this section we will review and extend the space--time picture for
hadron production, and present how this picture is used as a starting
point to trace collision vertices throughout the time evolution of
the event.

\begin{figure}[t]
\includegraphics[width=0.53\textwidth]{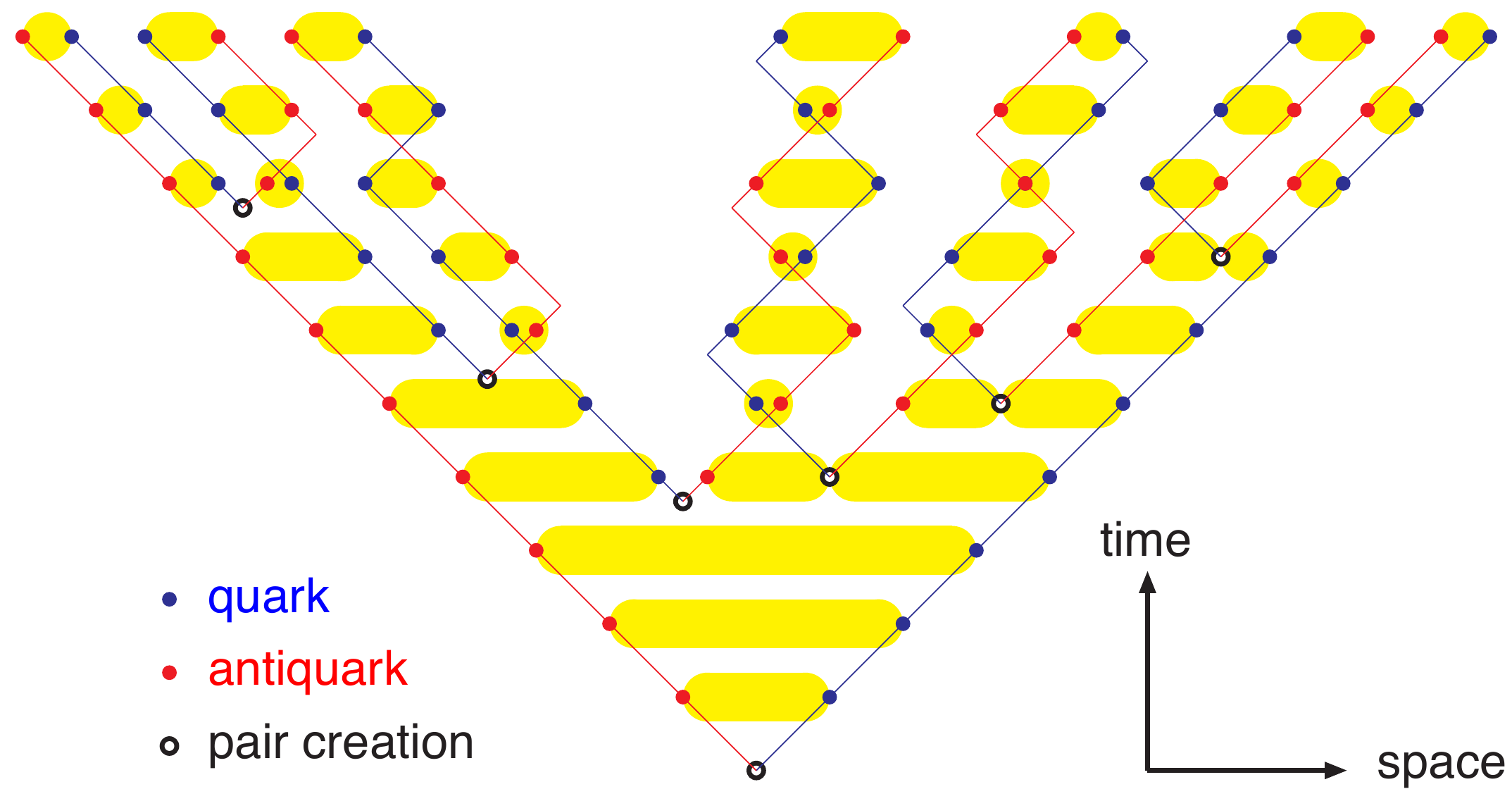}\hfill%
\includegraphics[width=0.40\textwidth]{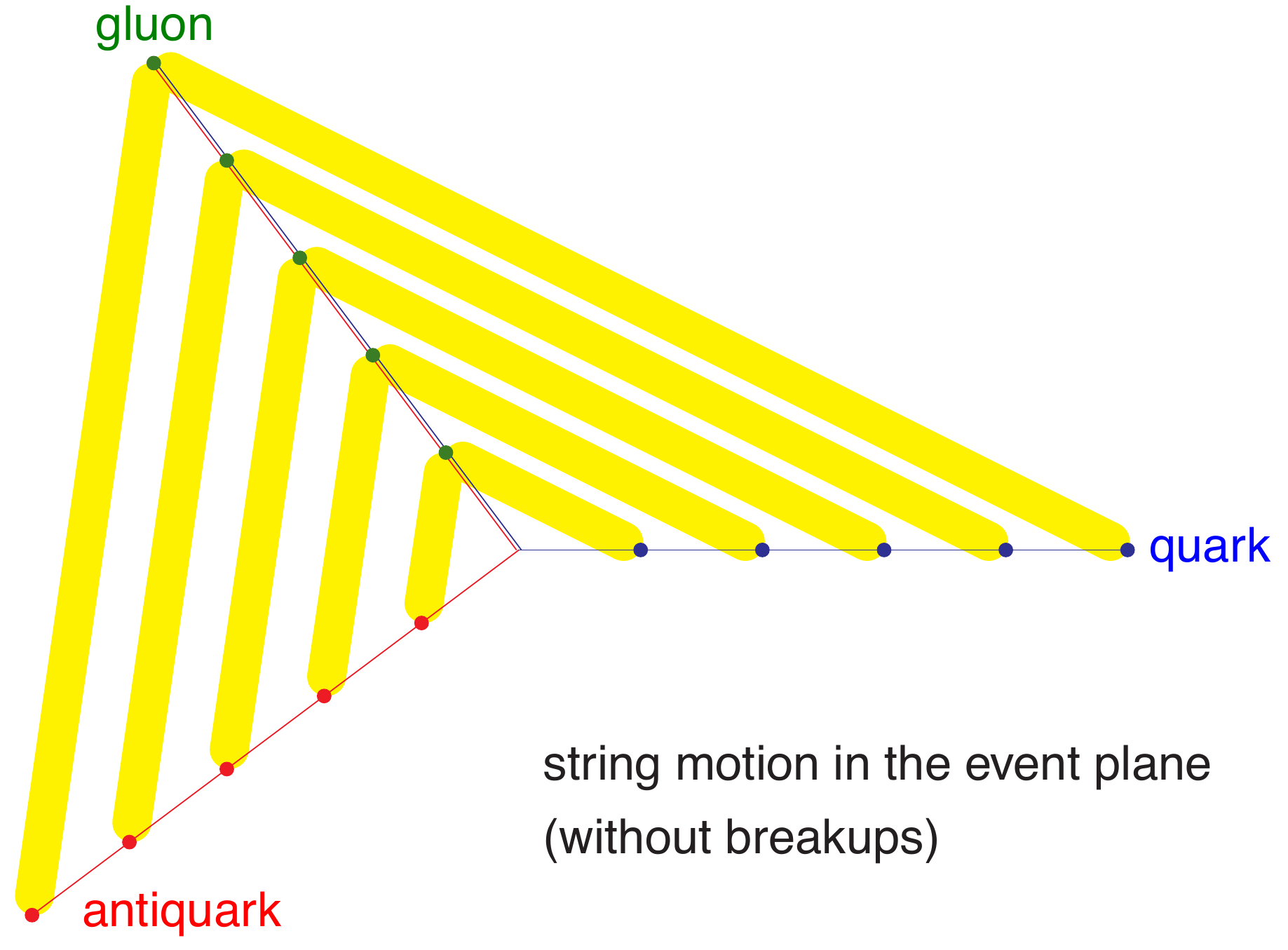}\\
\hspace*{0.25\textwidth}(a)\hspace{0.45\textwidth}(b)
\caption{(a) String breakup in a $\q\qbar$ event. The points
denote the location of quarks and antiquarks at snapshots in time, and
the yellow regions the string pieces then stretched out between them.
(b) String drawing in the plane of a $\q\qbar\g$ event.}
\label{fig:stringfrag}
\end{figure}

\subsection{Hadronization}
\label{subsec:hadronize}

The Lund string model is based on the assumption of linear
confinement, i.e.\ a string potential of $V = \kappa r$, where
the string tension $\kappa \approx 1$~GeV/fm and $r$ is the separation between a
colour triplet--antitriplet pair. For simplicity we may consider
the process $\e^+ \e^- \to \gamma^*/\Z^0 \to \q\qbar$, where the
quark--antiquark pair moves out along the $\pm z$ axis,
see \figref{fig:stringfrag}a. The linearity leads to a
straightforward relationship between the energy--momentum and
the space--time pictures:  
\begin{equation}
\left| \frac{\d p_{z,\q/\qbar}}{\d t} \right| =
\left| \frac{\d p_{z,\q/\qbar}}{\d z} \right| =
\left| \frac{\d E_{\q/\qbar}}{\d t} \right| =
\left| \frac{\d E_{\q/\qbar}}{\d z} \right| = \kappa ~.
\label{eq:xplinearity}
\end{equation}
It is necessary to keep track of signs: as the $\q$-to-$\qbar$
separation increases their energies decrease, with more and more of
the energy instead stored in the intermediary string. At the maximal
separation there would be no energy left for the quarks, and the string 
tension would then start to pull them together again, so that they would
perform an oscillatory motion often referred to as a ``yo-yo'' motion.

If there is enough energy, the string between an original
$\q_0 \qbar_0$ pair may break by producing new
$\q_i \qbar_i$ pairs, where the
intermediate $\q_i$ ($\qbar_i$) are pulled towards the $\qbar_0$ ($\q_0$)
end, such that the original colour field is screened. This way the
system breaks up into a set of $n$ colour singlets
$\q_0\qbar_1 - \q_1\qbar_2 - \q_2\qbar_3 - \ldots - \q_{n-1}\qbar_0$,
that we can associate with the primary hadrons. Each $\q_i\qbar_i$
pair is produced with zero energy and momentum at its common vertex,
since the string does not contain any local concentrations of energy.
The energy and momentum of a hadron $\hrm_i = \q_i\qbar_{i+1}$
therefore is provided by the string intermediate to the $\q_i \qbar_i$
and $\q_{i + 1} \qbar_{i + 1}$ breaks. This gives
$E_{\hrm_i} = \kappa (z_i - z_{i + 1})$ and
$p_{z,\hrm_i} = \kappa (t_i - t_{i + 1})$.
Note that $z_i > z_{i + 1}$ since $\q_0$ is moving in the $+z$ direction.
If boosted to a frame where $t'_i = t'_{i + 1}$, i.e.\ where the
hadron is at rest, one obtains
$m_{\hrm_i} = E'_{\hrm_i} = \kappa (z'_i - z'_{i + 1})$.

Unlike the intermediate vertices, the $\q_0 \qbar_0$ pair starts with
non-vanishing energy at the origin. The equivalent vertex for the $\q_0$
instead is where it has lost its energy, which (in the massless approximation) occurs at $t = z = E_{\q_0}(t = 0) / \kappa$.
This vertex can be used as the starting point for a recursive
procedure, where the location of each consecutive vertex can be
reconstructed from the $E$ and $p_z$ of the intermediate hadron.
Knowing the momenta of all hadrons it is therefore possible to
reconstruct all $\q_i \qbar_i$ production vertices, or the other way
around. Hadrons do not have a unique definition of a production ``vertex'',
being extended objects, but a convenient choice is the average of the 
$\q_i \qbar_i$ ones on either side of it \cite{Ferreres-Sole:2018vgo}.
Alternatives include an early or late choice, where the backward or 
forward light cones of the two $\q_i \qbar_i$ vertices cross.

Several issues have here been swept under the carpet, since they do
not directly affect the key relationship between the energy--momentum
and the space--time pictures. One issue is that quarks with
non-vanishing mass or $\pT$ should move along hyperbolae
$E^2 - p_z^2 = m^2 + \pT^2 = \mT^2$. When produced inside a string
they have to tunnel out a distance before they can end up on mass
shell. This tunnelling process gives a suppression of heavier quarks,
like $\s$ relative to $\u$ and $\d$ ones, and an (approximately)
Gaussian distribution of the transverse momenta. Effective
equivalent massless-case production vertices can be defined,
e.g.\ by replacing $m$ by $\mT$ in relations between $E$ and $p_z$.
Another issue is that the above notation only allows for meson
production. Baryons can be introduced e.g.\ by considering
diquark--antidiquark pair production, where a diquark is a colour
antitriplet and thus can replace an antiquark in the flavour chain.

Having simultaneous knowledge of both the energy--momentum
and the space--time picture of hadron production violates the 
Heisenberg uncertainty relations. In this sense the string model should 
be viewed as a semiclassical one, and there is no perfect way around that.
Smearing factors will be introduced to 
largely remove the tension for the transverse degrees of freedom,
and somewhat reduce it for the other ones. 
Either way, this semiclassical model does not introduce any clear systematic biases. Hence, there is no big problem in practice, since we are interested in average effects obtained by Monte Carlo sampling over a wide range of possible early histories. 

The real practical hurdle is to go on from a simple straight string to  
a larger string system. Consider e.g.\ 
$\e^+ \e^- \to \gamma^*/\Z^0 \to \q\qbar\g$.
In the limit where the number of colours is large, the $N_C \to \infty$
approximation \cite{tHooft:1973alw}, a string will be stretched from
the colour of the $\q$ to the anticolour of the $\g$, and then on
from the colour of the $\g$ to the anticolour of the $\qbar$,
Fig.~\ref{fig:stringfrag}b. To first approximation the two
string pieces each could be viewed as a boosted copy of a simple
$\q\qbar$ system. The problems arise around the gluon kink, as
follows. We already noted that a $\q/\qbar$ turns around when it has
lost its energy. When the same thing happens for a gluon, however,
it is instead replaced by a new expanding string region made out of 
inflowing momentum from the $\q$ and $\qbar$. Therefore there are
actually three string regions in which breaks can occur, and the third
one is especially important in the limit of a low-energy gluon.
Note that QCD favours the emission of
soft gluons, and that additionally a gluon is pulling out two string
pieces and therefore loses energy twice as fast as a quark, so such
third regions contribute a fair fraction of all hadron production.
For systems with more than one intermediate gluon the string motion 
becomes even more complicated. 

A framework to handle energy and momentum sharing in such complicated
topologies was developed in Ref. \cite{Sjostrand:1984ic}, and was then
extended to reconstruct matching space--time production vertices in
\cite{Ferreres-Sole:2018vgo}. (An earlier extension in
\cite{Gallmeister:2005ad} included several of the same main features, but
could not handle as complicated systems as required for LHC applications.) 
Again it can be described as a recursive procedure, starting from one end
of the string system, but now with additional rules how to pass from
one string region to the next. The reader is referred to Ref. \cite{Ferreres-Sole:2018vgo} for details. 

In addition to the main group of open strings stretched between $\q\qbar$
endpoints, there are two other common string topologies. One is a closed 
gluon loop, which can be viewed as an open string (with at least one 
intermediate gluon) where the $\q$ and $\qbar$ endpoints are fused into 
a single gluon, which closes the colour flow. Once an initial 
$\q_0\qbar_0$ breakup has been picked somewhere along the string, 
at random (within given rules), the further handling devolves back into 
the open string framework. The other is the junction topology,
represented by three quarks moving out in a different directions,
each pulling out a string behind itself. These strings meet at a common
junction vertex, to form a Y-shaped topology. The junction moves by the net 
pull of the string, and is at rest only in a frame where the opening angle
between each quark pair is $120^{\circ}$. Also in this case there may
be gluons on the string between a quark and the junction. Each of the
three legs may be hadronized according to the same basic rules as
above, with some special care needed where they meet at the junction,
around which a baryon is formed to carry the net baryon number
of the system.

There is one further aspect added to the framework presented so far. 
For the energy--momentum picture in a $\q\qbar$ system we started out
with a pure two-dimensional representation in $(E, p_z)$ space, but
then added random Gaussian $\pT$ kicks motivated by the tunnelling
mechanism. Alternatively we could have motivated such fluctuations
by the uncertainty relationship: a string could be expected to have
a radius roughly $\sqrt{2/3}$ that of the proton, since if
$r_{\p}^2 = \langle x^2 + y^2 + z^2 \rangle$ then
$\langle x^2 + y^2 \rangle = (2/3) r_{\p}^2$. Either argument gives
$\pT$ kicks of the order 0.3~GeV for each $\q_i \qbar_i$ pair,
consistent with data. By contrast, the basic machinery sets all
$\q_i \qbar_i$ production vertices to have $x = y = 0$, which gives
an unreasonably perfect lineup of the hadrons. For the studies in
\cite{Ferreres-Sole:2018vgo} we therefore introduced a Gaussian
$(x, y)$ smearing with a width according to the expressions above,
and will continue to do so. By the additional smearing to be introduced
in the next section, which partially might overlap, some reduction
of the width would be motivated, however.

Unfortunately, complications may arise in multiparton systems, notably
for those hadrons that have their two defining $\q_i \qbar_i$ vertices
in two different string regions, meaning there is no unique separation
between transverse and longitudinal degrees of freedom. Occasionally
this may give unreasonably large positive or negative
$\tau^2 = t^2 - x^2 - y^2 - z^2$. A few safety checks have been
introduced to catch and correct such mishaps as well as possible. 

\subsection{Multiparton interaction vertices}
\label{subsec:MPIvertices}

The framework described above assumes that all partons start out from
the same space--time production vertex, as would be the case e.g.\ in 
$\e^+\e^- \to \Z^0  \to \q \qbar$. In $\p\p$ the colliding hadrons are
extended objects, however. The Lorentz-contracted hadrons pass through
each other at a fairly well-defined time, conventionally $t = 0$, but
over a transverse region of hadronic sizes. In the overlap region
several parton-parton interactions can occur, as described by the MPI
framework in \textsc{Pythia} \cite{Sjostrand:1987su,Sjostrand:2017cdm}.

The probability for an interaction at a given transverse coordinate
$(x, y)$ can be assumed related to the time-integrated overlap of the
parton densities of the colliding hadrons in that area element. Let the
partons be described by a Lorentz contracted probability distribution
$P_{\mathrm{LC}}(x, y, z)$, which in its rest frame reduces to a spherically
symmetric $P(r)$ with $r^2 = x^2 + y^2 + z^2$. Setting the two incoming
beam particles $A$ and $B$ to move along the $z$ axis with velocity
$\pm v$, separated by $\pm b/2$ in the $x$ direction, where $b$ is the
impact parameter, this overlap (``eikonal'') reads
\begin{eqnarray}
  \mathcal{O}(x, y; b) & \propto &  \int \! \int
  P_{\mathrm{LC},A} \left( x -\frac{b}{2}, y, z - vt \right) \,
  P_{\mathrm{LC},B} \left( x +\frac{b}{2}, y, z + vt \right) \,
  \d z \, \d t  \nonumber \\
  & \propto & \int P_A \left( x -\frac{b}{2}, y, z_A \right) \, \d z_A \;
  \int P_B \left( x -\frac{b}{2}, y, z_B \right) \, \d z_B ~,
\end{eqnarray}
the latter by suitable variable transformation. The answer can be further
simplified in case of a Gaussian distribution
$P(r) \propto \exp( - r^2 / r_0^2)$:
\begin{eqnarray}
  \mathcal{O}(x, y; b) & \propto &  \int \exp \left(
  \frac{(x - b/2)^2 + y^2 + z_A^2)}{r_0^2} \right) \, \d z_A \;  
  \int \exp \left( \frac{(x + b/2)^2 + y^2 + z_B^2)}{r_0^2} \right) \, \d z_B  
  \nonumber \\ 
   & \propto & \exp \left( - \frac{2 \rT^2}{r_0^2} \right) \, 
   \exp \left( - \frac{\b^2}{2r_0^2} \right) ~,
   \label{eq:bMPIGaussian}
\end{eqnarray}
where $\rT^2 = x^2 + y^2$. That is, for a Gaussian proton the overlap
region is an azimuthally symmetric Gaussian, with no memory of
the collision plane, and the total overlap is a Gaussian in $b$.
The $r_0$ parameter can be approximately related to the proton radius
$r_{\p}$ by $\langle r^2 \rangle
= \langle x^2 + y^2 + z^2 \rangle = 3 r_0^2 / 2 = r_{\p}^2$.
The default in \textsc{Pythia} is a constant proton radius value $r_{\p} \approx 0.85$~fm
for the distribution of partons. With increasing energy, and a related
increase in the number of MPIs per collision, the effective edge of
interacting partons is pushed outwards and thus collision cross sections
can go up.

The Gaussian is a very special case, however. In general, the collision
region will be elongated either out of or in to the collision plane.
The former typically occurs for a distribution with a sharper proton
edge, e.g.\ a uniform ball, $P(r) \propto \Theta(r_0 - r)$, where
$\Theta$ is the step function, which gives rise to the almond-shaped
collision region so often depicted for heavy-ion collisions.
The latter shape instead occurs for distributions with a less pronounced
edge, such as an exponential, $P(r) \propto \exp( -r / r_0 )$.

In the \textsc{Pythia} MPI machinery the overlap distribution
$\mathcal{O}(b) = \int\!\int \mathcal{O}(x, y; b) \, \d x \, \d y$
can be chosen and tuned according to a few different forms.
The current default is $\mathcal{O}(b) \propto \exp((b/b_0)^p)$ with 
$p = 1.85$, i.e.\ close to but not quite Gaussian. A similar shape
and tune is obtained with a double Gaussian $P(r)$, where a smaller-radius
second Gaussian can be viewed as representing hot spots inside the proton.
In both cases a stronger-than-Gaussian peaking of $\mathcal{O}(b)$ at
$b = 0$ is required to get a sufficiently long tail out to largest charged
multiplicities in LHC and Tevatron minimum-bias events. 

The $P(r)$ and $\mathcal{O}(b)$ distributions as described so far are
likely to be significant simplifications, however. If one views the
evolution from a simple original parton configuration via initial-state
cascades into a set of interacting partons, then there are likely to
arise complicated patterns and correlations. One such framework is
presented in Ref. \cite{Bierlich:2019wld}, where an implementation of
Mueller's dipole model \cite{Mueller:1993rr,Mueller:1994jq} for the
two colliding hadrons are used to assign MPI production vertices.
These then turn out to give clearly non-isotropic distributions. 
In the future the relevant code for these assignments will be made
available, but using it comes at a cost in terms of a considerably
slower event generation.

For now, we have therefore settled for a simplified framework
with enough flexibility for our purposes. In it the MPIs locations 
by default are selected according to the Gaussian 
$\exp( - 2 \rT^2 / r_0^2)$, but optionally this can be modified 
in either of two ways. Either the $x$ coordinates 
are scaled by a factor $r_{\epsilon}$ and the $y$ ones by $1/r_{\epsilon}$,
or else the Gaussian is multiplied by a $\varphi$ modulation factor
\begin{equation}
\frac{\d N}{\d \varphi} \propto 1 + \epsilon \cos(2 \varphi) ~.  
\end{equation}
Here $r_{\epsilon} > 1$ or $\epsilon > 0$ means an enhancement in the 
collision plane and $r_{\epsilon} < 1$ or $\epsilon < 0$ out of it. 
Asymmetries in the spatial distribution also arise from the Monte Carlo 
sampling of a finite number of MPIs, and these may be even more important.

This machinery is used to select the $(x, y)$ coordinates of the MPI
vertices at $t = z = 0$. Only a fraction of the full beam-particle
momentum is carried away by the MPIs, leaving behind one or more beam
remnants \cite{Sjostrand:2004pf}. These are initially distributed
according to the basic $\exp( - \rT^2 / r_0^2)$ shape around the
center of the respective beam. By the random fluctuations, and by the
interacting partons primarily being selected on the side leaning
towards the other beam particle, the ``center of gravity'' will not
be located at the $x = \pm b/2, y = 0$ positions originally assumed. 
All the beam remnants will therefore be shifted so as to ensure that
the energy-weighted sum of colliding and remnant parton locations
is where it should be. As a small improvement on a uniform shift,
remnants located closer to the other remnant are shifted more, so as
to deplete the overlap region more. This is achieved by assigning
each remnant a weight
\begin{equation}
  \left( 1 + \frac{b}{r_{\p}} \exp\left( \frac{\pm x}{r_{\p}} \right)
  \right)^{-1}
\end{equation}
proportional to its eventual shift, where $x$ is relative to the
respective beam center with the other beam displaced $\mp b$ in the
$x$ direction. Shifts are capped to be at most a proton radius, 
so as to avoid extreme spatial configurations, at the expense of 
a perfectly aligned center of gravity.

Not all hadronizing partons are created in the collision moment
$t = 0$. Initial-state radiation (ISR) implies that some partons
have branched off already before this, and final-state radiation (FSR)
that others do it afterwards. These partons then can travel some
distance out before hadronization sets in, thereby further complicating  
the space--time picture, even if the average time of parton showers
typically is a factor of five below that of string fragmentation
\cite{Ferreres-Sole:2018vgo}. We will not trace the full shower
evolution, but instead include a smearing of the transverse location
in the collision plane that a parton points back to. Specifically,
a radiated parton is assigned a location at $t = 0$ that is smeared
by $\Delta \rT$ relative to its mother parton according to a
two-dimensional Gaussian with a width inversely proportional to its
$\pT$. The constant of proportionality can be set freely, but should
obviously be such that $\Delta \rT \, \pT \sim \hbar$. So as not
to obtain unreasonable $\Delta \rT$ shifts, the $\pT$ is set to be
at least 0.5~GeV in this context, comparable to the cut-off scale 
of the FSR showers. No attempt is made to preserve the center of 
gravity during these fluctuations.

The partons produced in various stages of the collision process
(MPIs, ISR, FSR) are initially assigned colours according to the
$N_C \to \infty$ approximation, such that
different MPI systems are decoupled from each other. By the beam
remnants, which have as one task to preserve total colour, these
systems typically become connected with each other. Furthermore,
colour reconnection (CR) is allowed to swap colours, partly to
compensate for finite-$N_C$ effects, but mainly that it seems like
nature prefers to reduce the total string length drawn out when
two nearby strings overlap each other. When such effects have been
taken into account, what remains to hadronize is one or more separate
colour singlet systems of the character already described in
\secref{subsec:hadronize}.

There is one key difference, however, namely that the strings now
can be stretched between partons that do not originate from the same
vertex. Even in the simplest case, a $\q$ connected with a $\qbar$
from a different MPI, there is a new situation not studied
previously, where the vertex separation should be equivalent to
a piece of string already at $t = 0$. For the energy--momentum picture 
it is traditionally assumed that its effects are sufficiently small 
that they can be neglected. If the effects of a 1~fm $\approx$ 1~GeV 
special term is to be spread over many hadrons, then the net effect 
on each hardly would be noticeable.

For the space--time picture we do want to be more careful about the
effects of the transverse size of the original source.
The bulk of the effects determining the hadronic production vertices do come from the framework of \secref{subsec:hadronize}, and therefore we will be satisfied if we can introduce a relevant amount of smearing on hadron production, without necessarily fully describe effects for the individual hadron. This is
achieved as follows.

For a simple $\q\qbar$ string, such as in  \figref{fig:stringfrag}a,
the relevant length of each hadron string piece is related to its energy.  
For a given hadron, define $E_{\hrm\q}$ ($E_{\hrm\qbar}$)
as half the energy of the hadron plus the full energy of all hadrons
lying between it and the $\q$ ($\qbar$) end, and use this as a measure
of how closely associated a hadron is with the respective endpoint. 
Also let $\mathbf{r}_{\perp\q}$ ($\mathbf{r}_{\perp\qbar}$) be the 
(anti)quark transverse production coordinates. Then define the hadron 
production vertex offset to be
\begin{equation}
  \Delta \mathbf{r}_{\perp\hrm} = \frac{E_{\hrm\qbar} \, 
    \mathbf{r}_{\perp\q} + E_{\hrm\q} \, \mathbf{r}_{\perp\qbar}}%
    {E_{\hrm\q} + E_{\hrm\qbar}}
  = \frac{(E_{\mathrm{tot}} - E_{\hrm\q}) \, \mathbf{r}_{\perp\q} 
    + E_{\hrm\q} \, \mathbf{r}_{\perp\qbar}}{E_{\mathrm{tot}}} ~,  
\end{equation}
relative to what a string motion started at the origin would have given.

This procedure is then generalized to more complicated string topologies.
In a $\q - \g_1 - \g_2 - \ldots - \qbar$ string, one may define $E_{\hrm\q}$
as above. If $E_{\hrm\q} < E_{\q} + E_{\g_1}/2$ the hadron is viewed as
produced between the $\q$ and $\g_1$, and the offset can be found as
above, only with $E_{\qbar}$ replaced by $E_{\g_1}/2$. If instead 
$E_{\q} + E_{\g_1}/2 < E_{\hrm\q} < E_{\q} + E_{\g_1} +  E_{\g_2}/2$ then
the excess energy $E_{\hrm\q} - E_{\q} - E_{\g_1}/2$ determines the
admixture of $\mathbf{r}_{\perp\g_1}$ and $\mathbf{r}_{\perp\g_2}$, 
and so on, stepping through region after region, for hadron after hadron, 
until the $\qbar$ end is reached. For junction topologies the same kind 
of approach can be used to iterate from each leg towards the central 
junction. The two lowest-energy legs are considered first, and an 
$\mathbf{r}_{\perp}$ towards which the third string is iterated is formed 
by the relative unused energy fractions of the first two. That way a 
junction baryon can receive contributions from all three legs. 

There are two obvious shortcomings. Firstly, the approach does not take
into account
the higher regions, handled in the complete string motion, e.g. made
up out of $\q$ and $\g_2$ momentum, where the hadron offset could be a more
complex combination of three different parton offsets. Secondly the
sharing according to energy is not Lorentz covariant. Nevertheless, we
believe this approach to provide a sensible approximation to the smearing
effects one may expect. There is also a third, less obvious problem, namely what to
do with closed gluon loops. There the hadronization is begun at a
random point, where the location of this point currently is not stored
anywhere. The algorithm as presented so far will start at another
point and therefore give a mismatch. We have not considered this a
big issue for now, since the default CR algorithm will dissolve almost
all such closed loops, and again the key issue is to provide some
relevant amount of smearing without attaching too deep a meaning to
each separate correction to the dominant hadronization picture.

\subsection{The space--time picture of hadronic rescattering}
\label{subsec:rescatterxyzt}

By the procedure outlined so far, each primary produced hadron has been
assigned a production vertex $x_0 = (t_0, \mathbf{x}_0)$ and a four-momentum $p = (E, \mathbf{p})$. The latter 
defines its continued motion along straight trajectories
$\mathbf{x}(t) = \mathbf{x}_0 + (t - t_0) \, \mathbf{p} /m$. Consider now two particles produced at $x_1$ and $x_2$ with 
momenta $p_1$ and $p_2$. Our objective is to determine whether these particles 
will scatter and, if so, when and where. To this end, the potential collision
is studied in the center-of-momentum frame of the two particles, with motion
along the $\pm z$ direction, i.e.\
\begin{equation}
\begin{split} 
    p_1 &= (E_1, 0, 0, p)~, \\
    p_2 &= (E_2, 0, 0, -p)~. 
\end{split}
\end{equation}
If they are not produced at the same time, the position of the earlier particle 
is offset to the creation time of the later particle. Particles moving away from 
each other already at this common time, i.e.\ with $z_1 > z_2$, are assumed unable 
to scatter.

Otherwise, the probability $P$ of an interaction is a function of the impact parameter $b$, the center-of-mass energy, and the two particle species. There is no solid theory for the $b$ dependence 
of $P$, so we will consider two different shapes. The default model is a Gaussian dependency,
\begin{equation}
    P(b) = P_0 e^{-b^2/b_0^2},
\label{eq:bGaussianEdge}    
\end{equation}
where $P_0$ is referred to as the opacity, a free parameter that is 0.75 by default, and the characteristic length scale is
\begin{equation}
    b_0 = \sqrt{\frac{\sigma}{P_0 \pi}}~,
\end{equation}
where $\sigma$ is the cross section. It is assumed that the only dependency on the energy and the particle species is through $\sigma$, which will be discussed in great detail in \secref{sec:rescattering}. Typical values of $b_0$ are around 1-2\ fm for the most common processes. An alternative model is a grey disk with interaction probability
\begin{equation}
    P(b) = P_0\, \Theta(b - b_0) , \label{eq:bSharpEdge}
\end{equation}
where $\Theta$ is the Heaviside step function. The $P_0 = 1$ case gives the often-used black disk limit. In both these cases, the parameter $b_0$ is chosen so that
\begin{equation} 
    \lint_0^\infty 2\pi b\, P(b)\,\d b = \sigma ~.
\end{equation}
This normalization ensures that if $b$ is chosen uniformly on a large disk, the 
total probability of an interaction is the same for both models. In reality, with a 
finite effective region, one may expect the Gaussian shape to give fewer scatterings.

If it is determined that the particles will interact, the interaction time is defined as the 
time of closest approach in the rest frame. The spatial component of the interaction vertex depends 
on the character of the collision. Elastic and diffractive processes can be viewed as
$t$-channel exchanges of a pomeron (or reggeon), and then it is reasonable to let each 
particle continue out from its respective location at the interaction time. For other 
processes, where either an intermediate $s$-channel resonance is formed or strings are 
stretched between the remnants of the two incoming hadrons, an effective common 
interaction vertex is defined as the average of the two hadron locations at the 
interaction time. In cases where strings are created, be it by $s$-channel processes 
or by diffraction, the hadronization starts around this vertex and is described in 
space--time as already outlined. This means an effective delay before the new hadrons 
are formed and can begin to interact. For the other processes, such as elastic scattering 
or an intermediate resonance decay, there is the option to have effective formation times 
before new interactions are allowed. One reason for why one would want this is that 
it takes some time for the new hadrons to break free from the volume formerly
occupied by the mothers and form their own new (spatial) wave functions. 

In actual events with many hadrons, each hadron pair is checked to see if it fulfils 
the interaction criteria and, if it does, the interaction time for that pair (in the CM frame of the event) is recorded
in a time-ordered list. During rescattering, unstable particles can decay, with the fastest-decaying ones having lifetimes comparable to the timescales of rescattering. 
For these particles, an invariant lifetime $\tau$ is picked at random according to 
an exponential $\exp(-\tau/\tau_0)$, where $\tau_0 = 1 / \Gamma$ is the inverse 
of the width. This is done for each short-lived hadron, and the resulting decay times are inserted into the same 
list. Then the scattering or decay that is first in time order is simulated unless the particles involved have already interacted/decayed.
This produces new hadrons that are checked for rescatterings or decays, and any such are
inserted into the time-ordered list. This process is repeated
until there are no more potential interactions.

There are some obvious limitations to the approach as outlined so far:

Firstly, the procedure is not Lorentz invariant, since the time-ordering of 
interactions is defined on the lab frame of the full collision, i.e.\ the CM frame 
for LHC events. We do not expect this to be a major issue: even if the time ordering 
would change depending on the frame chosen, it would not matter in choosing between 
two potential interactions with a spacelike separation, and only for a fraction of 
those with a timelike one. This has been studied and confirmed within existing
rescattering approaches \cite{Bass:1998ca,Xu:2004mz,Weil:2016zrk}. We will 
also present a check in \secref{subsec:lorentz}, where we confirm that the effect on observable quantities is negligible. More consistent time orderings 
have been proposed \cite{Peter:1994yq,Behrens:1994yr}, but are nontrivial to 
implement and have not been considered here.

Secondly, currently only collisions between two incoming hadrons are considered, even though 
in a dense environment one would also expect collisions involving three or more hadrons.
If one considers a closed system in thermal equilibrium, where $2 \to n$ processes
are allowed, indeed $n \to 2$ at commensurate rates would be a natural ingredient 
to maintain that balance. The system is rapidly expanding in $\p\p$ collisions, 
so for our current studies it should not be a big issue. One place where it could
make a difference is in baryon rates, where pair annihilation outweighs pair
creation within the current setup. In the future $3 \to n$ collisions could be
identified by isolating cases where a hadron has two very closely separated
potential $2 \to n$ interactions, which then could be joined into one. This would
also introduce an alternative argument for a formation time, as the borderline 
between separated and joined processes. 

Thirdly, introducing rescattering will change the shape of events, which of course is 
the point of the exercise, but it also affects distributions we do not want to change.
One example, related to the second limitation above, is that the charged multiplicity 
will increase, which has to be compensated by a tuning of other parameters. In this
article only a simple retune is made specifically for $\p\p$. More properly one 
should go back to $\e^+\e^-$ annihilation events and retune the fragmentation
of a simple string there, with rescattering effects included, before proceeding 
to $\p\p$. In $\e^+\e^- \to \Z^0 \to \q \qbar$ events, however, the bulk of 
rescattering should be related to nearest neighbours in rank, i.e.\ in order along 
the string. So, if such rescatterings are not simulated, then fragmentation 
parameters should not have to be changed significantly. A shortcut to avoid a 
bigger retune therefore is to forbid nearest-rank neighbours from rescattering 
also in $\p\p$ events, and this is one model variation we will consider.

Fourthly, all possible subprocesses are assumed to share the same impact-parameter
profile. In a more detailed modelling the $t$-channel elastic and diffractive 
processes should be more peripheral than the rest, and display an approximately 
inverse relationship between the $t$ and $b$ values.

Finally, the model only considers the effect of hadrons colliding with hadrons, 
not those of strings colliding/overlapping with each other or with hadrons.
The former is actively being studied within \textsc{Pythia}, as a shoving/repulsion 
of strings \cite{Bierlich:2016vgw,Duncan:2019poz}. Both shove and rescattering 
act to correlate the spatial location of strings/hadrons with a net push outwards,
giving rise to a radial flow. In reality the two could be combined, with shove
acting before hadronization and rescattering after. The two effects do not add 
linearly, however, since an early shove leads to a more dilute system of strings
and primary hadrons, and thereby less rescattering. Thus it will become a 
nontrivial task to distinguish the effects of the two possible phenomena, not
made any simpler if also string--hadron interactions were to be included in the mix.

\section{The hadronic rescattering model}
\label{sec:rescattering}

A crucial input for deciding whether a scattering can occur is the total cross section. Once a potential scattering is selected, it also becomes necessary to subdivide the total cross section into a sum of partial cross sections, one for each possible process, as these are used to represent relative frequencies for each process to occur. In this section, we discuss the possible processes we have implemented in our framework, including how their partial cross sections are calculated, and how those processes are simulated.

As we will see, a staggering amount of details enter in such a description,
owing to the multitude of incoming particle combinations and collision processes.
To wit, not only ``long-lived'' hadrons can collide, i.e.\ $\pi$, $\K$, $\eta$, $\eta'$, $p$, $n$, $\Lambda$, $\Sigma$, $\Xi$, $\Omega$, and their antiparticles, 
but also a wide selection of short-lived hadrons, starting with $\rho$, $\K^*$, $\omega$, $\phi$, $\Delta$, $\Sigma^*$ and $\Xi^*$. The possible processes that can occur depend heavily on the particle types involved.
In our model, the following types of processes are available:
\begin{itemize}
\item Elastic interactions are ones where the particles do not change species, i.e.\ $AB \to AB$. In our implementation, these are considered different from elastic scattering through a resonance, e.g. $\pi^+\pi^- \to \rho^0 \to \pi^+\pi^-$ (in reality there are likely to be interference terms that make this separation ambiguous). In experiments, usually all $AB \to AB$ events are called elastic because it is not possible to tell which underlying mechanism was involved. Therefore, when comparing with data for elastic cross sections, we do include contributions from resonance formation.
\item Resonance formation typically can be written as $AB \to R \to CD$, where $R$ is the intermediate resonance. This can only occur when one or both of $A$ and $B$ are mesons. It is the resonances that drive rapid and large cross-section variations with energy, since each (well separated) resonance should induce a Breit-Wigner peak.
\item Annihilation is specifically aimed at baryon--antibaryon collisions where the baryon numbers cancel out and gives a mesonic final state. This is assumed to require the annihilation of at least one $\q\qbar$ pair. This is reminiscent of what happens in resonance formation, but there the final state is a resonance particle, while annihilation forms strings between the outgoing quarks.
\item Diffraction of two kinds are modelled here: single $AB \to XB$ or 
$AB \to AX$ and double $AB \to X_1 X_2$. Here $X$ represents a massive excited 
state of the respective incoming hadron, and there is no net colour exchange
between the two sides of the event.
\item Excitation can be viewed as the low-mass limit of diffraction, where either one or both incoming hadrons are excited to a related higher resonance. It can be written as $AB \to A^*B$, $AB \to AB^*$ or $AB \to A^*B^*$. Here $A^*$ and $\B^*$ are modelled with Breit-Wigners, as opposed to the smooth mass spectra of the $X$ diffractive states. In our description, this has only been implemented in nucleon-nucleon interactions.
\item Nondiffractive topologies are assumed to correspond to a net colour 
exchange between the incoming hadrons, such that colour strings are stretched
out between them after the interaction.
\end{itemize}
All total and partial cross sections have a nontrivial energy dependence. Whereas we have made an effort to cover a fair amount of detail, it is not feasible to give all processes full attention in the first release of this framework, not even in the proportionately few cases where experimental data exist. Our hope is that since rescatterings will not be observable on an individual basis and instead the average effects they induce is what will be of interest, we can live with imperfections here and there so long as they do not generate non-negligible systematic biases. Refinements could be introduced over time without affecting the rescattering machinery as such. In \secref{subsec:rescatterRates} we will study the rates of different particle types participating in rescattering and at which energies most interactions occur, giving an indication of which cross sections are the most important for future refinement.

In the continued discussion, some common simplifications should be noted.
\begin{itemize}
\item Cross sections are invariant when all particles are replaced by 
their antiparticles. Whenever we talk about any particular cross section for two particles, it is always implicit that the exact same procedure is used to calculate the cross section for their antiparticles.
\item Many measured cross sections approximately scale in accordance 
with the Additive Quark Model (AQM) \cite{Levin:1965mi,Lipkin:1973nt},
i.e. like the product of the number of valence quarks in the two incoming hadrons.
The contribution of heavier quarks is scaled down relative to that of a $\u$ or 
$\d$ quark, presumably by mass effects giving a narrower wave function. 
Assuming that quarks contribute inversely proportional to their constituent masses,
this gives an effective number of interacting quarks in a hadron of approximately
\begin{equation}
    n_{\q,\mathrm{AQM}} = n_{\u} + n_{\d} + 0.6 \, n_{\s} + 0.2 \, n_{\c} +
    0.07 \, n_{\b}~. 
\label{eq:nqAQM}    
\end{equation}
For lack of alternatives, many unmeasured cross sections are assumed to scale 
in proportion to this.
\item The neutral Kaon system is nontrivial, with strong interactions
described by the $\K^0/\Kbar^0$ states and weak decays by the 
$\K^0_{\mathrm{S}}/\K^0_{\mathrm{L}}$ ones. The oscillation time is of the
order of the $\K^0_{\mathrm{S}}$ lifetime, far above the rescattering scales 
of interest in this article. Therefore an intermediate ``decay'' invariant time 
of $10^9$~fm has been introduced for 
$\K^0/\Kbar^0 \to \K^0_{\mathrm{S}}/\K^0_{\mathrm{L}}$, well above 
hadronization scales but also well below decay ones. While the bulk of 
Kaon production is into the strong eigenstates, a fraction is into the weak ones,
such as $\phi \to \K^0_{\mathrm{S}} \, \K^0_{\mathrm{L}}$. Cross sections for $\K_S^0/\K_\mathrm{L}^0$ with a hadron are given by the mean of the cross section for $\K^0$ and $\Kbar^0$ with that hadron. When the collision occurs, the $\K_\mathrm{S,L}$ is converted into either $\K^0$ or $\Kbar^0$, where the probability for each is proportional to the total cross section for the interaction with that particle.
\end{itemize}
Finally, keep in mind that we here concern ourselves with cross sections for 
collisions at low CM energies, with most rescatterings occurring
below 2~GeV, and very few above 5~GeV, as we will see.

\subsection{Total cross sections}

The total cross section is needed by the rescattering algorithm to determine how close two hadrons need to be to interact. In the rescattering algorithm, each hadron pair (including the products of rescatterings) is checked for potential interactions, and thus naively $\mathcal{O}(n_\mathrm{primary}^2)$ total cross sections must be calculated. Quick checks that can exclude a fair fraction of all pairs at an early stage are essential to keep time consumption at a manageable level. In particular, we have made an effort to ensure that total cross sections can be calculated efficiently, and that partial cross sections are only calculated for a hadron pair when it has been determined that they should interact.

\begin{figure}[tbp]
\begin{minipage}[c]{\linewidth}
\centering
\includegraphics[width=0.48\linewidth]{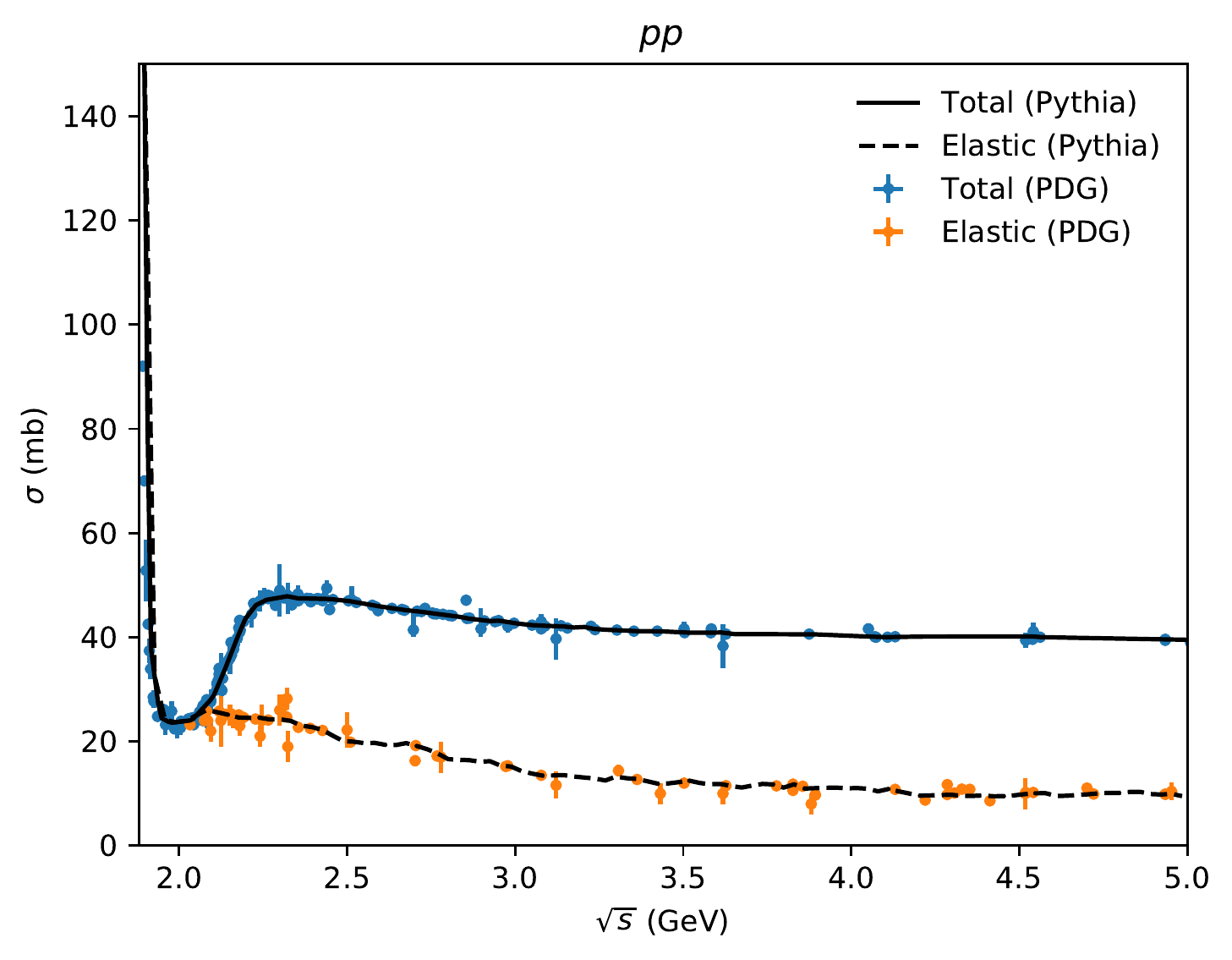}
\includegraphics[width=0.48\linewidth]{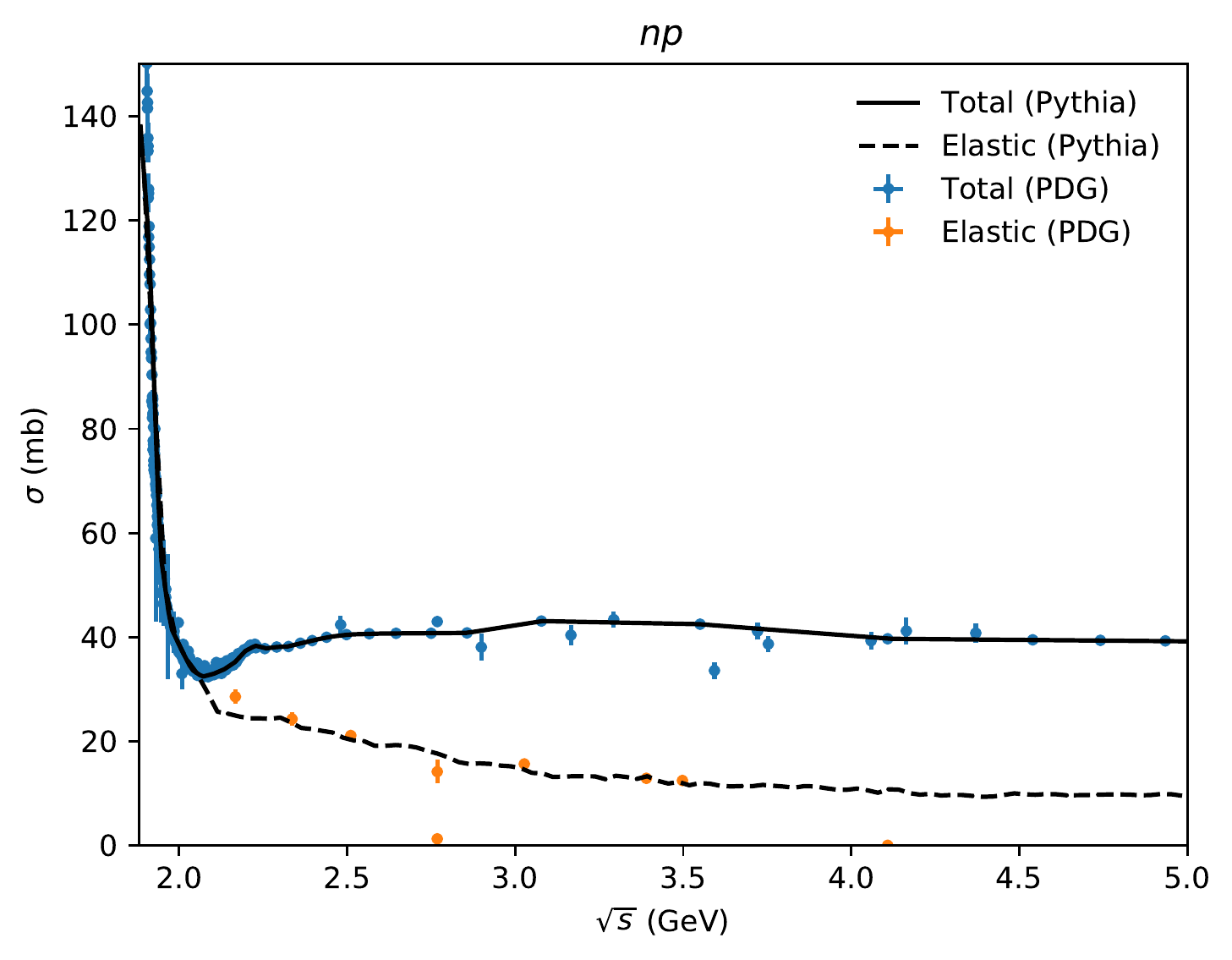}
\end{minipage}

\begin{minipage}[c]{\linewidth}
\centering
\includegraphics[width=0.48\linewidth]{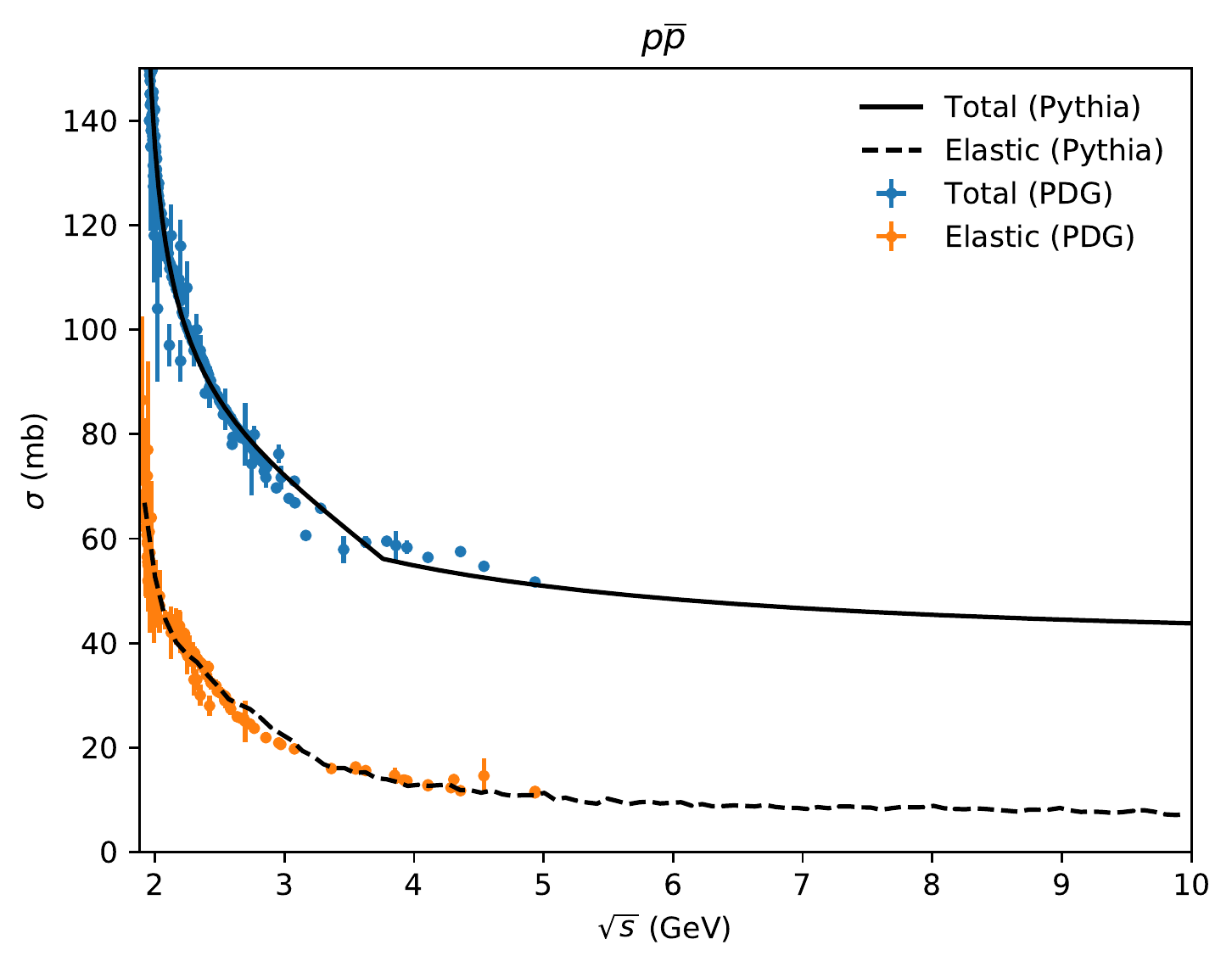}
\includegraphics[width=0.48\linewidth]{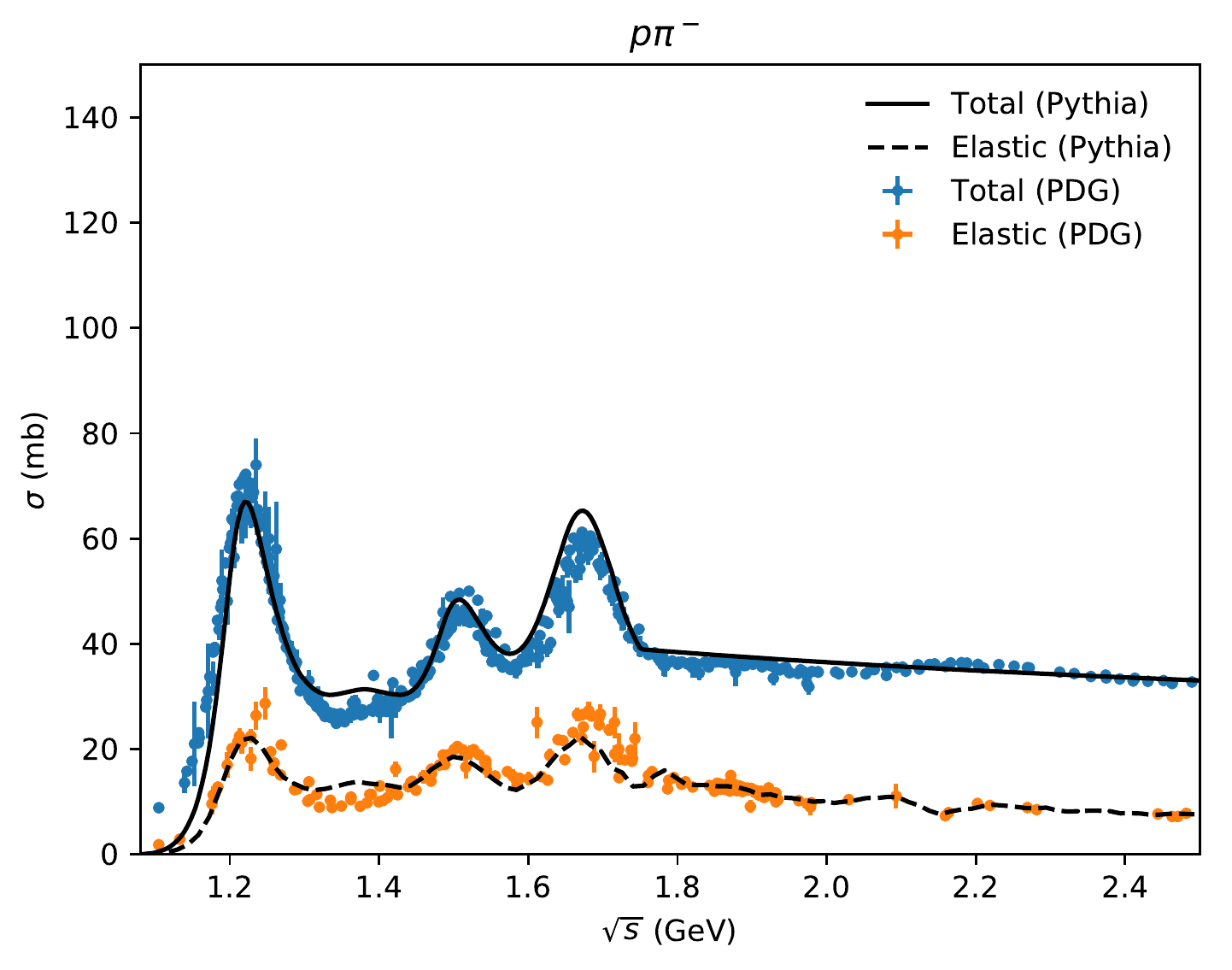}
\end{minipage}

\begin{minipage}[c]{\linewidth}
\centering
\includegraphics[width=0.48\linewidth]{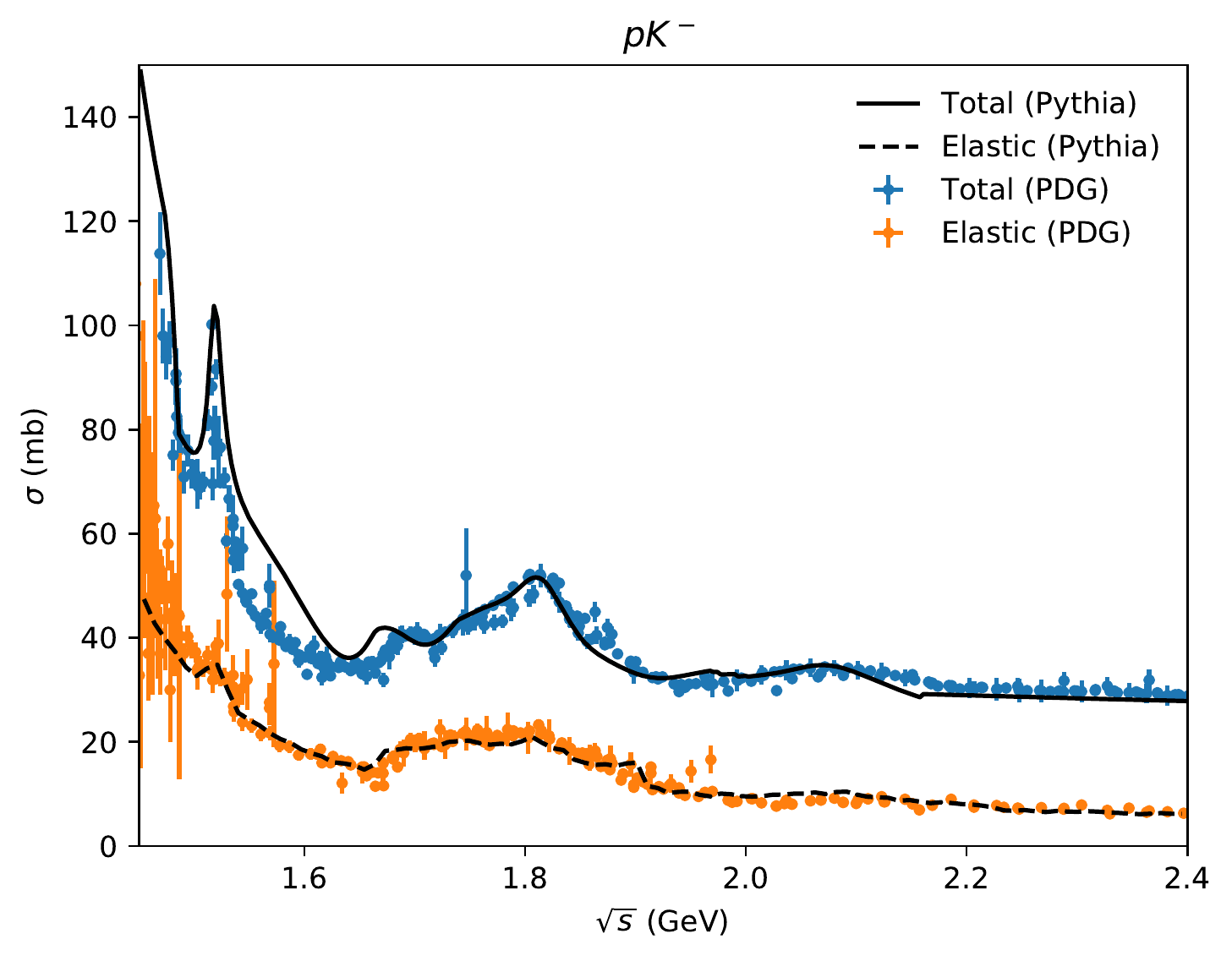}
\includegraphics[width=0.48\linewidth]{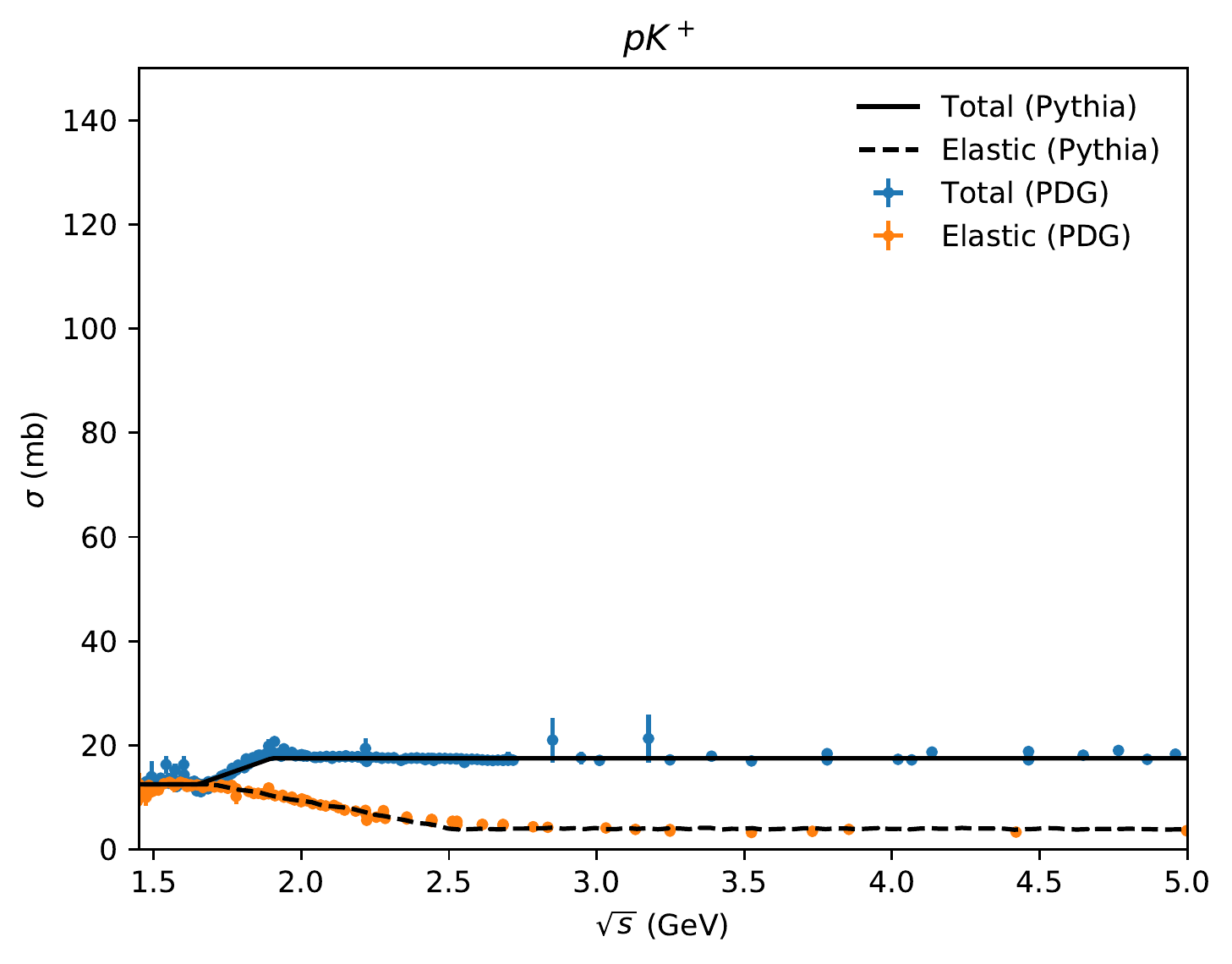}
\end{minipage}

\caption{Total and elastic cross sections for some important processes. The elastic cross sections for $\p\pi^-$ and $\p\K^-$ include elastic scattering through a resonance, $AB \to R \to AB$, which notably do not correspond to the elastic cross sections calculated in \secref{subsec:sigmaEl}.}
\label{fig:sigmaTotEl}
\end{figure}

A brief summary of total cross sections is provided in \tabref{tab:sigmaTotal}. \figref{fig:sigmaTotEl} shows the total and elastic cross sections for some important processes where PDG data is available \cite{Tanabashi:2018oca}.

\begin{table}[ht]
    \centering
    \begin{tabular}{|c | c|}
        \hline
        Case & Method \\
        \hline
        $\N\N$, $<5$ GeV & Fit to data \\
        $\N\N$, $>5$ GeV & $HPR_1R_2$ parameterization \\
        Other $\B\B$ & AQM (UrQMD) parameterization \\
        \hline
        $\p\pbar$, $<5$ GeV & Ad hoc parameterization \\
        $\p\pbar$, $>5$ GeV & $HPR_1R_2$ parameterization \\
        Other $\B\Bbar$ & AQM rescaling of $\p\pbar$ \\
        \hline 
        $\pi\pi$ and $\K\pi$ & Parameterization based on \cite{GarciaMartin:2011cn,Pelaez:2019eqa} and \cite{Pelaez:2016tgi}\\
        $\N\K^-, \N\Kbar^0$ & Resonances + ad hoc parameterization \\
        $\N\K^+, \N\K^0$ & Ad hoc parameterization \\
        $\M\B/\M\M$ with resonances & Resonances + elastic \\
        Other $\M\B/\M\M$ & $HPR_1R_2$ if available, otherwise AQM \\
        \hline
    \end{tabular}
    \caption{Summary of total cross section descriptions. Here, $\N$ is used to denote a nucleon ($\p$ or $\n$), $\B$ a baryon and $\M$ a meson.}
    \label{tab:sigmaTotal}
\end{table}

\subsubsection{Baryon-baryon}

For $\N\N$ collisions below 5~GeV, the total cross section is found by 
an interpolation of experimental data \cite{Tanabashi:2018oca}. The $\n\n$ 
cross section is taken to be the same as the $\p\p$ one. Above 5~GeV, 
the cross section is found using the $HPR_1R_2$ parameterization 
\cite{Tanabashi:2018oca},
\begin{equation}
    \sigma_\total = P + H \log^2\left( \frac{s}{s_0} \right) 
    + R_1 \left( \frac{s}{s_0} \right)^{\eta_1} 
    + R_2 \left( \frac{s}{s_0} \right)^{\eta_2} ~,
\end{equation}
where:
\begin{itemize}
    \item $P$, $R_1$ and $R_2$ depend on the specific particle species, 
    as shown in \tabref{tab:HPR1R2}.
    \item $s_0$ depends on the masses of $A$ and $B$ and is given by 
    $(m_A + m_B + M)^2$, where $M = 2.1206$ GeV is a constant.
    \item $H = \pi (\hbar c)^2/M^2 = 0.2720$ mb, $\eta_1 = 0.4473$ and $\eta_2 = 0.5486$ are constants. 
\end{itemize}
In other baryon--baryon cases, the cross section is found using the AQM ansatz as
\begin{equation}
    \sigma_{\mathrm{AQM},AB} = (40~\mathrm{mb}) \, \frac{n_{\q,\mathrm{AQM},A}}{3}  
    \, \frac{n_{\q,\mathrm{AQM},B}}{3} ~.
    \label{eq:AQM}
\end{equation}

\begin{table}[tbp]
    \centering
    \begin{tabular}{|l|c c c|}
        \hline
        Process & $P$ & $R_1$ & $R_2$ \\
        \hline
        $\p\p/\n\n$  & 34.41 & 13.07 & -7.394 \\
        $\p\n$       & 34.71 & 12.52 & 6.66  \\
        $\pbar \p$   & 34.41 & 13.07 & 7.394 \\
        $\N\pi^\mp$ & 18.75 &  9.56 & $\pm$1.767 \\
        $\p\Kbar$ & 16.36 &  4.29 &     3.408 \\
        $\n\Kbar$ & 16.31 &  3.70 &     1.826 \\
        \hline
    \end{tabular}
    \caption{Parameters for the $HPR_1R_2$ parameterization, for processes used in our rescattering framework. All numbers are in units of mb. $\N$ stands for either $\p$ or $\n$ and $\Kbar$ stands for either $\K^-$ or $\Kbar^0$.}
    \label{tab:HPR1R2}
\end{table}

\subsubsection{Baryon-antibaryon}

For $\B\Bbar$, we parameterize the cross section as a function of the absolute value of the center-of-mass momentum $p_\CM$ of the colliding hadrons. For $\p\pbar$ below $p_\CM < 6.5$ GeV, we use the UrQMD parameterization \cite{Bass:1998ca}:
\begin{equation}
\sigma_\total(\p\pbar) = \begin{cases}
    271.6 e^{-1.1 \, p^2}~, & p < 0.3 ~,\\
    75.0 + 43.1 \, p^{-1} + 2.6 \, p^{-2} - 3.9 \, p~, & 0.3 < p < 6.5~,
\end{cases}
\end{equation}
For $p_\CM > 6.5$~GeV, we use $HPR_1R_2$. The boundary at 6.5~GeV has been chosen to give a smooth transition between the two regions, and is slightly different from the boundary at 5~GeV used by UrQMD. For all other baryon-antibaryon interactions, the total cross section is found using the same parameterization, but rescaling by an AQM factor,
\begin{equation}
    \sigma_\total(\B\Bbar) = \frac{\sigma_{\AQM,\B\Bbar}}{\sigma_{\AQM,\p\pbar}}
    \, \sigma_\total(\p\pbar) ~,
    \label{eq:AQMrescaling}
\end{equation}
where $\sigma_\AQM$ is given in eq.~\eqref{eq:AQM}. 

In some cases no quarks can annihilate, e.g.\ for 
$\Delta^{++}(\u\u\u) + \Deltabar^+(\dbar\dbar\dbar)$. In these cases, the 
annihilation cross section (see \secref{sec:Annihilation}) is subtracted 
from the total one.

\begin{figure}[tbp]
\begin{minipage}[c]{\linewidth}
\centering
\includegraphics[width=0.48\linewidth]{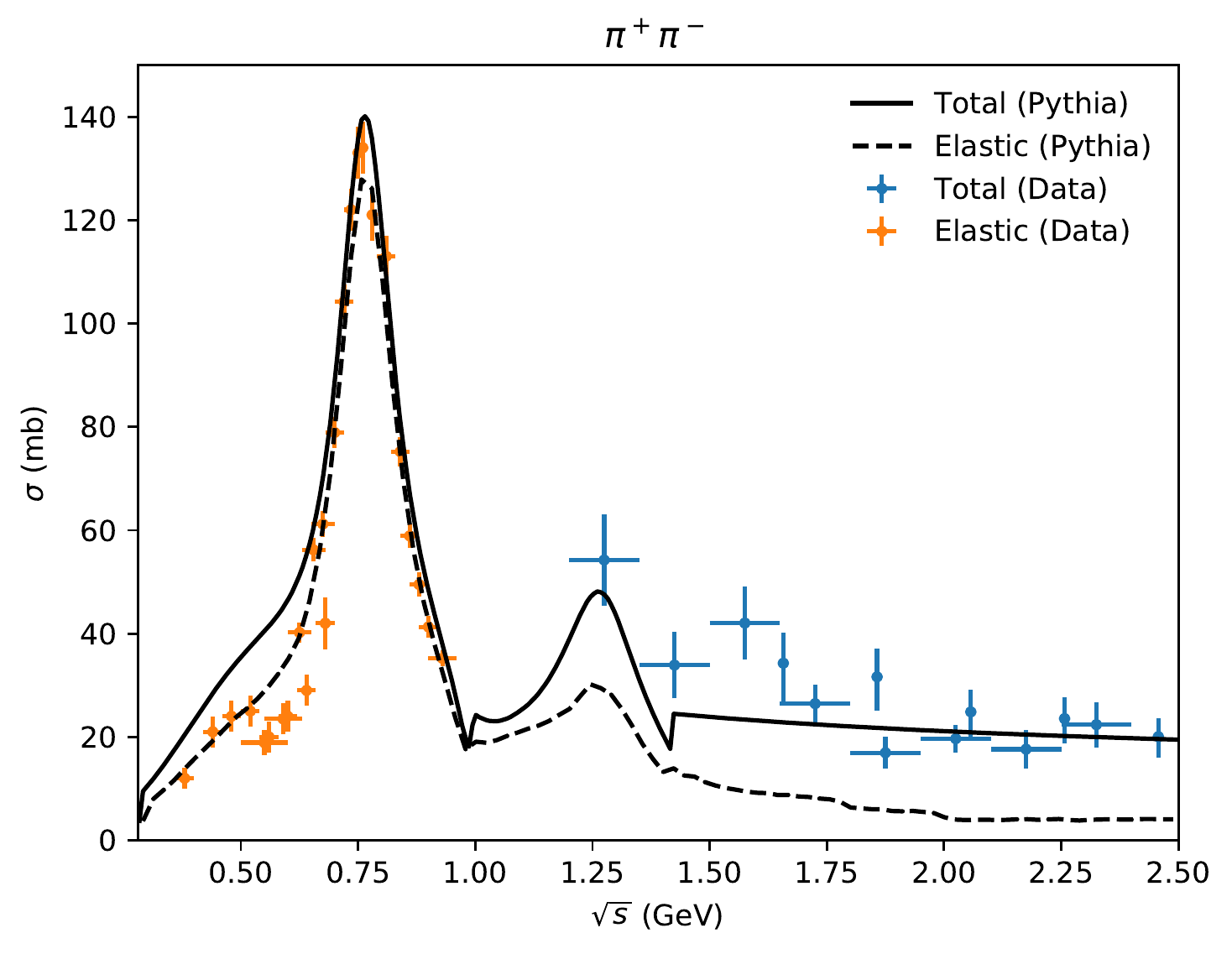}
\includegraphics[width=0.48\linewidth]{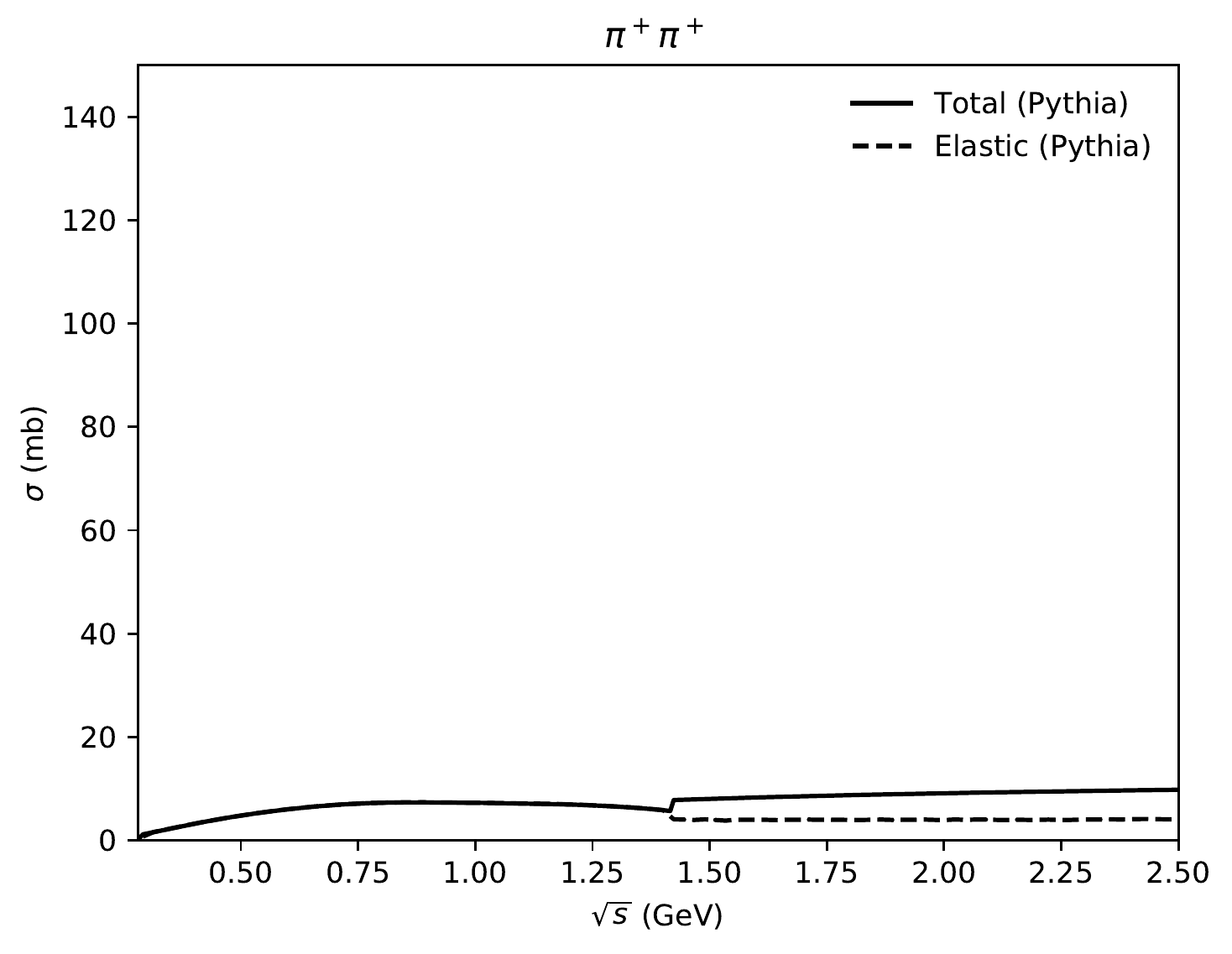}
\end{minipage}
\begin{minipage}[c]{\linewidth}
\centering
\includegraphics[width=0.48\linewidth]{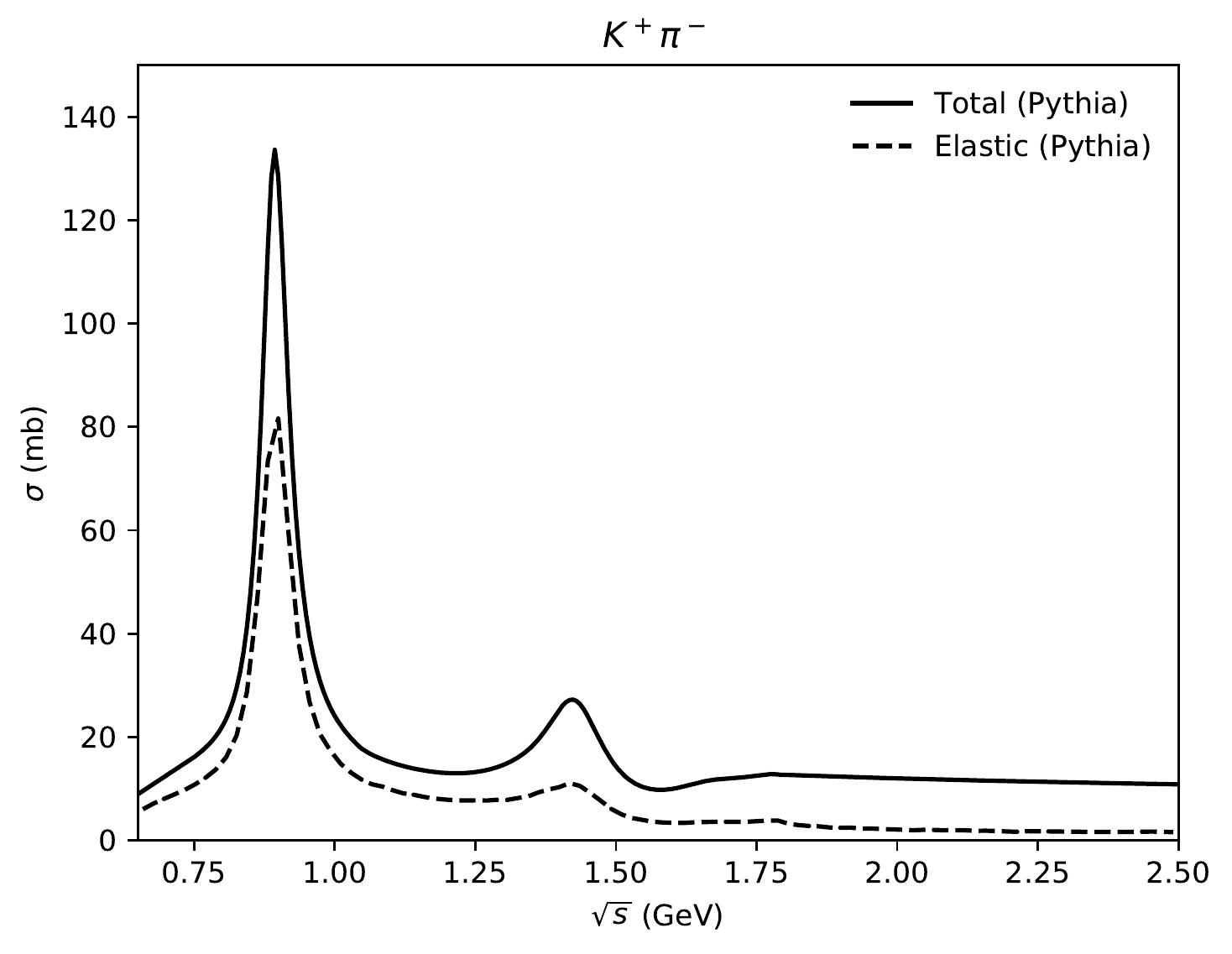}
\includegraphics[width=0.48\linewidth]{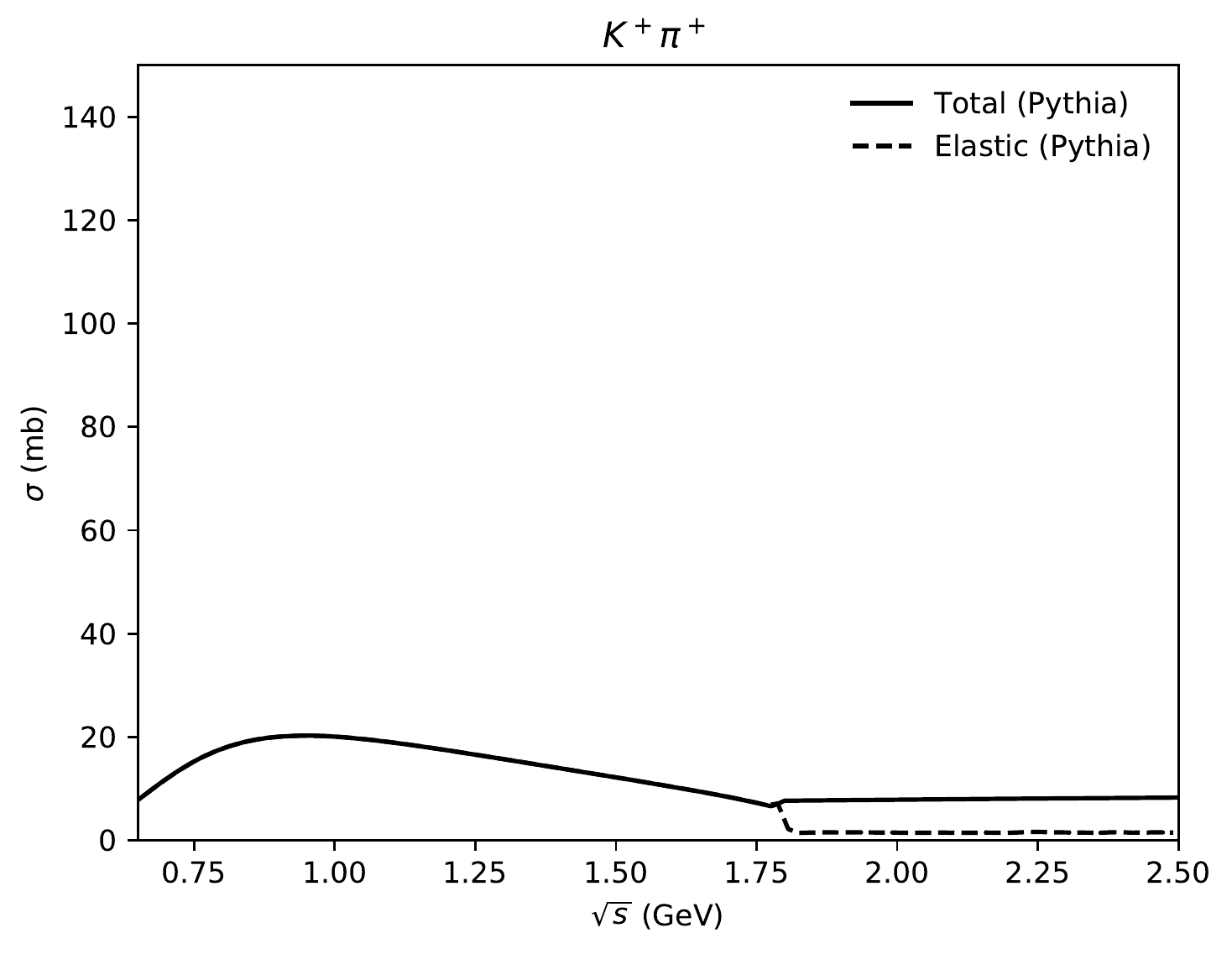}
\end{minipage}
\caption{Total and elastic cross sections for $\pi\pi$ and $\K\pi$ interactions. We see that resonances exist for $\pi^+ \pi^-$ and $\K^+\pi^-$, but not for $\pi^+\pi^+$ and $\K^+\pi^+$. The elastic cross sections include cross sections for elastic scattering through a resonance. For $\pi^+ \pi^-$, the elastic data comes from \cite{Srinivasan:1975tj,Protopopescu:1973sh} and total data comes from \cite{Biswas:1967mpl, Robertson:1973tk}. Note that in some theory calculations the concept of elastic is extended to related processes, e.g.\ $\pi^+ \pi^- \to \pi^0 \pi^0$ may count as part of a broader $\pi \pi \to \pi \pi$ ``elastic'' process. If we had taken that viewpoint, the elastic cross sections for $\pi^+\pi^-$ and $\K^+\pi^-$ would have equalled the total cross section at low energies.}
\label{fig:sigmaTotElmeson}
\end{figure}

\subsubsection{Meson-hadron}

The most common meson-meson interactions are $\pi\pi$ and $\K\pi$. In these two cases,
the total cross sections are found using the calculations of Pel{\'a}ez et al.
\cite{GarciaMartin:2011cn,Pelaez:2019eqa, Pelaez:2016tgi}. Below $1.42$ GeV for $\pi\pi$ 
and below $1.8$ GeV for $\K\pi$, values of the total cross sections have been tabulated 
and are found using interpolation, for the sake of efficiency. Above these thresholds, the cross
section is parameterized as
\begin{equation} 
    \sigma_\total(AB) = \frac{4\pi^2\left( \beta_P s 
    + \beta_\rho s^{\alpha_{\rho}} 
    + \beta_2 s^{\alpha_{R2}} \right)}{\sqrt{(s - (m_A - m_B)^2) (s - (m_A + m_B)^2)}},
\label{eq:pipiHE}    
\end{equation}
where, $\alpha_\rho = 0.53$, $\alpha_{R2} = 2\alpha_\rho - 1 = 0.06$, and the $\beta$ parameters depend on the exact process as given in \tabref{tab:pipiParameters}. Total and elastic cross sections for $\pi\pi$ and $\K\pi$ interactions are shown in \figref{fig:sigmaTotElmeson}.

\begin{table}[ht]
    \centering
    \begin{tabular}{|c|c|c|c|}
        \hline
        Case & $\beta_P$ & $\beta_\rho$ & $\beta_2$ \\
        \hline
        $\pi^\pm \pi^\mp$ & 0.83 & 1.01  &  0.013  \\
        $\pi^\pm \pi^0$   & 0.83 & 0.267 & -0.0267 \\
        $\pi^0 \pi^0$     & 0.83 & 0.267 &  0.053  \\
        $\pi^\pm \pi^\pm$ & 0.83 & -0.473 & 0.013  \\
        \hline
        $\K\pi^\pm, I=1/2$ & 6.9032 & 8.2126 & 0.0 \\
        $\K\pi^0, I = 1/2$ & 3.4516 & 4.1063 & 0.0 \\
        $\K\pi, I = 3/2$ & 10.3548 & -5.76786 & 0.0 \\
        \hline
    \end{tabular}
    \caption{Parameter values for the $\pi\pi$ and $\K\pi$ cross sections, as used in
    eq.~\eqref{eq:pipiHE}. In the case of $\K\pi$, $I$ refers to the sum of the third isospin components for the incoming particles. The two $I = 1/2$ cases are equivalent, except for Clebsch-Gordan coefficients.}
    \label{tab:pipiParameters}
\end{table}

For some of the remaining meson-hadron interactions, explicit resonances are implemented.
In these cases, at low energies (below $\sim2$ GeV, depending on the specific interaction),
the total cross section is given by the elastic cross section plus the sum of resonance
cross sections,
\begin{equation}
    \sigma_\total = \sigma_{\elastic} + \lsum_{\mathrm{resonances}} \sigma_{\resonance} 
    ~,
\label{eq:sumresel}    
\end{equation}
where $\sigma_\elastic$ and $\sigma_\resonance$ will be described in the following sections.
There is an option in Pythia to also calculate the $\pi\pi$ and $\K\pi$ cross sections 
this way instead of using the default methods of Ref. \cite{Pelaez:2019eqa, Pelaez:2016tgi}, but there are two drawbacks of using this approach. In terms of physics, it is less accurate because it does not take into account interference effects between
resonances. And in terms of computational efficiency it is slower, which can have a significant impact on performance that is exacerbated by how common these interactions are.

One important case with a lot of data is $\p/\n + \K^-/\Kbar^0$. Summing resonances does 
not accurately match data at low energies, so an additional contribution has been added,
based on formulae from UrQMD. Furthermore we add an explicit elastic contribution
not present in UrQMD in order to get an even better fit. Above $2.16$ GeV, we use the $HPR_1R_2$ parameterization. 
The case $\p/\n + \K^+/\K^0$ is also important and much data exists, but in this case
resonances cannot form since there are no common quark--antiquark pairs to annihilate. 
We use an ad hoc parameterization to fit these cross sections to data at low energies.
Specifically, the total cross section is given by 12.5~mb below 1.65~GeV and 17.5~mb 
above 1.9~GeV, with a linear transition in the intermediate range. The total and 
elastic cross sections for both these $\N\K$ cases are shown in \figref{fig:sigmaTotEl}. 

The last special case is $N\pi$ which uses the $HPR_1R_2$ parameterization above the resonance region. All other cases use the AQM parameterization above the resonance region. For those processes where resonances are not available, AQM is instead used at all energies.

\subsection{Elastic scattering} 
\label{subsec:sigmaEl}

In this section we discuss the directly elastic processes $AB \to AB$, leaving aside
scattering through a resonance, $AB \to R \to AB$. A summary of $\sigma_{\mathrm{el}}$
descriptions is provided in \tabref{tab:sigmaElastic}.

\begin{table}[ht]
    \centering
    \begin{tabular}{|c | c|}
        \hline
        Case & Method \\
        \hline
        $\p\p/\n\n/\p\n$, $<5$ GeV & Fit to data \\
        $\p\p/\n\n/\p\n$, $>5$ GeV & CERN/HERA parameterization \\
        Other $\B\B$  &  AQM parameterization \\
        \hline 
        $\p\pbar$  & UrQMD parameterization \\
        Other $\B\Bbar$  & Rescaling $\p\pbar$ \\
        \hline 
        $\pi\pi$, $<1.42$ GeV  & Parameterization by Pel{\'a}ez et al. \cite{Pelaez:2019eqa} \\
        $\pi\pi$, $>1.42$ GeV & Constant 4 mb \\
        $\K\pi$, $I=1/2$, $<1.8$ GeV & No scattering except through resonances \\
        $\K\pi$, $I=3/2$, $<1.8$ GeV & Parameterization by Pel{\'a}ez et al. \cite{Pelaez:2016tgi} \\
        $\K\pi$, $>1.8$ GeV & Constant 1.5 mb \\
        $\N\pi$, $<4$ GeV  & Fit to data \\
        $\N\pi$, $>4$ GeV  & CERN/HERA parameterization \\
        $\N\K$  & Ad hoc parameterization \\
        Other $\M\B/\M\M$  & AQM parameterization \\
        \hline
    \end{tabular}
    \caption{Summary of elastic cross section descriptions. Here, $\N$ is used to denote a nucleon, $\B$ a baryon and $\M$ a meson. For $\K\pi$ below 1.8 GeV, $I$ refers to the sum of the third isospin component of the incoming particles.}
    \label{tab:sigmaElastic}
\end{table}

\begin{table}[ht]
    \centering 
    \begin{tabular}{|c|c c c c c|}
        \hline
        Case & $a$ & $b$ & $n$ & $c$ & $d$ \\
        \hline
        $\N\N$ & 11.9 & 26.9 & -1.21 & 0.169 & -1.85 \\
        $\p\pbar$ & 10.2 & 52.7 & -1.16 & 0.125 & -1.28 \\
        $\N\pi$ & 0 & 11.4 & -0.4 & 0.079 & 0 \\
        \hline
    \end{tabular}
    \caption{CERN/HERA parameters}
    \label{tab:HERAparameters}
\end{table}

For $\p\p$, $\n\n$, and $\p\n$, the elastic cross section is fitted to PDG data below 5~GeV \cite{Tanabashi:2018oca}, which is assumed to be the same as the total cross section up to 2.1 GeV. Above 5~GeV, $\sigma_{\mathrm{el}}$ is parameterized as a function of laboratory momentum $p_\lab$, according to the CERN/HERA parameterization \cite{Montanet:1994xu} with the general form
\begin{equation}
    \sigma_\mathrm{HERA}(p) = a + b \, p^n + c \log^2 p + d \log p ~,
\end{equation}
with parameters given in \tabref{tab:HERAparameters}. For all other $\B\B$ cases, 
the elastic cross section is given by an elastic AQM-style parameterization
\cite{Bass:1998ca},
\begin{equation}
    \sigma_{\AQM,\elastic} =  0.039 \, \sigma_{\AQM,\total}^{3/2}~. \label{eq:AQMel}
\end{equation}
The CERN/HERA parameterization is also used for $\p\pbar$ for $p_\lab > 5$ GeV, albeit with different parameters. Below this lab momentum, we use another ad hoc parameterization from UrQMD \cite{Bass:1998ca},
\begin{equation}
\sigma_{\mathrm{el}}(\p\pbar) = \begin{cases}
    78.6 ~, & p < 0.3 ~, \\
    31.6 + 18.3 \, p^{-1} - 1.1 \, p^{-2} - 3.8 \, p ~, & 0.3 < p < 5~.
\end{cases}
\end{equation}
For all other baryon-antibaryon cases, the elastic cross section is found by 
rescaling the $\p\pbar$ cross section, using an AQM factor in the same way as 
for total cross sections.

For elastic cross sections involving mesons, there are several special cases. For $\pi\pi$, we separate our calculation into two regions, below and above 1.42~GeV, as for the total cross section. Below, the purely elastic cross section is found by parameterizing the d-wave contribution from Pel{\'a}ez et al. \cite{GarciaMartin:2011cn,Pelaez:2019eqa}. This parameterization can be seen in \figref{fig:sigmaTotElmeson}, where it is equal to the total $\pi^+\pi^+$ cross section since no resonances can be formed in that case. The other $\pi\pi$ cases get the same contribution, except with a scale factor that depends on the exact case. Above 1.42~GeV, a constant elastic cross section of 4~mb is consistent with the parameterization of Ref. \cite{Pelaez:2019eqa} when the contribution from resonances is taken into account. For $\K\pi$, we divide the region into below and above 1.8~GeV. Below this threshold, for total isospin $I = 1/2$, the whole elastic cross section is well described by scattering through a resonance. For total isospin $I = 3/2$, resonances cannot form, and we instead use a parameterization by Ref. \cite{Pelaez:2016tgi}. Above 1.8~GeV, we use a constant 1.5~mb for all cases.

In $\N\pi$ interactions, the non-resonant elastic cross section vanishes below around $1.8$ GeV. Between this energy and up to 4~GeV, we add a non-resonant contribution by interpolating data. Above 4~GeV, we use the CERN/HERA parameterization.

The last special case is $\N\K^+/\N\K^0$. This uses a simple fit to data, using 12.5~mb below 1.7~GeV and 4.0~mb above 2.5~GeV, with a linear transition in between. In all remaining cases, the AQM parameterization given in eq. \eqref{eq:AQMel} is used.

The angular distribution for non-resonant $AB \to AB$ is specified by the 
selection of the $t$ value according to an exponential $\exp(B_{\mathrm{el}}t)$, 
where the slope is given by
\begin{equation}
B_{\mathrm{el}} = 2 b_A + 2 b_B + 2 \alpha' \, \ln \left( \frac{s}{s_0} \right) ~.
\end{equation}
Here $b_{A,B}$ is 2.3~GeV$^{-2}$ for unflavoured baryons and 1.4~GeV$^{-2}$ for mesons, $\alpha' = 0.25$~GeV$^{-2}$ is the slope of the pomeron trajectory, and 
$s_0 = 1 / \alpha' = 4$~GeV$^2$  \cite{Donnachie:1985iz,Schuler:1993wr}. 
The $b_{A,B}$ values are rescaled by AQM factors for strange or heavier hadrons, while
$\alpha'$ is assumed universal.

Note that, strictly speaking, the 
$\sigma_{\total}$, $\sigma_{\elastic}$, $B_{\elastic}$ and $\rho$ 
(the ratio of the real to imaginary parts of the forward scattering amplitude) 
should be connected by the optical theorem. Here we make no attempt to model
$\rho$ or to exactly fulfil the optical theorem, which would have been quite
messy in the low-energy resonance region. Note that an $L = 0$ resonance would
decay isotropically, meaning a more complicated overall angular distribution
when interference between elastic and resonance contributions is considered.
We have checked, however, that the optical theorem is approximately obeyed
above the resonance region, assuming that $\rho$ is not giving large effects.

\subsection{Resonance formation} 
\label{subsec:sigmaRes}

\begin{figure}[tbp]
\begin{minipage}[c]{\linewidth}
\centering
\includegraphics[width=0.48\linewidth]{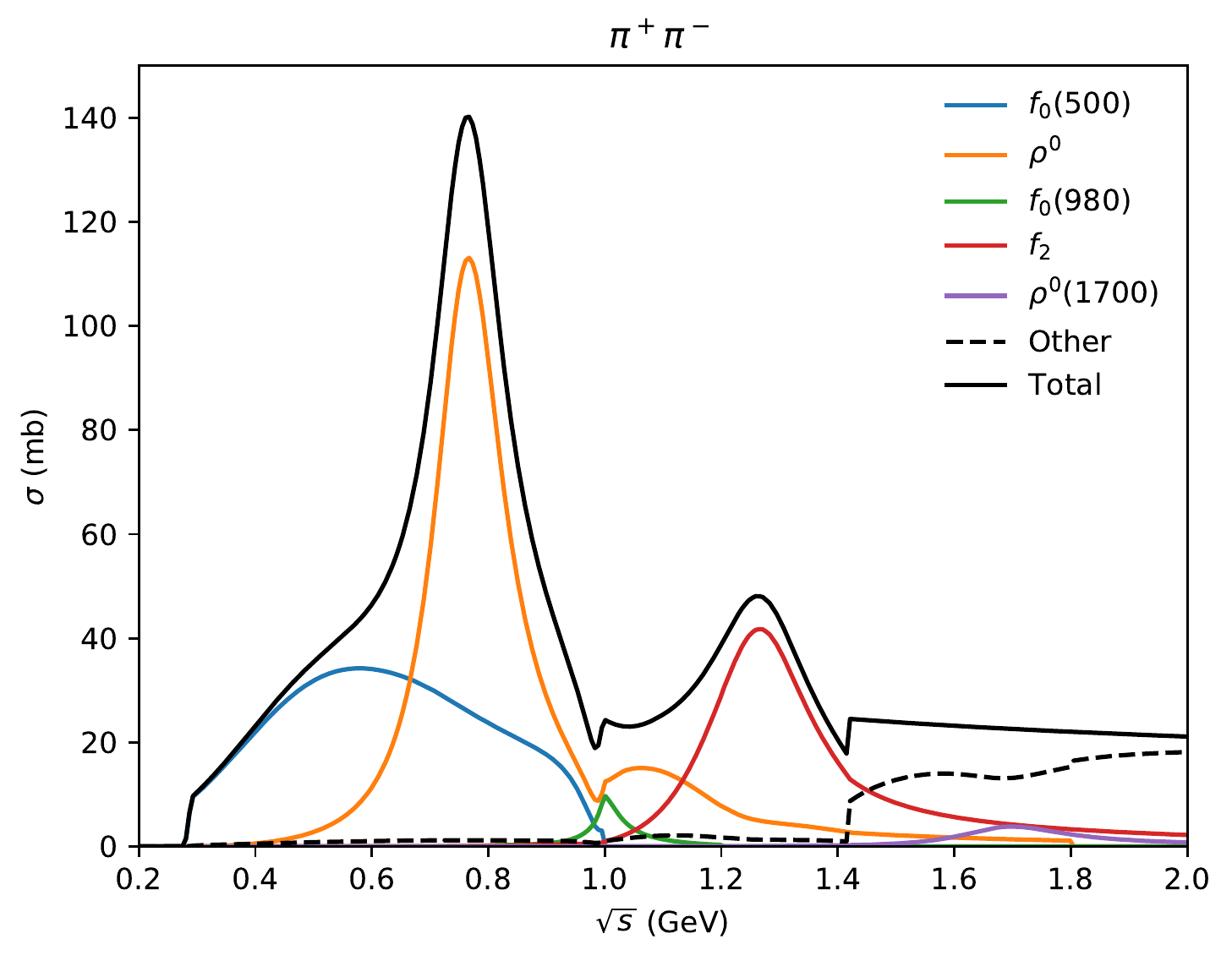}
\includegraphics[width=0.48\linewidth]{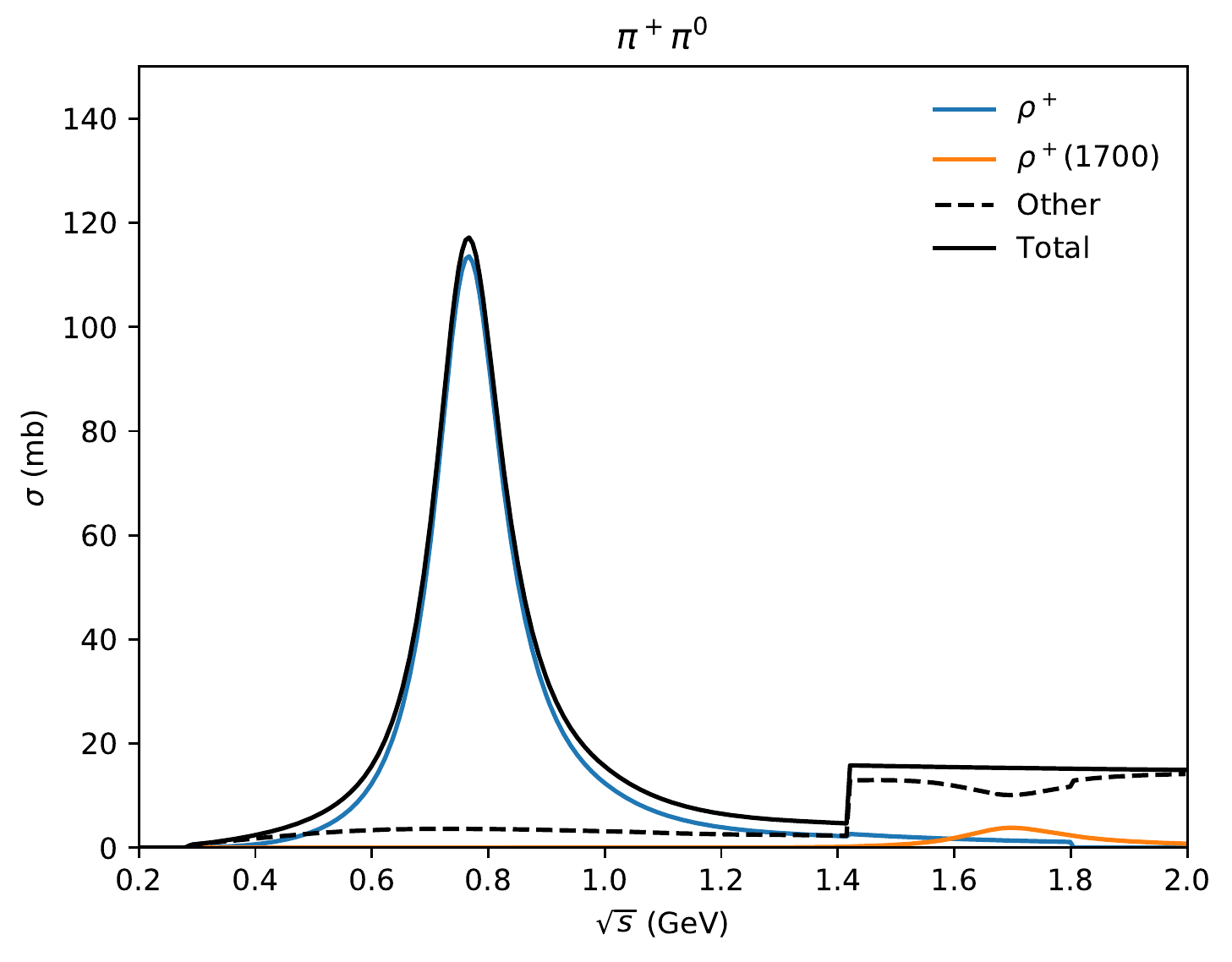}
\end{minipage}

\begin{minipage}[c]{\linewidth}
\centering
\includegraphics[width=0.48\linewidth]{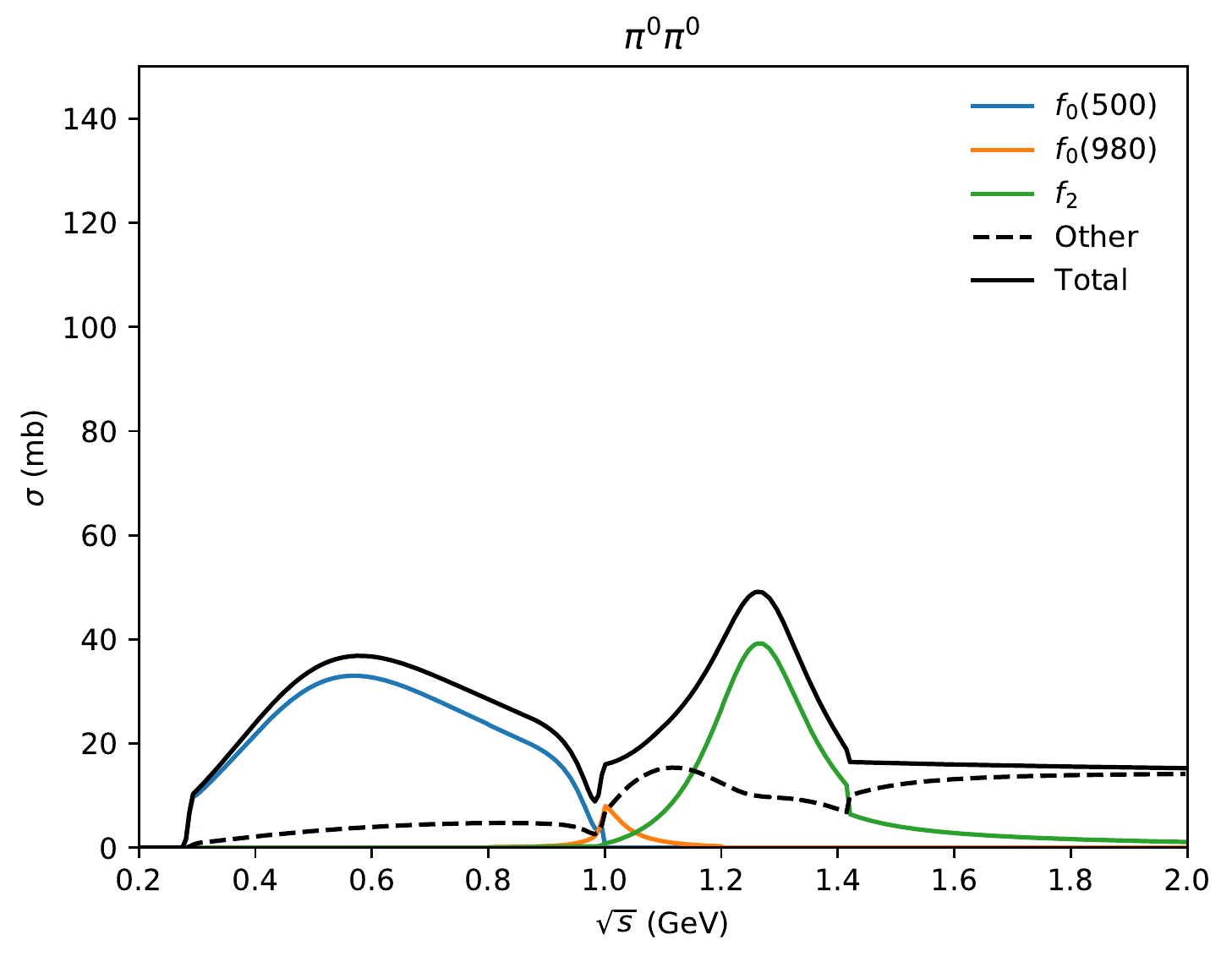}
\includegraphics[width=0.48\linewidth]{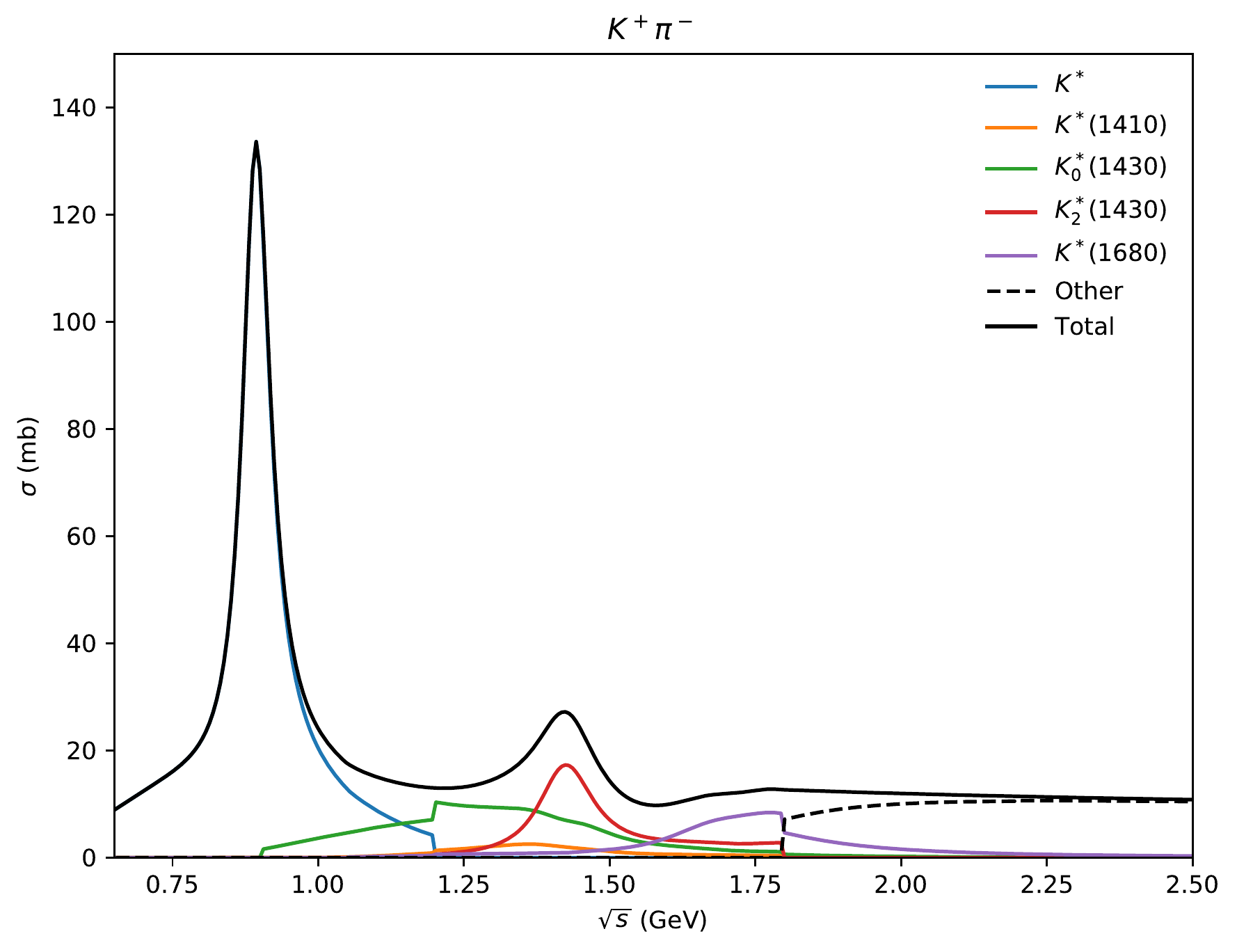}
\end{minipage}

\begin{minipage}[c]{\linewidth}
\centering
\includegraphics[width=0.48\linewidth]{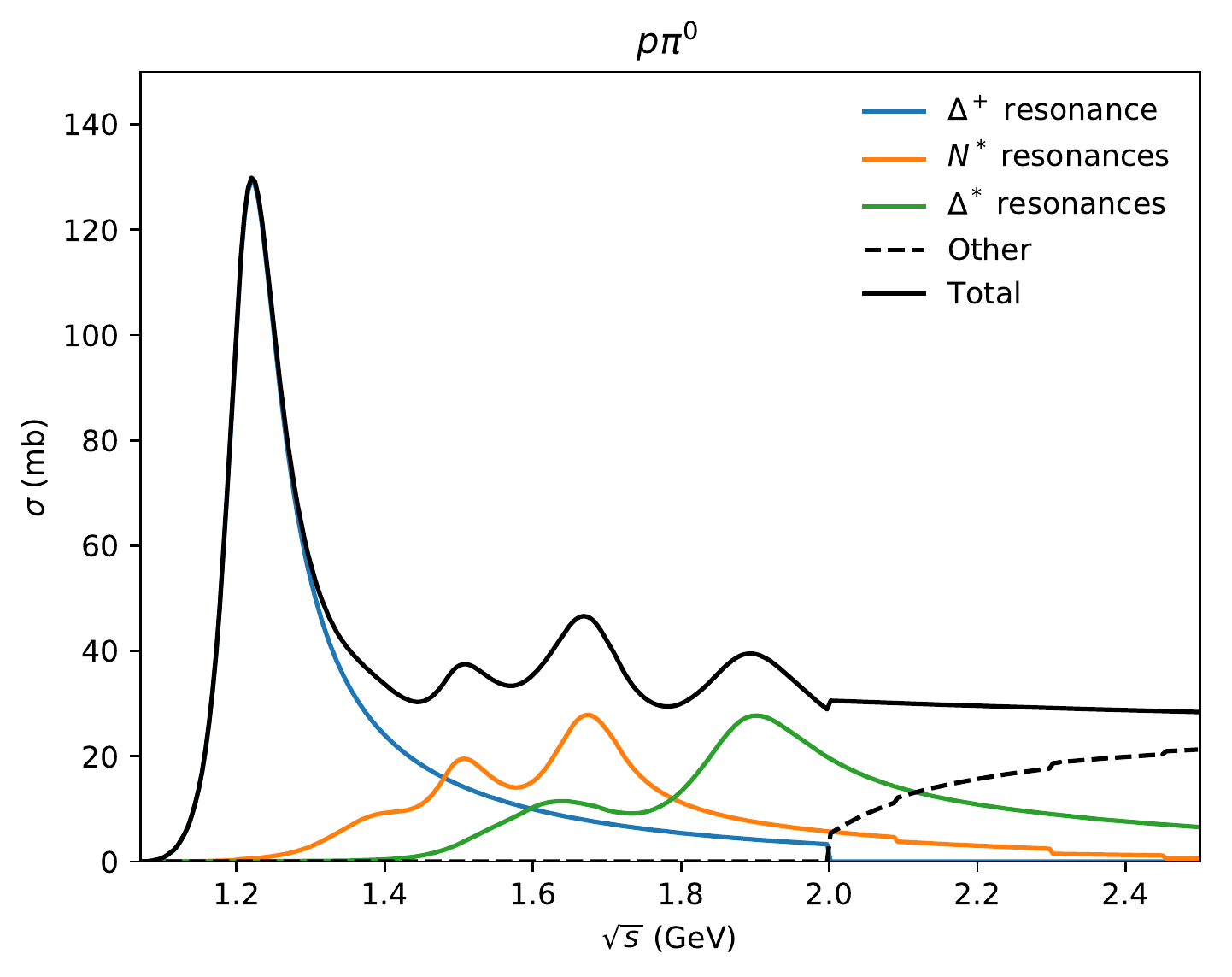}
\includegraphics[width=0.48\linewidth]{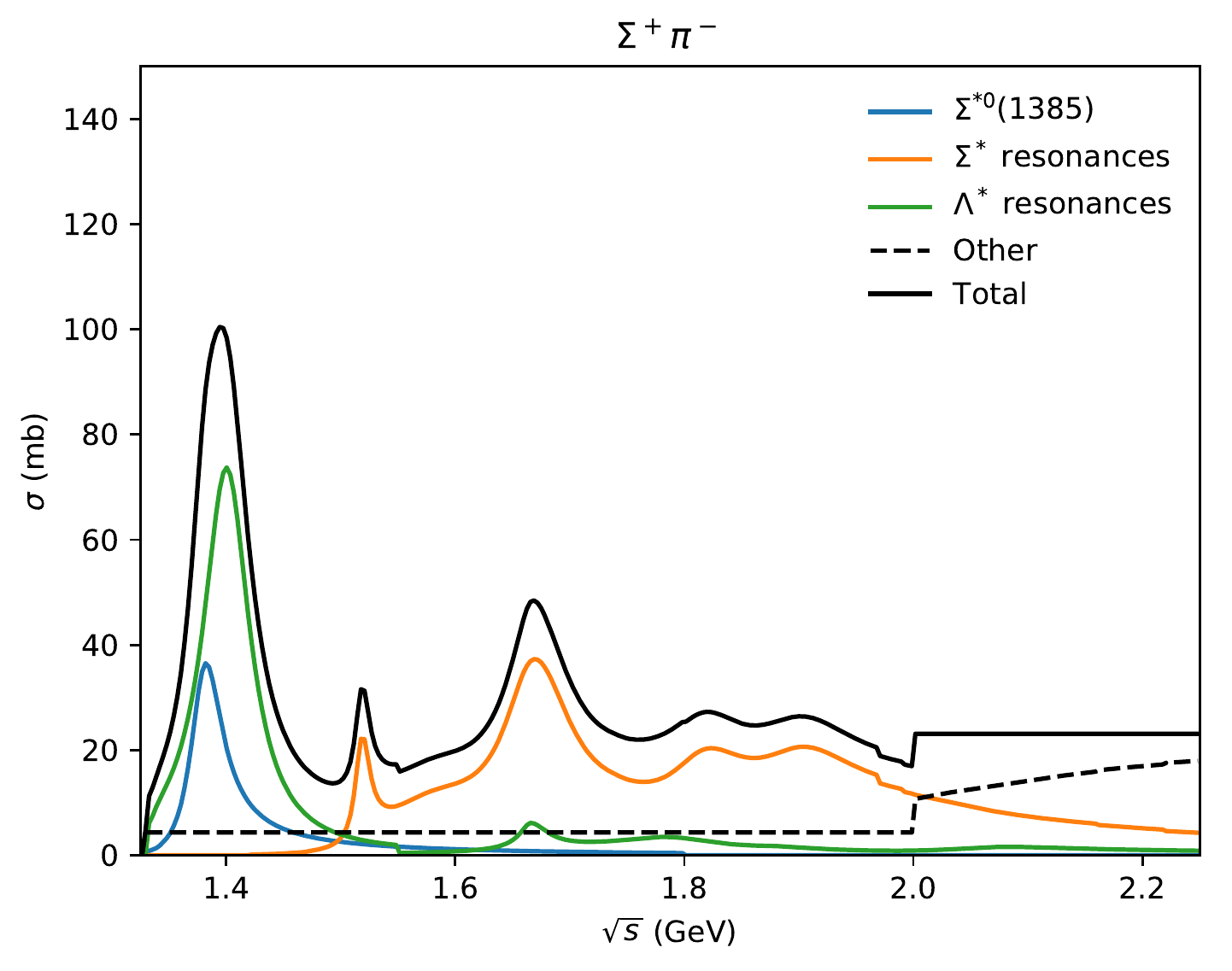}
\end{minipage}

\caption{Resonant cross sections for some important cases, with partial cross sections for each resonance. For $p\pi^0$ and $\Sigma^+\pi^-$ there are many resonances, and we have divided them into groups for readability. The "other" cross sections include elastic, diffractive and non-diffractive.}
\label{fig:sigmaRes}
\end{figure}

Explicit resonance formation has been implemented for $\pi\pi$, $\K\pi$, $\N\pi$, $\N\eta$, $\N\omega$, $\Sigma \pi$, $\Sigma \K$, $\Lambda\pi$, $\Lambda\K$, and $\Xi\pi$. This includes all isospin configurations of these particles where resonances exist (e.g. $\Sigma^+ \pi^-$, but not $\Sigma^- \pi^-$). For the formation of a particular resonance $R$ the cross section is given by a nonrelativistic Breit--Wigner \cite{Tanabashi:2018oca}
\begin{equation}
    \sigma_{AB \to R} = \frac{\pi}{p_\CM^2} \frac{(2S_R+1)}{(2S_A+1)(2S_B+1)} 
    \frac{\Gamma_{R \to AB} \Gamma_R}{(m_R - \sqrt{s})^2 + \frac14 \Gamma_R^2} ~,
    \label{eq:resonanceFormation}
\end{equation}
where $S$ is the spin of each particle, $p_\CM$ is the CM momentum of the incoming particles, $\Gamma_{R \to AB}$ is the mass-dependent partial width, and $\Gamma_R$ is the total mass-dependent width of $R$, found by summing the partial widths. The partial widths of a particle at mass $m$ are given by UrQMD as
\begin{equation}
    \Gamma_{R \to AB}(m) = \Gamma_{R \to AB}(m_0) \frac{m_0}{m} 
    \frac{\left< p^{2l+1}(m) \right>}{\left< p^{2l+1}(m_0) \right>} 
    \frac{1.2}{1.0 + 0.2 \frac{\left< p^{2l}(m) \right>}{\left< p^{2l}(m_0) \right>}} ~, \label{eq:massDependentWidth}
\end{equation}
where $m_0$ is the nominal mass of the particle and $\Gamma_{R \to AB}(m_0)$ is the nominal width, both known from experiment, and $l$ is the angular momentum of the outgoing two-body system. The final factor ensures that widths do not blow up at large masses. The phase space factors are given by
\begin{equation}
    \left< p^{2l+1}(m) \right> = \iint p_\CM^{2l+1}(m, m_A, m_B)\, A(m_A) \, A(m_B)
    \, \d m_A \, \d m_B ~,  
    \label{eq:phaseSpace}
\end{equation}
where 
\begin{equation} 
    p_\CM(m, m_A, m_B) = \frac{\sqrt{(m^2 - (m_A + m_B)^2)(m^2 - (m_A - m_B)^2)}}{2m}
    \label{eq:KallenForRes}
\end{equation}
and $A(m)$ are the mass distribution functions, given by a Breit--Wigner,
\begin{equation} 
    A(m) = \frac{1}{2\pi} \frac{\Gamma(m)}{(m^2 - m_0^2)^2 + \frac14 \Gamma^2(m)},
\end{equation}
which reduces to $A(m) = \delta(m - m_0)$ for particles with zero width. Note that although the mass distribution depends on mass-dependent widths, which again depend on the mass distribution of other particles, there is no circular dependency since particle widths 
can only depend on the widths of lighter particles.

\figref{fig:sigmaRes} shows the resonant cross sections for some important cases. For the $\pi\pi$ cases there is a small elastic cross section below 1.42 GeV, corresponding to a d-wave contribution. For $\K^+\pi^-$  there is no direct elastic cross section at low energies, but a significant fraction of the resonances formed will decay back to the initial state particles, cf. \figref{fig:sigmaTotElmeson}. We also observe a discontinuous behaviour at some points. One reason for this is that resonance particles are assigned a restricted mass range outside which they cannot be formed, which is particularly noticeable for example for $\p\pi^0 \to \Delta^+$ at 2.0 GeV. Another reason for a non-smooth behaviour is the fact that the total cross section is parameterized using the more sophisticated machinery of \cite{GarciaMartin:2011cn, Pelaez:2019eqa, Pelaez:2016tgi} and the resonance cross sections are scaled to sum to this value. This is especially noticeable for $\pi^+ \pi^- \to \rho^0$, where the total cross section is significantly larger than the sum of resonance cross sections in the range around 1.0-1.2~GeV, and is why the cross section for $\pi^+ \pi^- \to \rho^0$ has a second peak in that region instead of looking like a regular Breit-Wigner. Both these kinds of discontinuities are visible in the $\K^+\pi^-$ cross sections, at the $K^*$ cutoff at 1.2~GeV.

One exceptional case is the formation of $f_0(500)$ resonances in $\pi^+\pi^-$ or $\pi^0\pi^0$ interactions. The nature of the $f_0(500)$ meson is not fully understood and it has certain exotic properties, notably its width is about the same as its mass. For this reason, eq.~\eqref{eq:resonanceFormation} does not describe its formation well. We find the relevant cross sections by interpolating values calculated based on the work by Pel{\'a}ez et al. \cite{GarciaMartin:2011cn, Pelaez:2019eqa}. After the $f_0(500)$ has been produced, it is treated as any other meson, including in its decay.

The formula for mass-dependent partial widths works only for two-body decays. These are the dominant ones for most resonances we consider, but some hadrons have three- or four-body decays, for instance $\rho^0 \to \pi^+ \pi^- \pi^+ \pi^-$. For such particles, we calculate the mass-dependent partial widths for the two-body channels according to eq. \eqref{eq:massDependentWidth}, but assume that the multibody channels have a constant width for the purposes of calculating the total width needed in eq. \eqref{eq:resonanceFormation}.

In the space--time description, the resonance is created at the average location of the two incoming hadrons at the interaction time in the collision CM frame. The resonance is then treated as any unstable particle with a mean lifetime that is assumed to be $\tau = 1/\Gamma(m_0)$, even if the resonance is off-shell. If all decay channels of the resonance are two-body decays, then eq. \eqref{eq:massDependentWidth} is used to calculate the branching ratios. In this case, the masses of the outgoing particles are picked according to
\begin{equation}
    \d\Gamma_{R \to AB} \sim p_\CM^{2l+1}(m, m_A, m_B)\,A(m_A)\,A(m_B)\,\d m_A\,\d m_B~.
\end{equation}
If there is one or more multibody decay channels, the particle is instead decayed using the existing \textsc{Pythia} machinery.

\subsection{Annihilation} 
\label{sec:Annihilation}

\begin{figure}[t!]

\begin{minipage}[c]{\linewidth}
\centering
\includegraphics[width=0.48\linewidth]{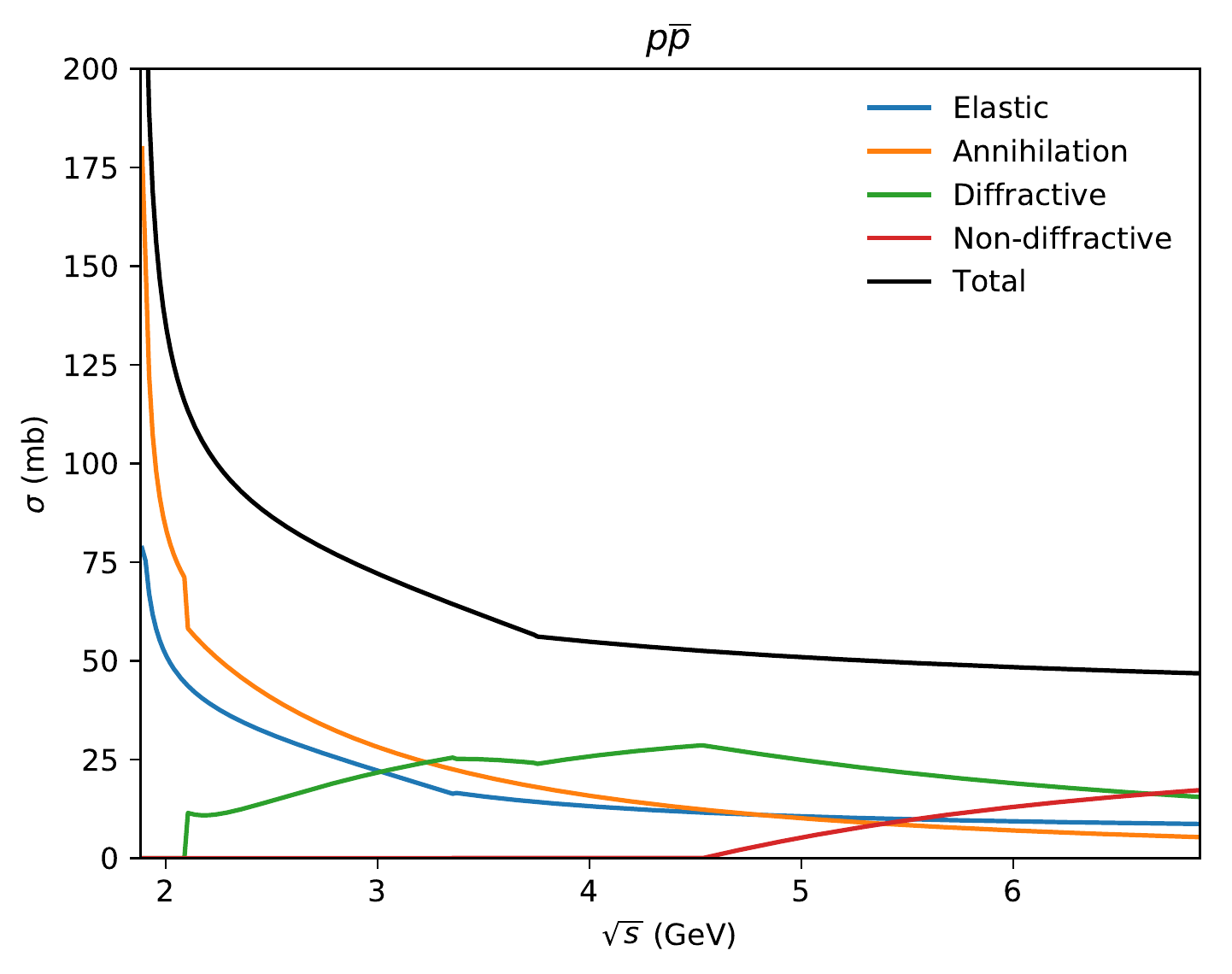}
\end{minipage}

\begin{minipage}[c]{\linewidth}
\centering
\includegraphics[width=0.48\linewidth]{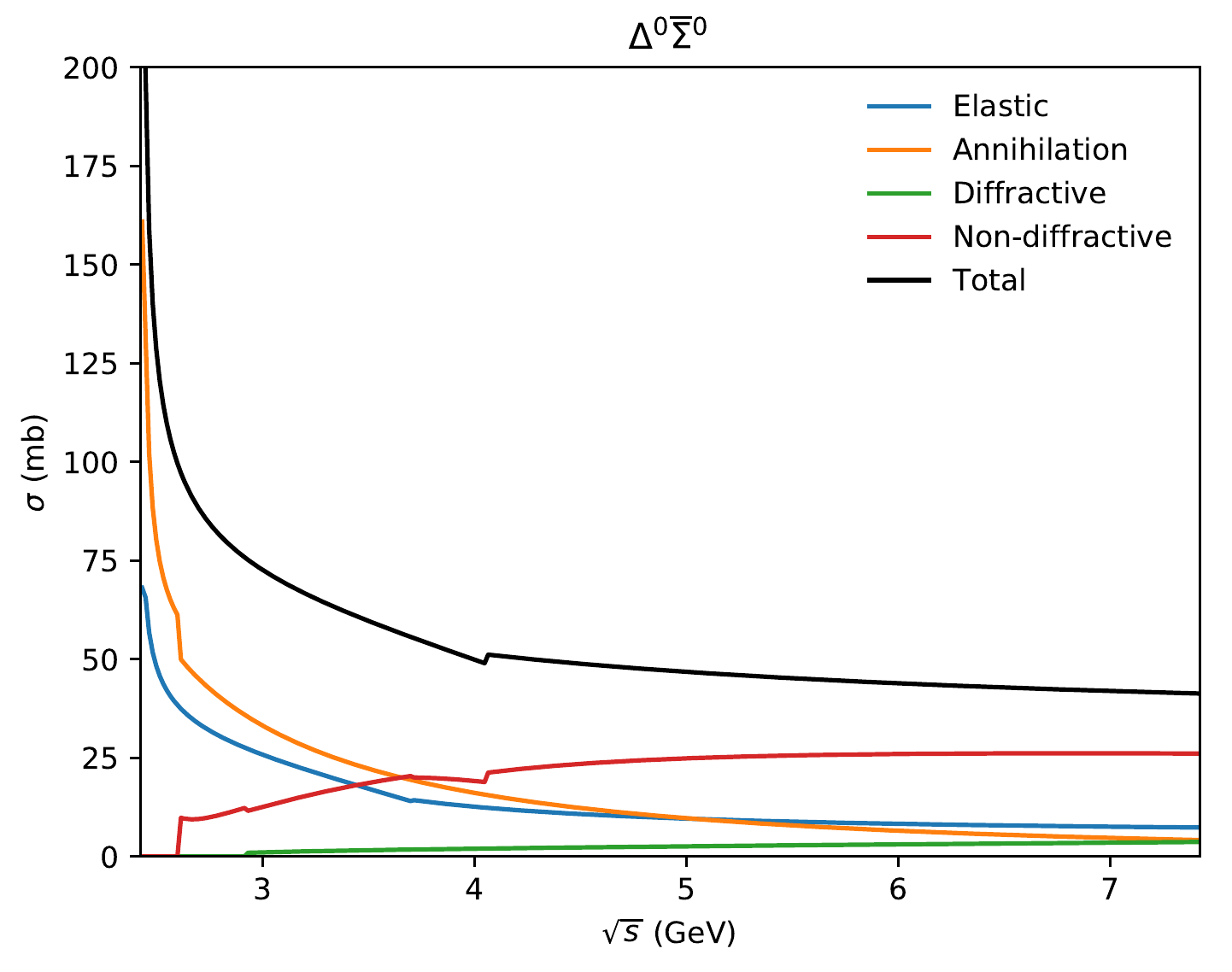}
\includegraphics[width=0.48\linewidth]{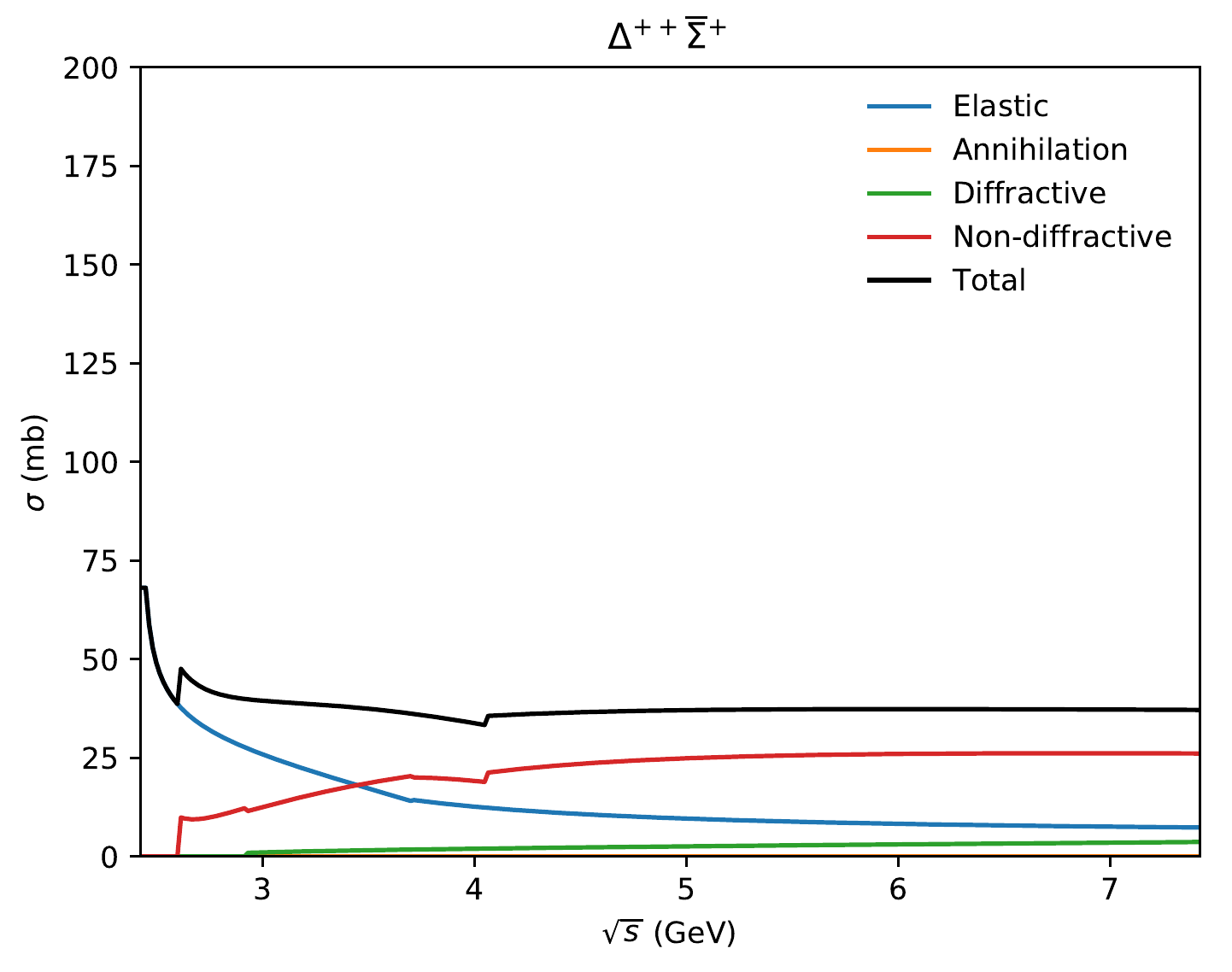}
\end{minipage}

\caption{Partial cross sections for $\p\pbar$, $\Delta^0\Sigmabar^0$ and $\Delta^{++}\Sigmabar^+$. We see that $\Delta^0\Sigmabar^0$ is simply a rescaling of the $\p\pbar$ case, except it gets different diffractive and non-diffractive contributions because $\p\pbar$ implements explicit resonances. For $\Delta^{++}\Sigmabar^+$ annihilation is not possible, so the annihilation cross section is subtracted from the total, significantly changing its shape.}
\label{fig:sigmaAnn}
\end{figure}

In $\B\Bbar$ collisions the baryon number can be annihilated, so that only mesons
remain in the final state. For $\p\pbar$, below 2.1 GeV, annihilation counts for all inelastic processes, so below this threshold,
\begin{equation} 
    \sigma_\annihilation = \sigma_\total - \sigma_\elastic. 
\end{equation}
Above the threshold, it is given by a parameterization by Koch and Dover \cite{Koch:1989zt},
\begin{equation}
    \sigma_\text{ann} = 120 \frac{s_0}{s} \left( \frac{A^2 s_0}{(s - s_0)^2 + A^2 s_0}
    + 0.6 \right)~,
\end{equation}
where $s_0 = 4 m_p^2$ and $A = 0.05$ GeV. For other $\B\Bbar$, this is rescaled in the same way as for the total cross section. Note that the cross section is taken to be the same regardless of whether the baryons have one, two or three quarks in common, but if there are none then currently no annihilation is assumed, even though in principle it would be possible to decompose a $\B\Bbar$ system with no $\q\qbar$ pairs in common into three separate $\q\qbar$ strings. \figref{fig:sigmaAnn} shows the cross sections for $\p\pbar$, $\Delta^0 \Sigmabar^0$ and $\Delta^{++} \Sigmabar^+$. 

When an annihilation process occurs, one or two quark-antiquark pairs are annihilated. 
If two or more pairs are available, the probability for a second annihilation is given 
by a free parameter, by default 0.2, to represent a small but existing rate. No complete
annihilation of all three pairs is performed, since the rate presumably is small and
since it then would be necessary to recreate a new pair, making little net difference. 
The pair(s) to be annihilated is (are) chosen uniformly among all possible combinations.
If only one quark pair remains, a single string is stretched between the $\q$ and
$\qbar$, along the original collision axis. If two pairs remain, a random pairing 
is done to form two separate strings. The procedure for sharing momentum is similar 
to the one described below in \secref{sec:Nondiff}. The possibility of having a single
string stretched between a diquark--antidiquark pair is omitted, since then a new
baryon--antibaryon pair would be produced.

\subsection{Diffractive processes}

Diffractive cross sections in the continuous regime are calculated using SaS ansatz 
\cite{Schuler:1993wr,Schuler:1996en}. The basic version of SaS is designed to deal 
only with processes involving $\p$, $\pbar$, $\pi$, $\rho$, $\omega$ and $\phi$ 
(as needed for $\p\p/\gamma\p/\gamma\gamma$ collisions), and only for collision energies
above 10~GeV. It is here extended to all baryons by applying an AQM rescaling factor
to the corresponding $\p$ cross sections. For mesons a similar rescaling to $\pi$
($= \rho$) cross sections is performed, except that here $\phi$ is retained as the 
template for $\s\sbar$ interactions. The $\eta$ and $\eta'$ cross sections thus are
the appropriate mixes of $\pi$ and $\phi$ ones.

The differential cross section for single diffraction $AB \to XB$ is taken to be of 
the form 
\begin{equation}
    \d\sigma_{XB} \propto \frac{\d M_X^2}{M_X^2} \, \left( 1 - \frac{M_X^2}{s} \right)
    \, \exp( B_{XB} \, t) \, \d t~, 
    \label{eq:sigSD}
\end{equation}
where 
\begin{equation}
    B_{XB}(s) = 2 b_B + 2 \alpha' \ln \left( \frac{s}{M_X^2} \right)~,
\end{equation}
with $b_B$ and $\alpha'$ as for elastic scattering. The constant of proportionality 
involves hadron--pomeron and triple-pomeron couplings, specified for the few 
template processes and then multiplied by AQM factors. The diffractive mass spectrum
is taken to begin at $M_{X,\mathrm{min}} = m_A + 2 m_{\pi} = m_A + 0.28$~GeV and extend 
to the kinematical limit $M_{X,\mathrm{max}} = E_{\mathrm{CM}} - m_B$, while $t$ can 
take values within the full allowed range \cite{Sjostrand:2006za}. 
Above $E_{\mathrm{CM,min}} = 10$~GeV the integrated cross section has been 
parameterized. Below this scale, our studies show that a shape like
\begin{equation}
    \sigma_{XB}(E_{\mathrm{CM}}) = \sigma_{XB}(E_{\mathrm{CM,min}}) \,
    \left( \frac{E_{\mathrm{CM}} - M_{X,\mathrm{min}} - m_B}%
    {E_{\mathrm{CM,min}} - M_{X,\mathrm{min}} - m_B} \right)^{0.6}
    \label{eq:sigSDinterpol}
\end{equation}
provides a good representation of the behaviour down to the kinematic threshold.
Note that $m_A$ and $m_B$ are the actual masses of the colliding hadrons, not 
those of the corresponding template process.

Single diffraction $AB \to AX$ is obtained by trivial analogy with $AB \to XB$.
For double diffraction $AB \to X_1 X_2$ the cross section reads
\begin{equation}
    \d\sigma_{XX} \propto \frac{\d M_1^2}{M_1^2} \, \frac{\d M_2^2}{M_2^2} \, 
    \left( 1 - \frac{(M_1 + M_2)^2}{s} \right) \, 
    \left( \frac{s \, m_{\p}^2}{s \, m_{\p}^2 + M_1^2 \, M_2^2} \right)
    \, \exp( B_{XX} \, t) \, \d t~, 
\end{equation}
where 
\begin{equation}
    B_{XX}(s) = 2 \alpha' \ln \left( e^4 + \frac{s \, s_0}{M_1^2 \, M_2^2} \right)~,  
\end{equation}
again with $s_0 = 1 / \alpha'$. For the behaviour below 10 GeV, our studies suggest that
\begin{equation}
    \sigma_{XX}(E_{\mathrm{CM}}) = \sigma_{XX}(E_{\mathrm{CM,min}}) \,
    \left( \frac{E_{\mathrm{CM}} - M_{X_1,\mathrm{min}} - M_{X_2,\mathrm{min}}}%
    {E_{\mathrm{CM,min}} - M_{X_1,\mathrm{min}} - M_{X_2,\mathrm{min}}} \right)^{1.5}
   \label{eq:sigDDinterpol}
\end{equation}
is a suitable form. 

So far we only considered the continuum production, which dominates for large 
diffractive masses. For small masses, diffractive cross sections can also include 
the formation of explicit resonances, and the contribution from these should be 
added to the continuum contribution. In our framework, this can occur as 
$\N\N \to \N\N^*$ or $\N\N \to \N\Delta^*$ (single diffractive), or 
$\N\N \to \Delta \N^*$ or $\N\N \to \Delta\Delta^*$ (double diffractive),
and similarly when one baryon is replaced by its antibaryon. Higher excitations 
are implicitly part of the continuum diffractive treatment and not 
considered here. The cross section for $A B \to C D$ is given by Ref. \cite{Bass:1998ca}
\begin{equation}
    \sigma_{AB \to CD} = (2S_C + 1) \, (2S_D + 1) \, 
    \frac1s \frac{\left< p_{CD} \right>}{\left< p_{AB} \right>} \, |\mathcal{M}|^2~,
    \label{eq:sigmaExcitePartial}
\end{equation}
where $S$ is the spin of each particle, $\mathcal{M}$ is the matrix element, and $\left< p_{ij} \right>$ are phase space factors given by eq.~\eqref{eq:phaseSpace} (assuming $l = 0$). In practice, this expression will sometimes lead to the sum of partial cross sections being larger than the total one. In those situations, we rescale the excitation cross sections (leaving other partial cross sections unchanged) so that the sum of partial cross sections is equal to the total.

For the matrix elements, we use the same as UrQMD \cite{Bass:1998ca}. For $\N\N \to N\Delta$ it is given by
\begin{equation}
    |\mathcal{M}|^2 = A \, \frac{m_{\Delta}^2 \Gamma_{\Delta}^2}%
    {(s - m_{\Delta}^2)^2 + m_{\Delta}^2 \Gamma_{\Delta}^2}~,
\end{equation}
where $m_{\Delta} = 1.232$~GeV and $\Gamma_{\Delta} = 0.115$~GeV are the nominal mass and width of $\Delta$, and the coefficient is $A = 40000$. For $\N\N \to \Delta\Delta$, the matrix element is a constant $|\mathcal{M}|^2 = 2.8$. Finally, for the remaining classes, the matrix element takes the form
\begin{equation}
    |\mathcal{M}|^2 = \frac{A}{(m_C - m_D)^2 (m_C + m_D)^2}~,
\end{equation}
where $m_C$ and $m_D$ are the nominal masses for the outgoing particles (which will  never be the same for these classes, so the matrix element cannot diverge), and the  coefficient $A$ is $A = 6.3$ for $\N\N \to \N\N^*$, $A = 12$ for $\N\N \to N\Delta^*$,  and $A = 3.5$ for $\N\N \to \Delta N^*$ and $\N\N \to \Delta \Delta^*$.

In eq.~\eqref{eq:sigmaExcitePartial}, the only dependence on outgoing masses comes 
from the phase space term. Thus, the masses of the outgoing particles are distributed 
according to
\begin{equation}
    \frac{\d\sigma_{AB \to CD}}{\d m_C \d m_D} \sim p_\CM(E_\CM, m_C, m_D)\,A(m_C)\,A(m_D)~,
\end{equation}
from eq.~\eqref{eq:phaseSpace}. The $t$ behaviour is assumed to be given by an
exponential slope with the same $B_{XB}/B_{XX}$ as in the continuum single/double 
diffraction for the given diffractive masses.

Calculating the integrals in eq. \eqref{eq:phaseSpace} during event generation would be debilitatingly slow. Therefore, we tabulate the cross sections for each process up to 8 GeV and use interpolation to get the total and partial excitation cross sections. For energies above this threshold, the expansion
\begin{equation} 
p_\CM(E_\CM, m_C, m_D) = \frac12 E_\CM \left( 1 - \frac{m_C^2 + m_D^2}{E_\CM^2} + \mathcal{O}(E_\CM^{-3}) \right)
\end{equation}
shows that $p_\CM$ is approximately constant with respect to $m_C$ and $m_D$ when $E_\CM \gg m$. At the same time, the mass distributions $A(m)$ vanish at large $m$. Thus, in this limit, the phase factor can be approximated as
\begin{equation}
    \left<p_{CD}\right> \approx p_\CM(E_\CM, m_{C,0}, m_{D,0}) \int \d m_C\, A(m_C) \int \d m_D\, A(m_D),
\end{equation}
By integrating $A$ ahead of time, the cross sections can be calculated efficiently during run-time also above the tabulated region.

For other incoming hadron combinations, we fall back on the simpler smooth low-mass
enhancement implemented in SaS to compensate for the lack of explicit resonances. 
For $AB \to XB$ the differential cross section in eq.~\eqref{eq:sigSD} is multiplied 
by a factor
\begin{equation}
    c_{\mathrm{res}} \, \frac{(m_A + M_{\mathrm{res},0})^2}%
    {(m_A + M_{\mathrm{res},0})^2 + M_X^2}~.
    \label{eq:diffenhance}
\end{equation}
Here $c_{\mathrm{res}} = 2$ and $M_{\mathrm{res},0} = 2~\mathrm{GeV} - m_{\p}$ have been
chosen to provide a decent description of the low-mass enhancement in $\p\p$ collisions
at medium-high energies.
For energies below 10~GeV this part of the cross section can be described in the same 
spirit as the continuum part in eq.~\eqref{eq:sigSDinterpol}, but the power is changed 
from 0.6 to 0.3. Double diffraction can be handled in the same spirit. Three terms
contribute, where either side $A$, side $B$ or both are enhanced by a factor like
eq.~\eqref{eq:diffenhance}. In eq.~\eqref{eq:sigDDinterpol} the power is changed 
from 1.5 to 1.25 for the first two and to 1.0 for the last one.

The kinematics of events is provided by the mass and $t$ selections outlined above. 
The decays of the explicit low-mass resonances are assumed to be isotropic. In the other
cases a diffractive system is handled as a string stretched between two parts of the
incoming hadron. A baryon is split into a diquark plus a quark at random, where the 
former/latter is moving in the forwards/backwards direction in the rest frame of the 
hadron. Here forwards is the direction the hadron will be moving out along, once 
boosted to the collision CM frame. A meson is correspondingly split into a quark plus
an antiquark, but here the choice of which is moving forwards is taken to be random.  
The two string ends are given relative $\pT$ kicks of nonperturbative size, however, 
such that the string alignment along the collision axis is smeared. 

\begin{figure}[t!]
\begin{minipage}[c]{0.49\linewidth}
\centering
\includegraphics[width=\linewidth]{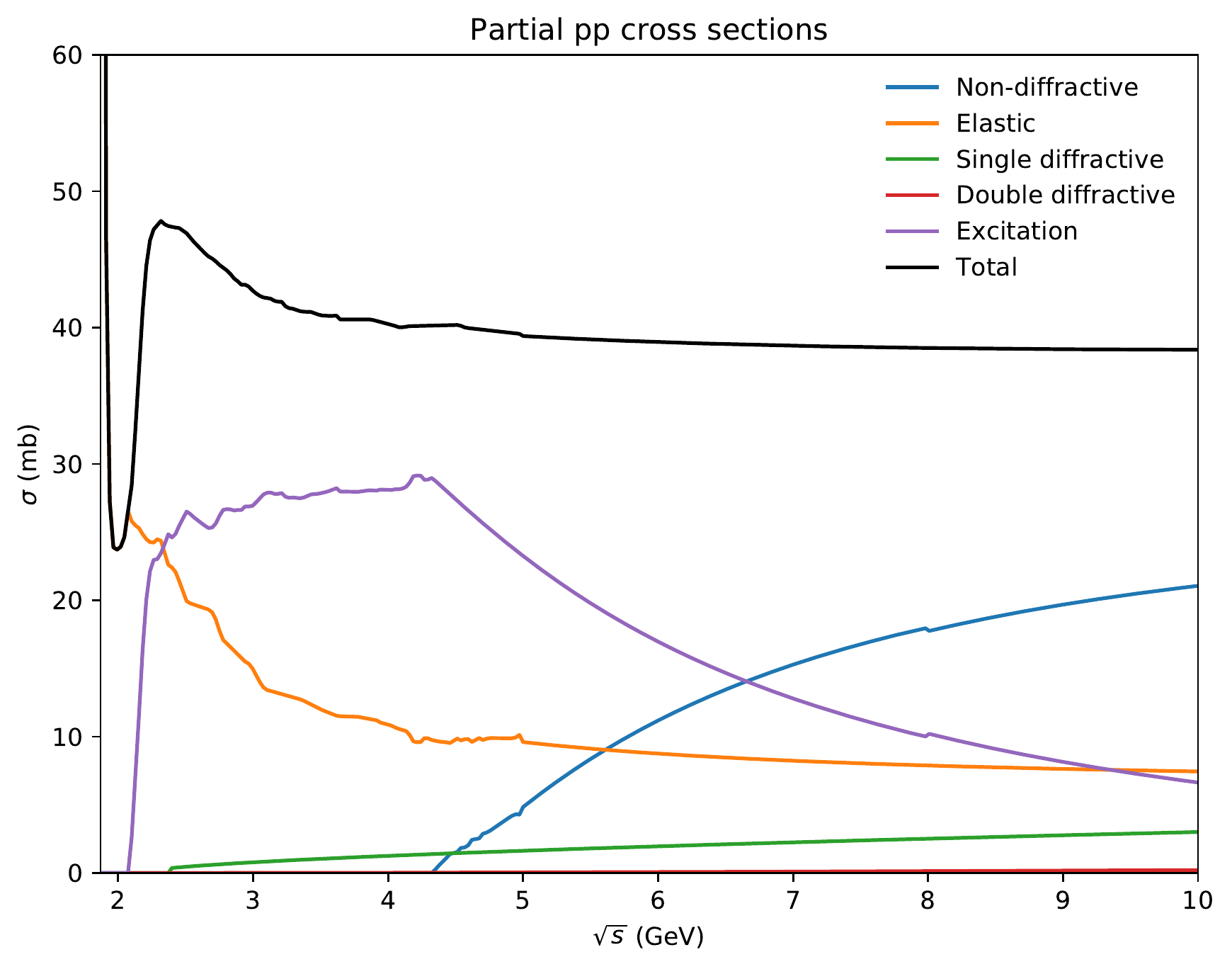}\\
(a)
\end{minipage}
\hfill
\begin{minipage}[c]{0.49\linewidth}
\centering
\includegraphics[width=\linewidth]{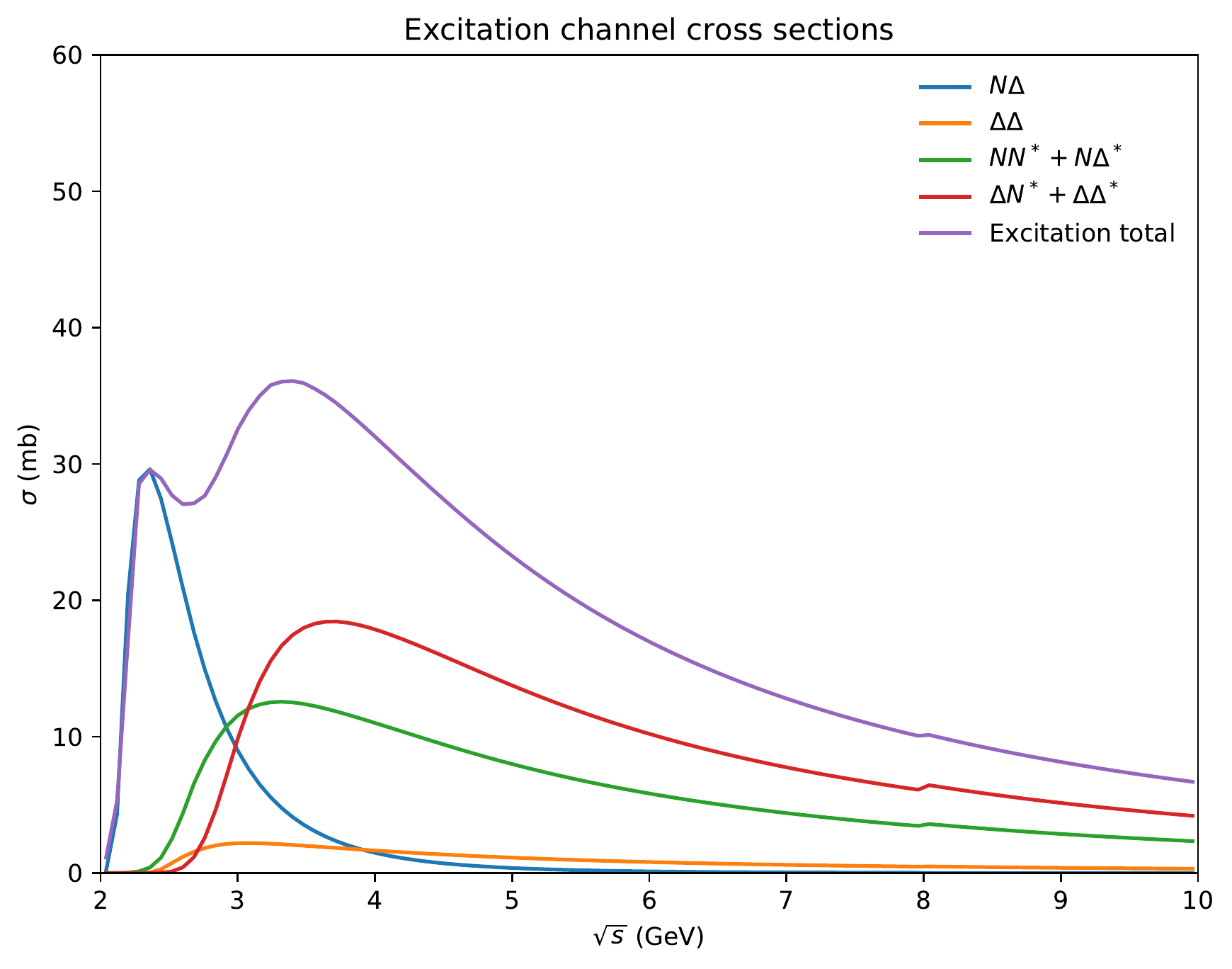}\\
(b)
\end{minipage}
\caption{
(a) All partial cross sections for $\p\p$ interactions. (b) Excitation cross sections, according to eq. \eqref{eq:sigmaExcitePartial}. Note the small jump at 8 GeV, at the boundary between the tabulated and parameterized regions.
}
\label{fig:ppPartial}
\end{figure}

\figref{fig:ppPartial}a shows all partial cross sections for $\p\p$ collisions. We see that the single diffractive cross section is very small compared to other cross sections, and the double diffractive one almost vanishes. The excitation cross section is here shown separately from the cross sections describing diffraction in the continuous region. Note that below around 4.5 GeV, the excitation cross section is set equal to the difference $\sigma_\total - \sigma_\elastic$ instead of following the form given by eq. \eqref{eq:sigmaExcitePartial}. The full shape of the excitation cross sections are shown in \figref{fig:ppPartial}b.

\subsection{Nondiffractive processes}
\label{sec:Nondiff}

Nondiffractive cross sections are found by subtracting all other partial cross sections 
from the total cross section,
\begin{equation}
    \sigma_{\nondiff} = \sigma_{\total} - \sigma_{\elastic} - \sigma_{\diff}
    - \sigma_{\resonance} - \sigma_{\annihilation}.
\end{equation}
At large energies the nondiffractive processes dominate the total cross section,
but at low energies they can have a small or even vanishing cross section. Since it is defined as the difference between the total and the other partial cross sections, it can sometimes have a fluctuating energy dependence with no clear physics explanation. 

A nondiffractive event is associated with the exchange of a gluon between 
the two incoming hadrons, where the gluon carries negligible momentum but leads 
to a rearranged
colour topology. To this end, each initial hadron is separated into a colour 
(a quark or an antidiquark) part and an anticolour (an antiquark or a diquark) part. 
For a baryon the selection of the diquark part is done according to the $SU(6)$ 
decomposition (in three flavours times two spins), while the meson subdivision is 
trivial. After the colour-octet gluon exchange, the colour end of one hadron forms a colour singlet with the anticolour end of the other hadron, and vice versa.
(Cases with more complicated colour-charge topologies are
suppressed and are neglected here.) This leads to two strings being stretched out 
between the two octet-state ``hadrons''.

Consider the collision in its rest frame,
with hadron $A$ ($B$) moving in the $+z$ ($-z$) direction. In that frame, the colour 
and anticolour objects of each hadron are assumed to have an opposite and compensating 
$\pT$. This is chosen according to a Gaussian with the same width as used to describe 
the $\pT$ smearing in string breakup vertices. In the breakup context a width of 
$\langle \pT^2 \rangle \approx (0.35~\mathrm{GeV})^2$ is motivated by a tunnelling 
mechanism, but a number of that magnitude for the parton motion inside a hadron
could equally well be viewed as a consequence of confinement in the transverse 
directions by way of the Heisenberg uncertainty relations.

Including (di)quark masses, the transverse masses $m_{\perp A 1}$ and $m_{\perp A 2}$ 
of the two $A$ hadron constituents are defined. Next a $z_A$ value is picked that 
splits the $A$ lightcone momentum $p_A^+ = E_A + p_{z A}$ between the two, 
$p_{A 1}^+ = z_A p_A^+$ and $p_{A 2}^+ = (1 - z_A) p_A^+$ \cite{Sjostrand:2004pf}.
For a meson $z = x_1 / (x_1 + x_2)$, where the $x_i$ are picked at random according to 
$(1 - x_i)^{0.8} / \sqrt{x_i}$. For a baryon first each of the three quarks are 
assigned an $x_i$ according to $(1 - x_i)^{2.75} / \sqrt{x_i}$. If $z_A$ is associated 
with the diquark, made out of the first two quarks, then 
$z_A = 2(x_1 + x_2) / (2(x_1 + x_2) + x_3)$. Note that here the diquark tend to take 
most of the momentum, not only because it consists of two quarks, but also by an 
empirical enhancement factor of 2. The $p_{A i}^-$ can now be obtained from 
$p^+ p^- = m_{\perp}^2$, and combined to give an effective mass $m_A^*$ that the 
$A$ beam remnant is associated with: 
$m_A^{* 2} = m_{\perp A 1}^2 / z + m_{\perp A 2}^2 / (1 - z)$.
The same procedure can be repeated for the $B$ hadron, but with 
$p^+ \leftrightarrow p^-$. Together, the criterion $m_A^* + m_B^* < E_{\mathrm{CM}}$
must be fulfilled, or the whole selection procedure has to be restarted. (Technically,
some impossible values can be rejected already at earlier stages.) Once an acceptable
pair $(m_A^*, m_B^*)$ has been found, it is straightforward first to construct the
kinematics of $A^*$ and $B^*$ in the collision rest frame, and thereafter the
kinematics of their two constituents.

Since the procedure has to work at very small energies, some additional aspects should be mentioned.
At energies very near the threshold, the phase space for particle production is limited.
If the lightest hadrons that can be formed out of each of the two new singlets 
together leave less than
a pion mass margin up to the collision CM energy, then a simple two-body production of 
those two lightest hadrons is (most likely) the only option and is thus performed.
There is then a risk to end up with an unintentional elastic-style scattering.
For excesses up to two pion masses, instead an isotropic three-body decay is 
attempted, where one of the strings breaks up by the production of an intermediate 
$\u\ubar$ or $\d\dbar$ pair. If that does not work, then two hadrons are picked as 
in the two-body case and a $\pi^0$ is added as third particle.

One reason why $m_A^* + m_B^* < E_{\mathrm{CM}}$ might fail is if
the constituent transverse masses are too big. Thus, after a number of failed attempts,
their values are gradually scaled down to increase the likelihood of success. 
This, on the other hand, increases the risk of obtaining two strings with low
invariant masses. A further check is therefore made that each string has a mass
above that of the lightest hadron with the given flavour content, and additionally 
that the mass excess is at least a pion mass for one of the two strings. 

The two strings can now be hadronized, but often one or both have small masses. 
To this end the ministring framework, used when at most two hadrons can be formed 
from a string, has been extended to try harder. Several different approaches are
used in succession, until one of them works. The order is as follows.\\
(1) Several attempts are made to produce two hadrons from the string by a traditional 
string break in the middle.\\ 
(2) If not, a hadron is formed consistent with the endpoint flavour content.
Four-momentum is shuffled between it and one of the partons of the other string, so as
to put the hadron on mass shell while conserving the overall four-momentum. Since the
string with lowest mass excess is considered first, the two partons of the other string
should normally be available.\\ 
(3) If no allowed shuffling is found, then a renewed attempt is made to 
produce two hadrons by a string break, but this time the two lightest hadrons of the 
given flavour content are chosen.\\ 
(4) If that does not work, one lightest hadron is formed from the endpoint flavours 
and the other is set to be a $\pi^0$.\\ 
(5) It still no success, then go back to forming one hadron, but the lightest 
possible, and again shuffle momentum to a parton.\\ 
(6) Finally, the problem may occur also for the string with higher mass excess, 
i.e.\ after the first string was hadronized, and possibly took some four-momentum in
the process. Then a collapse to one hadron (at random or eventually the lightest) 
with the recoil taken by another hadron is attempted.

\subsection{The transition to high-energy processes}
\label{sec:transitiontoHE}

We have now described a framework for low energy hadron-hadron interactions.
Our motivation for doing this has been to apply it to rescattering, but in principle, having this framework means that it is now possible to generate events in \textsc{Pythia} at these low energies.
Despite all the technical details, the structure of the resulting events is quite simple. At most two objects (either hadrons or strings) are created in the first step of the process. The strings are stretched out almost perfectly along the collision axis and fragment into hadrons with only small nonperturbative $\pT$ kicks. 

This is in contrast to the high-energy framework used to simulate the primary LHC 
$\p\p$ collision, e.g.\ in inelastic nondiffractive processes. Here the multiparton 
interactions machinery very much is based on perturbation theory, where each 
interaction requires the use both of hard matrix elements and parton distribution 
functions (PDFs), giving scattered partons over a wide range of $\pT$ scales, even if 
the lower scales dominate. Many string pieces are stretched criss-cross in the event,
and fragment into the high-multiplicity initial state that the rescattering framework
will be applied to. If one uses this perturbative framework at lower and lower
energies the average number of MPIs will decrease, as will their typical $\pT$ scale. Gradually the idea of applying a perturbative approach becomes less 
appealing. Technically the machinery can be applied down to 10~GeV CM energy, 
but is then highly questionable. 
Furthermore, many of the cross sections described here do not scale correctly at higher energies.
For a high-energy 
$\p\p / \p\pbar$ primary collision four different models are available
\cite{Rasmussen:2018dgo}. Only one of them explicitly covers some more collision
types, but extensions by AQM rescaling could be possible. 

\begin{figure}[tbp]
\begin{minipage}[c]{0.49\linewidth}
\centering
\includegraphics[width=\linewidth]{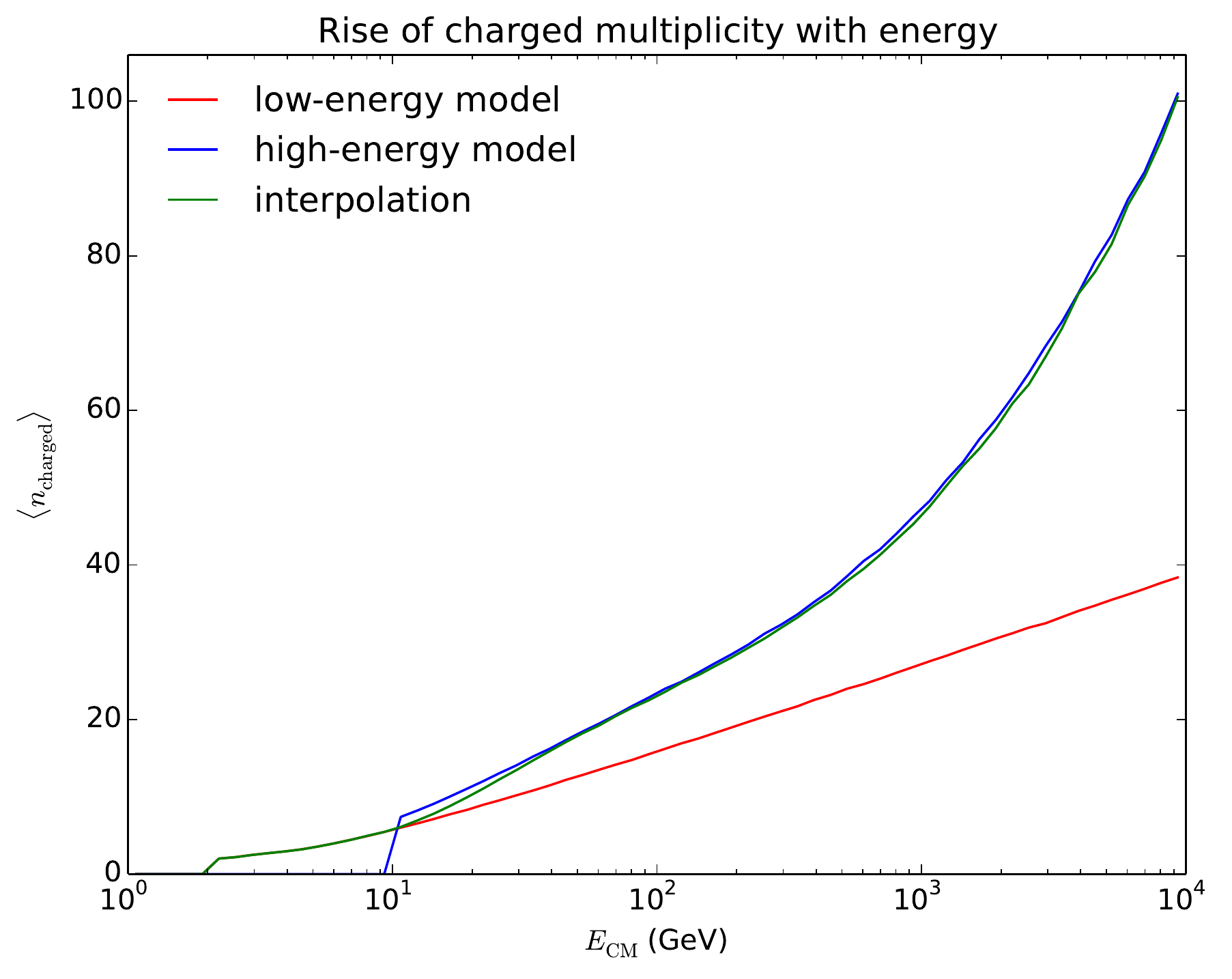}\\
(a)
\end{minipage}
\hfill
\begin{minipage}[c]{0.49\linewidth}
\centering
\includegraphics[width=\linewidth]{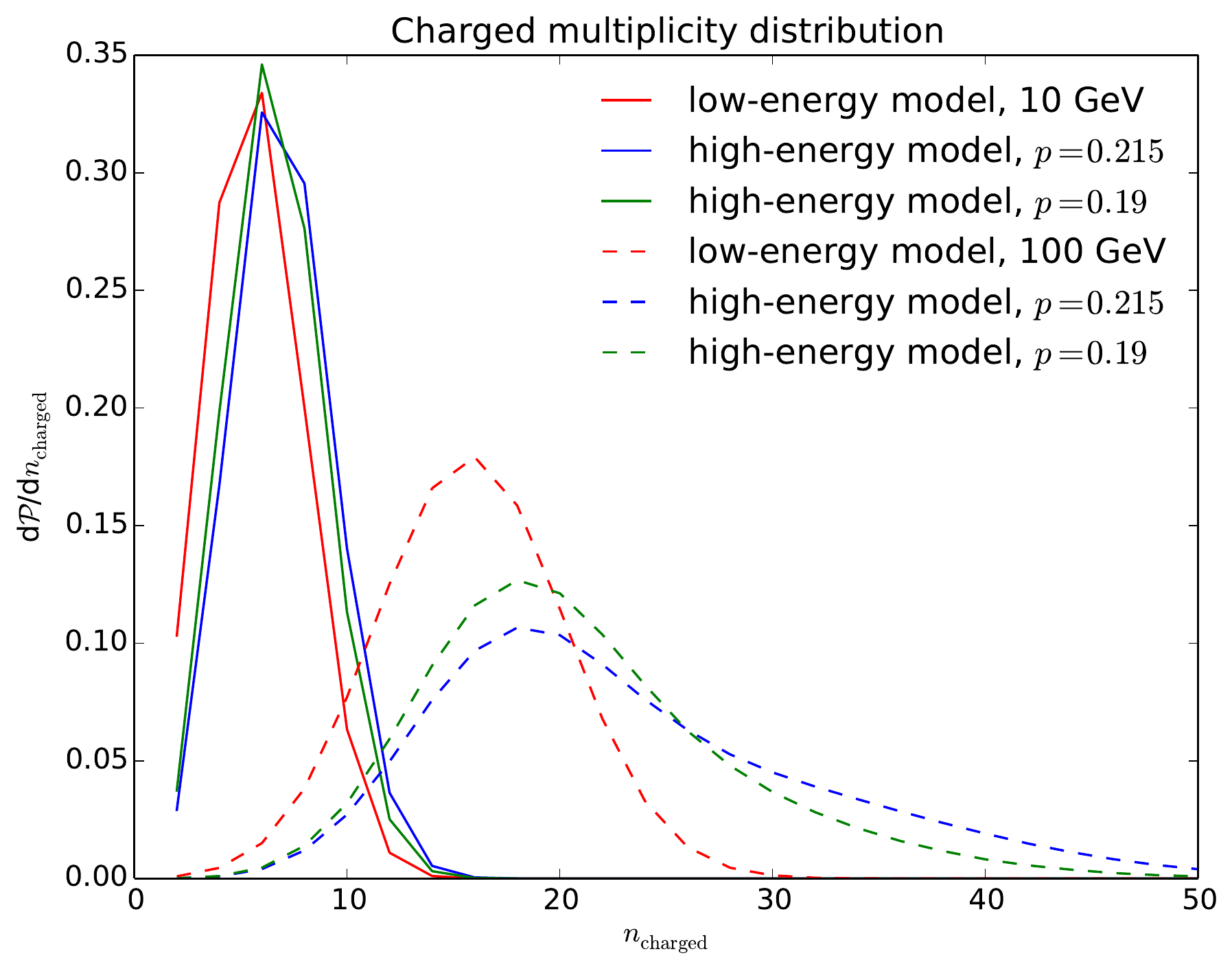}\\
(b)
\end{minipage}\\
\begin{minipage}[c]{0.49\linewidth}
\centering
\includegraphics[width=\linewidth]{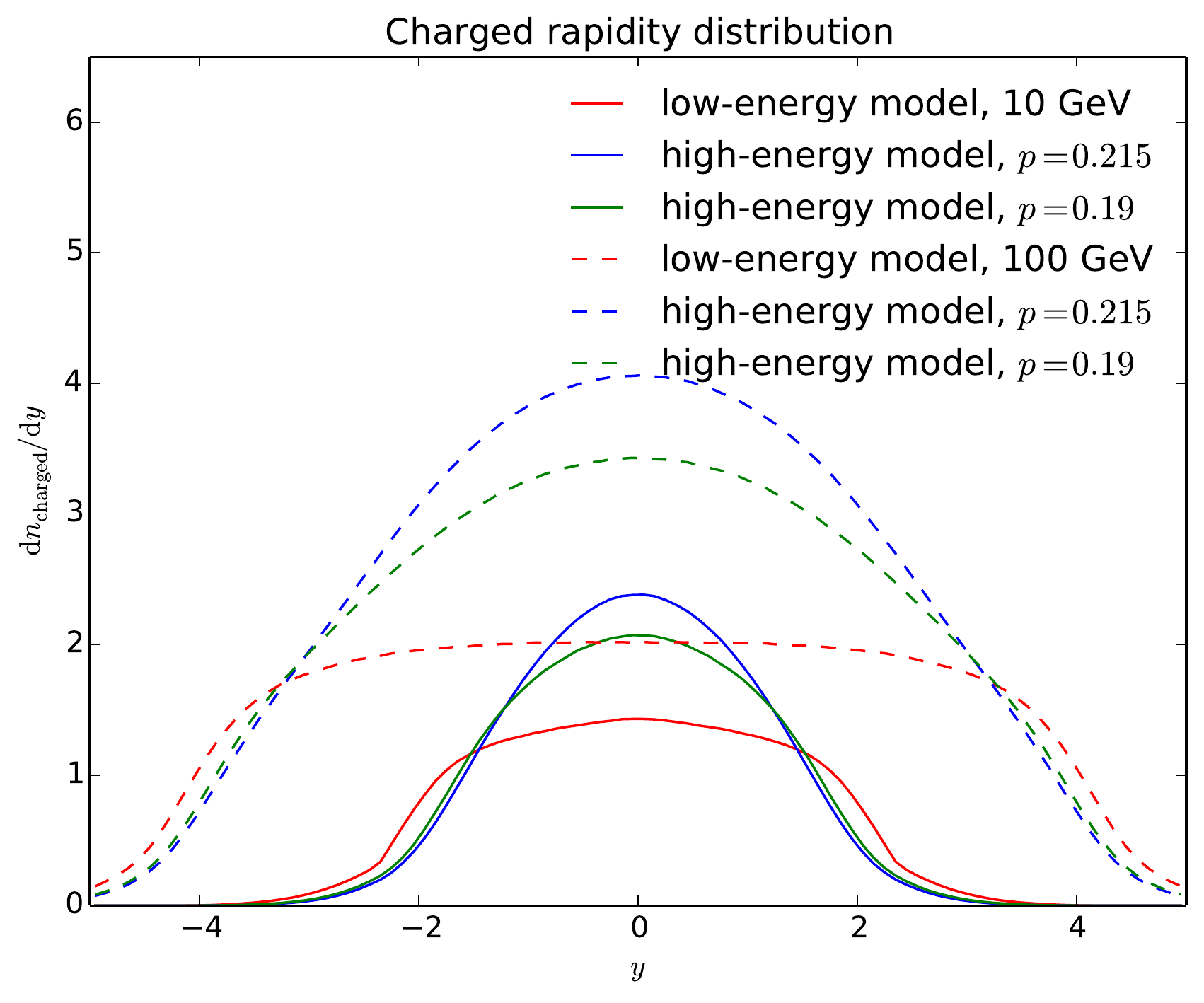}\\
(c)
\end{minipage}
\begin{minipage}[c]{0.49\linewidth}
\centering
\includegraphics[width=\linewidth]{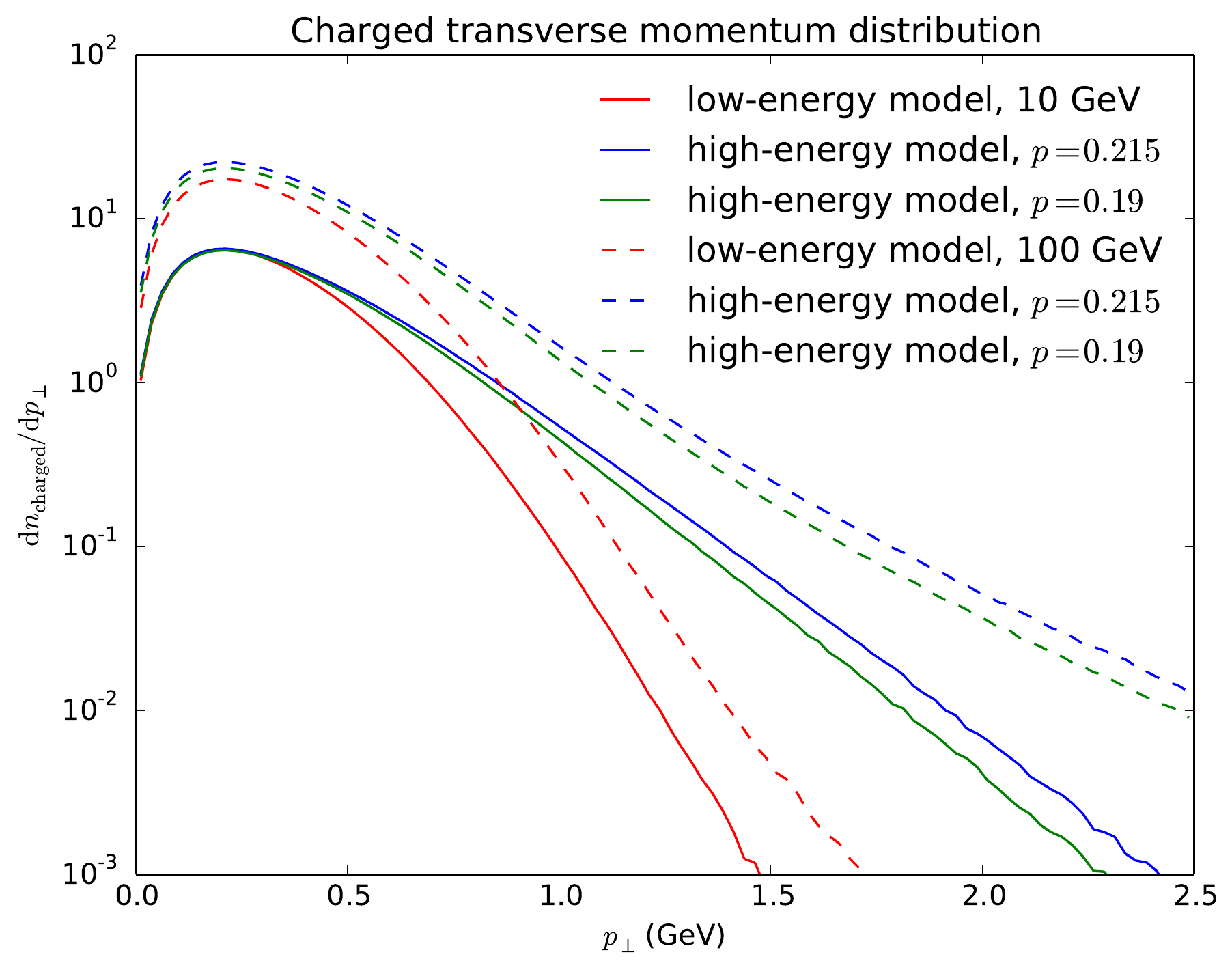}\\
(d)
\end{minipage}
\caption{
(a) Energy dependence of the average charged multiplicity 
in nondiffractive $\p\p$ collisions. (b,c,d) Comparison of 
charged multiplicity, rapidity and transverse momentum distributions 
for 10 and 100~GeV nondiffractive $\p\p$ collisions.
}
\label{fig:LEHE}
\end{figure}

Therefore it is tempting to interpolate between the two descriptions. There is 
now such an option available. In it, the fraction of perturbatively handled events
rises from the threshold energy $E_{\mathrm{thr}} = 10$~GeV as 
\begin{equation}
P_{\mathrm{pert}} = 1 - \exp \left( - 
\frac{E_{\mathrm{CM}} - E_{\mathrm{thr}}}{E_{\mathrm{wid}}} \right) ~,   
\end{equation}
where $E_{\mathrm{wid}} = 10$~GeV is a measure of the size of the transition 
region. This is actually the same form as already used previously to transition
between a nonperturbative and a perturbative description of diffraction, with
the diffractive system mass replaced by $E_{\mathrm{CM}}$ 
\cite{Navin:2010kk,Rasmussen:2018dgo}. 

How this transition works in practice is illustrated in 
\figref{fig:LEHE}a, for the energy dependence of the charged 
multiplicity in nondiffractive events. In this figure the difference 
between the low-energy and high-energy model multiplicities is not so large in
the transition range 10 -- 30~GeV, but the importance of the perturbative 
components obviously increases with energy. Zooming in on the behaviour at 
the 10~GeV threshold and at an energy above it, at 100~GeV, 
\figref{fig:LEHE}b,c,d show some differential distributions. At 10~GeV 
the limited phase space does not allow for high multiplicities, while a longer 
perturbatively-induced tail is apparent at 100~GeV. Nevertheless, the MPI 
activity is reflected in a shift towards central rapidities and the presence 
of a high-$\pT$ tail already at 10~GeV.

The perturbative model results have been obtained with the default Monash tune
\cite{Skands:2014pea}, which mainly is based on comparisons with LEP, Tevatron
and LHC data. One should therefore be aware that the extrapolation to lower 
energies is not without its problems. As an example, the key parameter of the
MPI framework is the $\pTo$ one, that regularizes the divergence of the 
perturbative $2 \to 2$ cross sections in the limit $\pT \to 0$. It is assumed
to have an energy dependence that scales like $\pTo \propto E_{\mathrm{CM}}^p$
(but more complicated forms could be considered).
The default values, with $p = 0.215$, gives $\pTo = 0.56$ and 0.91~GeV, 
respectively, at 10 and 100~GeV. If $p$ is changed to 0.19, then 
instead $\pTo = 0.66$ and 1.02~GeV, respectively, at the low energies, 
assuming a fixed $\pTo$ value at 7~TeV. The result of such a modest change 
is illustrated in Fig.~\ref{fig:LEHE}b,c,d. Qualitatively the 
difference to the low-energy model remains, but quantitatively it is 
visibly reduced.

One may also note that the string drawing can be quite different in the two cases.
In the nonperturbative model the $\p\p$ events always are represented by two strings,
each stretched between a quark and a diquark. When MPIs are included, it becomes
frequent that two quarks are kicked out of the same proton, more so
at low energies where the high-$x$ valence-quark part of PDFs is probed.
This leads to so-called junction topologies, where the baryon number can wander 
more freely in the event \cite{Sjostrand:2002ip}. Technically, this makes 
the hadronization of low-energy events more messy, and may require repeated 
attempts to succeed.

In diffraction, the excited masses $M_X$ vary between events, also for a fixed CM
energy. To handle perturbative activity inside the diffractive system then would 
seem to require a time-consuming re-initialization of the MPI framework for 
each new diffractive system. Instead, at the beginning of a run, an 
initialization is done for a set of logarithmically spaced diffractive 
masses, and numbers relevant for the future generation are saved in arrays. 
By interpolation, required numbers can then be found for any mass during the 
subsequent event generation. This approach has now been extended also to be
available for nondiffractive processes, if so desired. This means that $\p\p$ 
collisions can be simulated essentially from the threshold to LHC energies and 
beyond without any need to re-initialize. The prize to pay is a somewhat longer 
initialization step at the beginning of a run, but still in the range of seconds 
rather than minutes. One current limitation is that it is numbers for the MPI
generation that are stored, so it is not now possible to pick a specific hard 
process for handling in the same way. 

Another limitation is that the perturbative framework requires access to PDFs for the
colliding hadrons, which restricts us to $\p$, $\n$ and (with big uncertainties)
$\pi$. Additionally PDFs are available for the photon and the pomeron, the latter
used in diffraction, and in that sense they can be handled on equal footing with 
hadrons. A further restriction is that \textsc{Pythia} can only be set up for one 
combination of incoming beams at a time, so as to handle the perturbative processes. 
The simpler nonperturbative machinery used for rescatterings has no such restriction, 
of course.

\section{Model tests}

\label{sec:modelTests}

In this section, we will study the properties of the rescattering model.
We start with studying how rescattering affects simple observables such as $\pT$ spectra, charged multiplicity, jet structure, and the potential for collective flow. We also look at how event properties change when rescattering is performed in a Lorentz boosted frame, in order to verify that the frame-dependence described in \secref{subsec:rescatterxyzt} does not significantly alter the final state. 

Next, we look at the rates at which different particle types participate in rescattering and the rates at which the different types of processes occur. Finally, we consider the free parameters and model choices that have gone into the framework, and study the effect of changing those.

\subsection{Basic effects of rescattering}

\begin{figure}[t!]
\begin{minipage}[c]{0.49\linewidth}
\centering
\includegraphics[width=\linewidth]{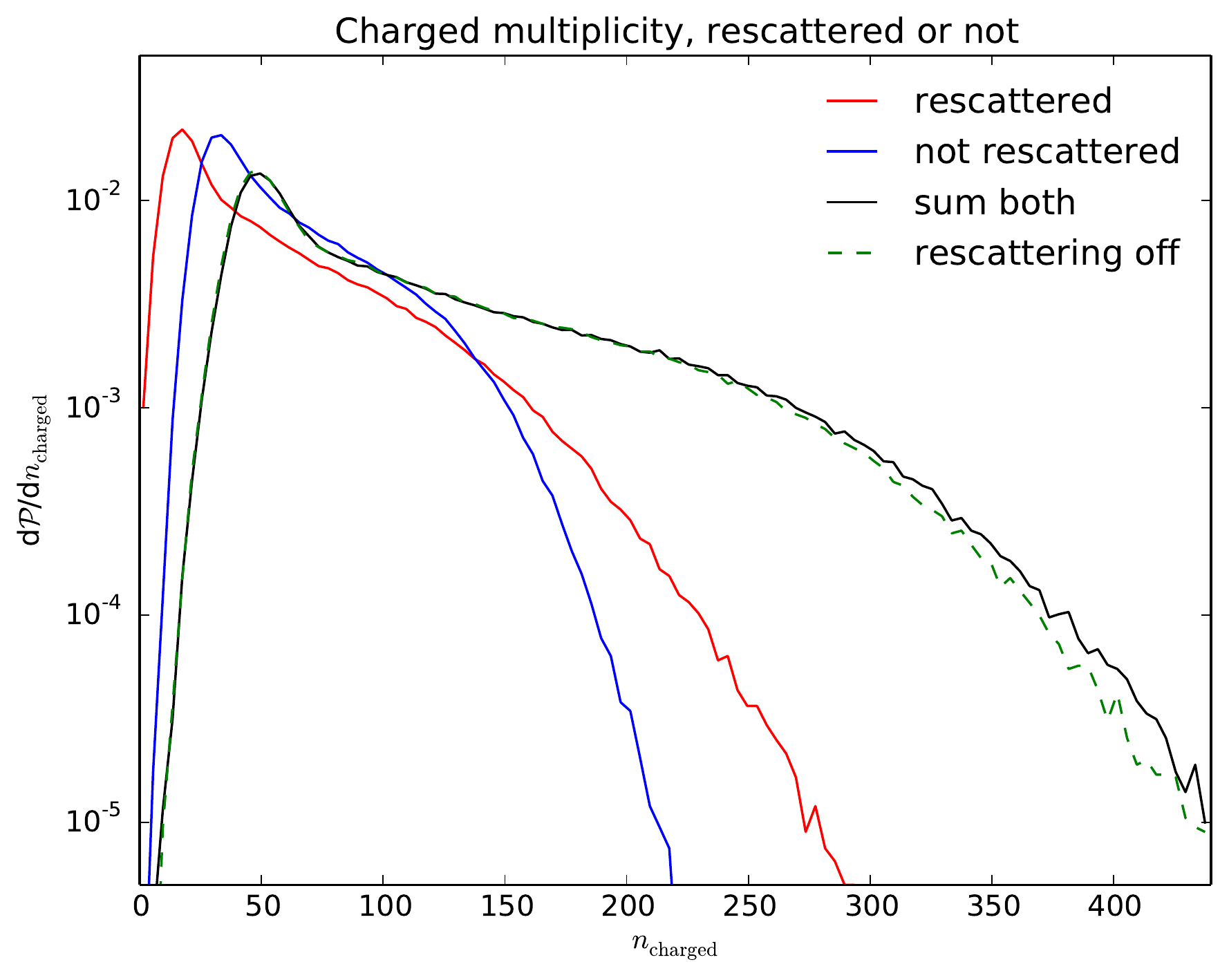}\\
(a)
\end{minipage}
\begin{minipage}[c]{0.49\linewidth}
\centering
\includegraphics[width=\linewidth]{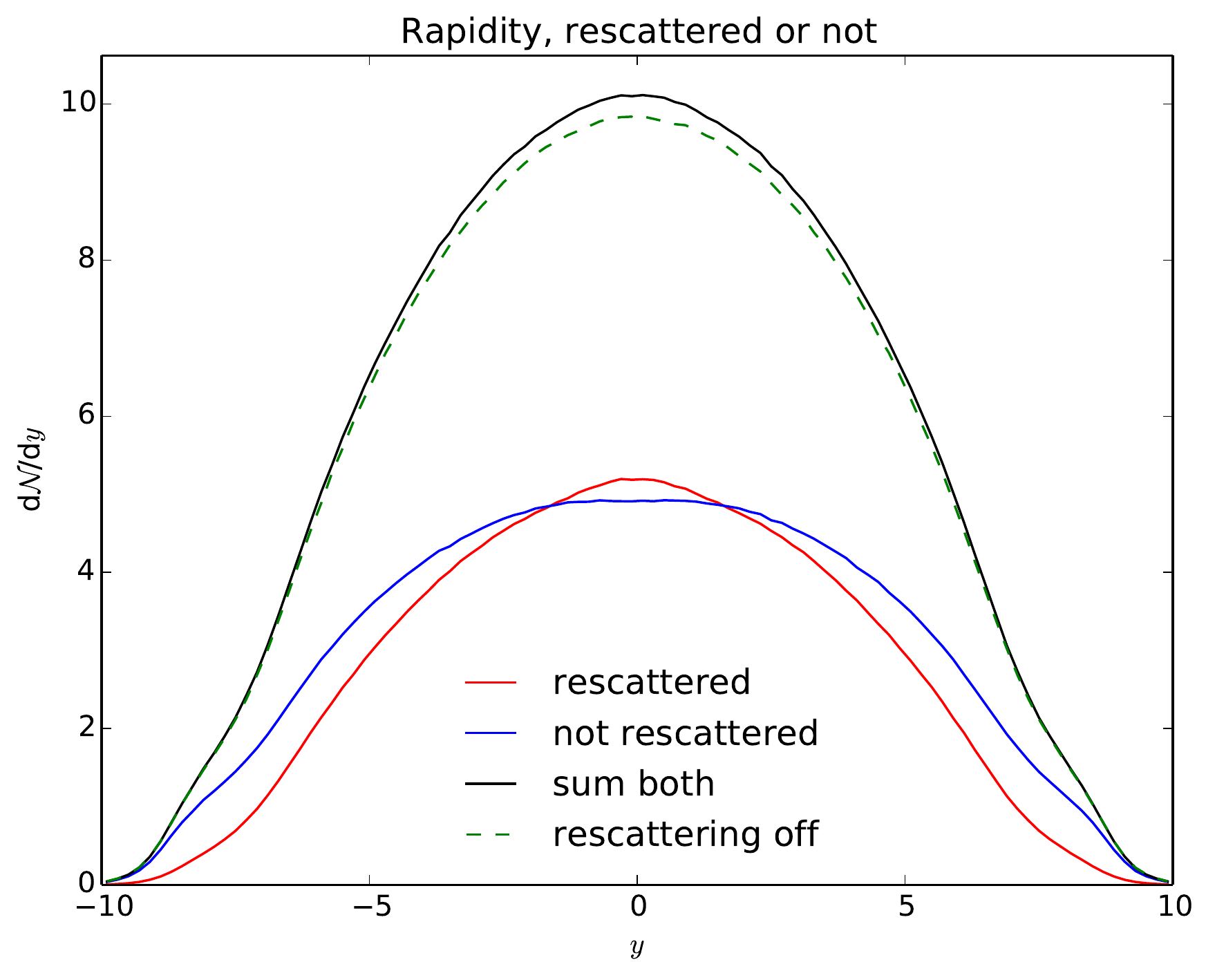}\\
(b)
\end{minipage}\\
\begin{minipage}[c]{0.49\linewidth}
\centering
\includegraphics[width=\linewidth]{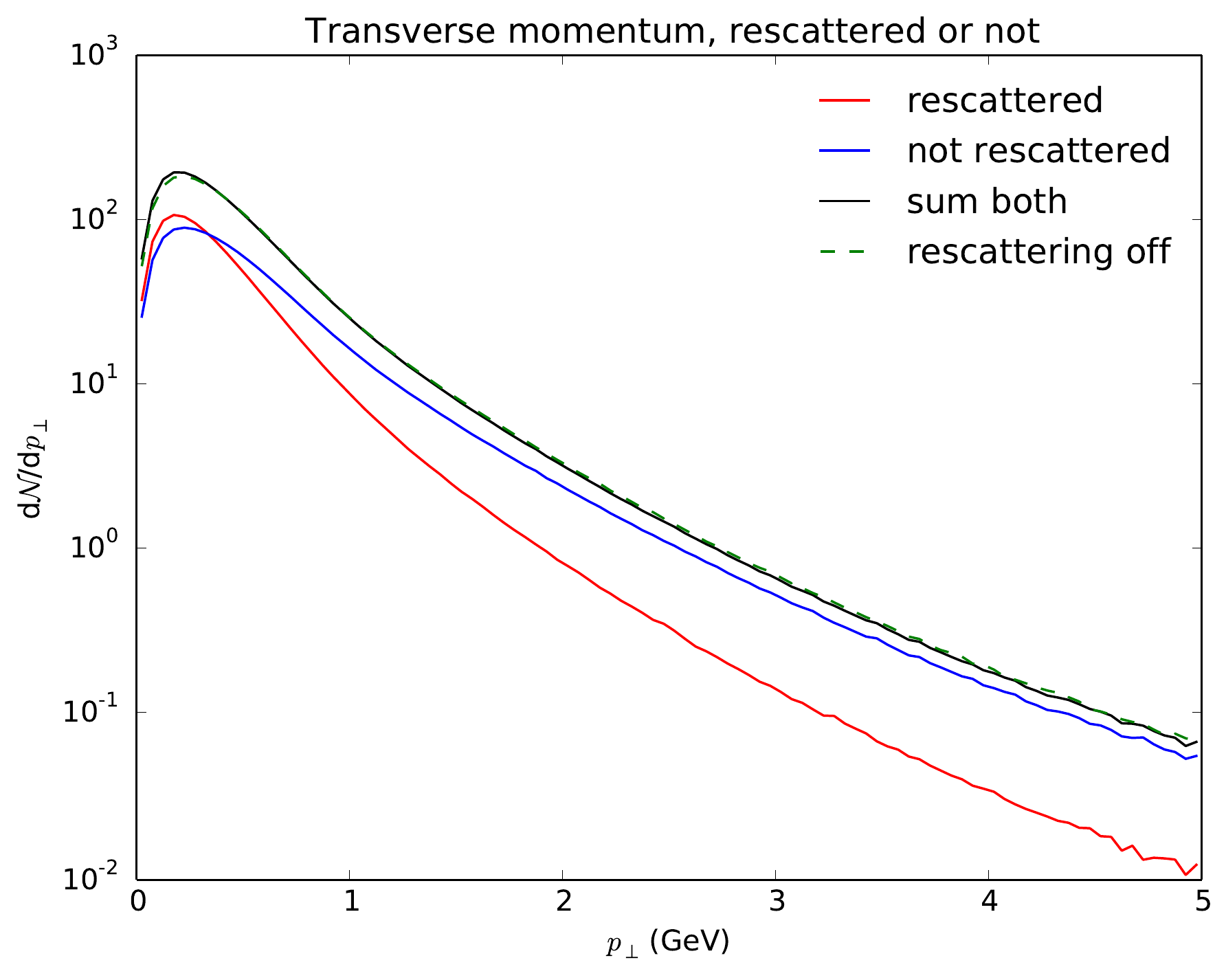}\\
(c)
\end{minipage}
\begin{minipage}[c]{0.49\linewidth}
\centering
\includegraphics[width=\linewidth]{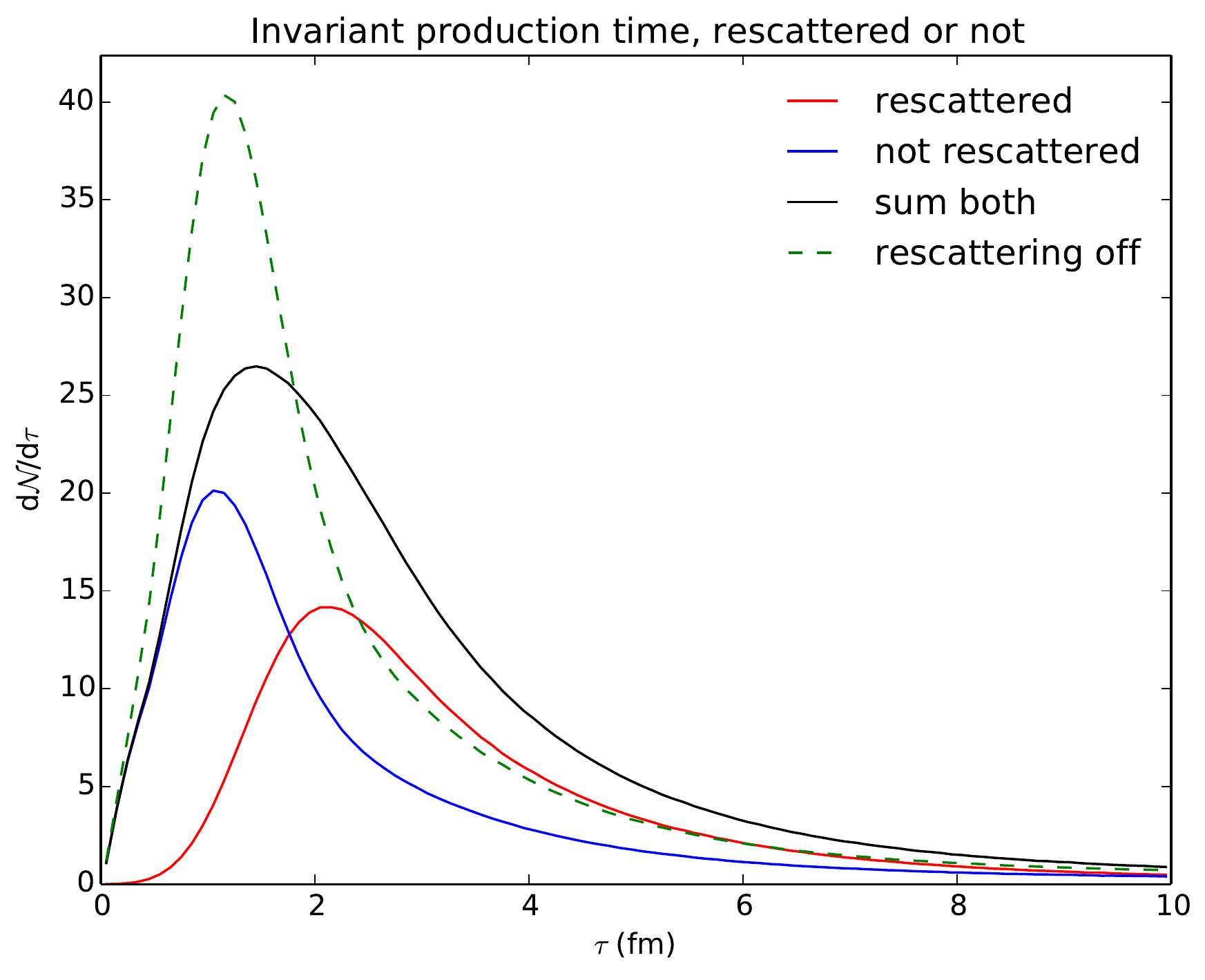}\\
(d)
\end{minipage}
\caption{(a) Multiplicity, (b) rapidity, (c) transverse momentum,
and (d) invariant production time spectra of charged
final-state hadrons, subdivided into those that have been involved in
rescatterings and those that have not, in 13~TeV nondiffractive $\p\p$
events. As reference a comparison is also made with events without
rescatterings.}
\label{fig:rescattersplit}
\end{figure}

As the most basic check, \figref{fig:rescattersplit} shows how charged multiplicity, rapidity spectra, transverse momentum spectra, and invariant production times are affected by rescattering. We see that 
rescattering increases charged multiplicity, which is obviously expected when one considers the fact that we have implemented $2 \to n$, $n \geq 3$ interactions, but not interactions involving multiple incoming particles. The rescatter-affected 
hadrons have a broader multiplicity distribution than those not 
involved: events that start out with a low number of primary 
hadrons have a smaller rescattering probability than average, and vice versa.

In the same vein, the rescattered fraction is larger for central 
rapidities, where there are more hadrons to begin with, and this is also 
where inelastic rescatterings give a multiplicity increase. An interesting observation 
is that higher-$\pT$ hadrons seldom participate in rescattering, 
\figref{fig:rescattersplit}c. The natural explanation is that these hadrons 
typically are produced at larger transverse distances by (mini)jet fragmentation,
where the particle density is reduced by having fewer overlapping MPI
systems than at small $r_{\perp}$. Notable is also the slight net 
decrease at high $\pT$ by rescattering, (over)compensated by the 
increase at small $\pT$. Finally, and quite logically, rescattering 
kicks in with some delay in invariant time, since a sufficient amount 
of primary hadrons have to be produced first. 

The point of introducing rescattering is to change some event properties, 
but not all changes are relevant rescattering signals, since some
could easily be compensated by a retuning of many other parameters. 
In particular, the average (charged) event multiplicity is such a signal. Indeed, the fact that it is changed by rescattering means that a retune is necessary in order to restore it to the experimentally well-known value. 
The MPI framework, which is the main 
driving force in generating the multiplicity spectrum, is 
sufficiently uncertain to easily absorb the rescattering effects
on the multiplicity. More specifically, when we study the effects of 
rescattering, the $p_{\perp 0}$ parameter of the MPI framework, 
\texttt{MultipartonInteractions:pT0Ref}, is adjusted to restore
the average charged multiplicity in the $\eta < 2.5$ range to the 
no-rescattering value. Its default value in \textsc{Pythia} is \texttt{pT0Ref~=~2.28}~(GeV), and we have found that setting it to \texttt{pT0Ref~=~2.345} restores charged multiplicity to the correct value. We will use this value in all subsequent studies, unless otherwise noted. In the future, a more detailed retune would be desirable.

\subsection{Jets}

\begin{figure}[t!]
\begin{minipage}[c]{0.49\linewidth}
\centering
\includegraphics[width=\linewidth]{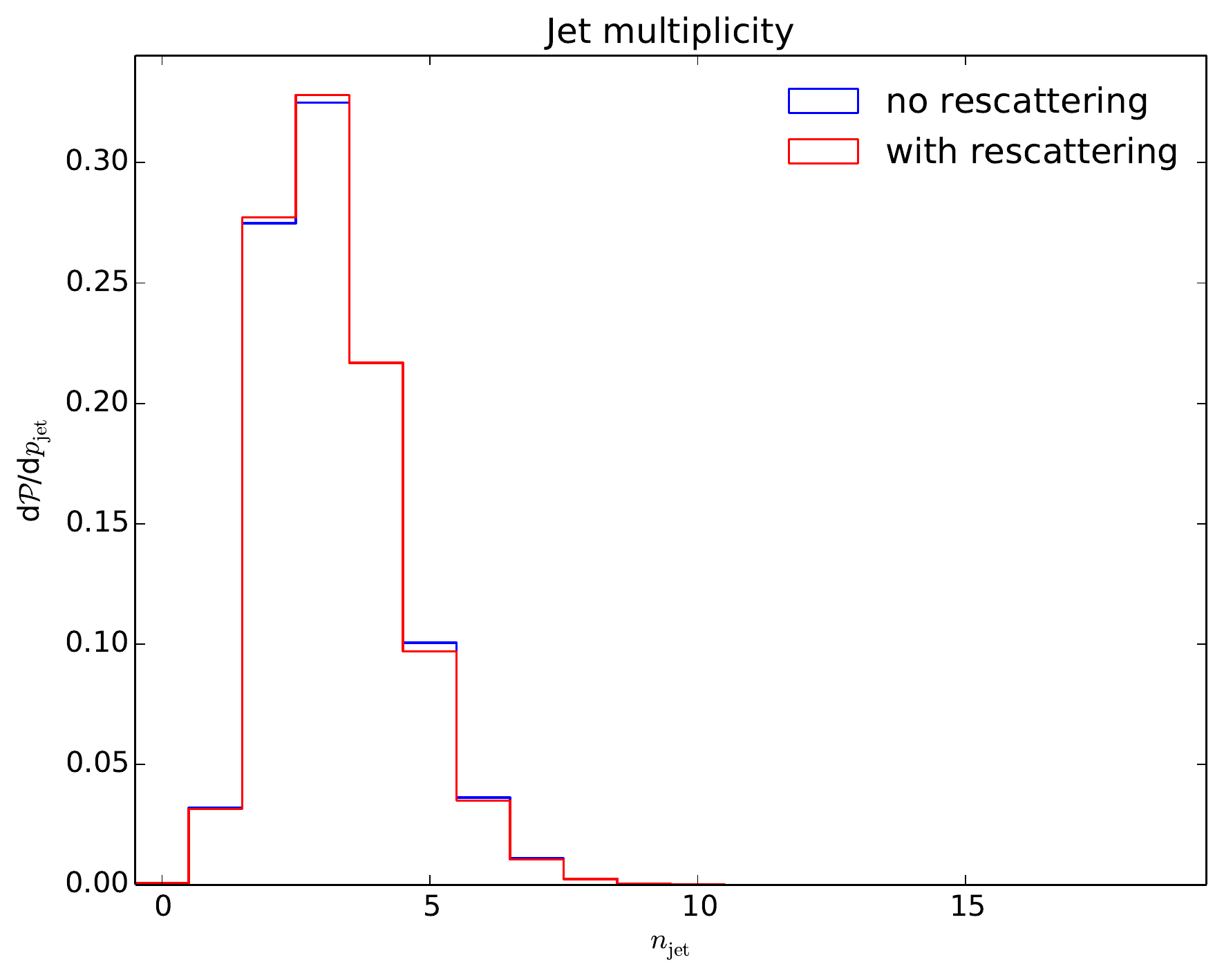}\\
(a)
\end{minipage}
\begin{minipage}[c]{0.49\linewidth}
\centering
\includegraphics[width=\linewidth]{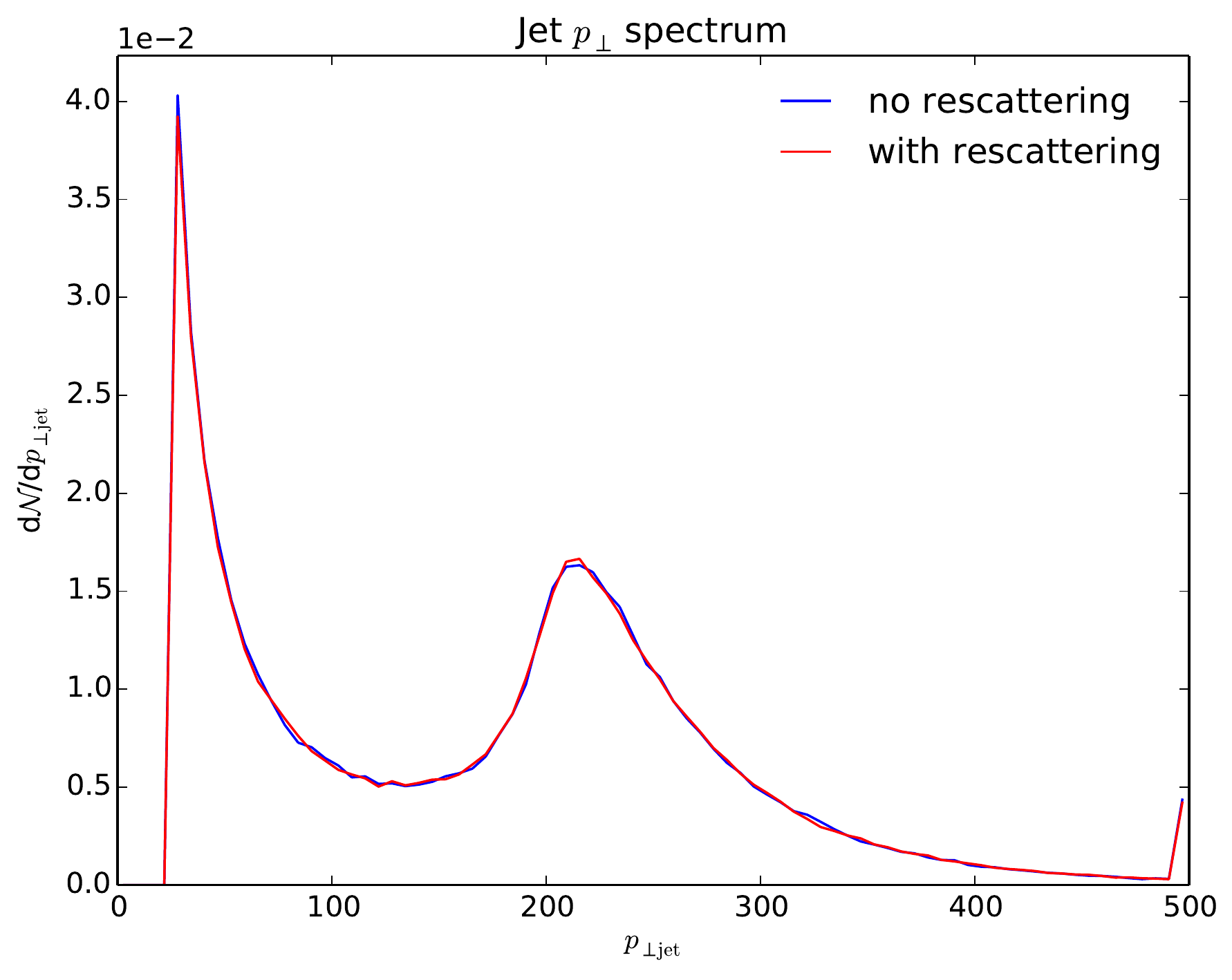}\\
(b)
\end{minipage}
\caption{
Production rates (a) and inclusive $\pT$ spectra (b) of jets in 13~TeV
$\p\p$ collisions, as further described in the text. The uptick in
the last bin of (b) is because all jets with $\pT > 500$~GeV have been
put there.}
\label{fig:jetrate}
\end{figure}

We have already argued that high-$\pT$ particles are less affected
by rescattering than low-$\pT$ ones, and hence jets should remain
essentially unchanged. This also turns out to be the case. As an example,
QCD two-jet production with $\pT > 200$~GeV hard collisions at 13~TeV was
studied, and anti-$k_{\perp}$ jets found for a 0.7 radius and a 25~GeV
lower cut-off \cite{Cacciari:2008gp}. We then find that the particle multiplicity inside a jet with rescattering on is about 2\% higher than with rescattering off.
This increase is almost uniformly spread from
the center to the periphery of the jet. The $\pT$-weighted jet profile
is almost identical, however. Studying the jet rate itself, there is a
small net reduction in the number of jets when rescattering is allowed, 
\figref{fig:jetrate}a.
The difference is too small to be visible in the jet $\pT$ spectrum,
\figref{fig:jetrate}b. A closer inspection shows that the jet rate
above 150~GeV, i.e.\ in the domain of the two hard jets, is unchanged
within statistics. Below that scale, however, i.e.\ mainly additional
jets from parton showering, there is a drop by about 2\% in the rate. 
This is most likely related to a slight leakage of hadrons out of the
jet cone, shifting jet energies ever so slightly downwards. Such tiny differences could easily be tuned away, so in the end we conclude that
jet properties are not measurably affected.

\subsection{Collective flow}

One of the telltale signs of collective behaviour is an anisotropy in the azimuthal angle of outgoing particle momenta. Here we perform a preliminary study to see whether rescattering can produce azimuthal flow at all.

\begin{figure}[t!]
\begin{minipage}[c]{0.49\linewidth}
\centering
\includegraphics[width=\linewidth]{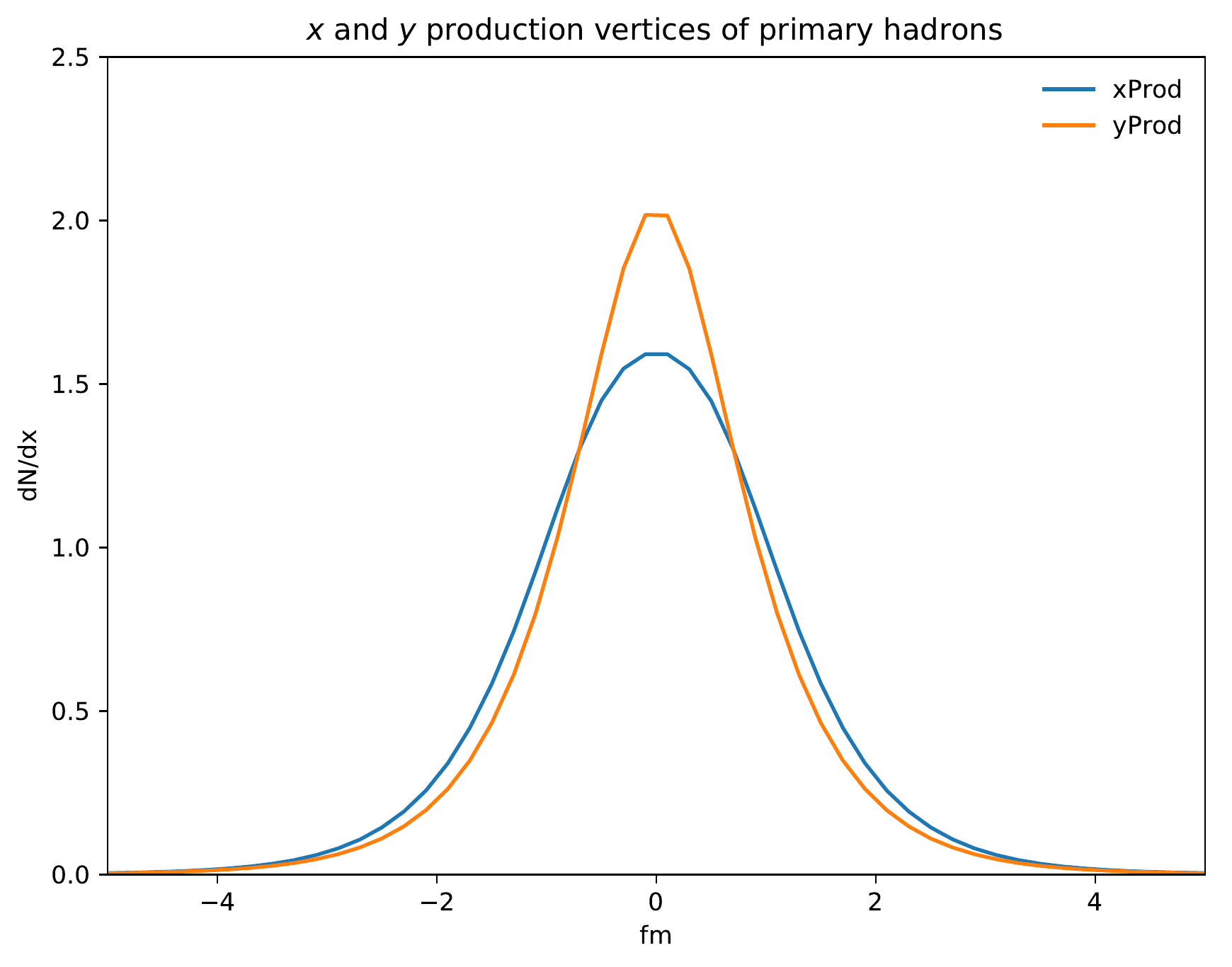}\\
(a)
\end{minipage}
\hfill
\begin{minipage}[c]{0.49\linewidth}
\centering
\includegraphics[width=\linewidth]{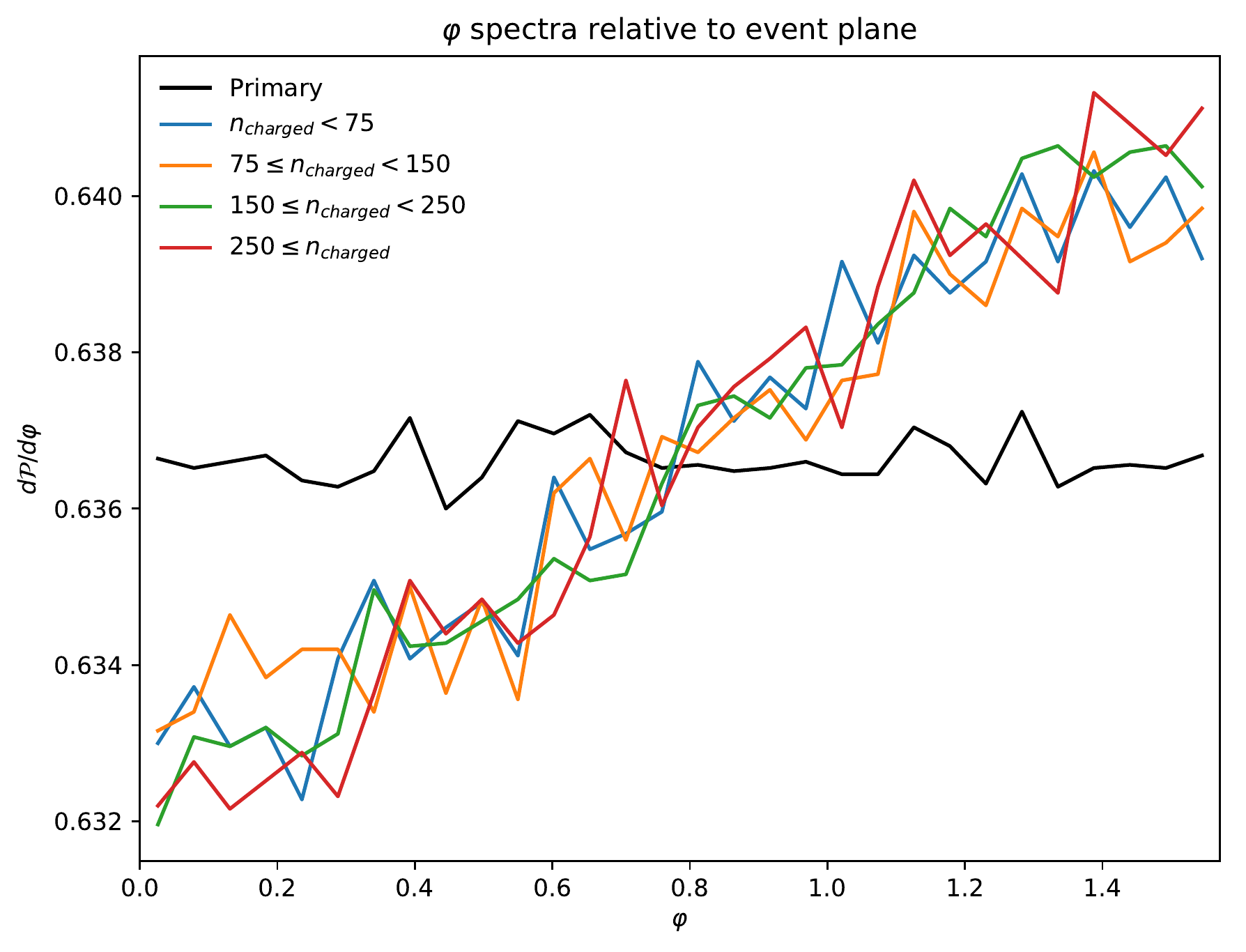}\\
(b)
\end{minipage}
\caption{
(a) $x$ and $y$ coordinates of primary hadrons, showing an initial anisotropy. (b) Azimuthal direction of momentum for outgoing hadrons, binned according to charged multiplicity. The angle is the acute angle to the event plane, $\varphi \in [0, \pi/2]$. The plot includes the spectrum for the primary hadrons, which illustrates that there is no flow before a collective behaviour has been induced by rescattering.
}
\label{fig:flow}
\end{figure}

In order to obtain a systematic flow, two things are required: an initial spatial 
anisotropy and a mechanism for collective behaviour. In this toy study an anisotropy 
is obtained by selecting the primary $\p\p$ collisions to have their impact parameter 
aligned along the $x$ axis, and choosing MPI vertices according to a Gaussian 
distribution multiplied by a $\varphi$ modulation factor with $\epsilon = 0.5$ 
(see \secref{subsec:MPIvertices}). The resulting $x$-$y$ anisotropy of primary hadron
production is illustrated in \figref{fig:flow}a. This causes an elliptic flow, as 
shown in \figref{fig:flow}b, where the $\varphi$ angle of final particle momenta is 
relative to the $x$ axis (which we know to be our event plane). By the symmetry of the initial anisotropy, the shape of the spectrum should depend only on the acute angle to the event plane, $0 < \varphi < \pi/2$, and we reduce the spectrum to this range to obtain better statistics.

The flow is aligned in the $y$-direction, 
consistent with the higher density gradient in this direction. Results are 
binned according to the charged multiplicity, which is correlated with the impact 
parameter. A low multiplicity is associated with peripheral events, for which the spatial anisotropy 
may be strong, but collective behaviour is suppressed by the low density. A high 
multiplicity, on the other hand, indicates a central event with much rescattering, 
but a low impact parameter so a less strict azimuthal alignment. In our simple study 
these two effects largely cancel to give comparable asymmetries independently of the 
multiplicity.

Unfortunately, the aforementioned study has been made under the unrealistic advantage
of a known event plane. In practice one would 
rather study e.g.\ two-particle azimuthal correlations. Furthermore, the initial anisotropy has been made implausibly large for illustratory purposes. 
When the simulation is repeated with more reasonable assumptions, we no longer observe any signs of flow. Therefore this brief 
study should be regarded as a proof of concept, and we hope to return to flow studies 
in the context of heavy-ion collisions, where a strong spatial anisotropy occurs 
naturally.

\subsection{Lorentz frame dependence}
\label{subsec:lorentz}

\begin{figure}[t!]

\begin{minipage}[c]{0.49\linewidth}
\centering
\includegraphics[width=\linewidth]{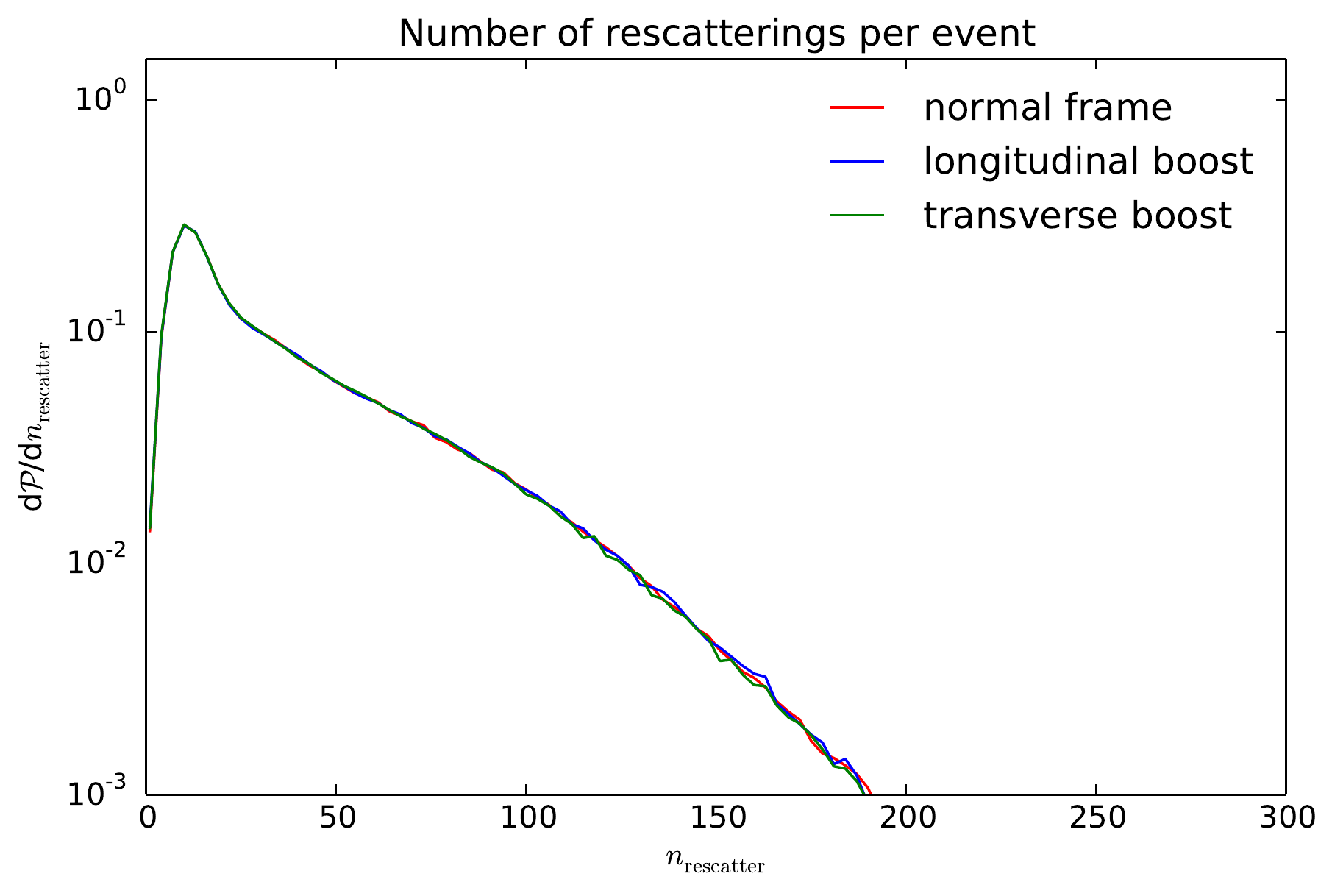}\\
(a)
\end{minipage}
\begin{minipage}[c]{0.49\linewidth}
\centering
\includegraphics[width=\linewidth]{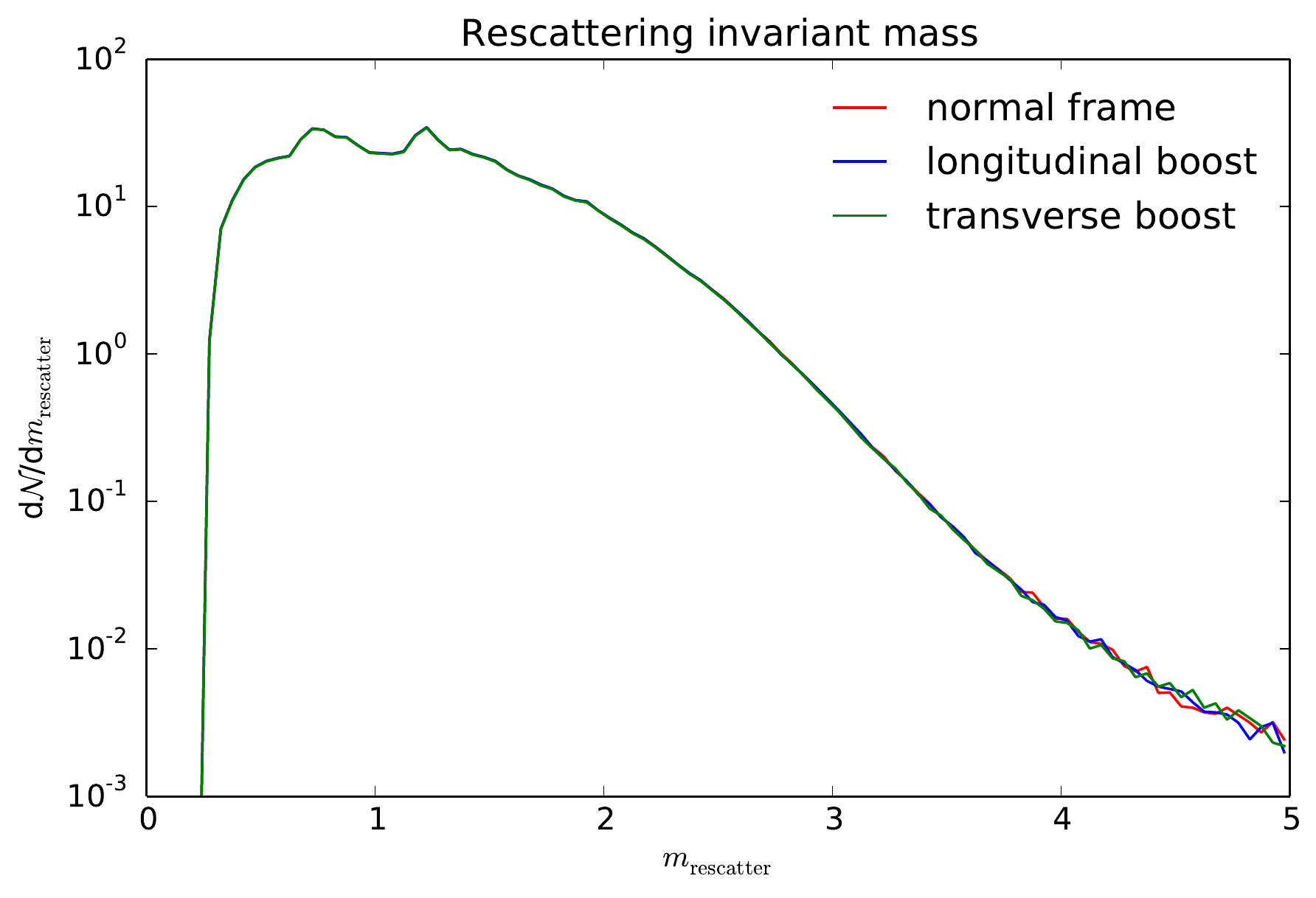}\\
(b)
\end{minipage}

\begin{minipage}[c]{0.49\linewidth}
\centering
\includegraphics[width=\linewidth]{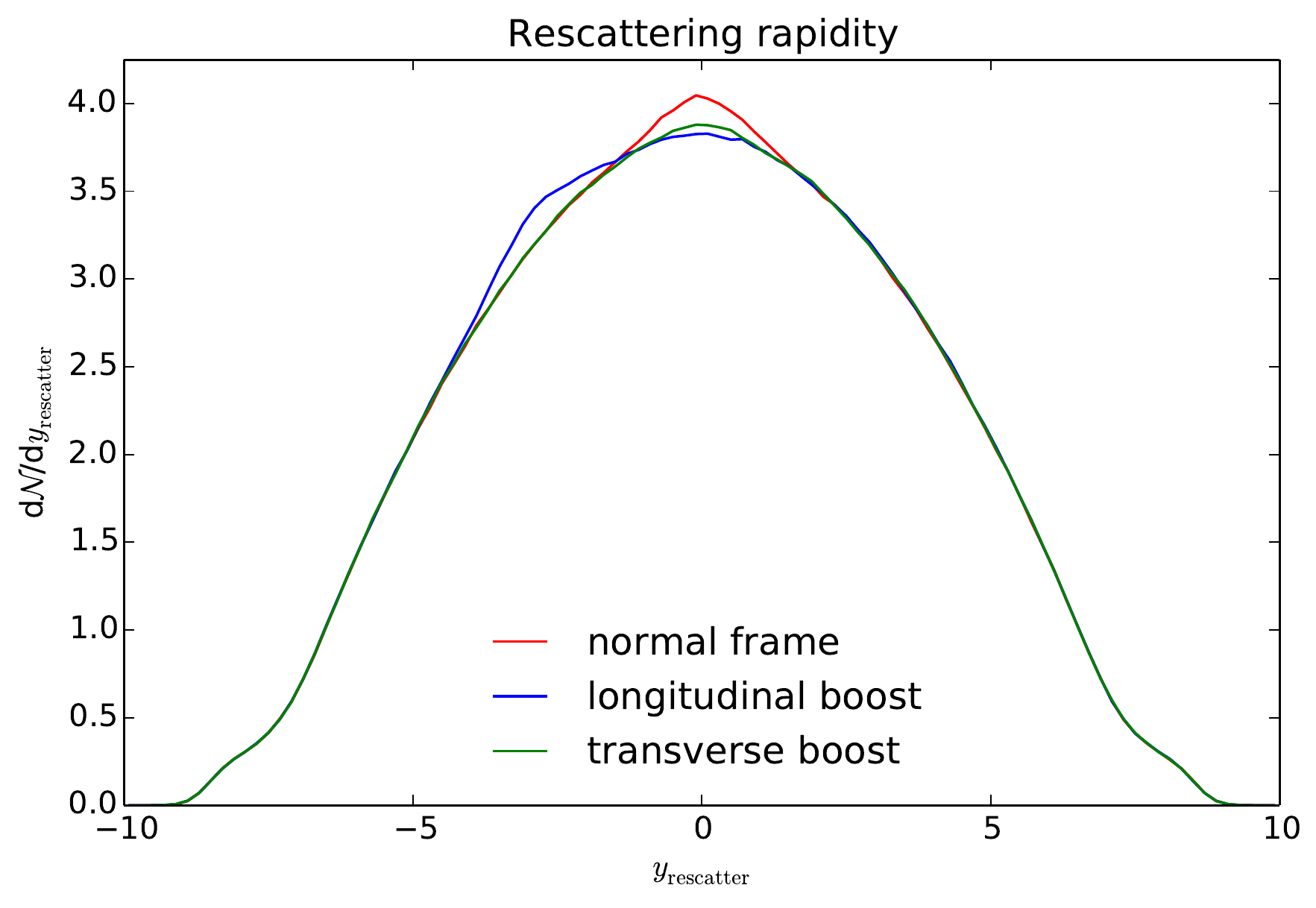}\\
(c)
\end{minipage}
\begin{minipage}[c]{0.49\linewidth}
\centering
\includegraphics[width=\linewidth]{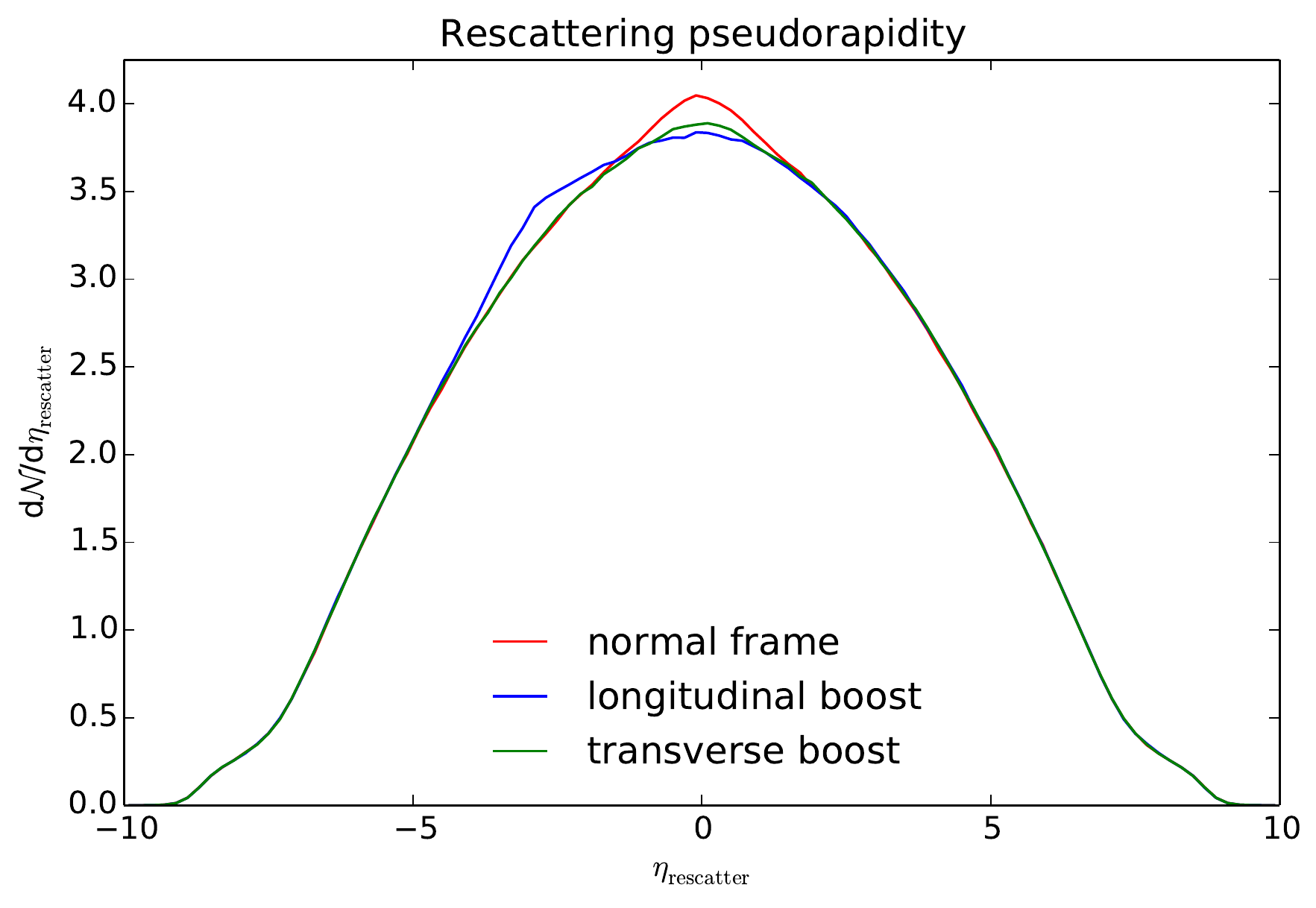}\\
(d)
\end{minipage}

\begin{minipage}[c]{0.49\linewidth}
\centering
\includegraphics[width=\linewidth]{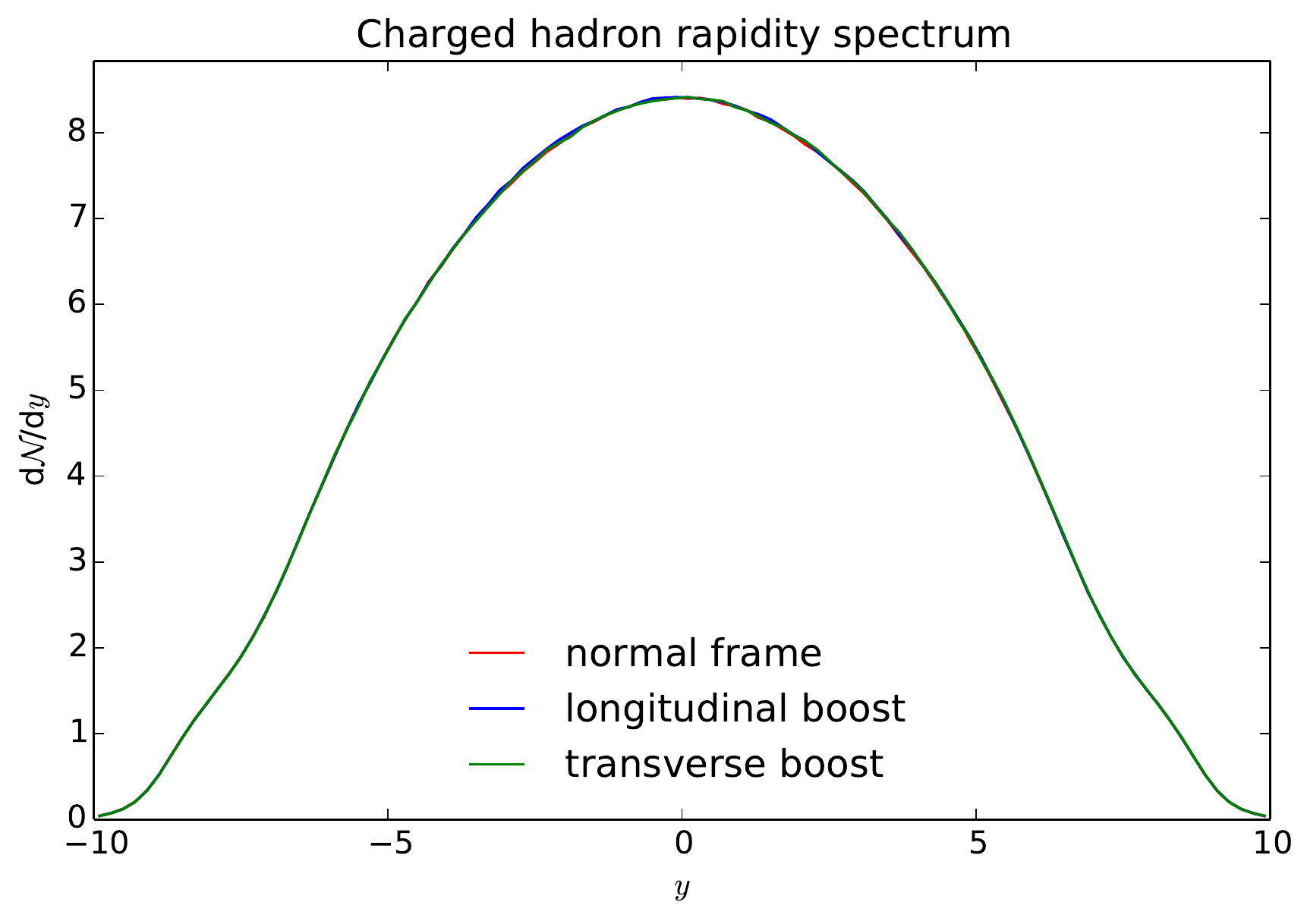}\\
(e)
\end{minipage}
\begin{minipage}[c]{0.49\linewidth}
\centering
\includegraphics[width=\linewidth]{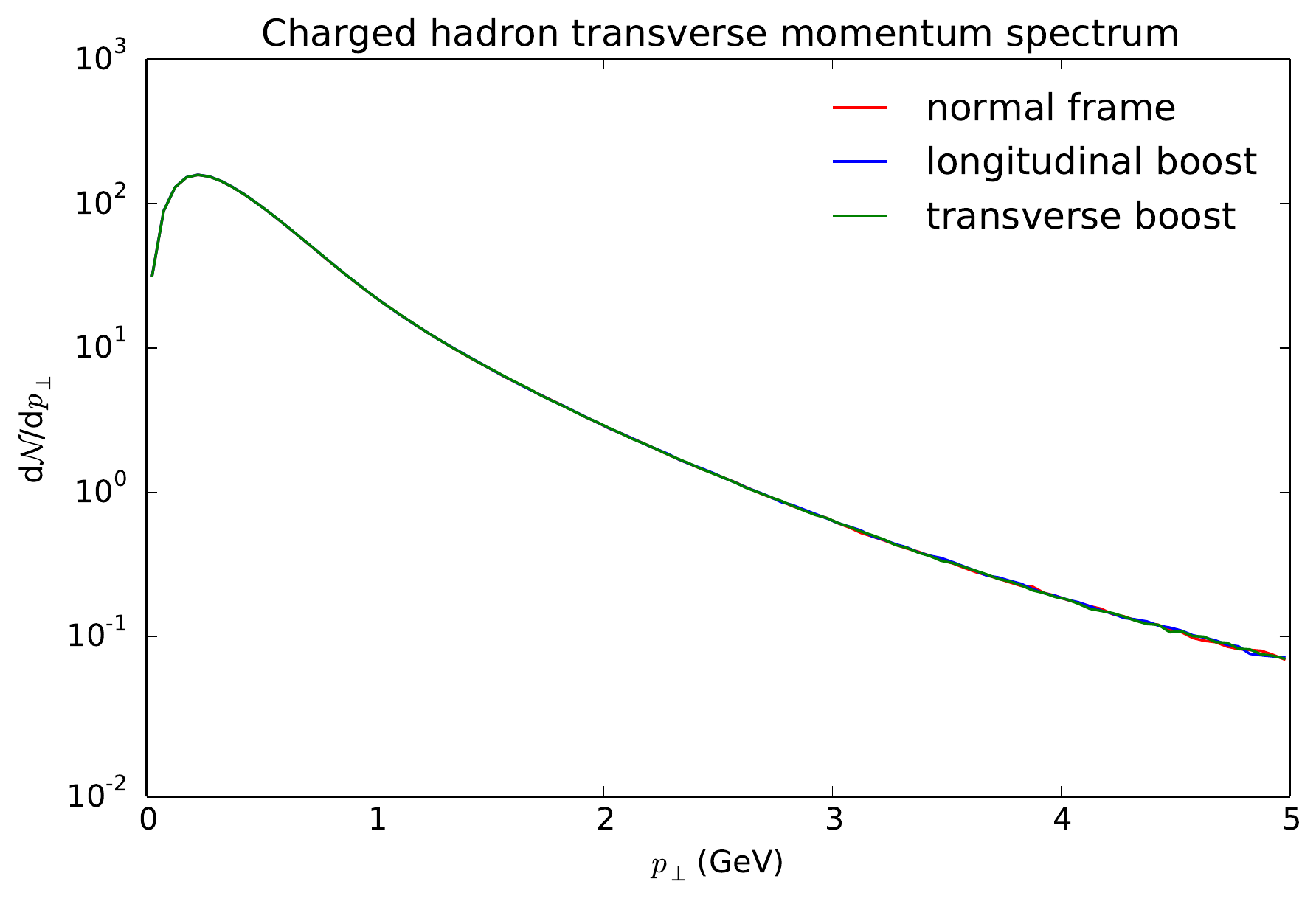}\\
(f)
\end{minipage}

\caption{
  Effects of modified time ordering on 13~TeV pp nondiffractive
  collisions, where ordering is either in the normal rest frame,
  or in a frame boosted either longitudinally or transversely by
  three units of rapidity. 
  (a) Number of rescatterings. 
  (b) Invariant mass distribution of rescatterings.
  (c) Rapidity distribution of rescatterings.
  (d) Distribution in $\eta = (1/2) \, \ln((t + z)/(t - z))$ of
  rescatterings.
  (e) Rapidity distribution of final charged hadrons.
  (f) Transverse momentum spectrum of final charged hadrons. 
}
\label{fig:boosteffect}
\end{figure}

The time ordering of rescatterings is not Lorentz invariant but, we do not expect this to be a major issue,
since most potential rescatterings cannot influence each other.
To confirm this more thoroughly, we boost the events by three units of rapidity either along or transverse to the
collision axis, perform rescattering in this boosted frame,
then boost back afterwards. Some results of performing this procedure, compared
with the ones in the normal CM frame, are shown in \figref{fig:boosteffect}.
One may first note that the number of rescatterings and their
invariant mass distribution are essentially unchanged. The rapidity spectrum of rescatterings however is somewhat
deformed by the forward boost, where rescatterings would begin at
around $y = -3$. Such rescatterings thus in part preempt ones at
larger times in that frame. The same applies for the space--time
pseudorapidity, $\eta = (1/2) \, \ln((t + z)/(t - z))$.
If instead the boost is transverse, the effects on the $y$ and
$\eta$ spectra are even smaller. Here collisions on the $-x$ side
of the event get an earlier start than those on the $+x$ one, giving a
$\pm 2$\% modulation in the azimuthal distributions of rescatterings
(not shown). These effects average out in other distributions, however, 
so that the $\pT$ and $r_{\perp} = \sqrt{x^2 + y^2}$  rescattering spectra 
are almost unchanged by transverse and longitudinal boost alike.

At the end of the day, the real test is whether observable properties
are affected or not. \figref{fig:boosteffect}e,f
show that the final-state charged-hadron rapidity and $\pT$ spectra
are almost completely insensitive to the choice of rest frame. The
same also applies for other distributions we have studied, such as
the azimuthal dependence, or the separate $\pi/\K/\p$ spectra. The
breach of Lorentz frame independence therefore is a negligible issue for
our studies.

\subsection{Rescatter rates} \label{subsec:rescatterRates}

\begin{figure}[t!]
\begin{minipage}[c]{0.49\linewidth}
\centering
\includegraphics[width=\linewidth]{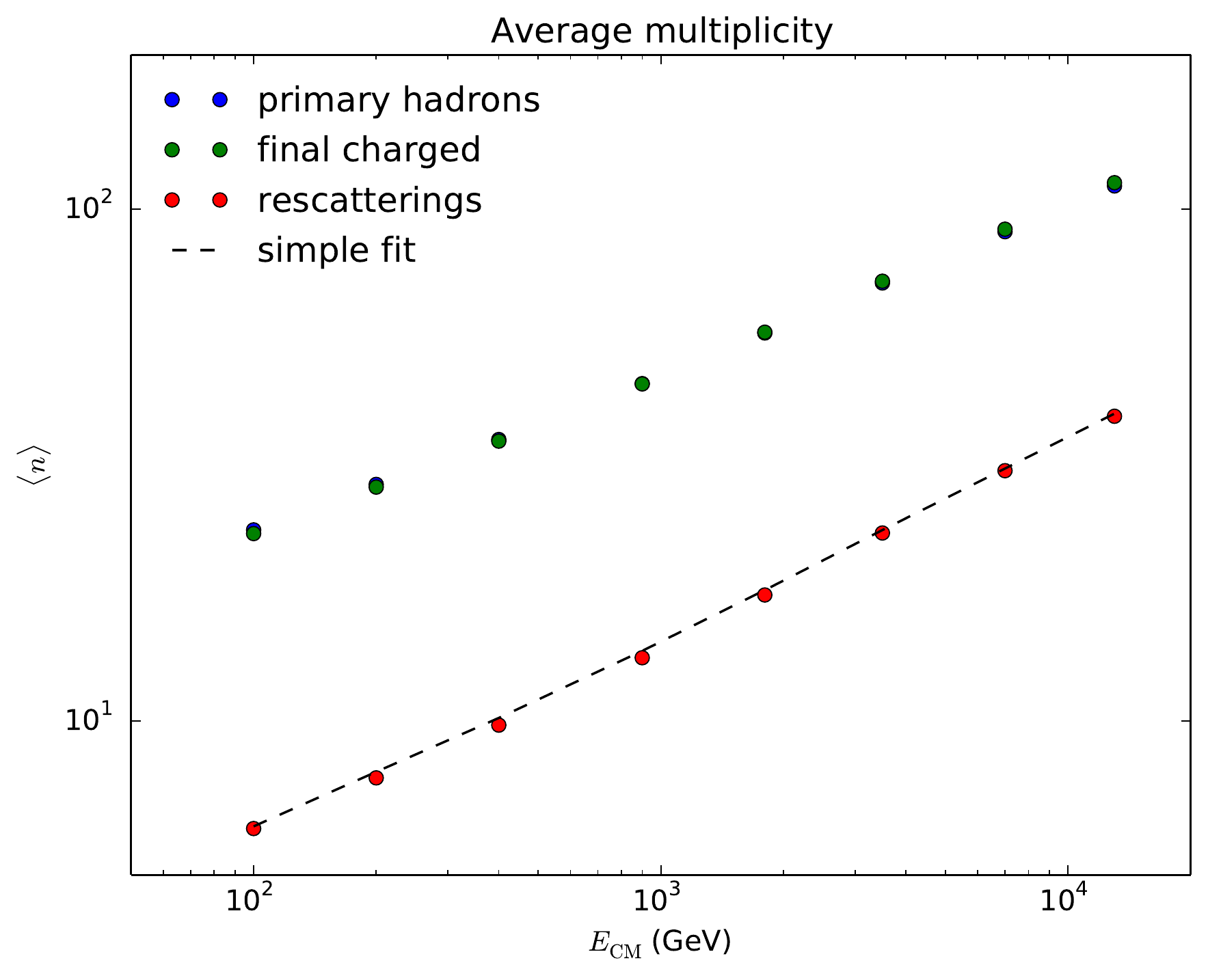}\\
(a)
\end{minipage}
\begin{minipage}[c]{0.49\linewidth}
\centering
\includegraphics[width=\linewidth]{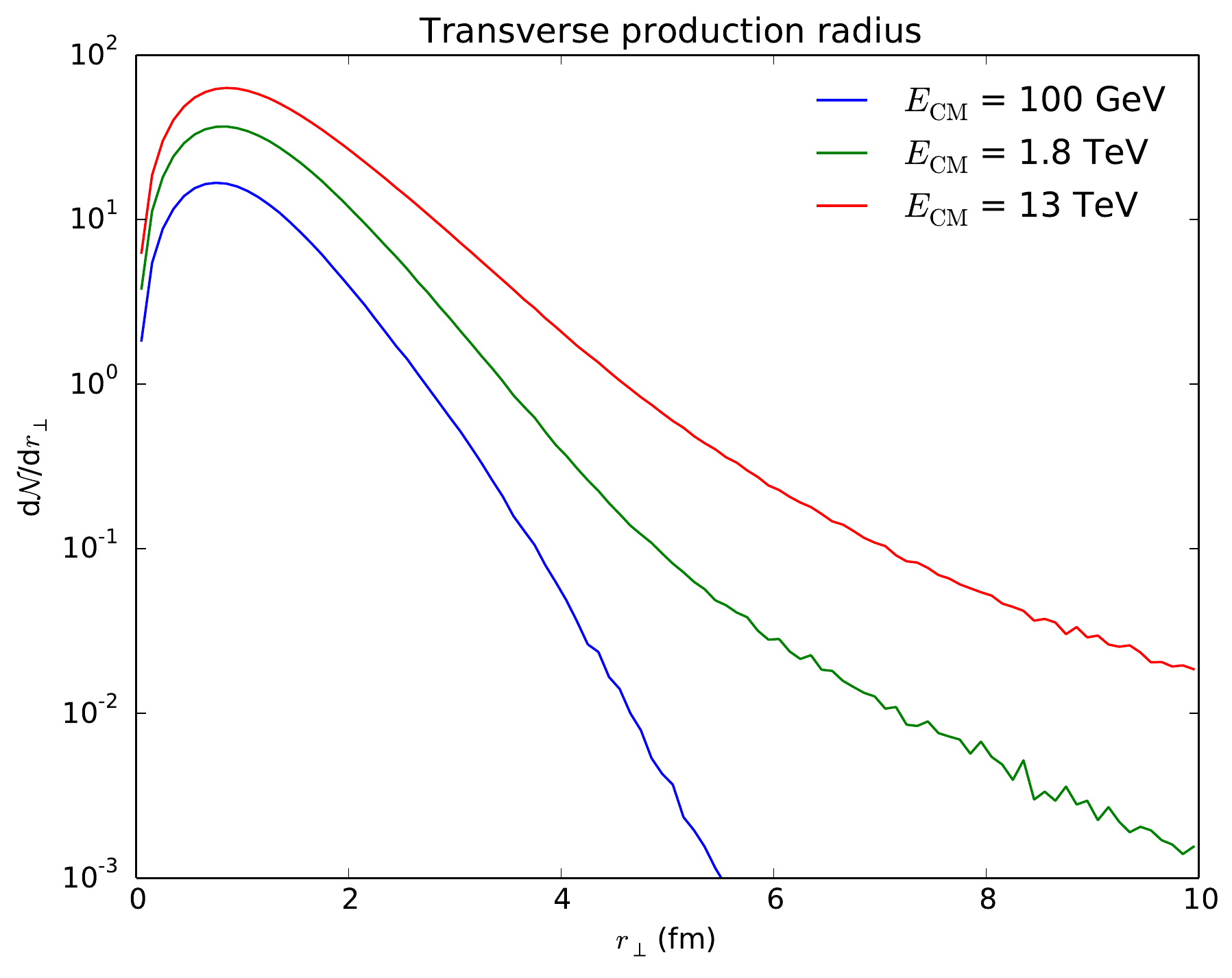}\\
(b)
\end{minipage}

\begin{minipage}[c]{0.49\linewidth}
\centering
\includegraphics[width=\linewidth]{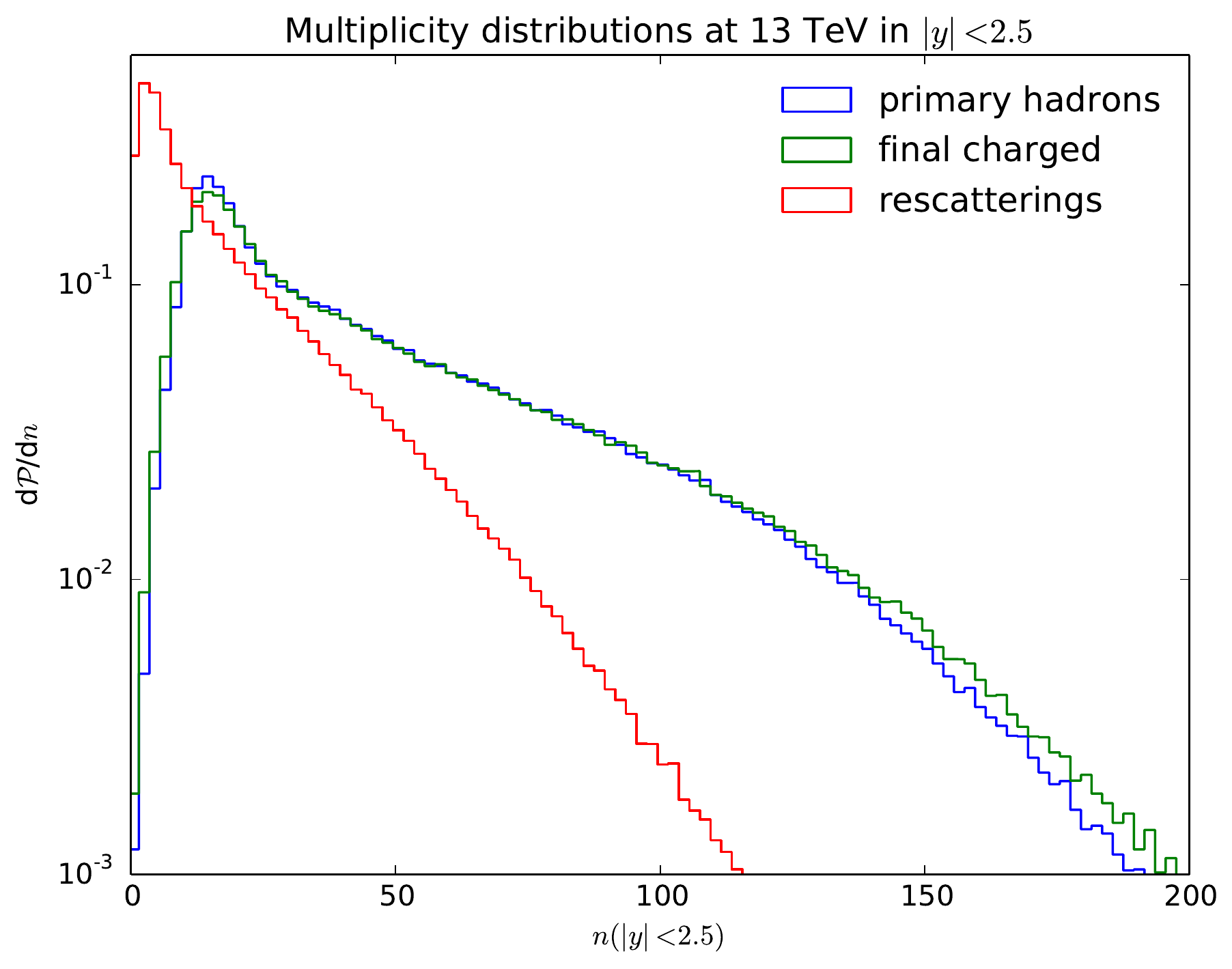}\\
(c)
\end{minipage}
\begin{minipage}[c]{0.49\linewidth}
\centering
\includegraphics[width=\linewidth]{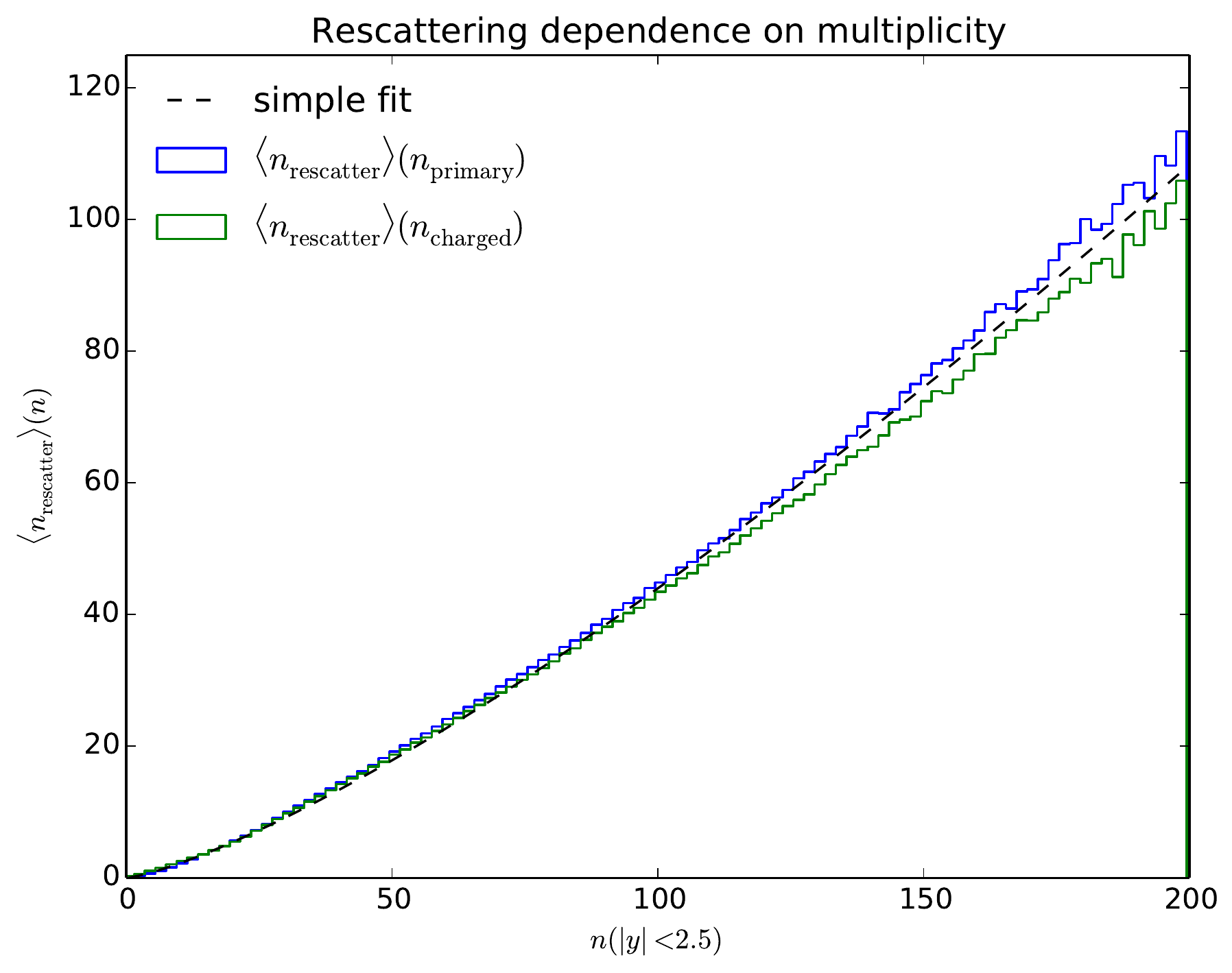}\\
(d)
\end{minipage}
\caption{\label{fig:rescatrate}
(a) Energy dependence of multiplicities in nondiffractive
$\p\p$ collisions.
(b) Primary hadron production in $r_{\perp} = \sqrt{x^2 + y ^2}$
at three energies.
(c) Distribution in the numbers of primary hadrons, charged hadrons
and rescatterings in the central $|y| < 2.5$ region of 13~TeV
nondiffractive $\p\p$ collisions.
(d) Multiplicity dependence of the number of rescatterings in 
events as above.}
\end{figure}

In this section we study how common rescatterings are, both overall 
and subdivided by hadron species and by process types. The average number of rescatterings per (inelastic) 
nondiffractive $\p\p$ event is shown as a function of the collision 
energy in \figref{fig:rescatrate}a. It is compared to the primary 
hadron multiplicity, i.e.\ the hadrons produced directly from the 
fragmenting strings, and to the final charged multiplicity. 
Note that these latter two are almost equal; the multiplicity 
increase from the decays of primary hadrons is compensated by
the decrease from the exclusion of neutral particles. This
largely holds also on an event-by-event level, so we may use 
the observable charged multiplicity as a simple measure of number of
primary hadrons that may rescatter. As an order-of-magnitude, 
the average number of rescatterings 
$\langle n_{\mathrm{rescatter}} \rangle$ is about half that of 
the primary multiplicity $\langle n_{\mathrm{primary}} \rangle$.
While the number of potentially colliding pairs increases like
$n_{\mathrm{primary}}^2$, the dashed line represents a fit according 
to a much slower $\langle n_{\mathrm{primary}} \rangle^{1.2}$.
The reason is that the system size also increases with energy. 
Obviously so in the longitudinal direction, but also in the 
transverse one, by an increasing MPI perturbative activity 
spreading production vertices over a larger transverse area,
\figref{fig:rescatrate}b. 

Zooming in on the central rapidity region of 13~TeV nondiffractive 
events, the different kinds of multiplicity distributions are 
displayed in \figref{fig:rescatrate}c, and the rescattering rate 
as a function of the primary or charged multiplicity in 
\figref{fig:rescatrate}d. In the latter, a simple fit 
$\langle n_{\mathrm{rescatter}} \rangle \propto n_{\mathrm{primary}}^{1.3}$
has been inserted to guide the eye, showing a similar scaling
as for the energy dependence. The power 1.3 also describes the 
dependence in the event as a whole, without the $|y| < 2.5$ restriction.

\begin{table}[tbp]
\centering
\begin{tabular}{|c|c||c|c||c|c|}
  \hline
  incoming & rate & incoming & rate & incoming & rate \\ \hline
  $\pi + \pi$         & 12.63 & $\K + \N$            & 0.39 & $\eta/\eta' + N$     & 0.19 \\
  $\pi + \rho$         & 4.59 & $\rho + \rho$        & 0.38 & $\pi + \B$           & 0.18 \\
  $\pi + \K$           & 3.84 & $\rho + \N$          & 0.36 & $N + \Delta$         & 0.16 \\
  $\pi + \N$           & 3.44 & $\rho + \omega/\phi$ & 0.34 &  $\pi + \Sigma^*$    & 0.15  \\
  $\pi + \omega/\phi$  & 2.08 & $\rho + \eta/\eta'$  & 0.30 & $\rho + \Delta$      & 0.14 \\ 
  $\pi + \eta/\eta'$   & 1.80 & $\pi + f_0(500)$     & 0.29 & $\eta/\eta' + \omega/\phi$ & 0.14 \\
  $\pi + \K^*$         & 1.33 & $\K + \omega/\phi$   & 0.27 & $\pi + \M$           & 0.12 \\
  $\pi + \Delta$       & 1.10 & $\K + \K$            & 0.26 & $\K + \Delta$        & 0.11 \\
  $\rho + \K$          & 0.54 & $\pi + \Lambda$      & 0.25 & $\K^* + \N$          & 0.11 \\
  $\pi + \Sigma$       & 0.46 & $\omega/\phi + \N$   & 0.24 &                      &      \\
  $\N + \N$            & 0.46 & $\K + \eta/\eta'$    & 0.23 &                      &      \\ 
  $\K + \K^*$          & 0.41 & $\rho + \K^*$        & 0.20 & other                & 1.87 \\ \hline
\end{tabular}
\caption{\label{tab:collisionkinds}
Number of collisions per 13~TeV nondiffractive $\p\p$ event, of different
incoming particle combinations, where particles have been grouped so as to
avoid too fragmented a view. $\M$ represents other meson species and $\B$
other baryon ones. All combinations with a rate below 0.1 have been summed
into the ``other'' group.}
\end{table}

\begin{table}[tbp]
\centering
\begin{tabular}{|l|c|}
  \hline
  Process type & rate \\ \hline
  resonant       & 17.80 \\
  elastic        & 14.08 \\
  nondiffractive & 6.92 \\
  annihilation   & 0.49 \\
  diffraction + excitation & 0.05 \\ \hline
\end{tabular}
\caption{\label{tab:collisiontypes}
Number of collisions of different types per 13~TeV nondiffractive $\p\p$ event.}
\end{table}
 
With well over a hundred different hadron species that can be 
produced, the number of different colliding hadron pairs are in 
the thousands, even if most of them are quite rare. To give some 
feel, \tabref{tab:collisionkinds} shows the most common groups of 
hadron pairs. Here $\pi$ represents all pions, $\K$ all Kaons
($\K^{\pm}$, $\K^0$, $\Kbar^0$, $\K_{\mathrm{S,L}}^0$),
$\N$ all nucleons ($\p$, $\n$, $\pbar$, $\nbar$), and so on.
As can be seen, $\pi\pi$ rescatterings dominate by far, constituting about a third of all rescatterings, while $\pi$ with anything else constitutes another third. This highlights the importance of accurate cross sections for processes involving pions.

Collisions are also characterized by which type of process occurs,
\tabref{tab:collisiontypes}. The resonant, elastic and nondiffractive
types dominate by far. Baryon--antibaryon annihilation is small but
not negligible for the baryon subclass of particles. Diffraction and 
excitation require more phase space to occur, and therefore become 
suppressed. 

\begin{figure}[t!]

\begin{minipage}[c]{0.49\linewidth}
\centering
\includegraphics[width=\linewidth]{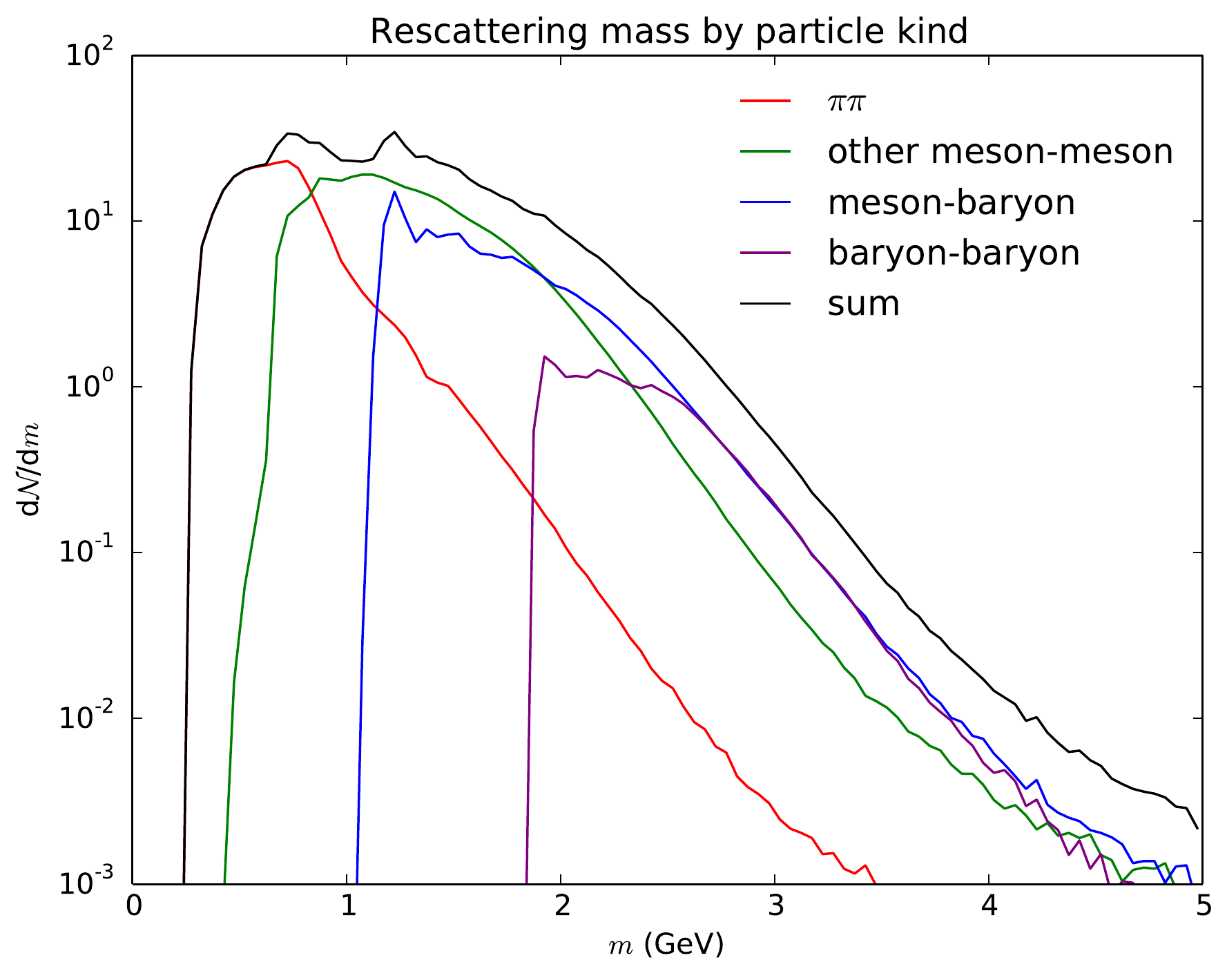}\\
(a)
\end{minipage}
\begin{minipage}[c]{0.49\linewidth}
\centering
\includegraphics[width=\linewidth]{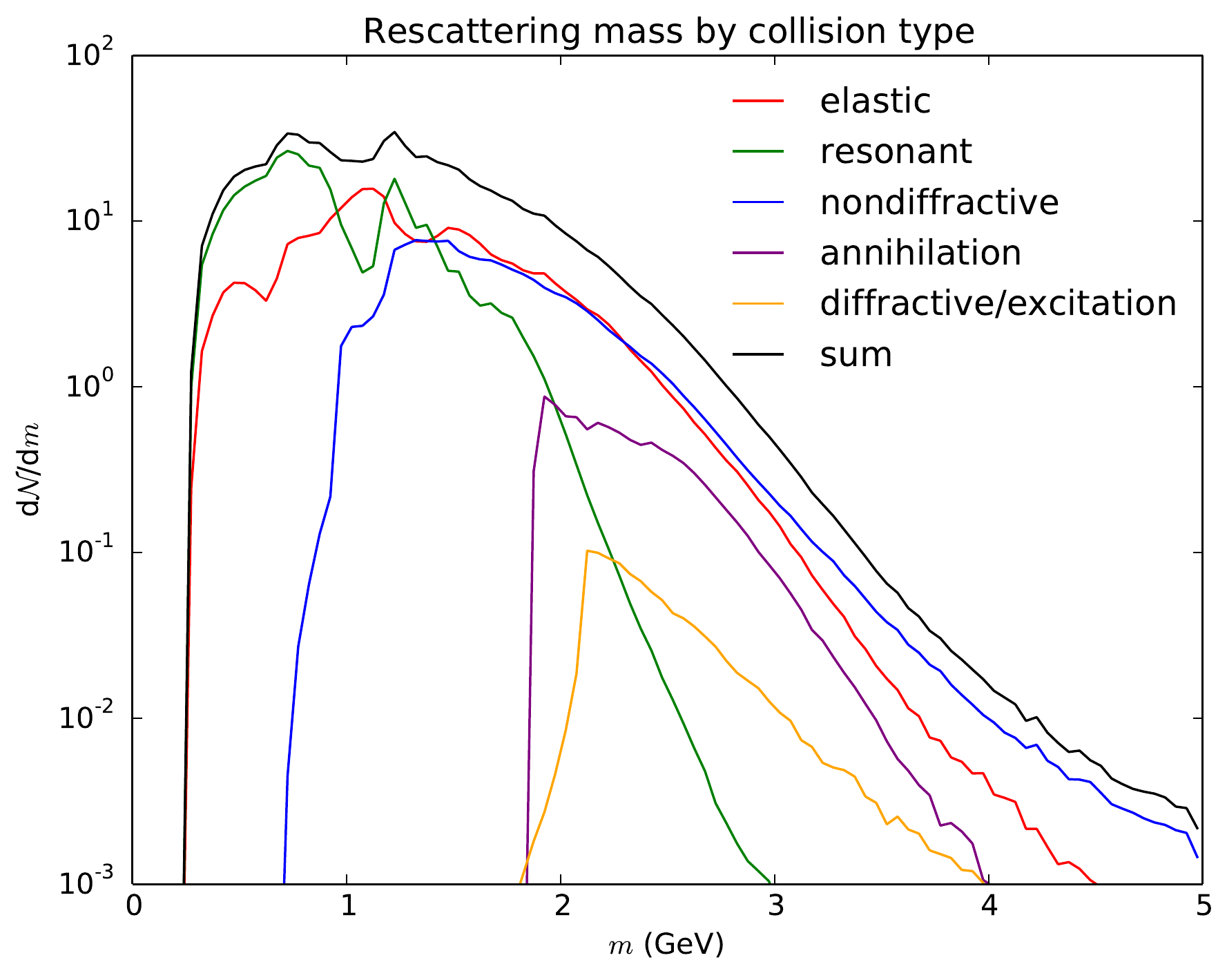}\\
(b)
\end{minipage}

\caption{Invariant mass distributions of rescattering pairs in 
13~TeV nondiffractive $\p\p$ events. (a) Grouped by incoming hadron
kinds. (b) Grouped by process type.}
\label{fig:masstypekind}
\end{figure}

It is also interesting to study the invariant mass spectrum of
collisions, \figref{fig:masstypekind}. There is a natural steep 
fall-off with mass for two particles to come close to each other, 
because of the way the fragmentation process correlates the 
space--time and energy--momentum pictures. Near each mass threshold 
there is also a phase-space suppression factor. On top of that 
the individual cross sections can give a more serrated shape for 
each collision type separately, mainly from resonance contributions, 
but these largely average out in the overall picture.

\subsection{Model variations}

\begin{table}[tbp]
\centering
\begin{tabular}{|l|r|l|}
    \hline
    Setting & Default & Effect on rescattering rate \\
    \hline
    \texttt{Rescattering:impactModel}       & 1 (Gaussian) & Black disk gives more \\
    \texttt{Rescattering:opacity}           & 0.9  & Larger values give more\\
    \texttt{Rescattering:quickCheck}        & on  & Turning it off gives more\\
    \texttt{Rescattering:nearestNeighbours} & on & Turning it off gives less\\
    \texttt{Rescattering:tauRegeneration}   & 1.  & Larger values give less\\ 
    \texttt{HadronVertex:mode}              & 0   & $\pm1$ gives much more/less\\
    \texttt{HadronVertex:kappa}             & 1. & Larger values give more\\
    \texttt{HadronVertex:xySmear}           & 0.5 & No significant effect\\
    \texttt{PartonVertex:modeVertex}        & 2 (Gaussian) & Has a small effect \\
    \texttt{PartonVertex:ProtonRadius}      & 0.85 & Larger value gives less\\ 
    \texttt{PartonVertex:EmissionWidth}     & 0.1 & No significant effect\\
    \hline
\end{tabular}
\caption{List of model choices and parameters used to study the range of possible 
rescattering effects, with their effect on the rescattering rate. Parameter names 
are as defined in the \textsc{Pythia} user interface. See the text for more detailed 
information.}
\label{tab:VaryParameters}
\end{table}

\begin{table}[tbp]
    \centering
    \begin{tabular}{|l | r|}
        \hline
        Setting & $n_\textrm{rescatter}$ \\
        \hline
        Default & 39.2 \\
        \hline
        \texttt{Rescattering:impactModel = 0} & 45.5 \\
        \hline 
        \texttt{Rescattering:opacity = 0.8} & 37.3 \\
        \texttt{Rescattering:opacity = 1.0} & 40.8 \\
        \hline
        \texttt{Rescattering:quickCheck = off} & 40.8 \\
        \hline
        \texttt{Rescattering:nearestNeighbours = off} & 25.4 \\
        \hline
        \texttt{Rescattering:tauRegeneration = 0.0} & 45.4 \\
        \texttt{Rescattering:tauRegeneration = 2.0} & 38.4 \\
        \hline 
        \texttt{HadronVertex:mode = -1} & 64.0 \\
        \texttt{HadronVertex:mode = 1}  & 21.7 \\
        \hline 
        \texttt{HadronVertex:kappa = 0.8} & 32.8 \\
        \texttt{HadronVertex:kappa = 1.2} & 44.4 \\
        \hline 
        \texttt{HadronVertex:xySmear = 0.3} & 40.2 \\
        \texttt{HadronVertex:xySmear = 0.7} & 39.1 \\
        \hline 
        \texttt{PartonVertex:modeVertex = 1} & 39.6 \\
        \hline 
        \texttt{PartonVertex:protonRadius = 0.7} & 39.3 \\
        \texttt{PartonVertex:protonRadius = 1.0} & 39.1 \\
        \hline
        \texttt{PartonVertex:EmissionWidth = 0.0} & 39.6 \\
        \texttt{PartonVertex:EmissionWidth = 0.2} & 39.2 \\
        \hline
    \end{tabular}
    \caption{Average number of rescatterings per event, when varying different settings individually.
    Events are \texttt{SoftQCD:nonDiffractive} processes at 13 TeV, using
    \texttt{MultipartonInteractions:pT0Ref = 2.345}.}
    \label{tab:nResc}
\end{table}

\begin{figure}[t!]
\begin{minipage}[c]{0.49\linewidth}
\centering
\includegraphics[width=\linewidth]{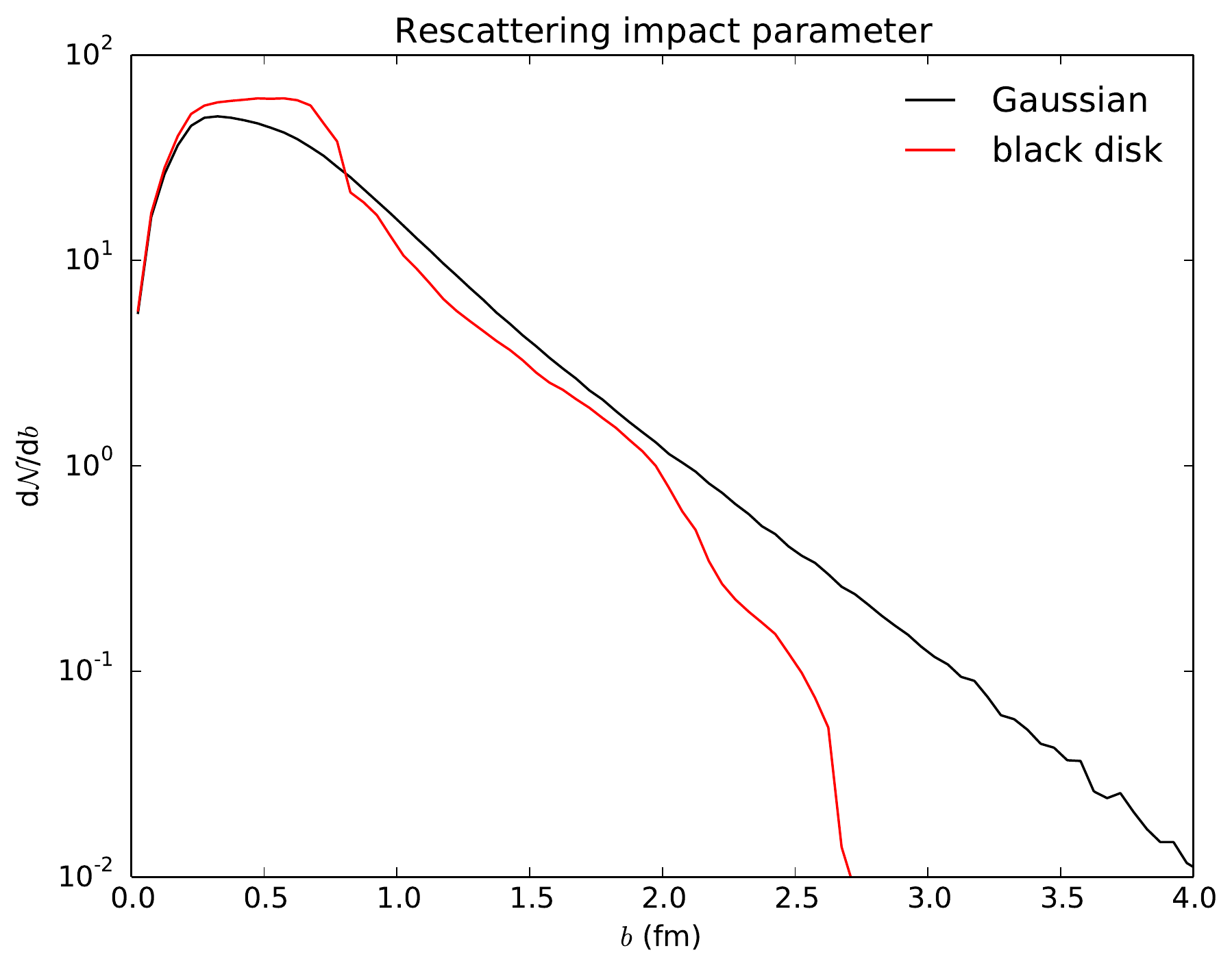}\\
(a)
\end{minipage}
\begin{minipage}[c]{0.49\linewidth}
\centering
\includegraphics[width=\linewidth]{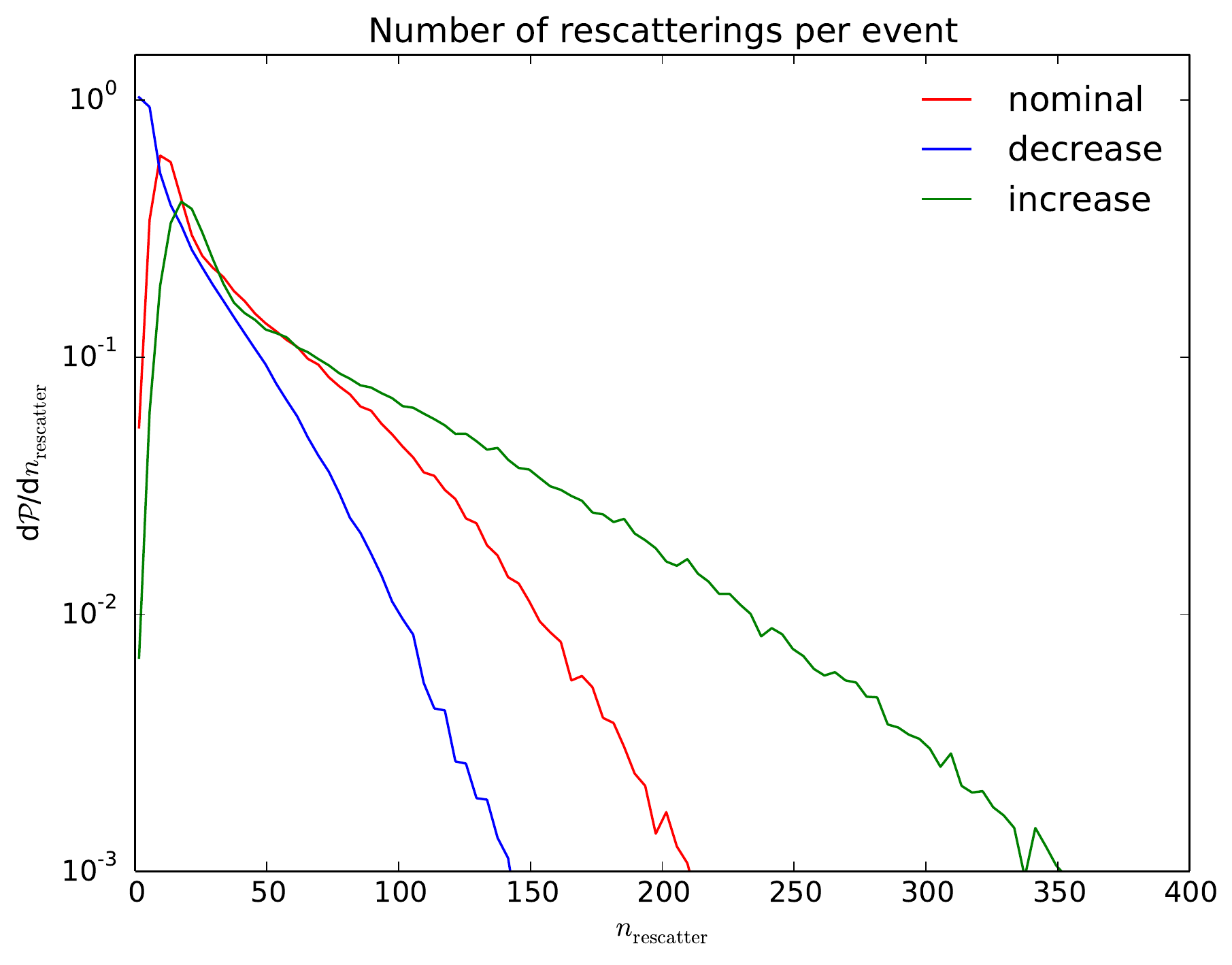}\\
(b)
\end{minipage}

\begin{minipage}[c]{0.49\linewidth}
\centering
\includegraphics[width=\linewidth]{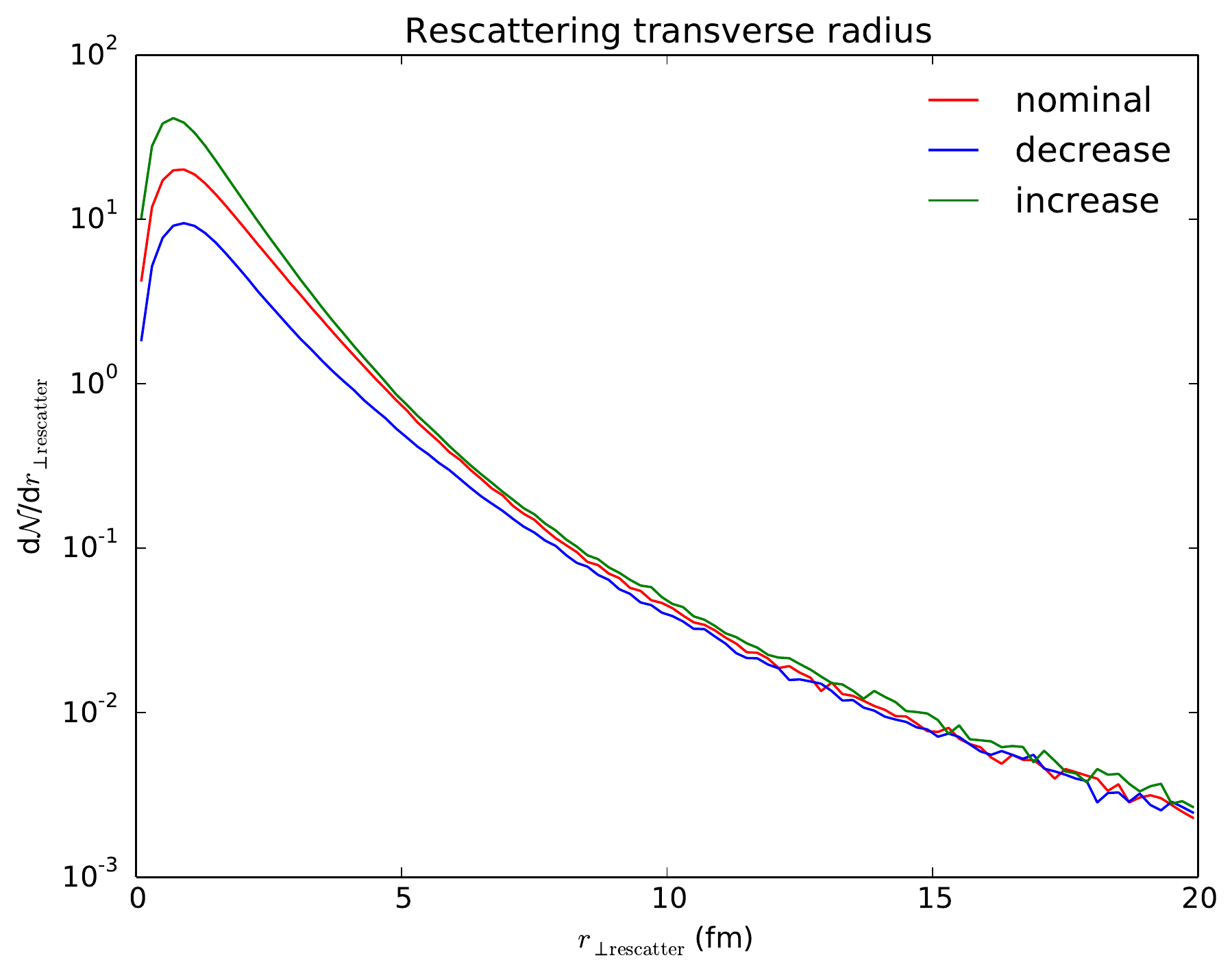}\\
(c)
\end{minipage}
\begin{minipage}[c]{0.49\linewidth}
\centering
\includegraphics[width=\linewidth]{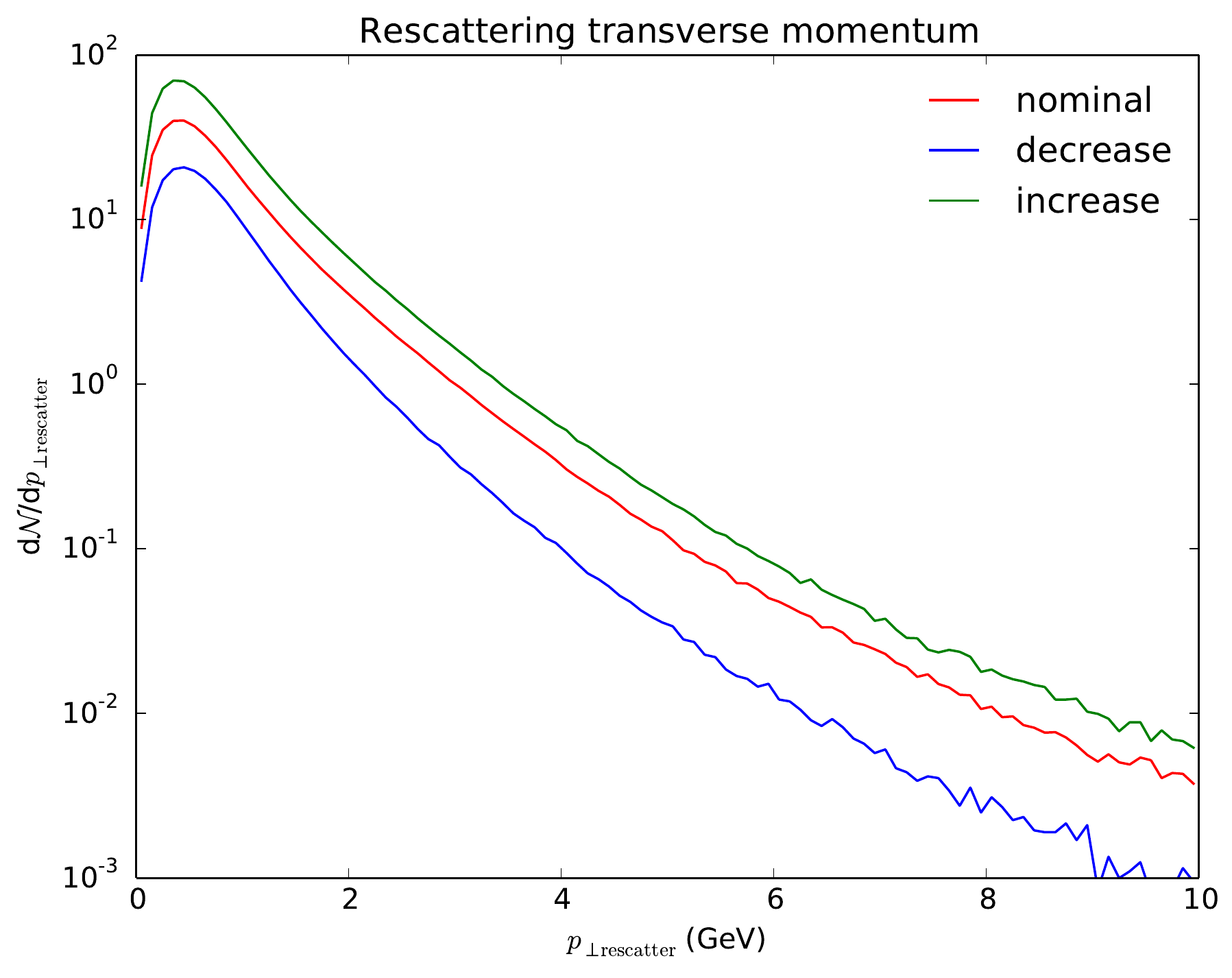}\\
(d)
\end{minipage}
\caption{\label{fig:VaryParameters}
(a) The impact-parameter distribution of rescatterings for the different impact models.
(b) Number of rescatterings per event.
(c,d) Distribution of rescatterings in $r_{\perp}$ and $\pT$.
Results are for 13 TeV nondiffractive $\p\p$ events.}
\end{figure}

As part of the new framework, several parameters and settings have been 
introduced. In this section, we study how changing these settings affects 
rescattering phenomenology. In particular, as a simple and direct test,
we present how each main model setting impacts the average number of rescatterings 
per event. In addition to these new settings, we also study existing settings that could have an effect 
on rescattering. A summary of settings and their overall effects is given in
\tabref{tab:VaryParameters}, with the average number of rescatterings for
different variations shown in \tabref{tab:nResc}. In more detail, the effect of the settings are as follows:
\begin{itemize}
\item \texttt{Rescattering:impactModel} describes how the rescattering probability
depends on the impact parameter $b$. The default (1) is a Gaussian fall-off, while the alternative (0) is a sharp edge,
see eqs. ~\eqref{eq:bGaussianEdge} and \eqref{eq:bSharpEdge}. In a uniform medium the two alternatives are 
normalized to result in  equal rescattering rates, as given by the cross section. 
In practice we see that the Gaussian option gives more long-range interactions, 
\figref{fig:VaryParameters}a, as expected,
but overall a somewhat reduced rescattering rate. 
This is because the particle density falls off from the central collision axis,
such that there are fewer pairs at large than at small impact parameter to begin 
with. The fact that the Gaussian option gives a lower rescattering rate means that the loss of events in the important 0.3--0.7 fm region
for the Gaussian model is not compensated for by including longer-range interactions.
\item \texttt{Rescattering:opacity} is the rescattering probability at $b = 0$, 
i.e.\ $P_0$ of eqs.~ \eqref{eq:bGaussianEdge} and \eqref{eq:bSharpEdge}. 
A lower opacity reduces the probability of close interactions, but increases 
the range of interactions. This gives fewer rescatterings, for the same reason 
as above.
\item \texttt{Rescattering:quickCheck} enables a simple check that tests whether two hadrons 
are moving away from each other at their respective time of creation in the CM 
frame of the event, and if so does not study further whether a rescattering is 
possible. This is faster than the more time-consuming full check, where the 
hadron pair is boosted to their common rest frame and the earliest particle is offset to a common time of creation before checking whether the hadrons
move away from each other. 
Performing the quick check first reduces the total execution time by about a factor of two, 
since the number of hadron pairs to consider in an LHC event may be of the order 
of 10\,000, whereof the vast majority are moving away from each other by any 
criterion (note that the full check is still performed on pairs that pass 
the simple check). The simple check rejects about 5\% of the collisions that 
would have been accepted by the full check, but these false rejections typically 
are close to the (unphysically sharp) accept/reject border, and do not make a 
significant impact on rescattering distributions. For these reasons the quick 
check is on by default.
\item \texttt{Rescattering:nearestNeighbours} allows hadrons that are produced 
as nearest neighbours along a string to rescatter against each other, see
\secref{subsec:rescatterxyzt}. The number of rescatterings goes up when on, 
but net effects do not change in proportion, since nearest-neighbour pairs are 
more likely to move in the same direction anyway. 
\item \texttt{Rescattering:delayRegeneration} and 
\texttt{Rescattering:tauRegeneration} are based on the assumption that it 
takes some formation time for a scattered hadron to build up a new wave function, 
and that during that time it has a reduced likelihood to scatter again. If
\texttt{delayRegeneration} is switched on, this time is chosen at random 
according to an exponential distribution with average proper time (in fm) given 
by the \texttt{tauRegeneration}. Hadrons produced from string fragmentation 
are not affected, since they get their time offset from the hadronization 
process itself, roughly corresponding to an average $\tau$ of 1.5~fm. 
Setting $\tau_{\mathrm{regen}} = 1$~fm reduces the number of rescatterings 
by about 10\% relative to an instantaneous regeneration. The effect seems to saturate however, and increasing it to 
2~fm does not make much further difference.
\item \texttt{HadronVertex:mode} defines where the hadron vertex is placed 
in string hadronization. By default, hadrons are defined to be produced at 
the average location of the two string breaks that define it (see
\figref{fig:stringfrag}). By setting \texttt{HadronVertex:mode = 1}, 
the production vertex is shifted forward in time to the point where the 
two colour endpoints meet for the first time, and setting it to $-1$ shifts 
it backwards in time by that same amount. These variations have a significant 
effect on the density of primary produced hadrons, changing the number of 
rescatterings by about 50\%. For this reason we do not 
vary this setting in our studies, but instead use \texttt{HadronVertex:kappa}, 
which gives similar but milder effects, as explained below.
\item \texttt{HadronVertex:kappa} is the string tension, by default $\kappa \approx 1$~GeV/fm,
eq.~\eqref{eq:xplinearity}. Increasing $\kappa$ compresses the production vertices 
and thus gives more rescattering. While the concept 
of a string tension is central in the hadronization framework, its exact value 
has not been relevant for the energy--momentum-related properties of an event. 
We allow for a generous $\pm$20\% variation to also cover some uncertainty in 
how to define the hadron production vertex, as described above.
\item \texttt{HadronVertex:xySmear} is the width of a Gaussian smearing of string 
breakup vertices in the plane perpendicular to the string, see \secref{subsec:MPIvertices}. Increasing this 
slightly increases the transverse offsets of the primary produced hadron 
vertices, but does not have significant overall effects on rescattering.
\item \texttt{PartonVertex:modeVertex} picks the shape of the overlap region 
between the two incoming protons, as used to pick the location of MPI vertices, 
see \secref{subsec:MPIvertices}. Different shapes give some variation in 
rescattering features, but they are small ones for most properties, and it 
is hard to quantify the difference between the various shapes. For this reason, 
we do not vary this setting in subsequent model tests. It is however a way to introduce 
spatial anisotropy in the primary hadron distribution, which is necessary for azimuthal flow. 
\item \texttt{PartonVertex:ProtonRadius} is the three-dimensional proton radius, 
which then gets converted to a two-dimensional one for the distribution of MPI 
production vertices, eq.~\eqref{eq:bMPIGaussian}. Increasing/reducing this by 
0.15~fm will increase/reduce the transverse radius of rescattering vertices 
by about 0.10~fm, and higher values give a slightly lower number of rescatterings. 
\item \texttt{PartonVertex:EmissionWidth} is the constant of proportionality for 
smearing of the transverse production vertices generated by partons showers, 
which are assumed to be inversely proportional to the $\pT$ of the parton. 
Varying this within a reasonable range has no significant effect on rescattering.
\end{itemize}

\begin{table}[tbp]
\centering
\begin{tabular}{|l|r|r|r|}
\hline
Setting                                 & decrease & nominal & increase \\
\hline
\texttt{Rescattering:impactModel}       & 1   &  1  & 0  \\
\texttt{Rescattering:opacity}           & 0.8  & 0.9  & 1.0   \\
\texttt{Rescattering:quickCheck}        & on  & on   &  off  \\
\texttt{Rescattering:nearestNeighbours} & off & on   &  on   \\
\texttt{Rescattering:tauRegeneration}   & 2.  & 1.   &  0.   \\ 
\texttt{HadronVertex:kappa}             & 0.8 & 1.   &  1.2  \\
\texttt{PartonVertex:ProtonRadius}      & 1.0 & 0.85 &  0.7  \\ 
\hline
\texttt{MultipartonInteractions:pT0Ref} & 2.305 & 2.345 & 2.385 \\
\hline
\end{tabular}
\caption{\label{tab:MinMaxParameters}List of model settings used to 
explore the range of possible rescattering
effects.  Here ``increase'' and ``decrease'' denote alternatives
with more or less amount of rescattering relative to the default
``nominal'' values.} 
\end{table}

For comparison purposes, one nominal scenario is defined as our best
assumption on relevant settings, and in addition two extremes with
decreased or increased rescattering rate, \tabref{tab:MinMaxParameters}.
For each case, \texttt{pT0Ref} has been tuned as shown in the table in order to restore charged multiplicity.

The resulting variations of rescattering rates are shown in 
\figref{fig:VaryParameters}b. The rate difference mainly arises around 
small transverse radii, \figref{fig:VaryParameters}c
(and early invariant times, not shown). By contrast, in properties 
such as the transverse momentum, \figref{fig:VaryParameters}d, 
or invariant mass of the collision systems (not shown), the variations
more affect the normalization than the shape of the distributions.
Comparisons to data will be given in \secref{subsec:compareMinMax}.

\section{Comparison with data}
\label{sec:comparisons}

While the standard \textsc{Pythia} generally gives a good description of LHC $\p\p$
data, there are some well-known discrepancies. One such is the shape of low-$\pT$ spectra of pions, Kaons and protons. Especially 
the poor description of the pion spectrum for $\pT < 0.5$~GeV has direct consequences 
for a number of other distributions \cite{Aad:2010ac}, e.g.\ when the pseudorapidity 
spectrum is studied either for $\pT > 0.1$~GeV or $\pT > 0.5$~GeV charged particles.
In this section, we study how these spectra are changed by rescattering, using Rivet \cite{Bierlich:2019rhm} to generate plots and comparisons to data. 
Results are shown initially for the default rescattering model, then for alternative parameter choices within this model, and eventually for model variations of the primary hadron production.    
Finally, we briefly consider the $\pT$ spectrum for the $\Lambda^0/\K_S^0$ ratio. 
As before, the \texttt{pT0Ref} parameter is retuned to ensure the same charged multiplicity in all scenarios studied.

\begin{figure}[t!]
\begin{minipage}[c]{\linewidth}
\centering
\includegraphics[width=0.48\linewidth]{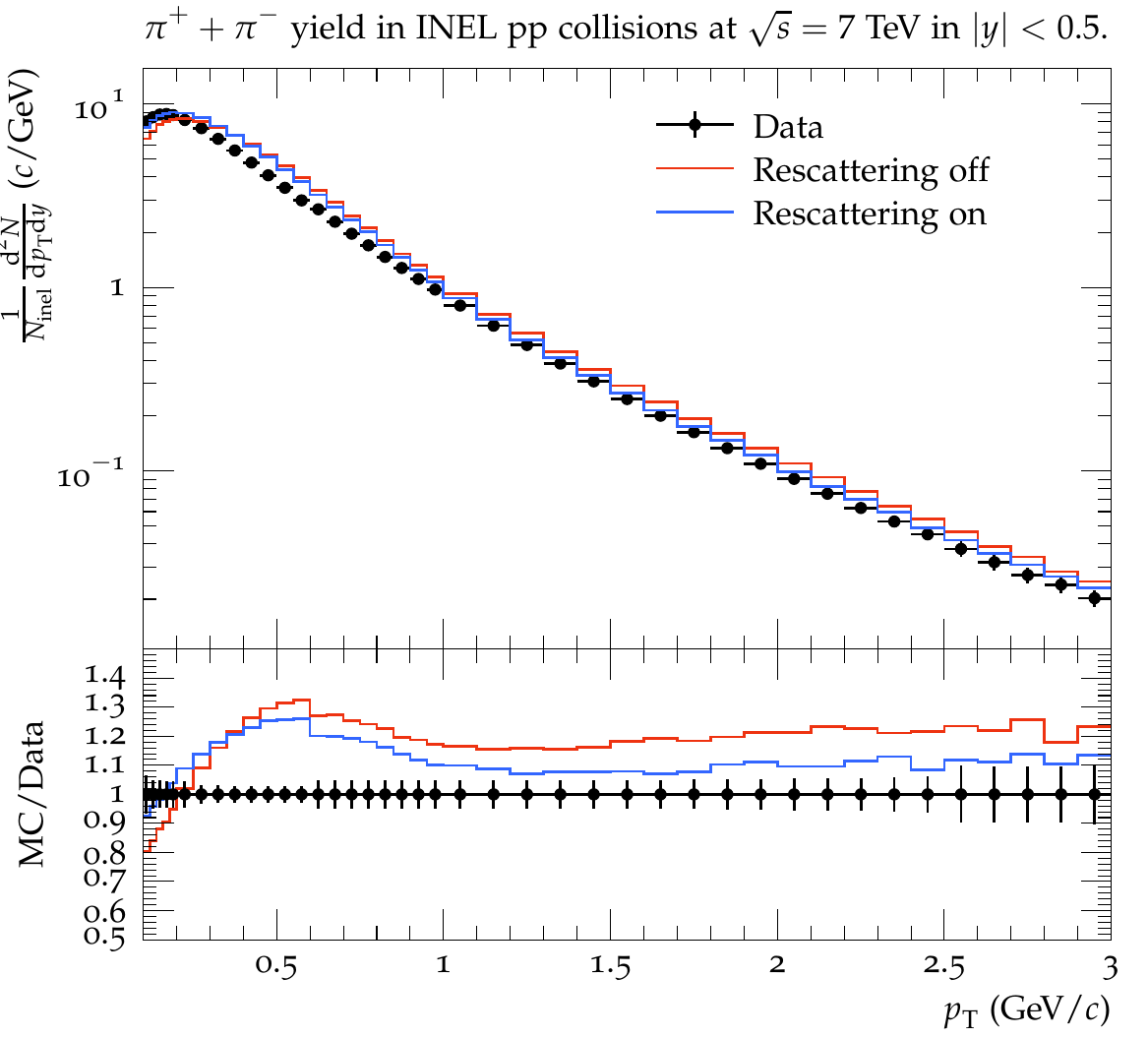}
\includegraphics[width=0.48\linewidth]{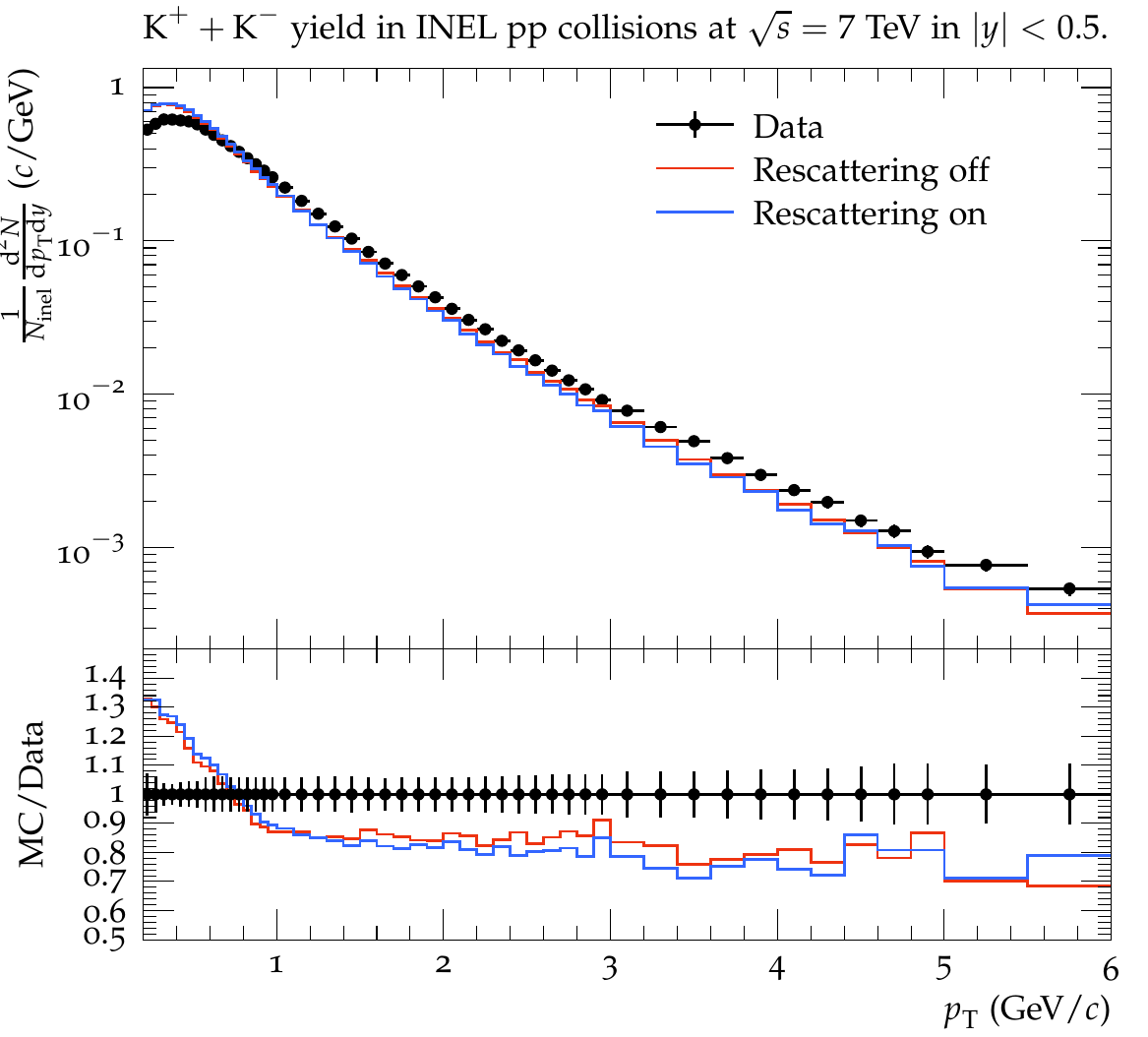}
\end{minipage}

\begin{minipage}[c]{\linewidth}
\centering
\includegraphics[width=0.48\linewidth]{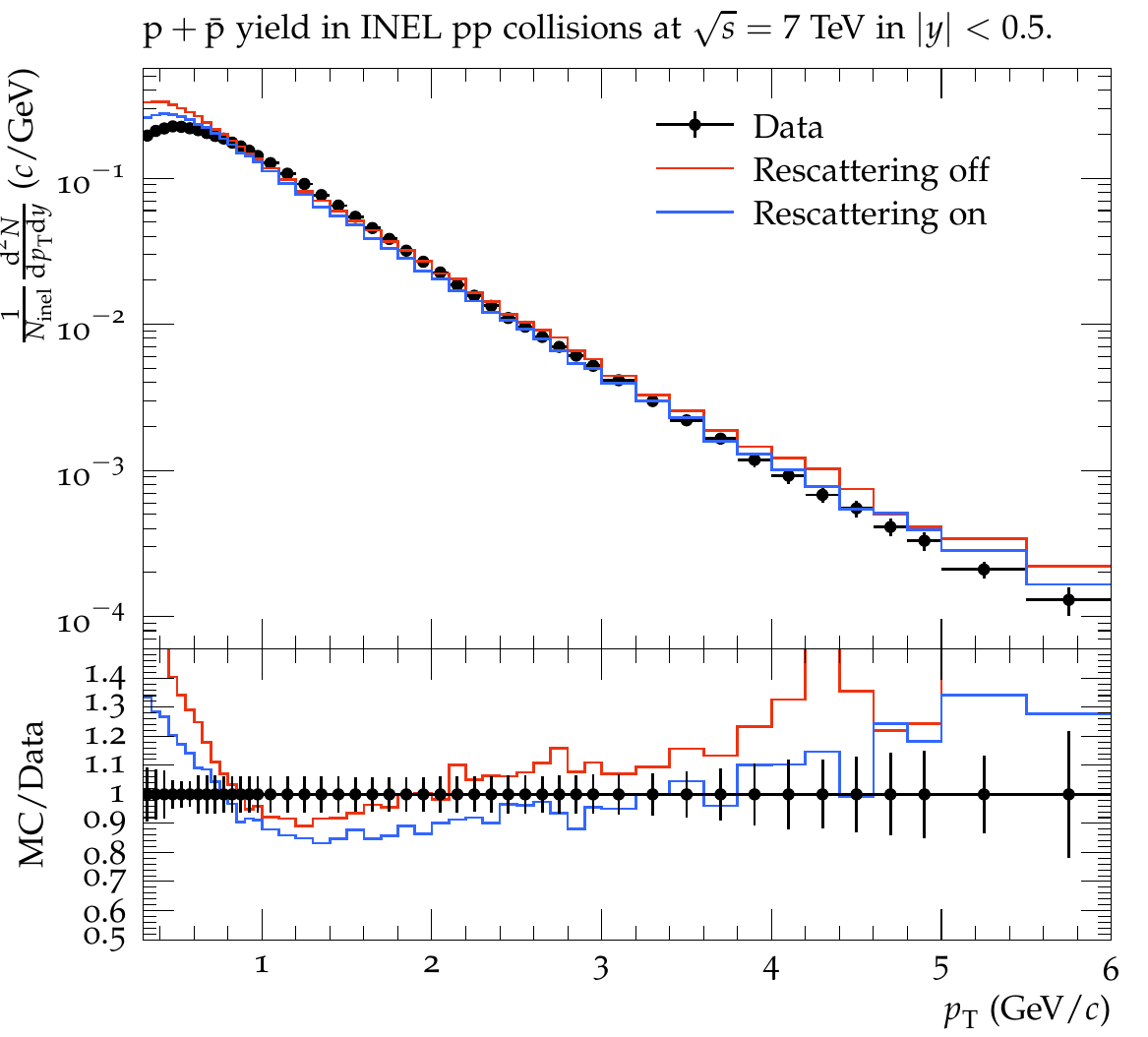}
\end{minipage}

\caption{$\pT$ spectra for $\pi^\pm$, $\K^\pm$ and $\p/\pbar$, compared with data from ALICE \cite{Abelev:2014qqa, Adam:2015qaa}.}
\label{fig:pTspectra}
\end{figure}

\begin{figure}[t!]
\centering
\begin{minipage}[c]{0.49\linewidth}
\centering
\includegraphics[width=\linewidth]{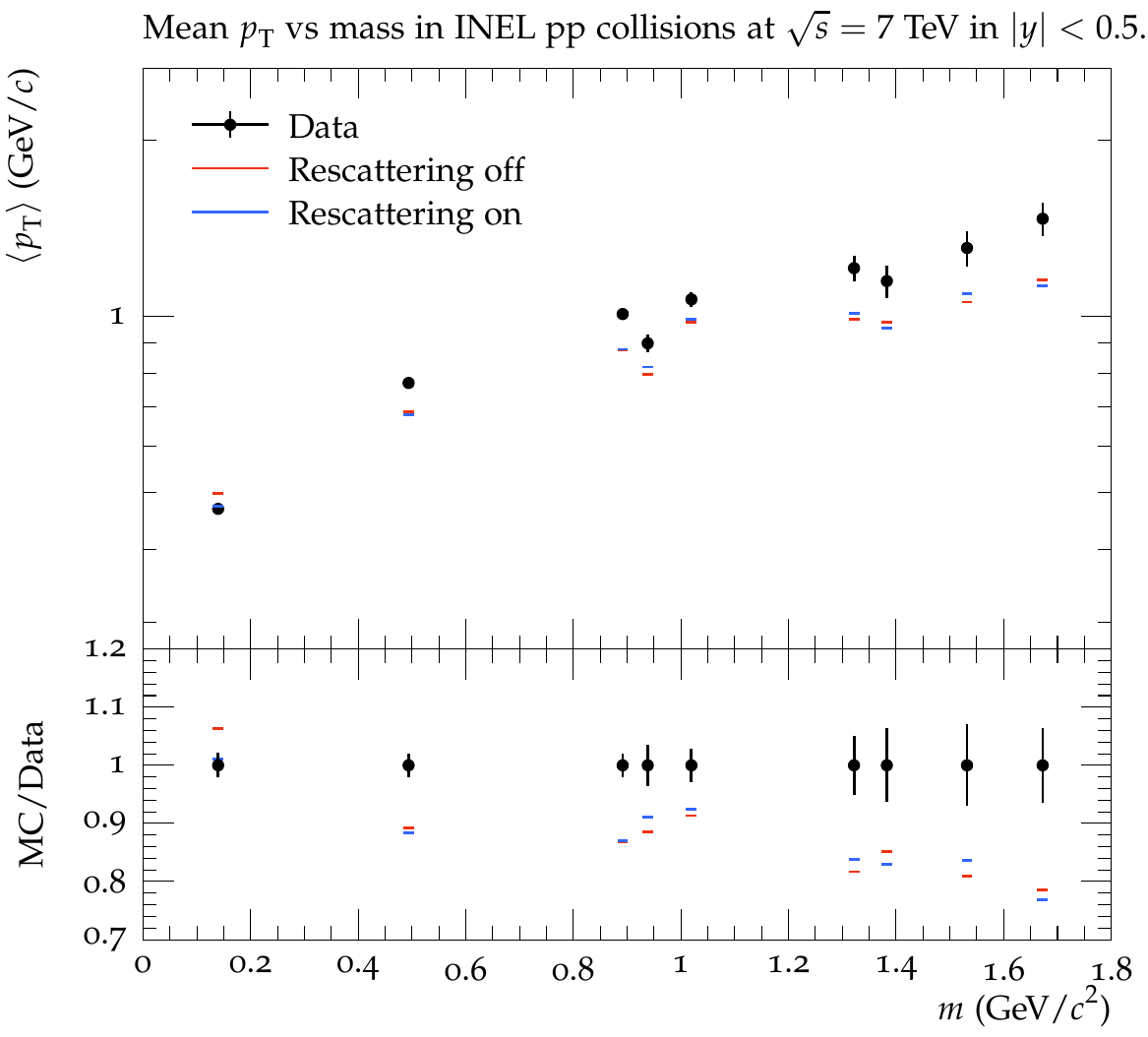}\\
(a)
\end{minipage}
\begin{minipage}[c]{0.49\linewidth}
\centering
\includegraphics[width=\linewidth]{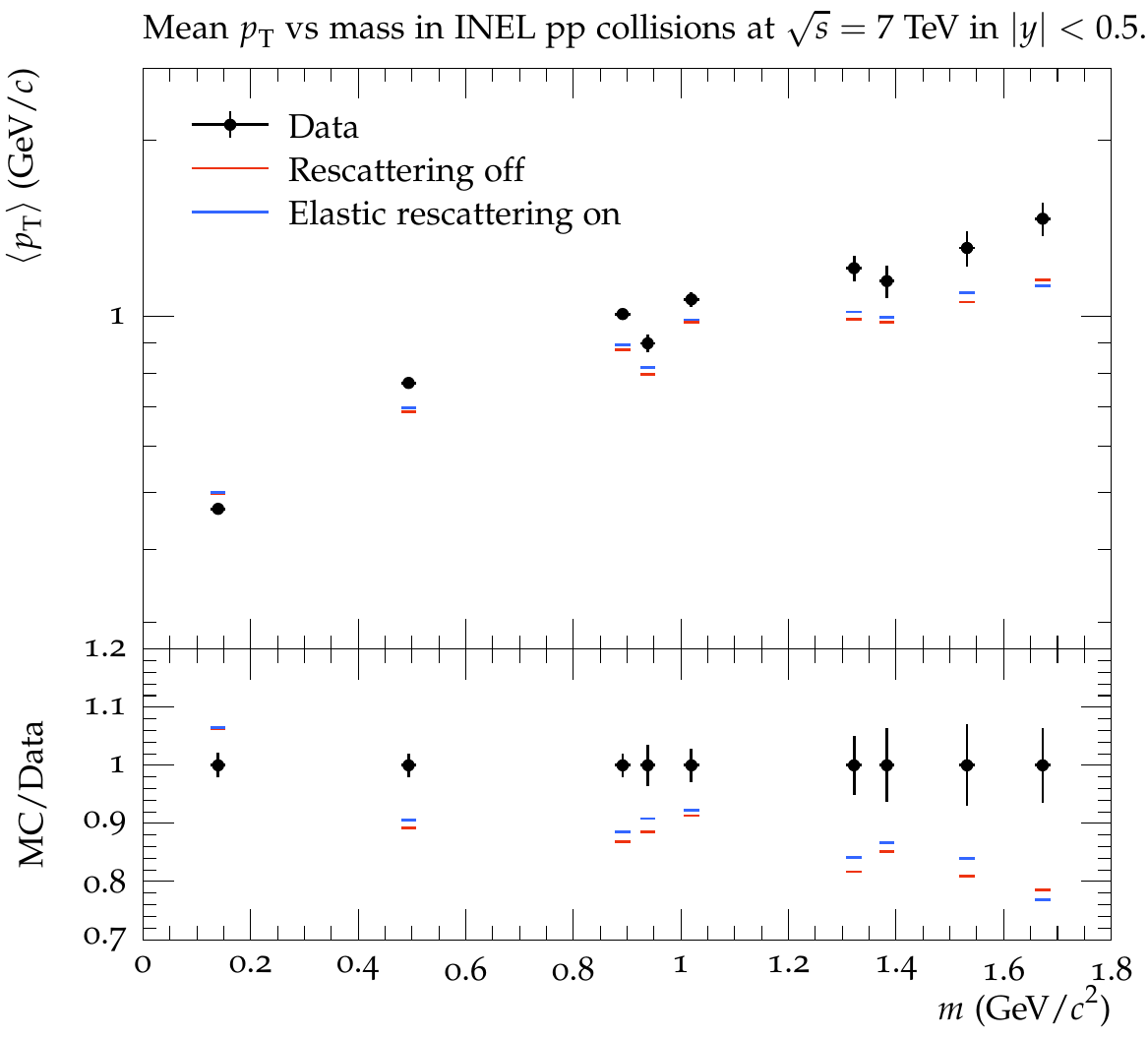}\\
(b)
\end{minipage}
\caption{Average $\pT$ for different particle species, ordered by mass, with data 
from ALICE \cite{Abelev:2014qqa, Adam:2015qaa}. The included particles are $\pi^\pm$, $\K^\pm$, $\K^*(892)^\pm$, $\p$, $\phi(1020)$, $\Xi^-$, $\Sigma^*(1385)^\pm$, $\Xi^*(1530)^0$ and $\Omega^-$. (a) Comparison of rescattering to no rescattering. (b) Comparison between the two when all rescatterings are forced to be elastic. Here we use the default \texttt{pT0Ref~=~2.28}, since elastic scattering does not change charged multiplicity.}
\label{fig:pTbyMass}
\end{figure}

\subsection{The effects of rescattering on transverse momentum speectra}

\figref{fig:pTspectra} shows the $\pT$ spectra for pions, Kaons and protons, with 
and without rescattering.
We see that rescattering gives a better fit to data for 
pions and protons, especially at low $\pT$, while for Kaons rescattering moves the $\pT$ spectrum away from data. The average $\pT$ for various particle species is shown 
in \figref{fig:pTbyMass}a, and here again there is an improvement for $\pi$ and $\p$, 
but a slight deterioration for $\K$. 

\begin{figure}[t!]
\begin{minipage}[c]{0.49\linewidth}
\centering
\includegraphics[width=\linewidth]{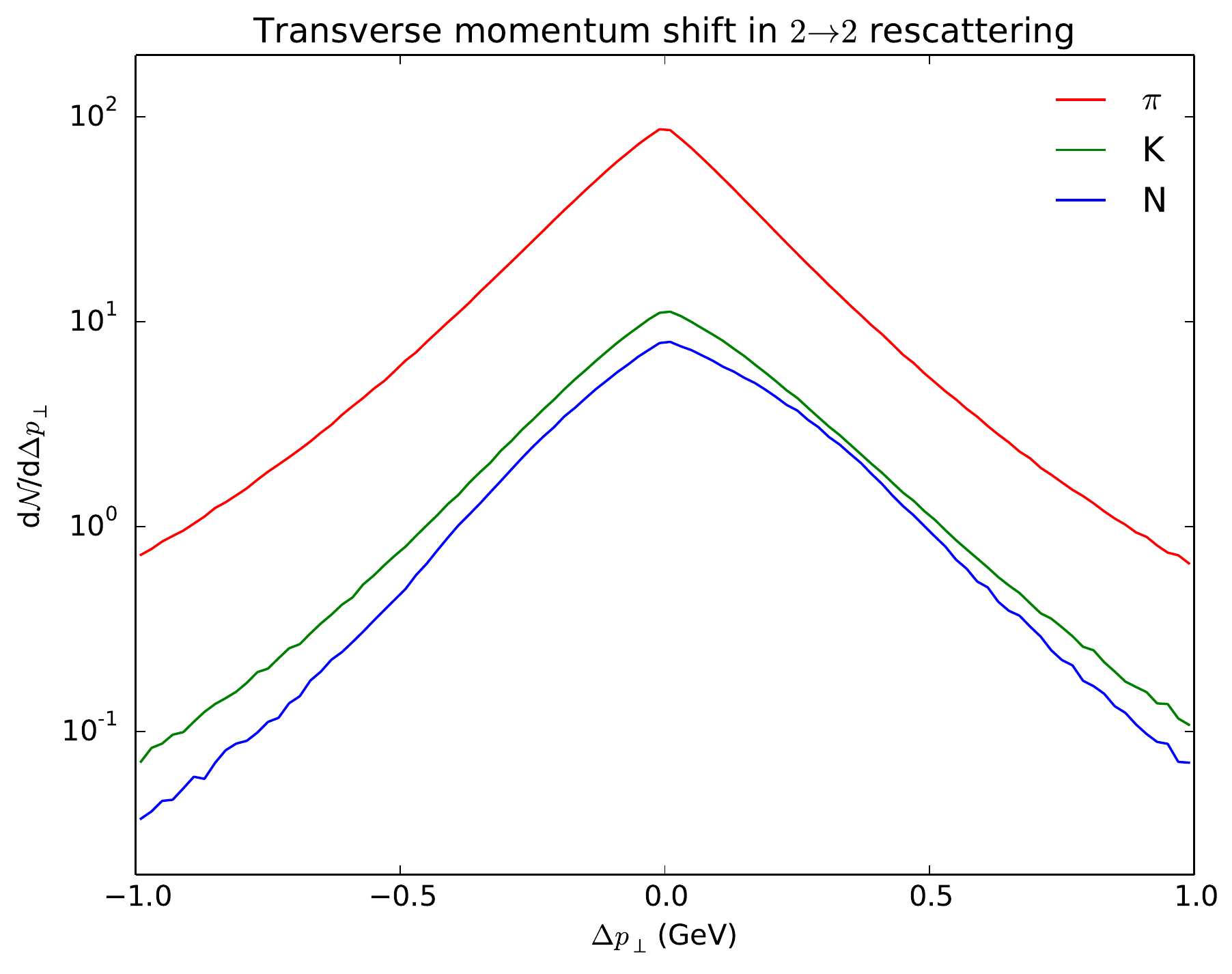}\\
(a)
\end{minipage}
\begin{minipage}[c]{0.49\linewidth}
\centering
\includegraphics[width=\linewidth]{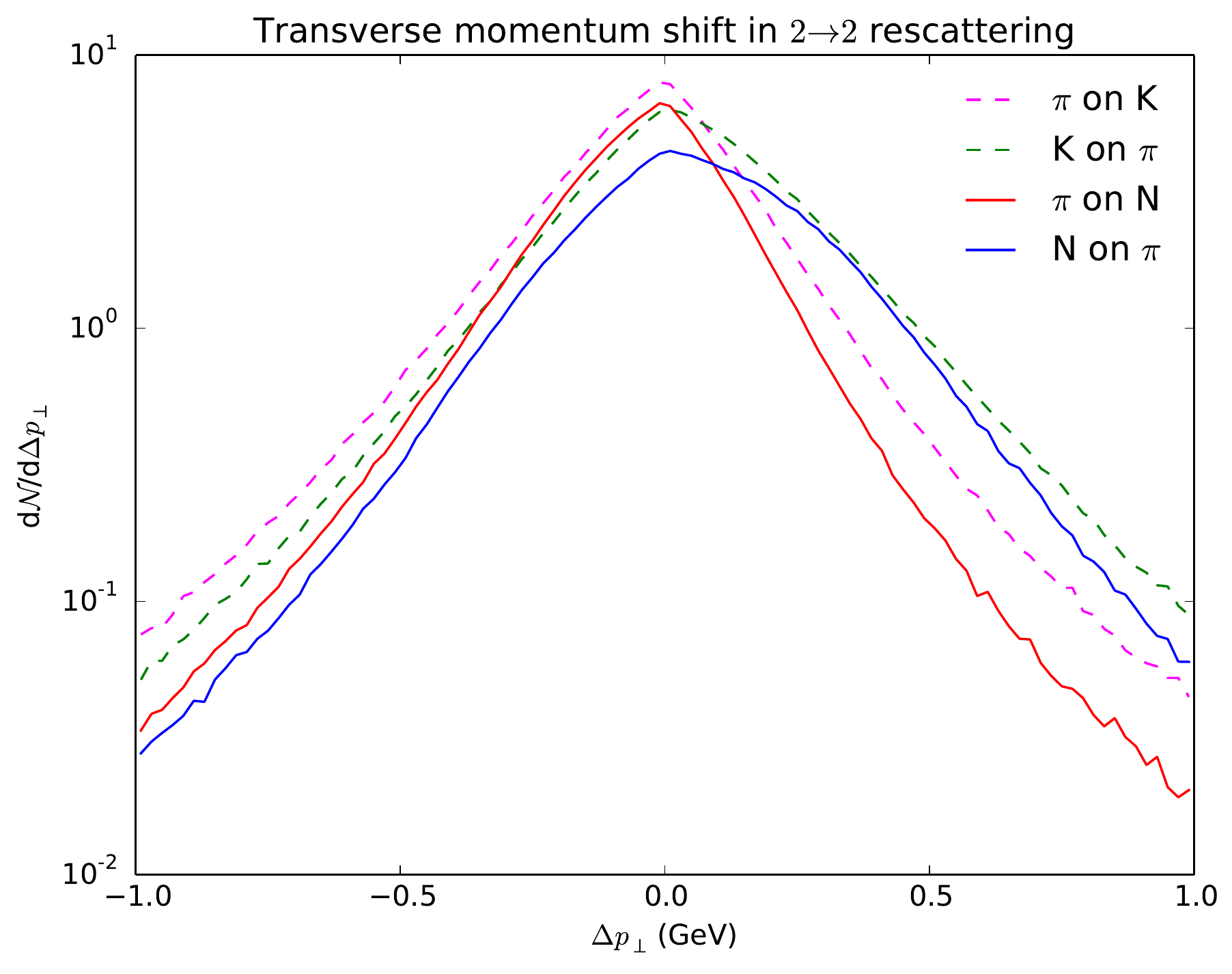}\\
(b)
\end{minipage}
\caption{Shift of transverse momentum by $2 \to 2$ elastic or resonant
processes, where positive numbers correspond to an increased $\pT$ in
the collision. (a) Inclusive shifts for $\pi$, $\K$ and $\N$ (including
antiparticles). (b) Shifts in $\K\pi \to \K\pi$ and $\N\pi \to \N\pi$ 
scatterings.}
\label{fig:pTshift}
\end{figure}

If we consider only elastic collisions, one would expect that rescattering should 
push lighter particles towards lower $\pT$ and heavier particles to higher $\pT$. This 
is because lighter particles generally move faster and will catch up with and push 
the heavier ones outwards, a phenomenon sometimes referred to as ``pion wind''. The actual momentum shifts in elastic rescatterings (including through resonances) is shown in \figref{fig:pTshift}.
Here we see a positive shift both for $\K$ and $\N$. This becomes more
apparent if one considers only $\K\pi \to \K\pi$ and $\N\pi \to \N\pi$
scatterings, \figref{fig:pTshift}b, where the heavier $\K / \N$ 
on the average gains $\pT$ at the expense of the lighter $\pi$.
A closer study reveals that the strongest $\pT$ shifts comes from resonance
production, i.e.\ $\K^*$ and $\Delta$ intermediate states. There are
two reasons for this. Firstly, these resonances give large cross 
sections in a mass range where the flux of colliding pairs is large
in the first place, and thus dominate over elastic scattering (in
the processes discussed here). Secondly, elastic scattering is peaked
in the forward direction, i.e.\ at small momentum transfers, while 
an $s$-channel spin 0 resonance decays isotropically in its rest frame.

In \figref{fig:pTbyMass}b, we look at $\left< \pT \right>$ shifts when only elastic scattering is permitted. 
Specifically, this is done by calculating each total cross section as before, but setting the elastic cross section equal to the total one (thus excluding elastic scattering through a resonance).
In this case, the $\langle \pT \rangle$ increases for all heavy particles except for $\Omega$, which is so rare so this can simply be explained by statistical fluctuations. 
For particles such as $\p$ and $\Sigma$, the change in $\left< \pT \right>$ is less than before, highlighting the fact that elastic scattering through a resonance gives the strongest momentum transfers.
(As a side note, an unexpected observation is that the average pion $\pT$ actually 
increases very slightly, which turns out to be a consequence of the narrow rapidity window 
$|y| < 0.5$ used in the experimental analysis; the average does decrease if all 
rapidities are included.)

So why then is the mean $\pT$ reduced for Kaons when inelastic interactions are allowed? The answer is that in processes classified as inelastic,
especially non-diffractive processes, we make a significant effort to ensure that at
least three particles are produced, so as to avoid the elastic channel.
Such interactions have to share the $\pT$ 
between more outgoing than incoming particles, which leads to a reduced average.
In principle, the opposite kind of interactions would be possible, where three (or more)
incoming particles could fuse to give two outgoing ones, presumably then with an 
increased $\pT$. We have not implemented these kinds of processes in the first version
of our framework, but their potential effect on the Kaon $\pT$ spectrum should make them a priority in future work.

Another observation from \figref{fig:pTbyMass}a is that the mean $\pT$ of $\Sigma^*$ is also reduced. In addition to the aforementioned effect of $2 \to n$ scattering, we have also observed that resonances formed during rescattering tend to have a lower $\pT$ than those produced directly from string fragmentation. 
From phase space considerations,
it is less likely for two random high-$\pT$ particles to have an invariant mass in the resonance
range than for two low-$\pT$ ones. The effect is especially large where the mass
difference between the resonance and the particles forming it is small, such as for
the $\Sigma^*$ baryons. These particles still tend to gain $\pT$ when they
themselves participate in rescattering, as we see in \figref{fig:pTbyMass}b.

\begin{figure}[t!]
\begin{minipage}[c]{0.49\linewidth}
\centering
\includegraphics[width=\linewidth]{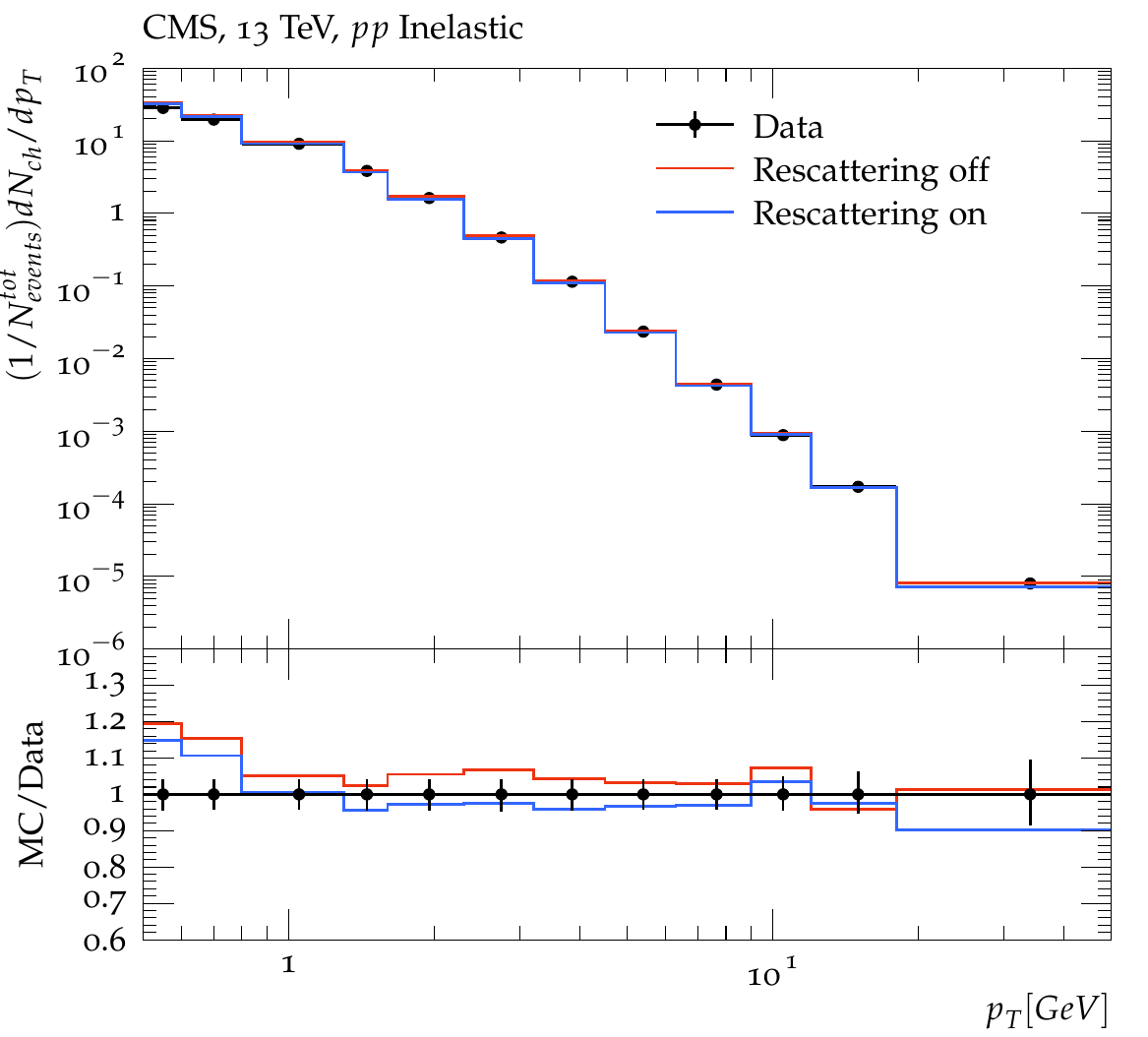}\\
(a)
\end{minipage}
\begin{minipage}[c]{0.49\linewidth}
\centering
\includegraphics[width=\linewidth]{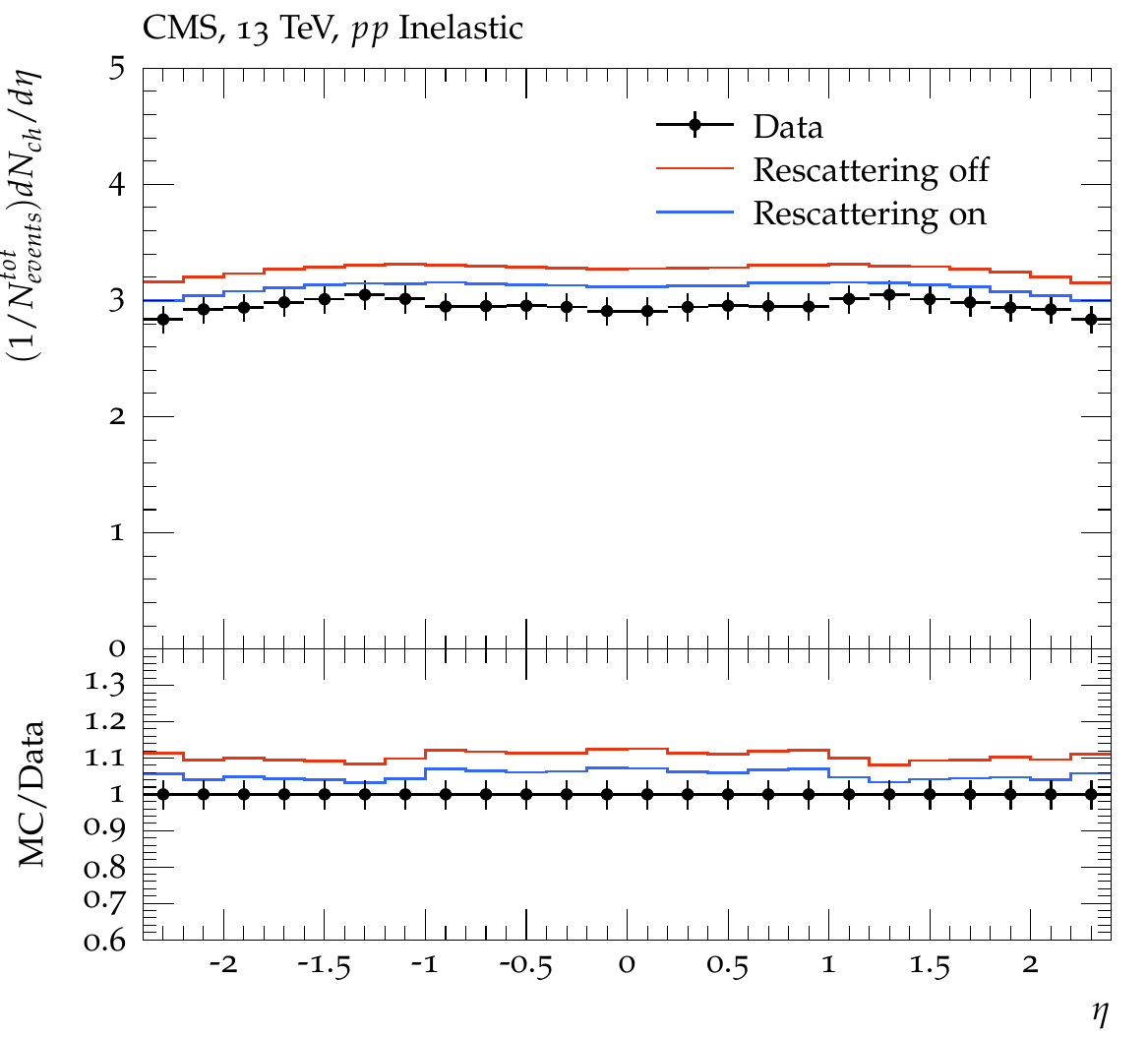}\\
(b)
\end{minipage}
\caption{$\pT$ and $\eta$ spectra compared with data from CMS \cite{Sirunyan:2018zdc}. Charged particles with $\pT > 500$ MeV and $|\eta| < 2.4$ are considered.}
\label{fig:pTetaCMS}
\end{figure}

\begin{figure}[t!]
\centering
\begin{minipage}[c]{0.49\linewidth}
\centering
\includegraphics[width=\linewidth]{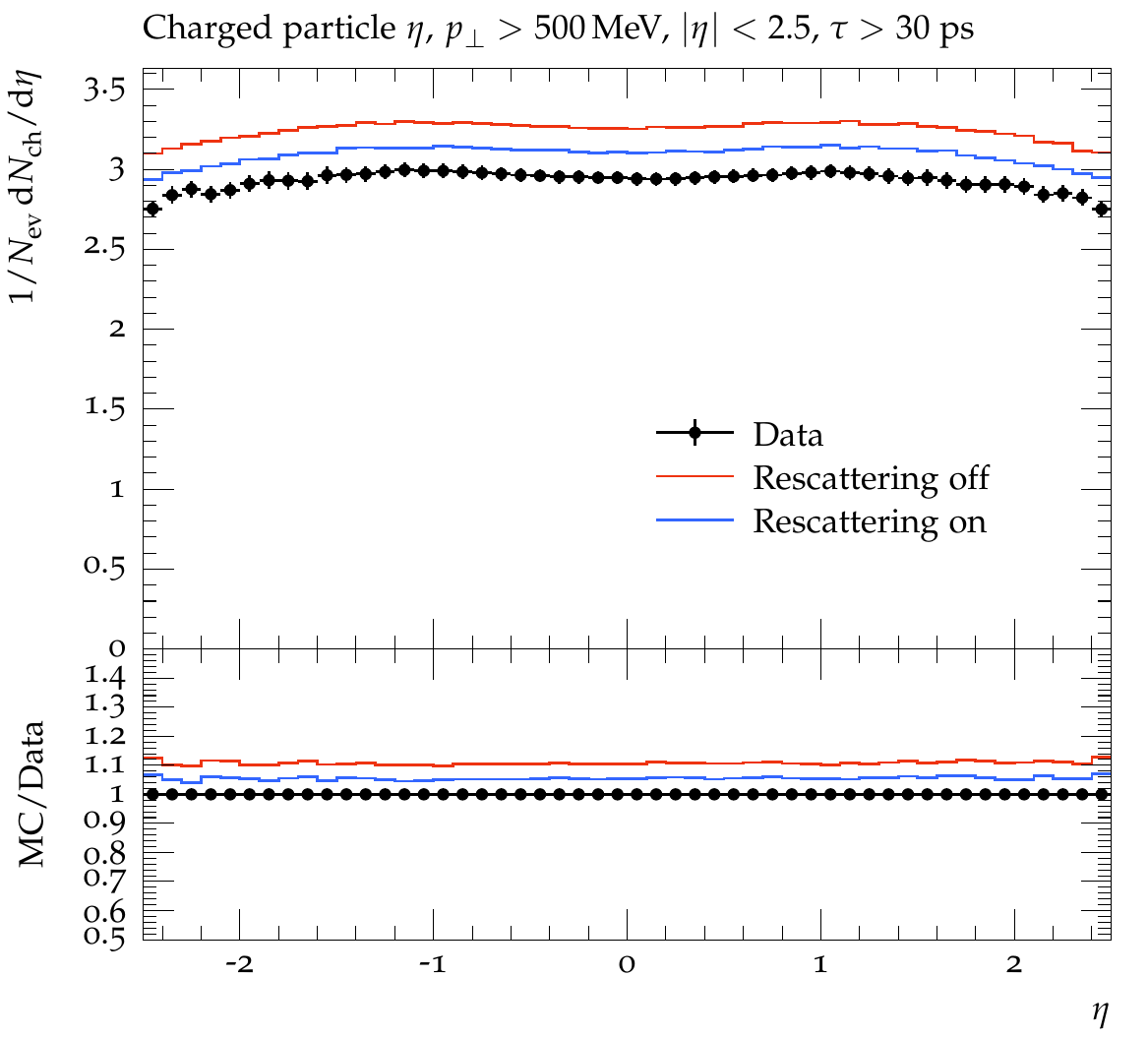}\\
(a)
\end{minipage}
\begin{minipage}[c]{0.49\linewidth}
\centering
\includegraphics[width=\linewidth]{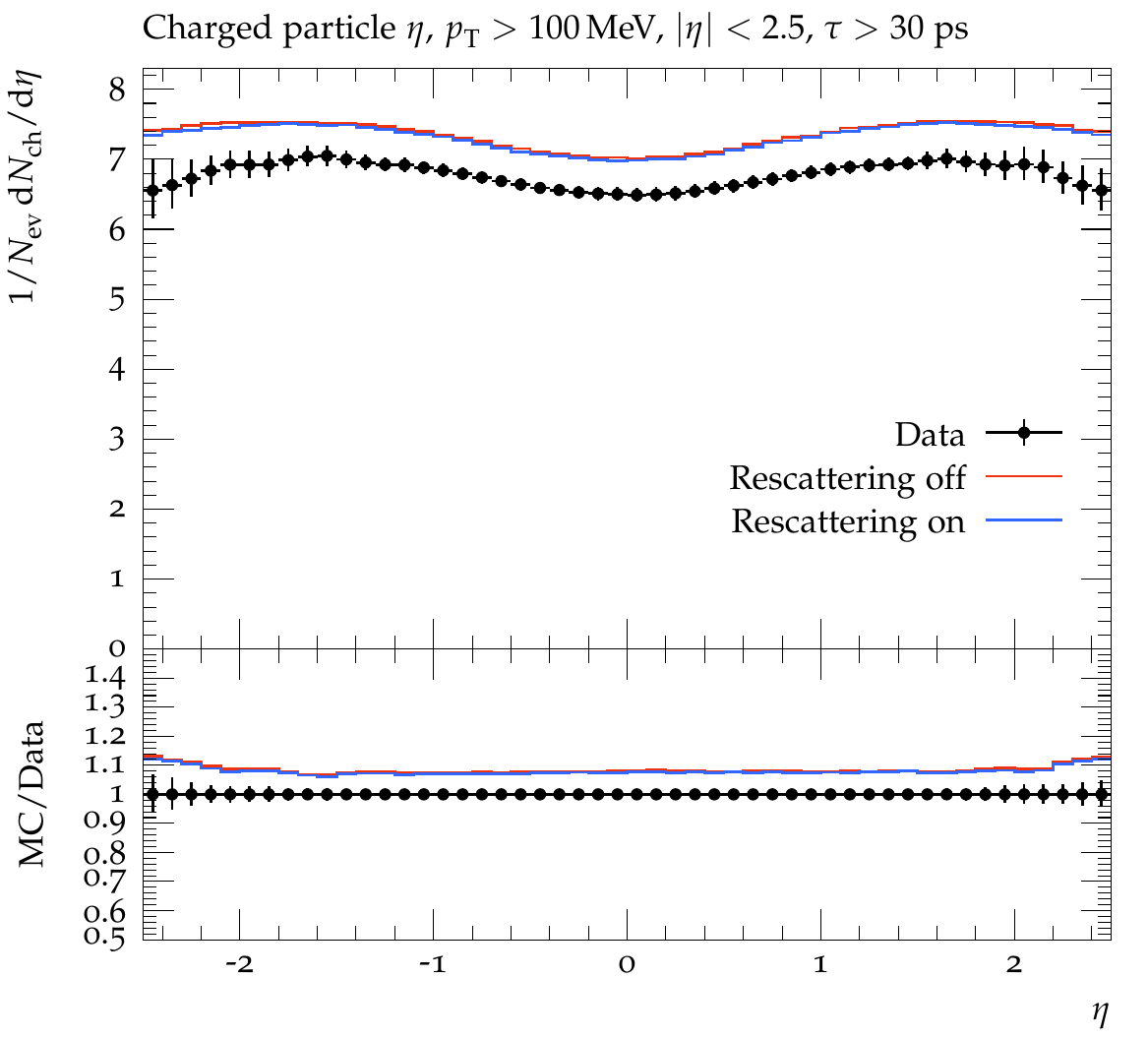}\\
(b)
\end{minipage}
\caption{Charged particle $\eta$ spectra compared with data from ATLAS \cite{Aad:2016mok,Aaboud:2016itf},  with cuts (a) $\pT > 500$ MeV, and (b) $\pT > 100$ MeV.}
\label{fig:etaATLAS}
\end{figure}

The total $\pT$ spectrum for all charged particles is shown in \figref{fig:pTetaCMS}a, and is 
improved overall by rescattering. The charged-particle pseudorapidity spectra in
\figref{fig:pTetaCMS}b and \ref{fig:etaATLAS}a show that when a cut $\pT > 500$~MeV is used, 
rescattering shifts the spectrum down by an approximately fixed amount, to a 
better agreement with data. However, this improvement is not visible in \figref{fig:pTetaCMS}b, where the cut is $\pT > 100$~MeV. 
This suggests that the ``true'' pseudorapidity spectra are mostly unaffected by rescattering, but because of $\pT$ shifts, rescattering has an indirect effect on the observed spectrum.
The takeaway from this is that data affected by low-$\pT$ particle production are 
likely to be better described when rescattering is included.

In summary, rescattering does what it is expected to in elastic scattering, 
i.e.\ slows down lighter hadrons and speeds up heavier ones. The disappointing aspect 
is that we have observed other mechanisms that work in the other direction. Finding 
ways to compensate for these effects should be addressed in future work.

\subsection{Model dependence of transverse momentum spectra} \label{subsec:compareMinMax}

Given the central role of the $\pT$ spectra, it is highly relevant to understand how 
sensitive they are to rescattering model variations. To this end, we can compare the 
default rescattering scenario with the two alternatives listed in \tabref{tab:MinMaxParameters}.
These two are selected to minimize or maximize the number of rescatterings, within 
reasonable extremes for each relevant setting. 

\begin{figure}[t!]
\begin{minipage}[c]{\linewidth}
\centering
\includegraphics[width=0.48\linewidth]{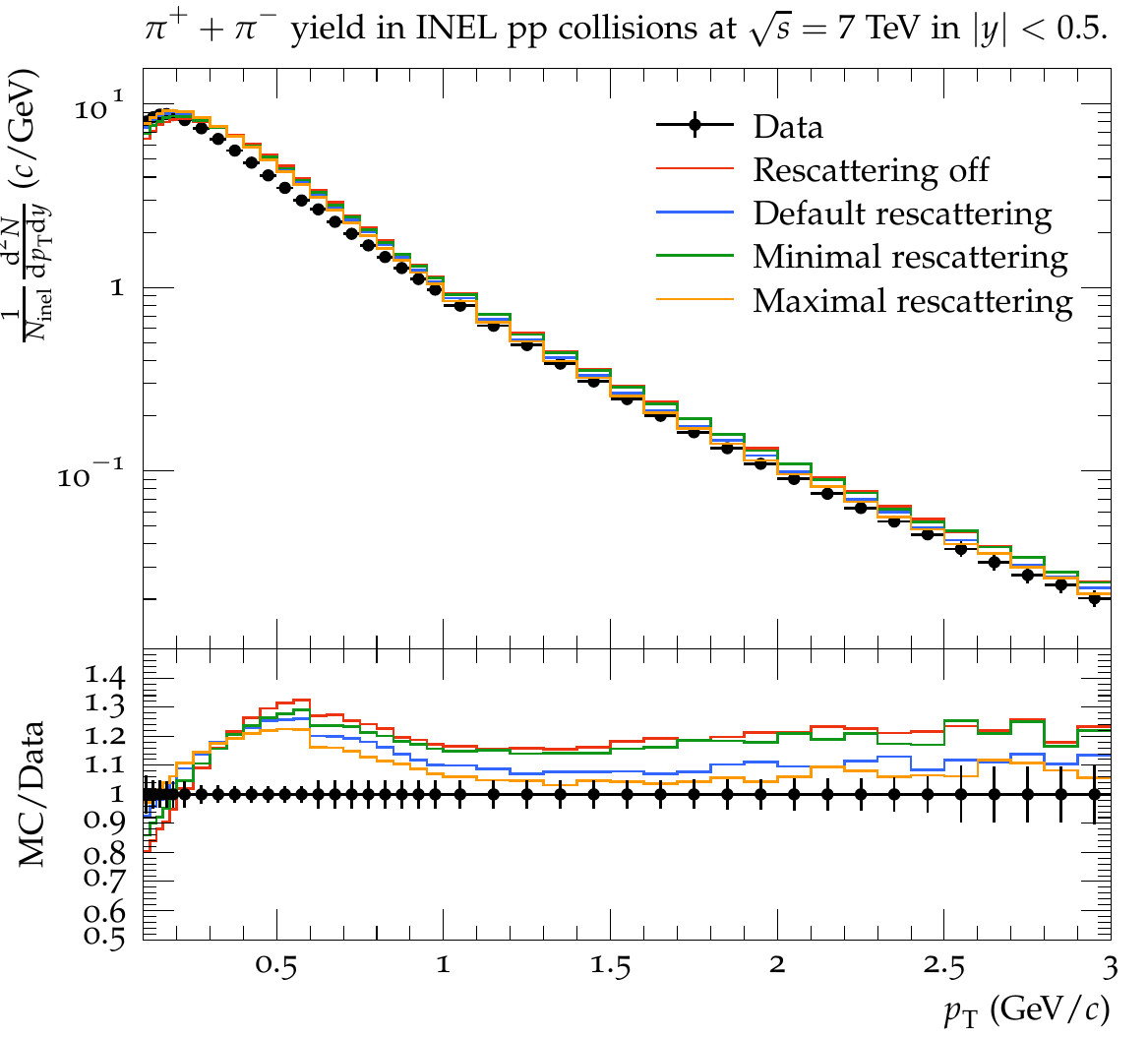}
\includegraphics[width=0.48\linewidth]{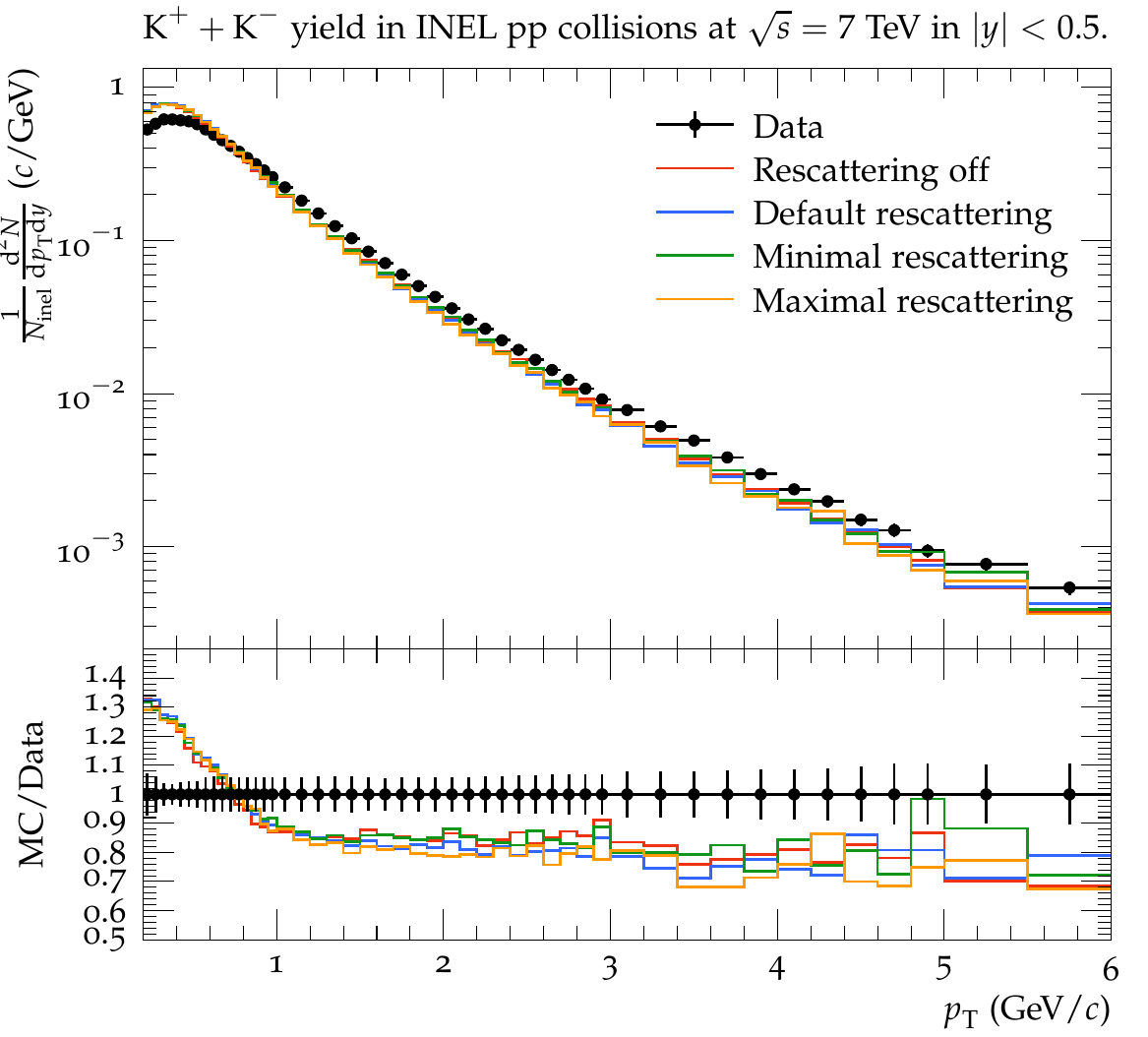}
\end{minipage}

\begin{minipage}[c]{\linewidth}
\centering
\includegraphics[width=0.48\linewidth]{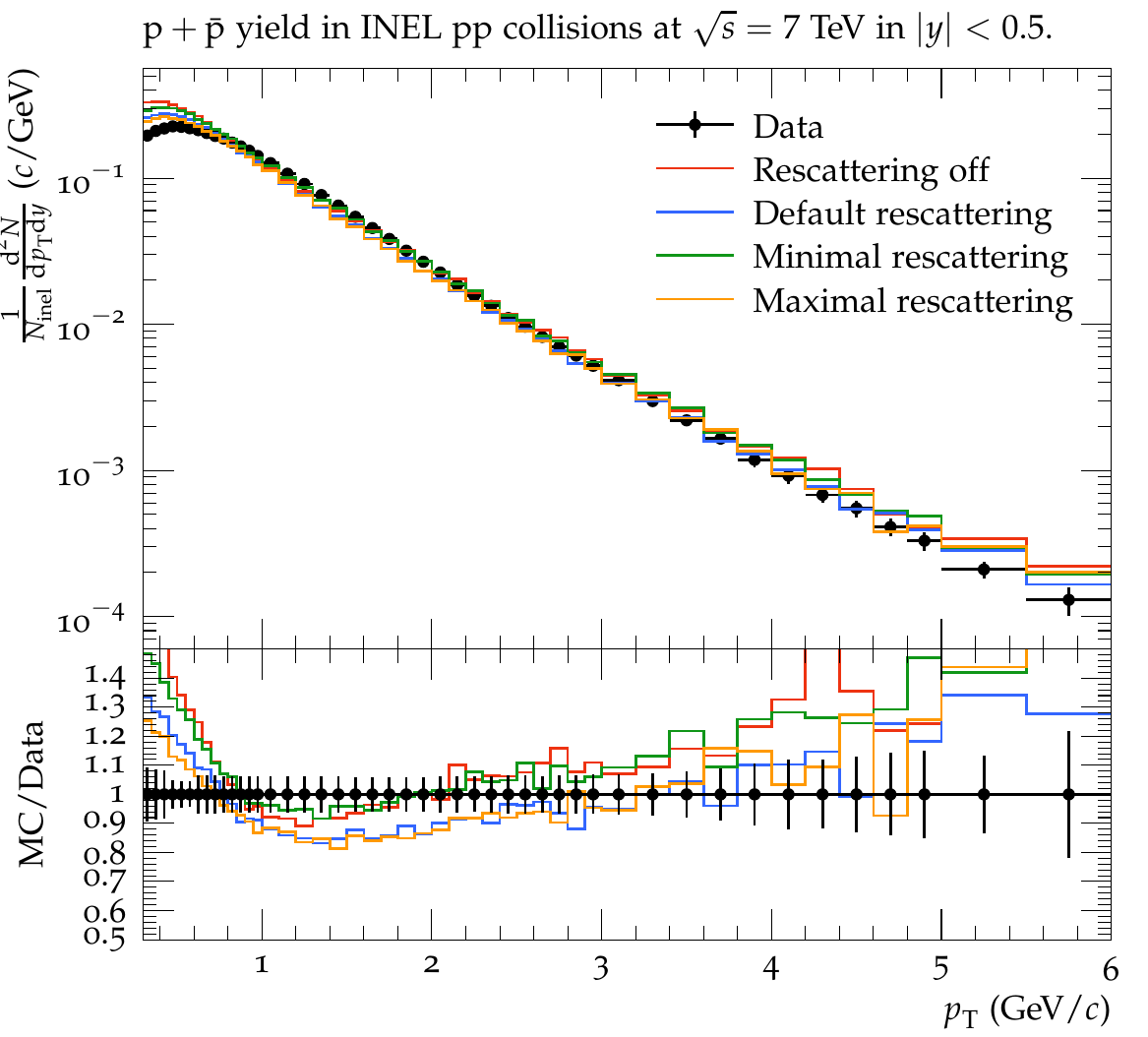}
\includegraphics[width=0.48\linewidth]{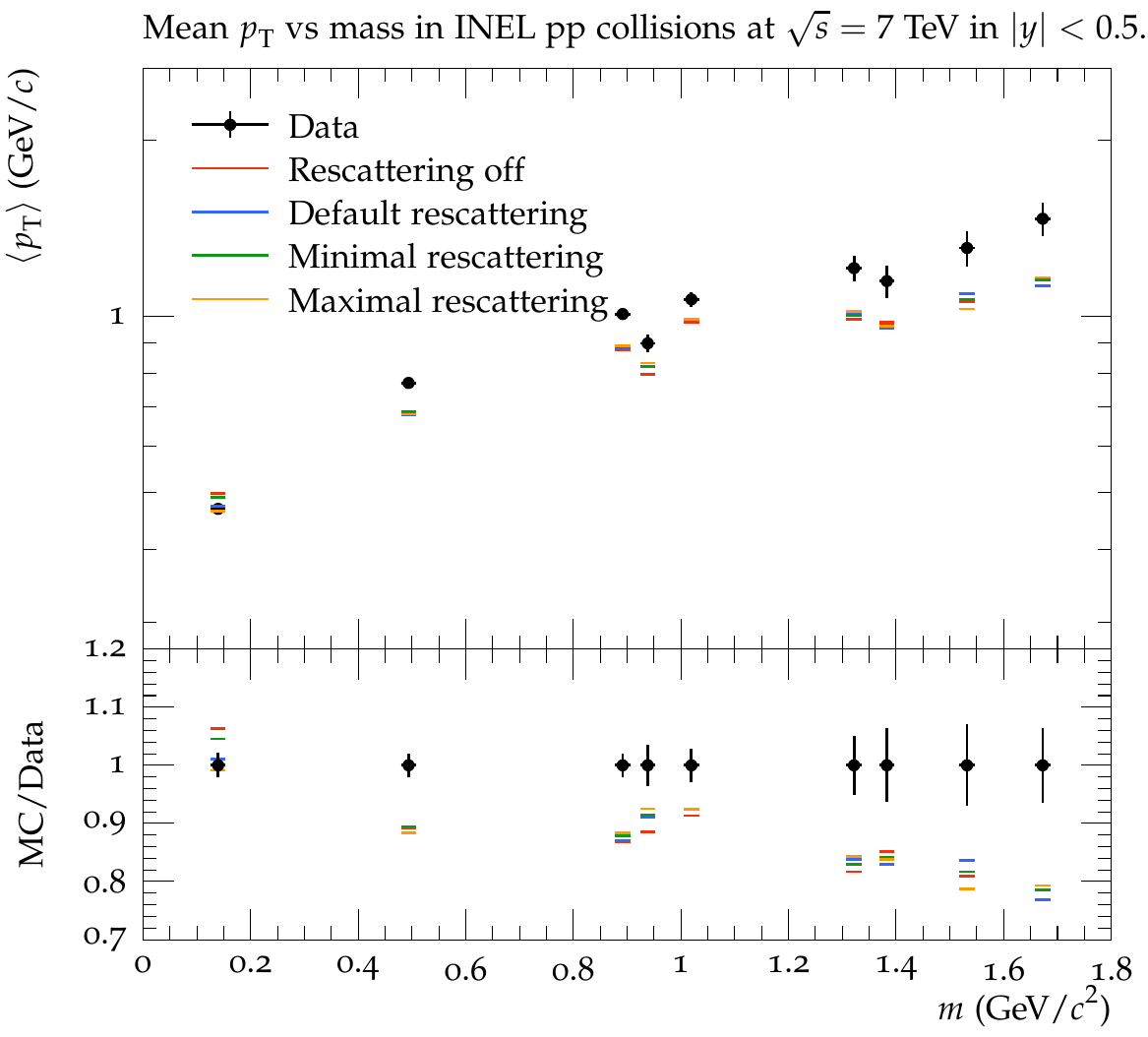}
\end{minipage}

\caption{$\pT$ spectra for $\pi^\pm$, $\K^\pm$ and $\p/\pbar$ and average $\pT$ for 
various particles, for different parameter configurations.}
\label{fig:pTspectra-minmax}
\end{figure}
 
The results are shown in \figref{fig:pTspectra-minmax}.
What we observe is that the effects on the $\pT$ spectra tend to scale with the 
amount of rescattering. This is especially clear for $\pi$ and $\p$, where the
minimum/maximum amount of rescattering give smaller/larger effects than the default 
values, respectively. At the same time, the maximum setup gives a relatively small 
further improvement over the default rescattering one. It is therefore meaningful to 
stay with the default scenario, rather than trying to use more extreme choices to
come closer to data.

\subsection{The thermal model alternative}

The rate of $\q\qbar$ string breaks is traditionally assumed to involve
a suppression factor $e^{-\pi m_{\perp \q}^2/\kappa}$: since the string 
does not contain any local concentrations of mass, a quark needs to tunnel 
out as a virtual particle until it has ``eaten up'' enough string length to 
correspond to its transverse mass \cite{Andersson:1983ia}. This gives a 
Gaussian $\pT$ spectrum to quarks and, by addition, to hadrons. 
The derivation is done for a single string in isolation, however, whereas 
the reality at hadron colliders is that the typical event contains several 
more-or-less overlapping strings. This may modify the primary particle 
production processes, which set the starting stage for the continued 
rescattering and decay processes we have considered in this article. 
Empirically, an exponential spectrum $\exp(-m_{\perp\mathrm{had}}/T)$ was 
early on proposed as a parameterization of hadron collision data, where $m_{\perp\mathrm{had}}$ is the transverse hadron mass
and $T$ could be associated with a temperature e.g.\ in the Hagedorn approach 
\cite{Hagedorn:1965st,Hagedorn:1970gh,Hagedorn:1983wk}. Interestingly, an 
effectively exponential fall-off could arise also starting from the Gaussian 
one, by assuming that the string tension is fluctuating along the string
length, also in the absence of other strings \cite{Bialas:1999zg}.

Based on such ideas, a ``thermal model'' option has been included as an 
alternative in \textsc{Pythia} \cite{Fischer:2016zzs}. Unlike purely 
statistical models, however, it is strictly based on the string model, with
local flavour and $\pT$ conservation. To this end, each $\q\qbar$ breakup
is associated with a (modified Bessel) $\pT$ distribution such that the
two-dimensional convolution results in an $\exp(-p_{\perp\mathrm{had}}/T)$ 
spectrum. In each fragmentation step, an old $\q$ flavour is always known 
when the new one is selected and a new hadron is formed out of the two. 
Each new quark and hadron possibility is assigned a relative weight 
$\exp(-m_{\perp\mathrm{had}}/T)$, times relevant spin and symmetry factors,
and these weights are used to make the random choice. The relative rate of
diquark/baryon production requires a free parameter, while an additional 
$\s$-quark suppression factor is needed to achieve better agreement with
observed production rates. The suppression of multistrange hadrons is
underestimated, however, whereas the standard string model overestimates it,
suggesting that ``the truth'' may lie somewhere in between.

\begin{figure}[t!]
\begin{minipage}[c]{\linewidth}
\centering
\includegraphics[width=0.48\linewidth]{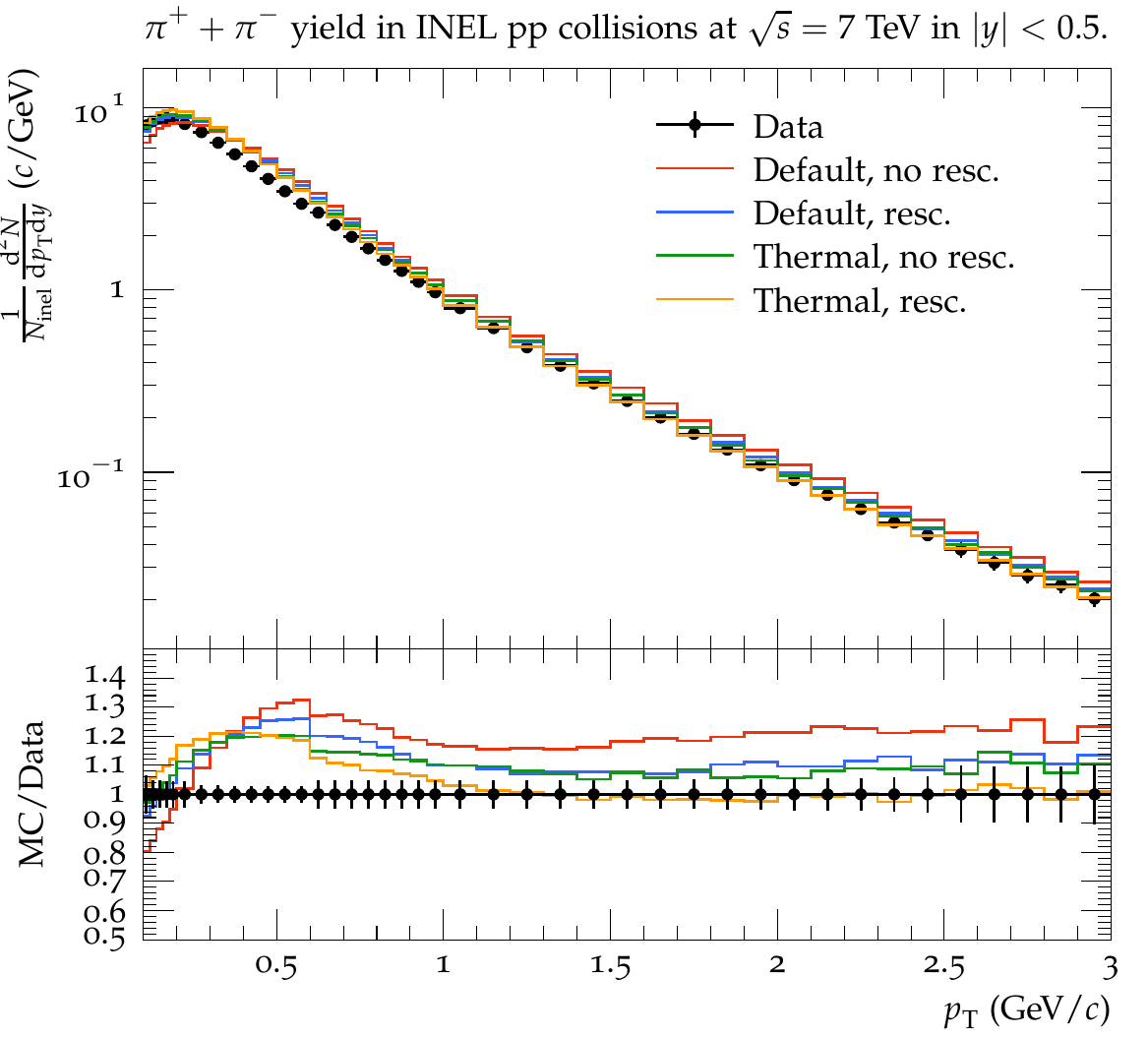}
\includegraphics[width=0.48\linewidth]{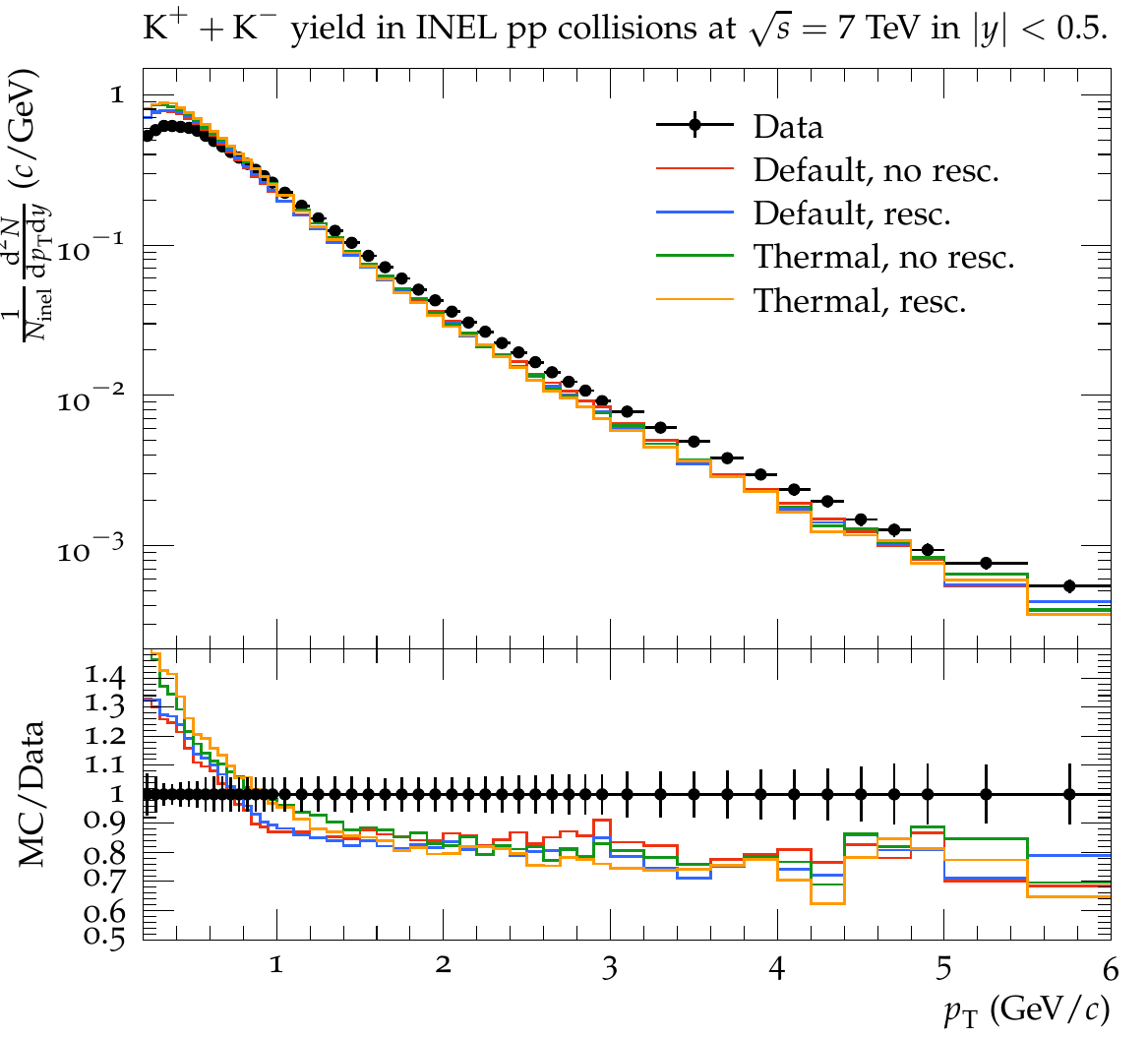}
\end{minipage}

\begin{minipage}[c]{\linewidth}
\centering
\includegraphics[width=0.48\linewidth]{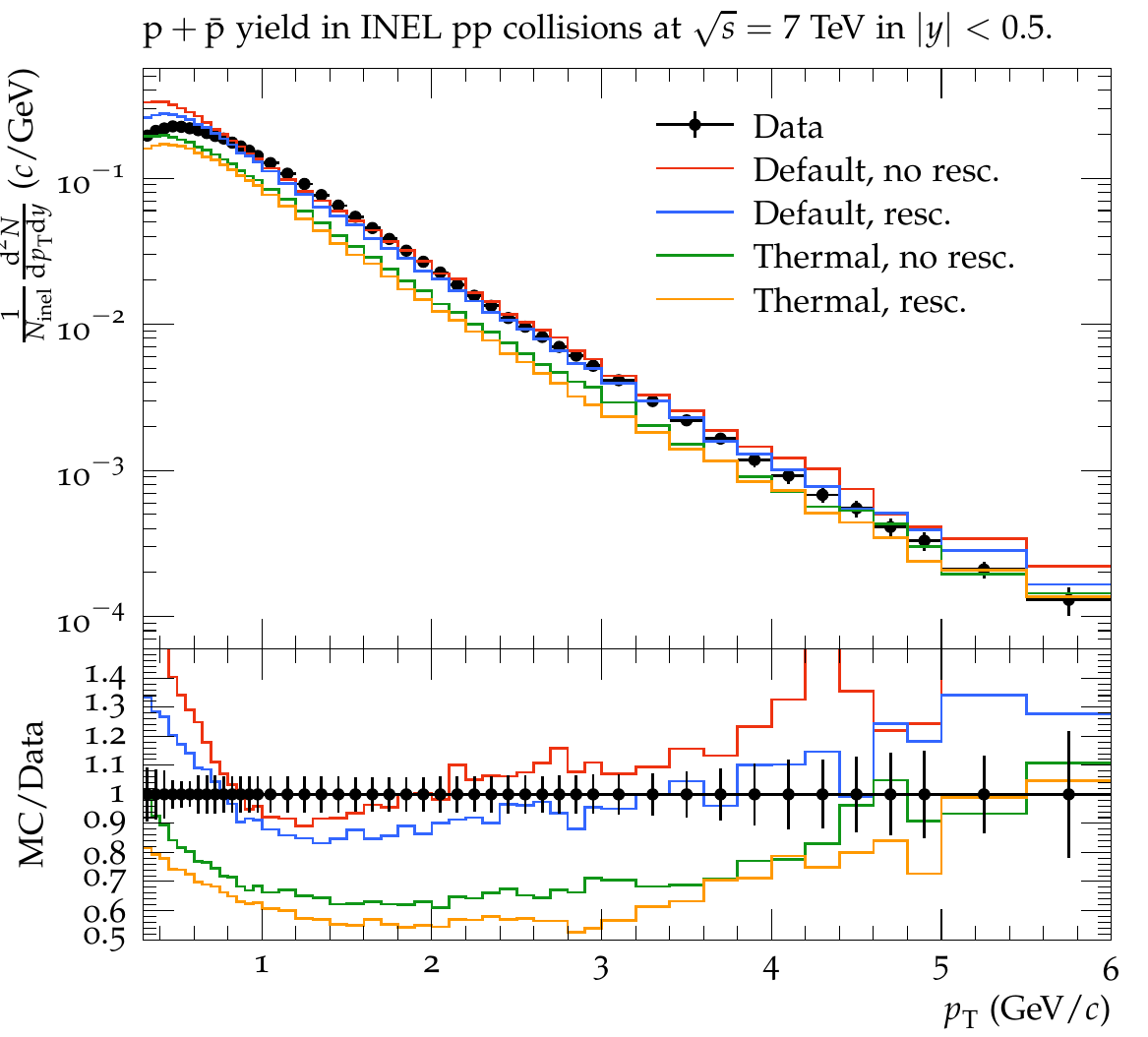}
\includegraphics[width=0.48\linewidth]{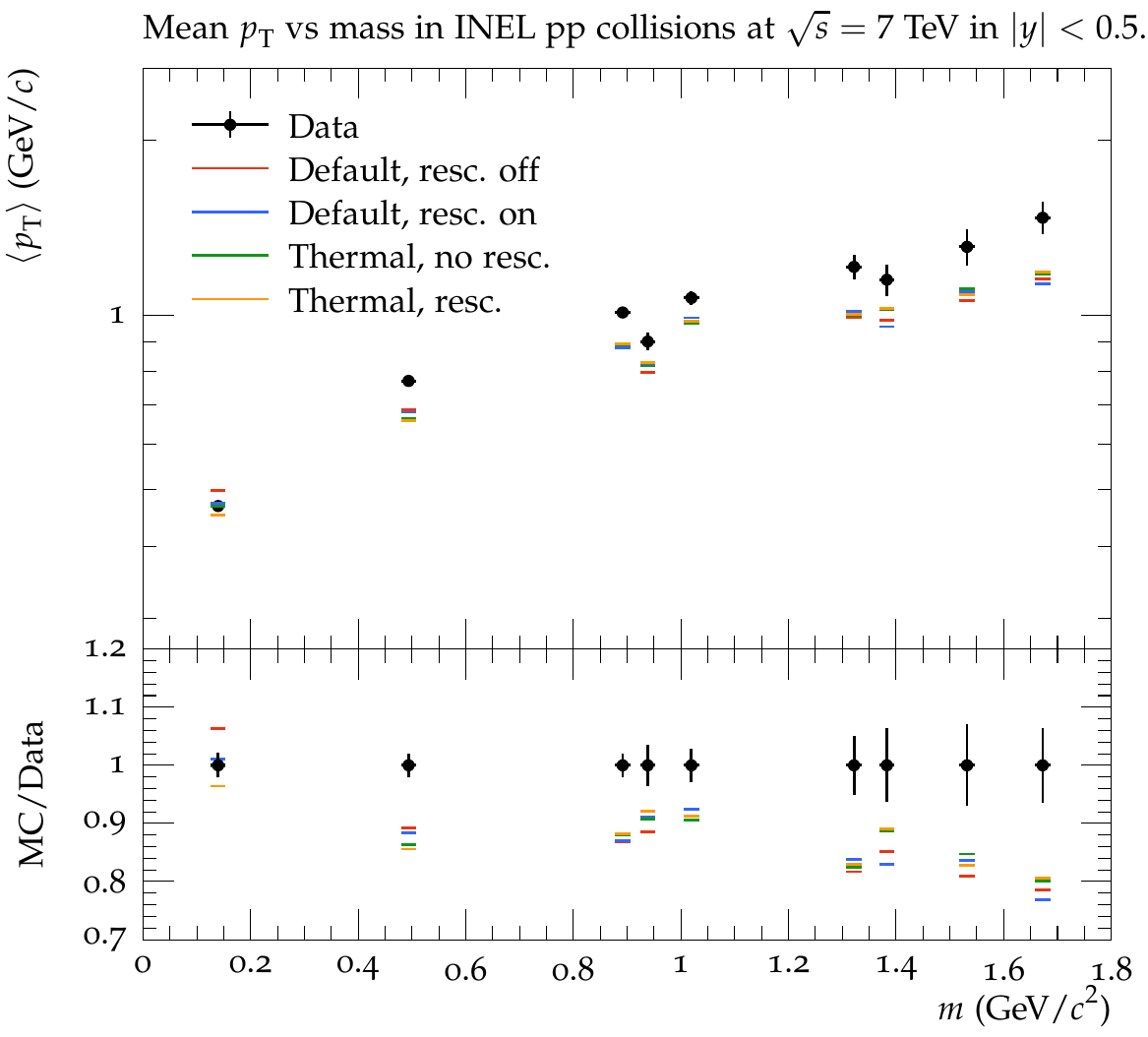}
\end{minipage}

\caption{$\pT$ spectra for $\pi^\pm$, $\K^\pm$ and $\p/\pbar$ and average $\pT$ 
for various particles, with comparing the Gaussian to the thermal model. When using the thermal model, \texttt{pT0Ref} has been tuned to 2.47 without rescattering and 2.52 with rescattering on, in order to maintain the correct $n_\mathrm{charged}$.}
\label{fig:pTspectra-thermal}
\end{figure}

A key aspect of the $\exp(-m_{\perp\mathrm{had}}/T)$ weight is that heavier 
primary hadrons obtain a larger $\langle \pT \rangle$ than lighter ones. 
While it does enhance low-$\pT$ pion production and deplete ditto baryon one, 
relative to the traditional string model, the effects are not large enough
to explain the data \cite{Fischer:2016zzs}. It is therefore interesting to
combine the thermal model with rescattering, to check whether the two 
together give a larger combined improvement than each individually. 
The results of this comparison are shown in \figref{fig:pTspectra-thermal}, 
where the Gaussian model is compared to the thermal model, both with and 
without rescattering. The effects of the thermal model are similar to the effects of rescattering, with an improvement for the mean $\pT$ of pions and protons and a deterioration for Kaons. For pions, the correction from the combination of the two in fact overshoots the $\langle \pT \rangle$ data, so that either of them individually gives a better result than the two combined, even if the pion $\pT$ spectrum itself looks rather reasonable. We also see that the $\pT$ spectrum for protons is less accurate, especially at higher $\pT$s. For these reasons, the results of using the thermal model are not particularly encouraging, at least not without a more thorough retuning.

\subsection{Close-packing}

Apart from the possibility of a randomly fluctuating string tension,
one may also expect systematic effects on the tension in a denser 
string environment, which can be modelled in different ways. 
One option implemented in \textsc{Pythia} is that
of colour ropes \cite{Bierlich:2014xba}, wherein several more-or-less
parallel strings can fuse into a ``rope''. The combined colour 
charge of this rope, as given by the Casimir operator, then gives a
scaling-up factor applied to the string tension. When the rope breaks, 
the difference in charge before and after the break gives the effective
charge involved in that $\q\qbar$ production step. The other option is 
based on the assumption that a close-packing of strings gives them a
smaller transverse area each, but preserves their separate identities
\cite{Fischer:2016zzs}. Also in this option the string tension is 
increased, but in principle as a smooth function of the amount of 
squeezing rather than in the discrete steps of the rope. In practice, 
there need not be any big difference between these two options, but in 
this study we choose the second one for simplicity. 

In this model, the creation of a new hadron is begun by an exploratory step
ahead, so that the number of strings overlapping the rapidity range 
of the intended next hadron can be estimated. This local string number 
is then raised to some (tuned) power to give a rescaling factor for the 
string tension. To this basic picture some damping is introduced for 
particle production at large $\pT$, which typically occurs at larger 
transverse radii, away from the denser region. Note that the current
implementation predates the introduction of space--time coordinates
for the hadronization process, such that there now is room for 
improvements, but not ones that are likely to give a qualitatively
changed behaviour for the properties studied here. 

\begin{figure}[t!]
\begin{minipage}[c]{\linewidth}
\centering
\includegraphics[width=0.48\linewidth]{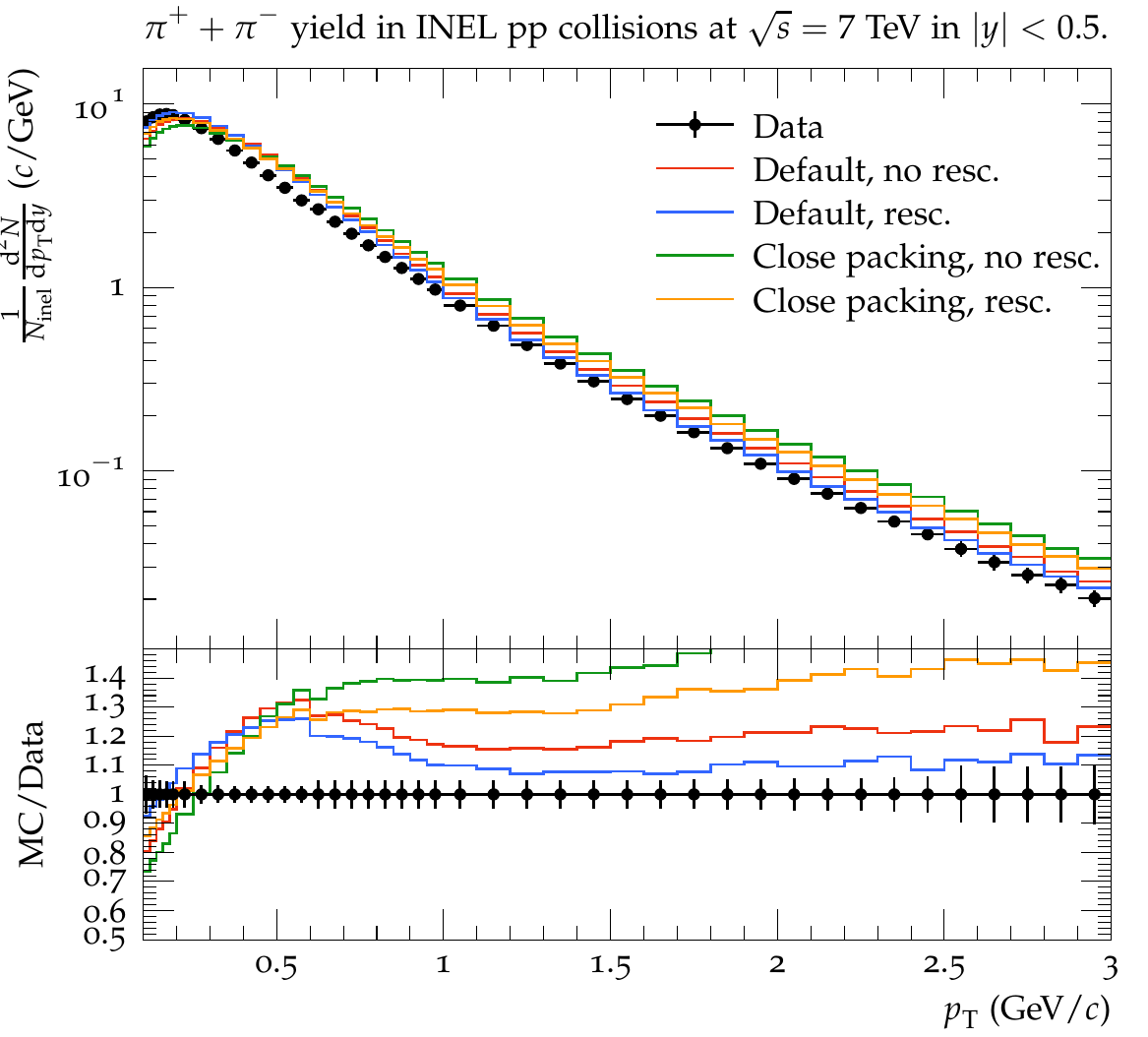}
\includegraphics[width=0.48\linewidth]{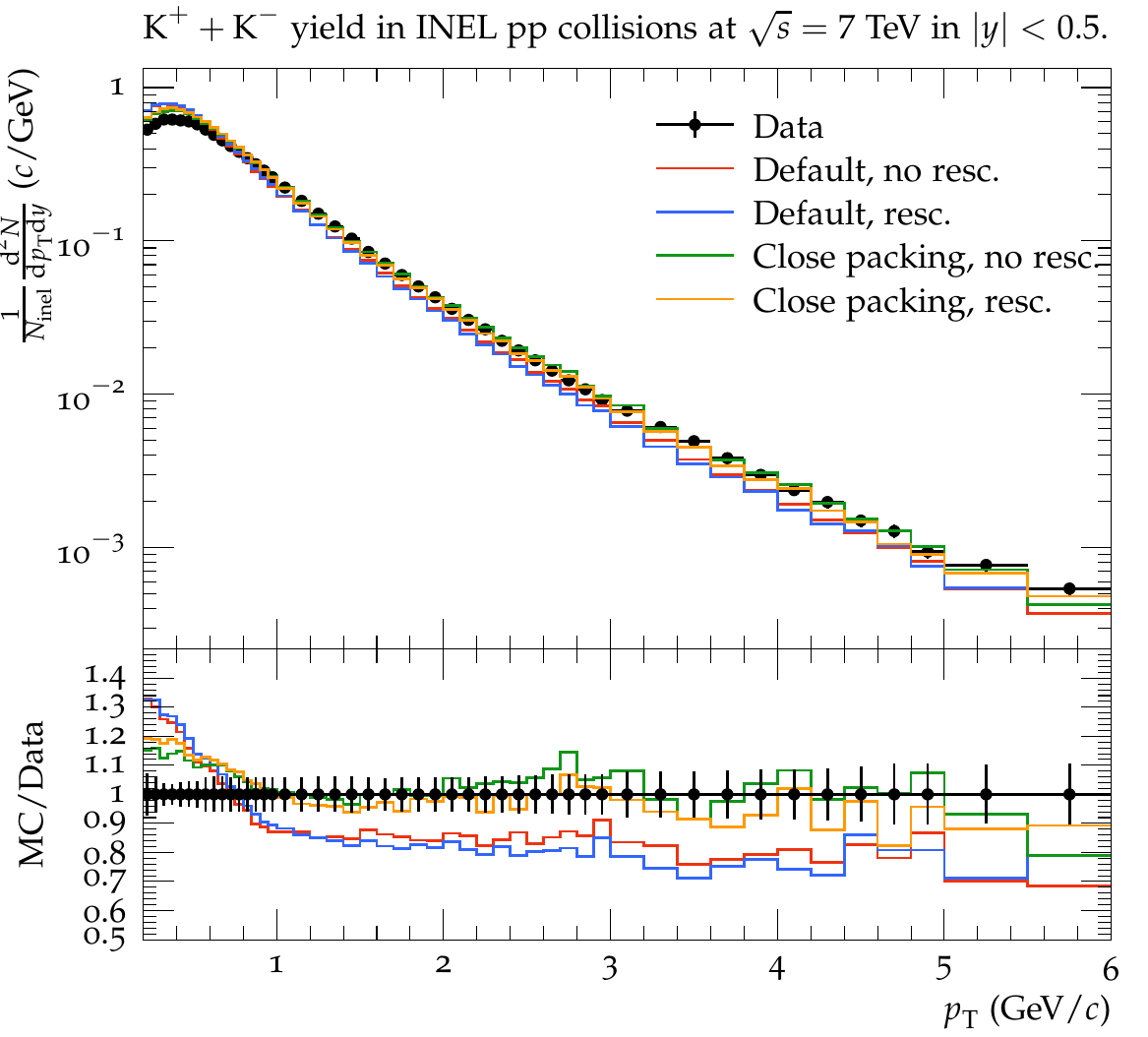}
\end{minipage}

\begin{minipage}[c]{\linewidth}
\centering
\includegraphics[width=0.48\linewidth]{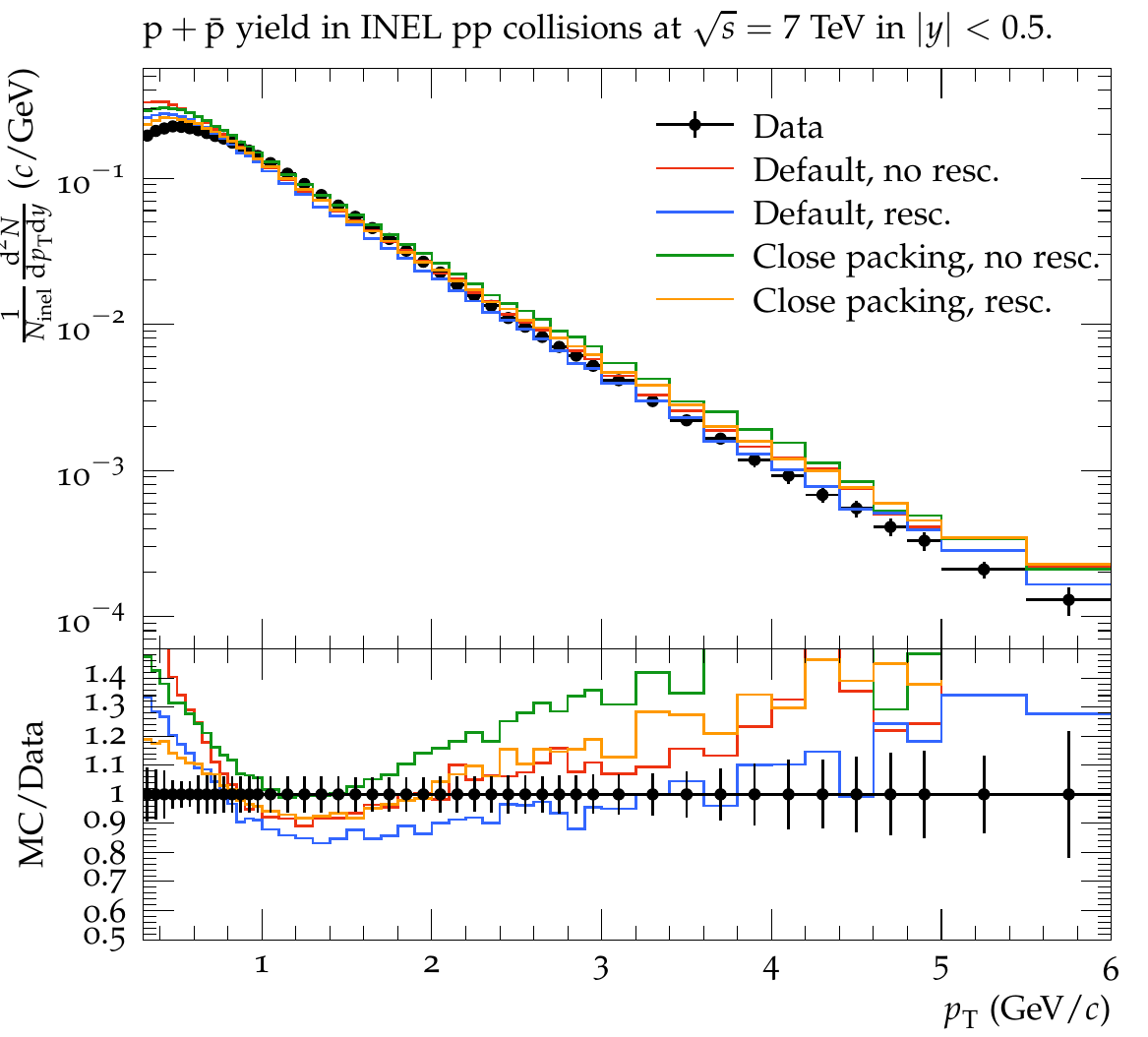}
\includegraphics[width=0.48\linewidth]{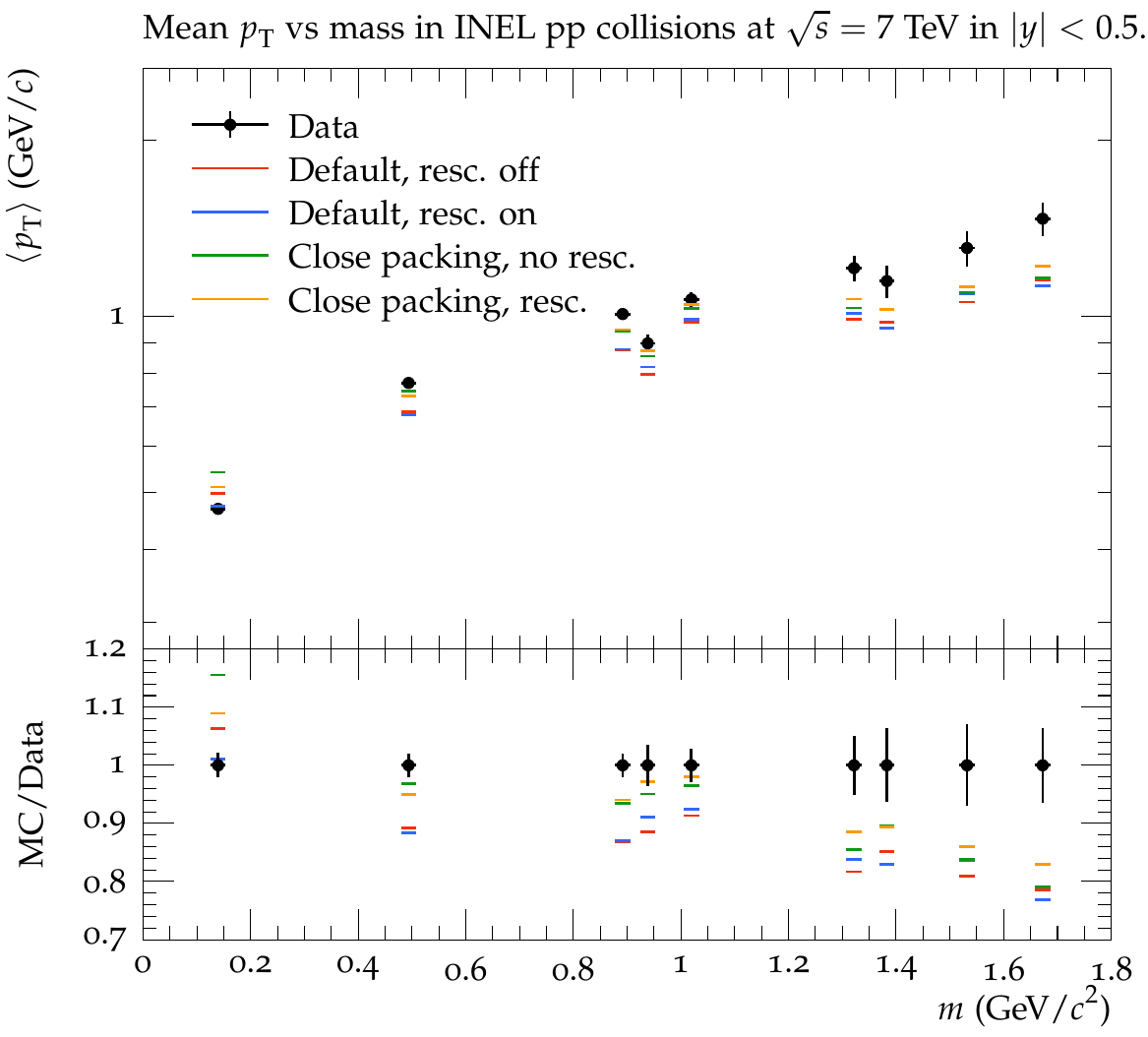}
\end{minipage}

\caption{$\pT$ spectra for $\pi^\pm$, $\K^\pm$ and $\p/\pbar$ and average 
$\pT$ for various particles, for the Gaussian model with and without 
close-packing corrections. When using the close-packing corrections, \texttt{pT0Ref} has been tuned to 2.18 without rescattering and 2.25 with rescattering on, in order to restore the correct $n_\mathrm{charged}$.}
\label{fig:pTspectra-close}
\end{figure}

The close-packing modification can be used either for the standard string
model or for the thermal alternative, by a rescaling either of $\kappa$
or of $T$. In \figref{fig:pTspectra-close}, we have used the former one. 
The trend here is that close-packing tends to increase $\pT$ for all particles, which means an improvement for all heavier hadrons, especially Kaons whose $\pT$ spectrum now follows data remarkably well above 1 GeV. However, this also means that the spectrum is worsened pions, and looking at their spectrum, the effect is quite severe. This deterioration is partially compensated for by rescattering, but not completely.
This makes the close-packing option unsuited as it stands. 
A retuning of fragmentation parameters might ameliorate the situation,
but that is beyond the scope of the current study.

\subsection{The role of vector mesons}

One of the standard assumptions is that the $\pT$ spectrum in $\q\qbar$ string
breaks is the same, independent of the quark species. This needs not be the case,
and higher-order corrections could well favour slightly different $\pT$ values 
for strange quarks \cite{Casher:1978wy,Skands:2012ts}, but for now we assume 
it to hold. Similarly, primary pseudoscalar and vector mesons are assumed to 
have the same $\pT$ spectra. The correct relative fraction of the two kinds of 
mesons is not known a priori, however, and for many hadrons it is difficult to 
measure their production rates, especially those with large widths. The prime 
example is the $\rho$, which we have seen contributes non-negligibly to the total 
rescattering rate. Since the $\rho$ has a higher mass than the Kaon, elastic 
$\rho\K$ collisions would tend to reduce the Kaon $\pT$, partially counteracting the gain from $\K\pi$ collisions.

\begin{figure}[t!]
\begin{minipage}[c]{\linewidth}
\centering
\includegraphics[width=0.48\linewidth]{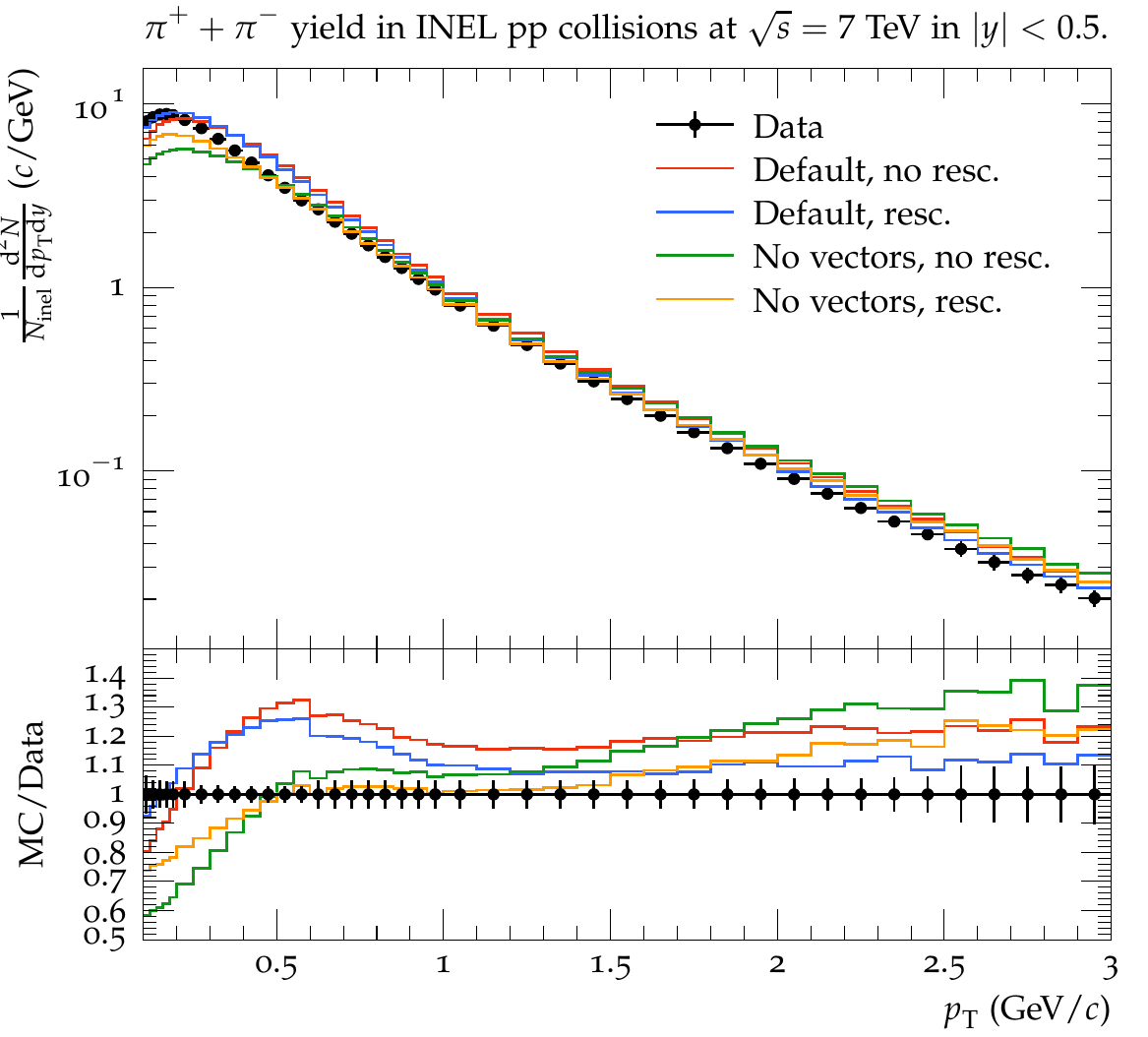}
\includegraphics[width=0.48\linewidth]{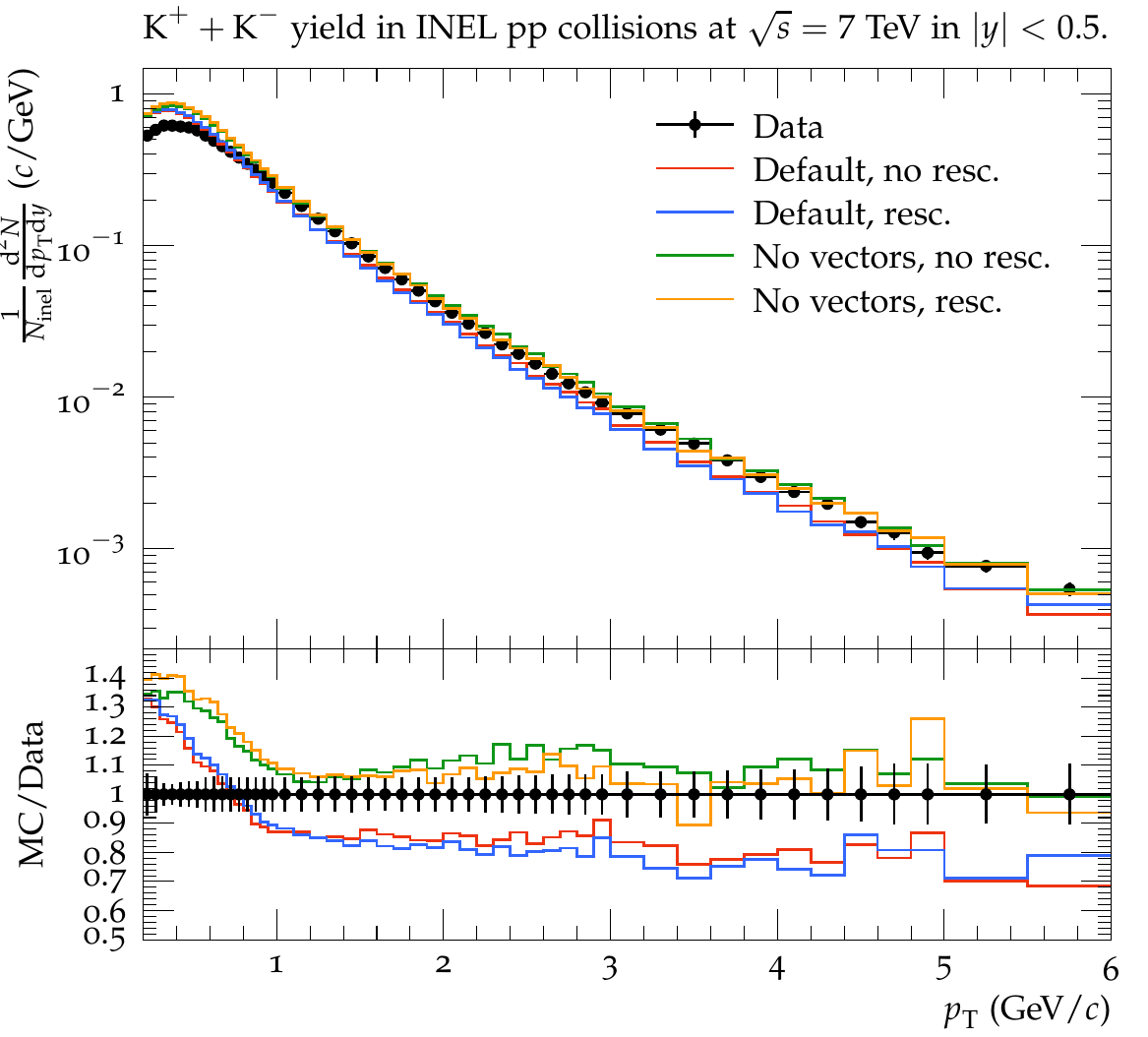}
\end{minipage}

\begin{minipage}[c]{\linewidth}
\centering
\includegraphics[width=0.48\linewidth]{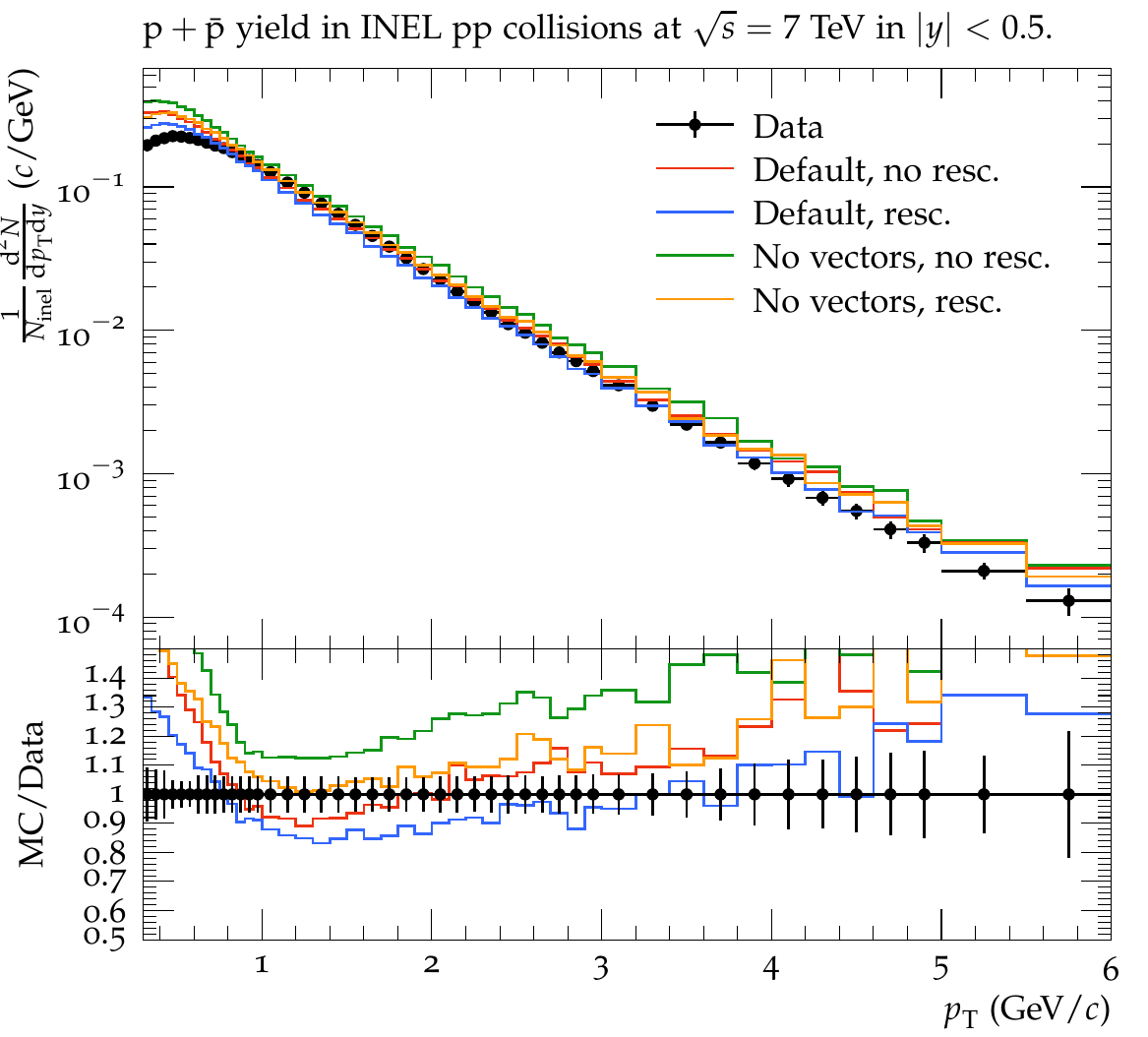}
\includegraphics[width=0.48\linewidth]{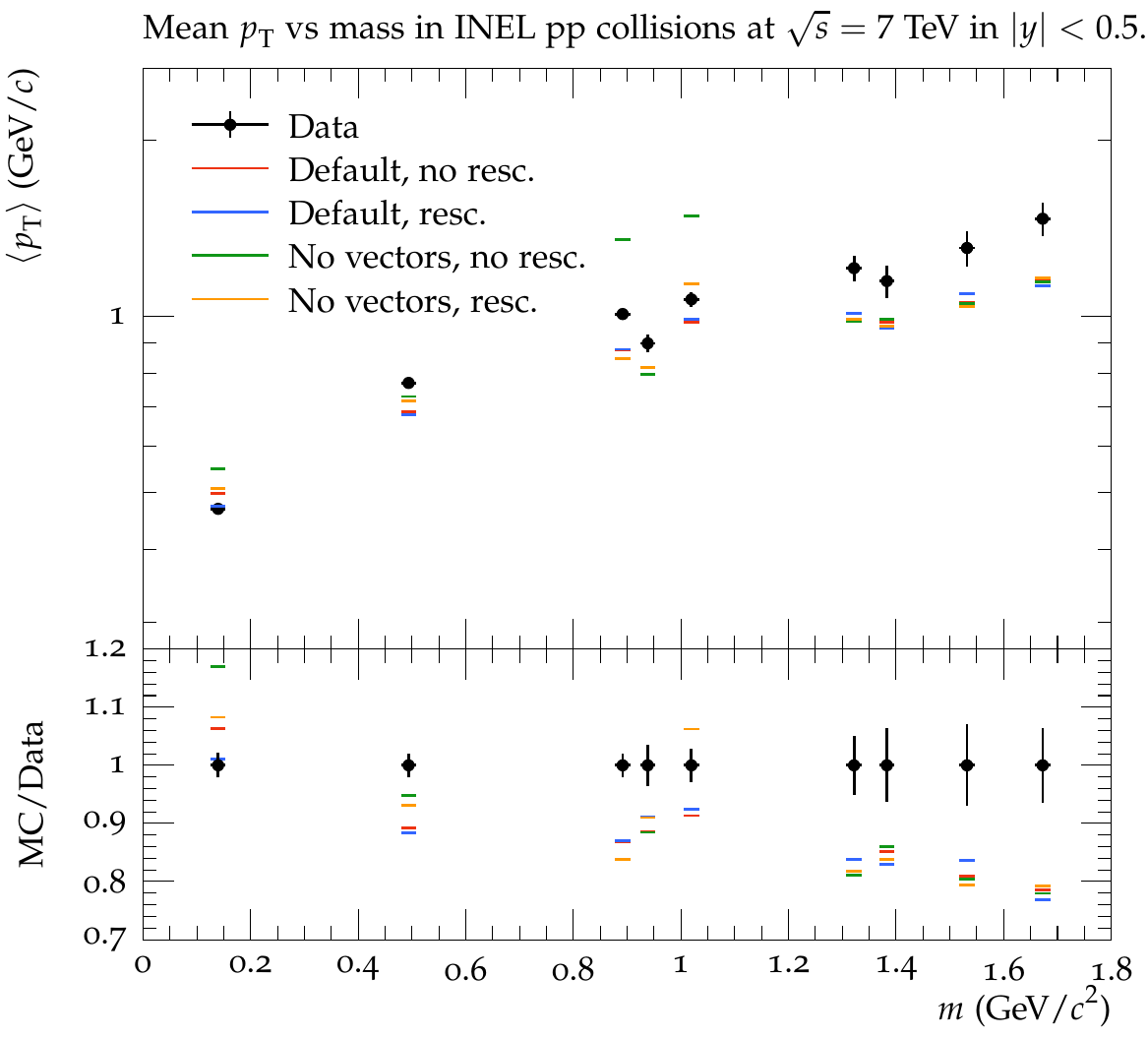}
\end{minipage}

\caption{$\pT$ spectra for $\pi^\pm$, $\K^\pm$ and $\p/\pbar$ and average $\pT$ for various particles, comparing rescattering to no rescattering, when no vector mesons are produced in the primary hadronization.}
\label{fig:pTspectra-noVMeson}
\end{figure}

As a simple test of the significance of heavy primary hadrons, we have studied
a toy scenario where no vector mesons at all are produced in the primary string 
fragmentation, but still can occur as intermediate states during rescattering. 
The resulting $\pT$ spectra are shown in \figref{fig:pTspectra-noVMeson}.
No attempt at a complete retune has been made, so it is the change by rescattering
that is most interesting, not the overall agreement. Not unexpectedly, the $\langle \pT \rangle$ is wildly off for $K^*(892)$ 
and $\phi(1020)$, which now cannot be produced in the primary process.
The ``pion wind'' effect is still there, in
that rescattering shifts pions to smaller $\pT$ and protons to larger. For Kaons
the $\langle \pT \rangle$ is still decreased by rescattering, providing further 
support that the primary mechanism for the Kaon $\pT$ loss is through $2 \to n$
processes, rather than from Kaon collisions with heavier particles.

It could have been informative also to go in the other direction, and include primary
production of higher resonances, with orbital or radial excitations. Measurements 
at LEP show that such mesons are produced at a non-negligible rate 
\cite{Tanabashi:2018oca}. And yet, their explicit inclusion tend to reduce the 
goodness of fit to many other properties, presumably because the assumed isotropic 
decay distributions do not represent the correct physics. Instead a higher-mass 
state could be viewed as a longer-than-normal string piece, with a decay along 
this string direction, just as if these products come directly from the string. 
Therefore we do not expect primary production of higher resonances to change 
$\pT$ properties appreciably, but currently do not have the full machinery 
necessary to test this assumption.

\subsection{Other transverse momentum spectra}\label{subsec:otherpTspectra}

\begin{figure}[t!]
\begin{minipage}[c]{0.49\linewidth}
\centering
\includegraphics[width=\linewidth]{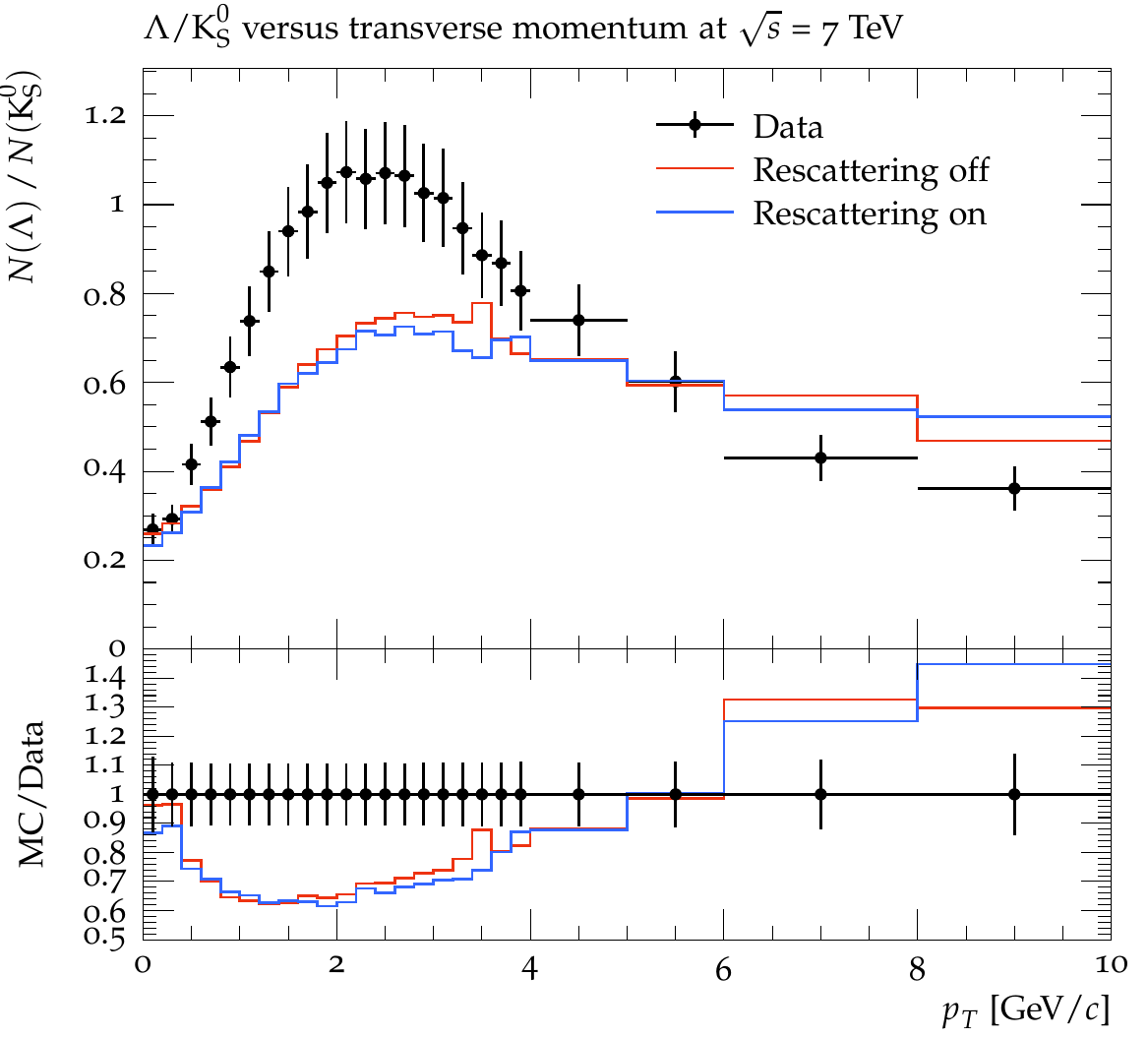}\\
(a)
\end{minipage}
\begin{minipage}[c]{0.49\linewidth}
\centering
\includegraphics[width=\linewidth]{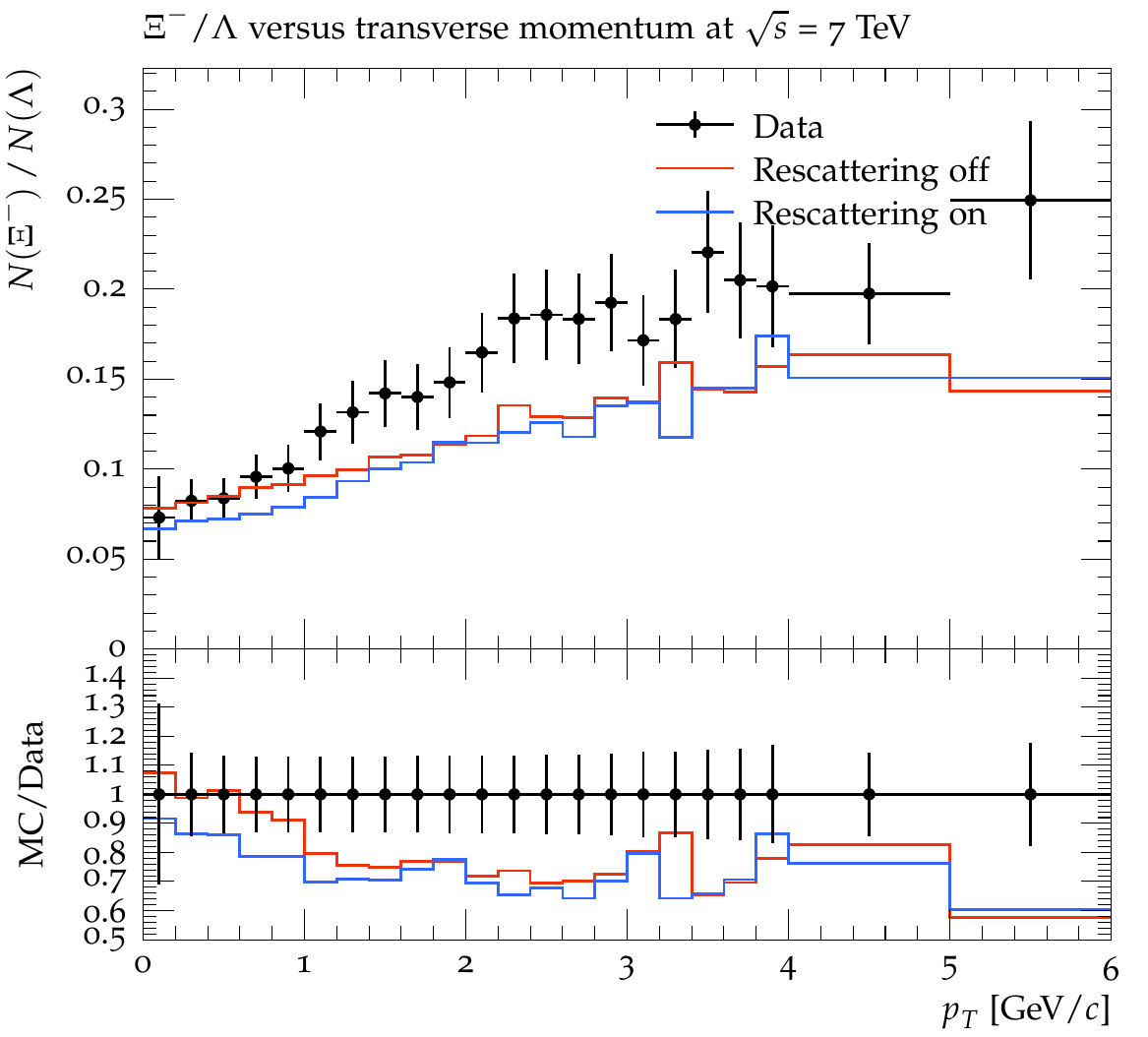}\\
(b)
\end{minipage}
\caption{$\pT$ ratios of (a) $\Lambda^0$ to $\K_\mathrm{S}$ and (b) $\Xi^-$ to $\Lambda^0$.}
\label{fig:strangeRatios}
\end{figure}

So far we have focused on $\pT$ spectra for pions, Kaons and protons. However, another experimental observation that pertains to collective behaviour is the peak for example in the $\Lambda^0 / \K^0_S$ ratio around $\pT \approx 2$~GeV. In \figref{fig:strangeRatios}, the ratios for $\Lambda^0 / \K^0_S$ and 
$\Xi^- / \Lambda^0$ are shown. Unfortunately rescattering 
does not provide an improvement. If anything it causes a deterioration, by reducing 
the relative number of $\Lambda^0$ and $\Xi^-$ baryons through the baryon-antibaryon 
annihilation mechanism. As before, an inclusion of $3 \to 2$ processes could help
alleviate the problem, but hardly give full agreement. In general, baryon production 
has been one of the more complicated and least successful aspects of the string 
fragmentation framework, already in the simpler $\e^+\e^-$ environment, and remains so.

\section{Summary and Outlook}
\label{sec:summary}

Hadronic rescattering is inevitable in the dense hadronic systems produced 
in high-energy $\p\p$ collisions. What less understood is the rate 
at which it happens, and the detailed modelling of the processes involved is open to discussion. 

In this article we have developed and studied a framework for hadronic
rescattering in $\p\p$ collisions. This involves three main aspects:
\begin{enumerate}
\item The space--time tracing of the motion of hadrons, with interleaved
scatterings and decays. The starting point here is our picture for the
space--time production of hadrons. Thereafter the motion of these hadrons 
is traced and possible crossings identified. The technical challenge is 
the fast growth of the number of hadron pairs to check, which can make 
have a significant impact on computing speed, even though most of these pairs never interact. 
\item The cross section for different collision processes. This is where 
most of the development effort has gone, and most of the new code can be
found. Much of the input has been from external sources, such as 
UrQMD ans\"atze, the calculations by Pel\'aez et al., the HPR$_1$R$_2$,
CERN/HERA and SaS parameterizations, and experimental data. We have tried to 
combine and extend these parts sensibly. For hadron pairs not described
in any other way, the Additive Quark Model is invoked to provide 
order-of-magnitude cross sections, also for charm and bottom hadrons.
\item The production of the new hadrons in these collisions. This is done either through explicit few-body channels, like elastic scattering or resonance 
formation, or through the existing string fragmentation
machinery. The typical collisions energies are so small, however, that 
extra efforts have to be made to translate these tiny strings into acceptable 
final states.
\end{enumerate}
Each of the three components are open to further refinements, but the new 
framework presented here should offer a good starting point for various 
studies as is. Other frameworks overlapping with ours already exist. To the 
extent feasible, one obvious future task would be to compare with other 
rescattering implementations, starting from the same initial hadron 
configuration.

Nevertheless, what we bring now is a cohesive implementation, 
where the full power of the traditional \textsc{Pythia} energy--momentum 
description is extended by the recent matching space--time picture and the 
new rescattering components, without the need to bridge disparate codes. 
This framework can then be applied to $\p\p$ collisions of any kind, from 
minimum-bias to high-$\pT$ physics. As far as we know, no other single 
program can offer as much.

The main emphasis in this study has been to develop and test the framework,
and to explore and understand how it behaves in general terms. Some applications
to LHC $\p\p$ studies have also been presented. In particular we note that
rescattering contributes to some aspects of collective flow, notably a
``pion wind'' that slows down pions and speeds up protons and (most) other
baryons. This helps remedy one of the glaring discrepancies of the 
traditional \textsc{Pythia} setup in comparisons with data. Unfortunately,
the effects are not large enough to fully resolve the discrepancies.
Worse, the Kaon $\pT$ spectrum is not modified appreciably, owing to a 
balance between speedup from the pion wind and slowdown from 
$2 \to n, n\geq 3$ processes. For this reason, one interesting topic for future study
is the modelling of $3 \to 2$ and related processes. 
There are also other phenomena, like azimuthal flow, where rescattering 
appears to give only a very small contribution. 

Thus it is obvious that further mechanisms will be needed to reach agreement
with a number of observables. We have here briefly explored some potential 
options, such as a randomly fluctuating string tension, i.e.\ the ``thermal'' 
model, and a larger string tension in a dense-string environment. Other 
ideas remain to be mixed in, such as string shoving. It may be disappointing
not to be in a situation where one simple model describes it all, but the
reality is that any physical process that can happen will also do so,
at some level.

The framework and its individual components have a higher applicability
than the one presented in this article, and we envisage several
follow-up studies. The most obvious one is to step up from $\p\p$ to
$\p$A and $\A\A$. This should be straightforward, since
\textsc{Pythia} already contains the Angantyr framework for heavy-ion
collisions \cite{Bierlich:2018xfw}. In a first step, we would study the
effects of rescattering on its own, without any other mechanisms for
collective flow. In a second step, one could combine it with other
effects, such as shove and rope formation, which also contribute to
flow effects. 

One relevant $\A\A$ study has already been done \cite{daSilva:2019and}, 
based on \textsc{Pythia}/Angantyr and its space--time picture, but 
interfacing UrQMD to handle the rescattering. Physics comparisons between 
the two approaches will be useful on its own, but additionally we hope 
that we can offer a more user-friendly framework, thereby simplifying 
the future experimental study of rescattering effects.

Although this article has mainly focused on rescattering, it should not be overlooked that the underlying framework, which allows for collisions for different beam 
particles and collision energies from the mass threshold and upwards, has other potential use cases. It could for example 
come in handy for other applications, such as the simulation of cosmic ray showers in the 
atmosphere and of hadronic 
showers in detectors. Currently this flexibility only works for soft
collisions, however. In order to fully include perturbative QCD
aspects, such as jets and MPIs, it is necessary to specify meaningful
PDFs for all colliding hadron species. Relevant combinations then
have to be stored such that it is easy to switch between them.
A special aspect is that, whereas collider physics mainly addresses
particle production at central rapidities, the evolution of hadronic
showers is especially sensitive to the production of the most forward 
hadrons, which therefore has to be carefully modelled.

In the current article, there has been no effort at a detailed retuning
of all model parameters, but only a modest revision of $\pTo$ to
retain the same total charged multiplicity as before when rescattering
is switched on. A future exercise would be to do a full-fledged
retuning. This could start with $\e^+\e^-$ annihilation events at LEP,
where no big effects are expected. Even small ones would be of interest,
however, since they could also add one more source of uncertainty in
$\W$ mass determinations \cite{Schael:2013ita}, in addition to colour
reconnection \cite{Sjostrand:1993hi} and Bose-Einstein
\cite{Lonnblad:1997kk}.

In conclusion, we hope that the current article and the new \textsc{Pythia} 
capabilities will be interesting for the experimental community, and also
open up for further developments and studies. By experience we know that 
new generator capabilities tend to inspire both expected and unexpected
applications.

\section*{Acknowledgements}

Thanks to J.R. Pel{\'a}ez, A. Rodas and J. Ruiz de Elvira for providing 
us with Mathematica code for their $\pi\pi$ and $\pi\K$ cross sections.
Work supported in part by the Swedish Research Council, contract number
2016-05996, and in part by the MCnetITN3 H2020 Marie Curie Innovative 
Training Network, grant agreement 722104.
This project has also received funding from the European Research
Council (ERC) under the European Union's Horizon 2020 research
and innovation programme, grant agreement No 668679.

\bibliographystyle{utphys}
\bibliography{lutp2011}

\end{document}